ЮГО-ЗАПАДНЫЙ ГОСУДАРСТВЕННЫЙ УНИВЕРСИТЕТ

На правах рукописи

Усатюк Василий Станиславович

# МЕТОД, АППАРАТНО-ОРИЕНТИРОВАННЫЙ АЛГОРИТМ И СПЕЦИАЛИЗИРОВАННОЕ УСТРОЙСТВО ДЛЯ ПОСТРОЕНИЯ НИЗКОПЛОТНОСТНЫХ КОДОВ АРХИВНОЙ ГОЛОГРАФИЧЕСКОЙ ПАМЯТИ

Специальность 05.13.05 – Элементы и устройства
вычислительной техники и систем управления

Диссертация на соискание ученой степени
кандидата технических наук

Доктор технических наук, доцент
Егоров Сергей Иванович

Курск – 2022

СОДЕРЖАНИЕ

# ВВЕДЕНИЕ

**Актуальность темы**

В связи с интенсивным увеличением объема информации, хранимой в электронном виде, в настоящее время актуальными являются вопросы развития систем голографического архивного хранения данных, в которых запись данных сопровождается процессами кодирования, а считывание данных – процессами их декодирования. Особенностью канала считывания информации голографического носителя является группирование ошибок и высокий уровень вероятности ошибки на бит − до $5{\cdot}10^{-2}$. Тогда как современные требования к надежности считывания архивной голографической памяти определяют вероятности ошибки на бит в диапазоне $10^{-8}$ − $10^{-10}$. Для достижения этого показателя надежности, ошибки, возникающие при хранении и считывании данных в голографической памяти, исправляют с помощью кодов с низкой плотностью проверок по четности (низкоплотных кодов).

Низкоплотностные коды были предложены Галлагером Р. в 60-х годах прошлого века и исследовались Таннером Р., Зябловым В.В. и Маргулисом Г.А. Всплеск интереса к применению этих кодов на практике возник в связи с появлением в 1997 году работы Маккея Д., посвященной декодированию с мягкими решениями этих кодов. В дальнейшем низкоплотностные коды исследовались Назаровым Л.Е., Габидулиным Э.М., Кудряшевым Б.Д., Трифоновым П.В., Фроловым А.А., Рыбиным П.С., Фоссорье М., Васичем Б. и другими. Тем не менее, в исследованиях этих ученых при построении кодов не уделялось достаточного внимания учету их дистантных свойств и спектров связности, что усложняет построение низкоплотностных кодов средней длины с высокой корректирующей способностью, требуемых в системах голографической памяти.

Приложениями этих кодов для голографической памяти занимаются компании AT&T (Alcatel-Lucent /NOKIA), Hitachi Maxell, Sony, Panasonic, Mitsubishi, Nichia, Alps Electric, Bayer Material Science, Sanyo, Lite-on, TrellisWare. Корректирующая способность низкоплотностного кода F-LDPC, предложенного компанией TrellisWare, для голографической памяти, далека от границы Полянского, что определяется ограниченной длиной кода (32000 бит), со скоростью кода 0.5, вызванной секторной организацией данных, и применением недостаточно совершенных методов построения этих кодов. Используемые в настоящее время программные реализации методов и алгоритмов построения низкоплотностных кодов имеют большую вычислительную сложность и недостаточную производительность, в них слабо учитываются комбинаторно-алгебраические свойства Таннер-графа и дистантные свойства кода, что приводит к появлению треппин-сетов и кодовых слов малого веса, отрицательно влияющих на корректирующую способность кодов.

Таким образом, объективно сложилось **противоречие** между необходимостью повышения надежности чтения в накопителях голографической памяти данных, за счет понижения вероятности ошибки на бит в области рабочих значений диапазона отношений сигнал-шум, устройствами коррекции ошибок низкоплотными кодами, путем разработки новых методов построения низкоплотностных кодов с использованием аппаратно-ориентированных алгоритмов и специализированных устройств для редукции числа циклов и определения дистантных свойств низкоплотностных кодов.

В связи с этим, **актуальной научно-технической задачей** является разработка методов, аппаратно-ориентированного алгоритма и специализированного устройства для построения низкоплотностных кодов для декодеров, обеспечивающих повышение надежности чтения в архивной голографической памяти.

**Целью диссертационной работы** является повышение надежности воспроизведения информации в накопителях архивной голографической памяти, за счет понижения вероятности ошибки на бит в области рабочих значений диапазона отношений сигнал-шум, устройствами коррекции ошибок низкоплотных кодов.

В соответствии с поставленной целью в диссертации решаются следующие **задачи**:

1. Анализ существующих методов построения низкоплотностных кодов, используемых в накопителях архивной голографической памяти, выбор и обоснование цели исследования;

2. Создание метода построения низкоплотностных кодов для накопителей архивной голографической памяти;

3. В рамках метода построения низкоплотностных кодов созданы частный метод оценки кодового расстояния и аппаратно-ориентированный алгоритм оценки кодового расстояния с использованием геометрии чисел;

4. Разработка специализированного устройства осуществляющего поиск кодового расстояния в подрешетке m-размерности для построения низкоплотностных кодов, и экспериментальная оценка надежности считывания данных из голографической памяти.

**Объект исследований –** вычислительные процессы, методы и аппаратно-ориентированный алгоритм в задаче кодирования-декодирования данных, считываемых из голографической памяти.

**Предмет исследований** - специализированное устройство для построения низкоплотностных кодов архивной голографической памяти.

**Методы исследования.** Для решения поставленных задач применялись методы: помехоустойчивого кодирования, геометрии чисел, абстрактной алгебры, теории графов, теории вероятностей, теории сложности вычислений, имитационного моделирования, параллельного программирования, теории проектирования ЭВМ.

**Научная новизна и положения, выносимые на защиту**:

1. Метод построения низкоплотностных кодов, состоящий из двух фаз построения и расширения протографа, отличающийся комбинированием жадного алгоритма запрещенных коэффициентов и стохастического алгоритма отжига, позволяющих улучшить дистантные свойства кодов и их спектры связности для фильтрации кодов кандидатов, обеспечивающих повышение надежности считывания информации в голографической памяти.

2. Метод оценки кодового расстояния, основанный на вложение кода в решетку, отличающийся применением для поиска кратчайших векторов параллельным перебором линейных комбинаций базисных векторов решётки, а также применением на этапе ортогонализации параллельных методов QR-разложения матриц, применением метода ветвей и границ в скользящем окне по подрешеткам $m$-размерности, позволяющий ускорить нахождение кодового расстояния.

3. Аппаратно-ориентированный алгоритм поиска кратчайшего вектора в решетке, отличающийся этапом распараллеливания вычисления координатных компонент с использованием зигзагообразного обхода Шнора элементов решетки, позволяющий оперативно получить необходимые индексы и кратчайший вектор нахождения кодового расстояния.

4. Специализированное устройство поиска кратчайшего вектора в решетке, включающее операции модификации координатных компонентов вектора и блоков вычисления частичных сумм совместно с блоком модификации/вычисления приращений координат и его границ, отличающееся использованием регистровых стеков и параллельным выполнением мультипликативных операции в одном временном интервале, позволяющее в подрешетки $m$-размерности сократить количество DSP процессоров в устройстве.

**Практическая ценность работы** состоит в следующем:

1. Комбинация метода построения низкоплотностных кодов для архивной голографической памяти, аппаратно-ориентированного алгоритма и специализированного устройства поиска кратчайшего вектора в решетке позволила построить новый низкоплотностный код для архивной голографической памяти, декодер которого обеспечивает повышение надежности воспроизведения информации от 8,9 раз при отношении значения сигнал-шум 1,1 дБ по сравнению с F-LDPC кодом, предложенным компанией TrellisWare для голографической архивной памяти.

2. Созданный метод оценки кодового расстояния линейных блочных кодов позволил дать оценки расстояний для низкоплотностных кодов длиной 32000 бит, используемых в голографической памяти.

3. Разработанное специализированное устройство поиска кратчайшего вектора в решетке обеспечивает выигрыш по быстродействию в сравнении с программной реализацией в 33.93 раза для подрешетки 512-размерности для низкоплотных кодов.

**Реализация и внедрение.**

Основные научные результаты и выводы диссертационной работы внедрены в ООО «Техкомпания Хуавей». Используемые результаты защищены компанией Huawei Technologies Co. тремя международными патентами. Также результаты диссертационной работы используются на кафедре вычислительной техники ЮЗГУ при преподавании дисциплин: «Защита информации» по направлению подготовки 09.03.01, «Схемотехника (элементная база перспективных ЭВМ)» по направлению подготовки 09.04.01. Внедрение подтверждается соответствующими актами.

**Достоверность** результатов диссертации обеспечивается обоснованным и корректным применением положений и методов математического аппарата алгебры и комбинаторики, теории вероятности, теории графов, теории помехоустойчивого кодирования, теории проектирования ЭВМ, а также подтверждается совпадением теоретических выводов с результатами имитационного моделирования.

**Соответствие диссертации паспорту научной специальности.**

Согласно паспорту специальности 05.13.05 – «Элементы и устройства вычислительной техники и систем управления» проблематика, рассмотренная в диссертации, соответствует пунктам 3, 4 паспорта специальности. 3. Разработка принципиально новых методов анализа и синтеза элементов и устройств вычислительной техники и систем управления с целью улучшения их технических характеристик, в части синтеза специализированного устройства поиска кратчайшего пути, необходимого для построения низкоплотностного кода, позволяющего в подрешетке $m$-размерности сократить количество DSP процессоров в устройстве. 4. Разработка научных подходов, методов, алгоритмов и программ, обеспечивающих надежность, контроль и диагностику функционирования элементов и устройств вычислительной техники и систем управления, в части создания метода и аппаратно-ориентированного алгоритма построения низкоплотностного кода, позволяющего повысить надежность воспроизведения данных голографической памяти ЭВМ.

**Апробация работы.** Основные теоретические положения и научные результаты диссертационной работы докладывались и обсуждались на следующих всероссийских и международных научных конференциях: 4-ой и 5-ой региональных научно-практических конференциях «Платоновские чтения» (г. Иркутск 2012, 2013), Всероссийской научной конференции «Наука. Технологии. Инновации» (г. Новосибирск, 2012), Всероссийских конференциях «Компьютерная безопасность и криптография» – «SIBECRYPT'12» в Институте динамики систем и теории управления СО РАН (г. Иркутск, 2012), «SIBECRYPT'13» (г. Томск, 2013), «XVI Всероссийском Симпозиуме по прикладной и промышленной математике» (г. Челябинск, 2015), 18-й Международной научно-технической конференции «Проблемы передачи в сетях и системах телекоммуникаций» (г. Рязань, 2015), II и III Международных конференциях «Инжиниринг & Телекоммуникации –En&T» (г. Москва/Долгопрудный, г. 2015, 2016), XIII Международной научно-технической конференции «Новые информационные технологии и системы» (г. Пенза, 2016), XIII Международной научно-технической конференции Оптико-электронные приборы и устройства в системах распознавания образов и обработки изображений «Распознавание 2017», (г. Курск, 2017), XII Международной научной конференции «Перспективные технологии в средствах передачи информации - ПТСПИ-2017» (г. Владимир-Суздаль, 2017), 15-й Международной конференции IEEE East-West Design & Test Symposium (г. Нови-Сад, Сербия, 2017), конференции «Applied Mathematics Day» в МИАН РАН (г. Москва, 22 сентября 2017), конференции «Машинное обучение и анализ алгоритмов» в ПОМИ РАН (г. Санкт-Петербург, 18-20 декабря 2017 г.), 41-й Международной конференции

«Telecommunications and Signal Processing» (г. Афины, Греция, 4-6 июля 2018 г.), 5-й Международной конференции по матричным методам в математике и приложениях, «The 5th International Conference on Matrix Methods in Mathematics and Applications (MMA 2019)»(19-23 Августа 2019 г. г. Москва), 43-й Международной конференции «Telecommunications and Signal Processing» (г. Милан, Италия, 2020 г.).

**Публикации.** По теме диссертации опубликовано 29 научных работ, в их числе 5 статей в научных рецензируемых изданиях, входящих в перечень ВАК Минобрнауки России, 8 работ проиндексированы в международной базе Scopus. Оригинальность технических решений, предложенных автором, подтверждена тремя Международными патентами на изобретения.

**Личный вклад соискателя.** Все выносимые на защиту научные результаты получены соискателем лично. В опубликованных в соавторстве работах по теме диссертации лично соискателем предложено: в [95, 96] метод поиска кратчайшего вектора в решетке, в [98, 104] методы оценки кодового расстояния на основе геометрии чисел, в [98, 101, 103-106, 143] методы построения низкоплотностных кодов, в [99] быстродействующее устройство поиска кратчайшего вектора в решетке.

**Объем и структура работы.** Диссертационная работа состоит из введения, четырех разделов, заключения, списка литературы и приложений. Работа содержит 160 страниц текста (с учетом приложений) и иллюстрируется 60 рисунками и 13 таблицами; список литературы включает 147 наименований.

# 1 АНАЛИЗ СУЩЕСТВУЮЩИХ МЕТОДОВ ПОСТРОНИЯ НИЗКОПЛОТНОСТНЫХ КОДОВ, ИСПОЛЬЗУЕМЫХ В НАКОПИТЕЛЯХ АРХИВНОЙ ГОЛОГРАФИЧЕСКОЙ ПАМЯТИ, ВЫБОР И ОБОСНОВАНИЕ ЦЕЛИ ИССЛЕДОВАНИЯ

## 1.1 Характеристики ошибок в каналах записи-воспроизведения архивной голографической памяти

Под голографической памятью понимают набор технологий, позволяющий осуществлять запись, воспроизведение и переформирование волновых полей оптического электромагнитного излучения. При этом с помощью лазера регистрируются, а затем восстанавливаются изображения трехмерных объектов. Процессы записи информации на голографический носитель и считывания ее представлены на рис. 1.1-1.2.

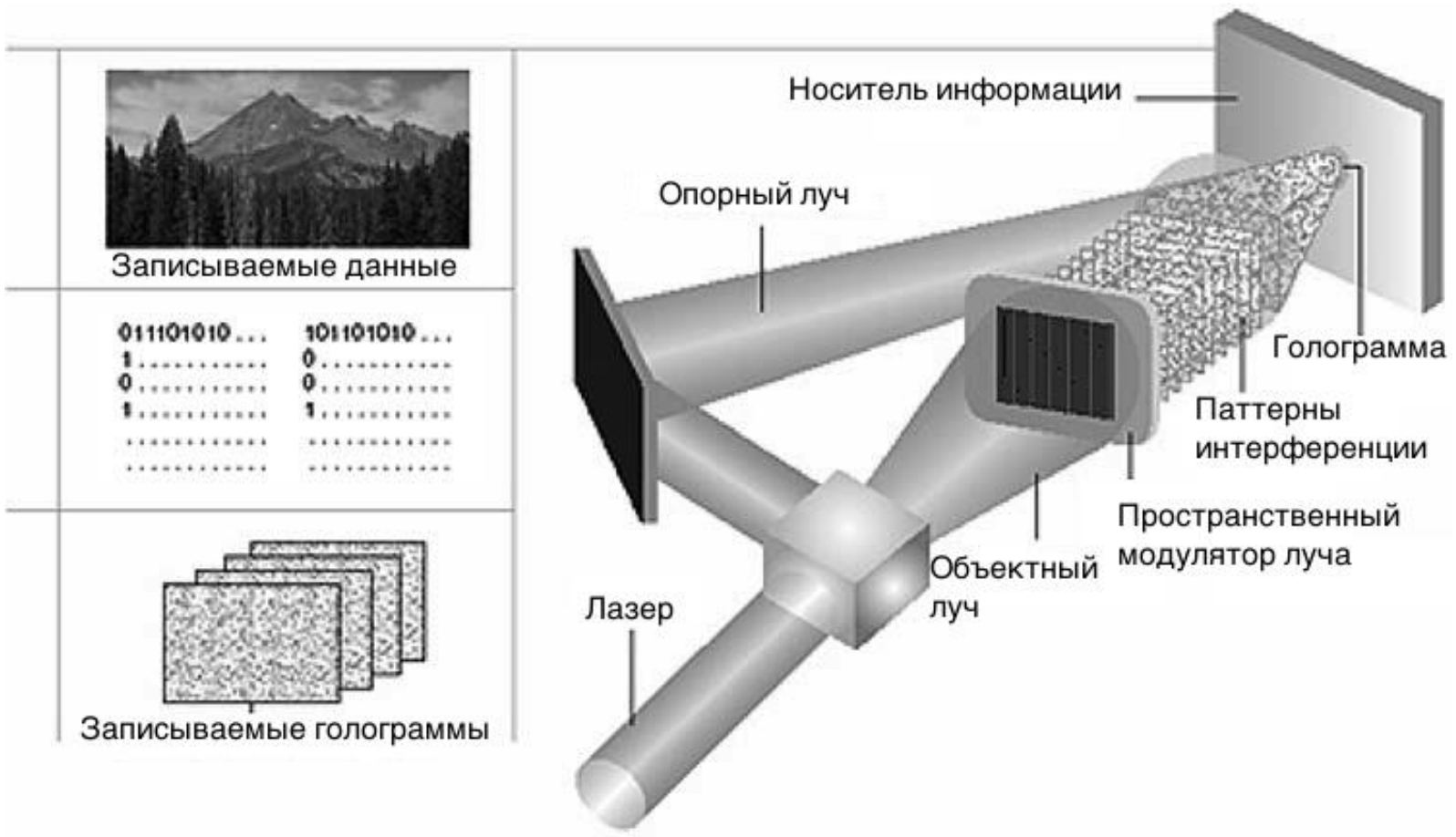

Рис. 1.1 - Процесс записи информации на голографический носитель

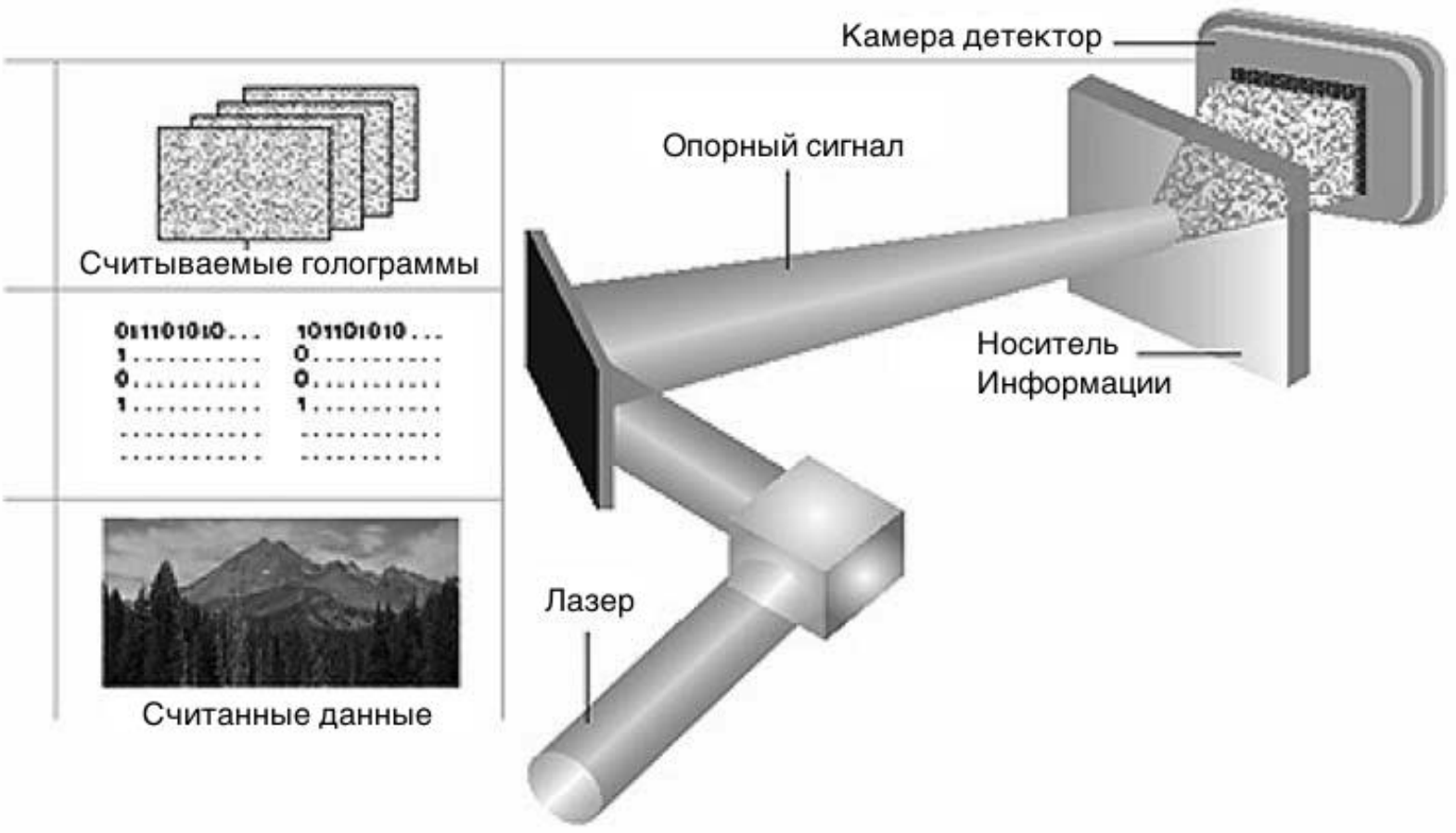

Рис. 1.2 - Процесс считывания информации с голографического носителя

К достоинствам оптических носителей голографической памяти относятся:

- высокая плотность записи информации на квадратный сантиметр;
- низкое потребление энергии;
- произвольный доступ к любому из секторов носителя;
- низкая стоимость носителей информации.

В случае применения носителя толщиной 1.5 мм и лазера с длиной волны $\lambda = 405$нм предельная плотность записи информации составляет 133 Тбайта на квадратный сантиметр [1-3].

К недостаткам оптических носителей голографической памяти относятся однократная запись носителя и относительно высокий уровень ошибок в считанных данных.

Основные узлы системы чтения и записи голографического носителя информации изображены на рис. 1.3.

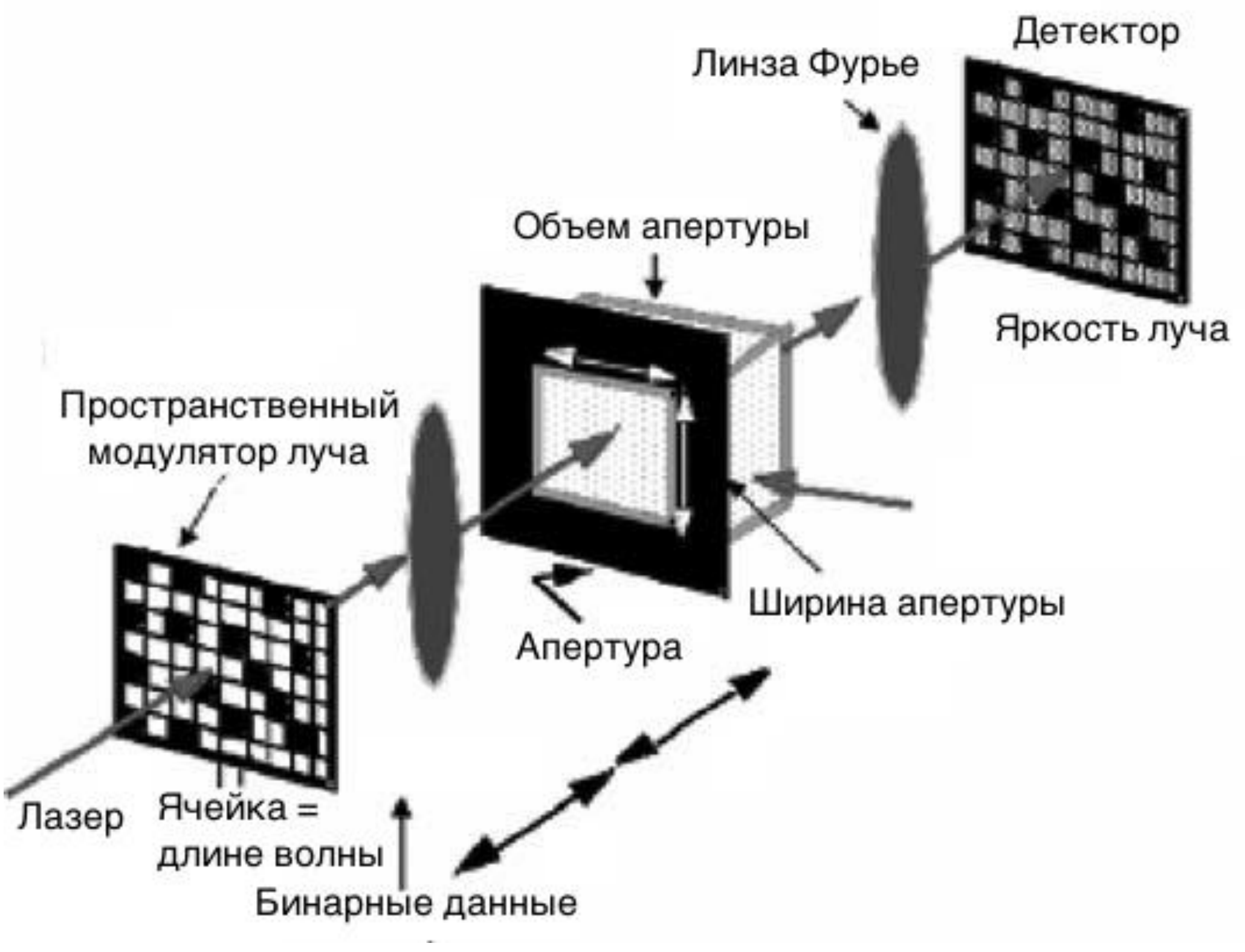

Рис. 1.3 – Основные узлы системы чтения и записи голографического носителя информации: пространственный модулятор луча, апертура голографического носителя информации и камера детектора

На вход пространственному модулятору луча подается битовый массив, который разбивается на страницы, записываемые в виде голограмм.

В процессе модуляции данных модулятор вносит ряд ошибок:

- Ошибки, обусловленные особенностями (искажениями) волнового фронта луча, проходящего через пространственный модулятор луча (ПМЛ). Волновой фронт, проходящий через ПЛМ, далек от плоского.

- Ошибки динамического диапазона. Фазовая маска, располагаемая после ПЛМ, осуществляющая усреднение интенсивности света в частотной области, ограничивает динамический диапазон.

- Ошибки вследствие температурной нестационарности среды.

- Ошибки, обусловленные когерентным шумом источника света.

- Прочие виды ошибок (электронный шум фазовой маски, влияние зеркал и масок).

Апертура формируется пространственным модулятором луча за счет оптической системы, обеспечивающей высокую предельную угловую разрешающую способность на участке траектории движения голографического носителя.

Детектор представляет собой детектирующую камеру высокой разрешающей способности, считывающую интенсивность излучения голографического носителя.

Интенсивность излучения, детектируемого камерой с голографического носителя, можно описать уравнением:

$$I_A = \left|(g \cdot a) * h_A + n_i + in_q\right|^2 + n_e, \quad (1.1)$$

где $g$ —двухмерный массив, характеризующий амплитуду света, $a$ —данные, $a \in \{0,1\}$, $h_A$ – функция распределения пикселей по амплитуде, $n_0 = n_i + in_q$ оптический (рассеянный) шум, заданный аддитивным и круговым нормальным распределением $\mathrm{N}(0, \sigma^2)$ и $n_e$ - аддитивный белый гауссовой шум.

Мощность сигнала превосходит мощность шума $n_0$ в 10 раз.

Сигнальная обработка голографического канала предусматривает деперемежение сигнала, оценку канала, обеление принятого сигнала путем устранения корреляции шума. Это позволяет в качестве модели голографического канала использовать набор параллельных АБГШ-каналов, представляющих 2-мерные массивы пикселей (слои в апертуре), называемых страницами.

Отношение Сигнал-Шум (SNR) для канала чтения голографической памяти вычисляется по формуле:

$$SNR_{hol} \equiv \frac{\mu_1-\mu_0}{\sqrt{\sigma_1^2-\sigma_0^2}}, \tag{1.2}$$

где $\mu_1, \mu_0$ средние, $\sigma_1$ и $\sigma_0$ стандартные отклонения детектирования значений бита 1 и 0, соответственно.

Для АБГШ-канала входная вероятность ошибки на бит вычисляется по формуле:

$$InBER = \frac{1}{2} erfc\left(\frac{Q}{\sqrt{2}}\right), \tag{1.3}$$

где $erfc$ –функция ошибки $erfc(x) = \frac{2}{\sqrt{\pi}}\int_0^x e^{-t^2}dt$ и $Q = \frac{\mu_1-\mu_0}{\sigma_1+\sigma_0}$.

Формула для вычисления отношения Сигнал-Шум в децибелах имеет вид:

$$SNR_{Дб} \equiv 20log_{10}\frac{\mu_1-\mu_0}{\sigma_1+\sigma_0}. \tag{1.4}$$

Для АБГШ-канала отношение Сигнал-Шум можно преобразовать в усредненную величину битовой ошибки:

$$InBER_{avg} = \frac{1}{2N}\sum_{i=1}^{N} erfc\left(\frac{Q_i}{\sqrt{2}}\right), \tag{1.5}$$

$Q_i$ –отношения Сигнал-Шум для каждой из голограмм, $N$ – число голограмм.

Соответственно, усредненное по голограммам отношение Сигнал-Шум вычисляется по формуле:

$$Q_{avg} = \sqrt{2}erfc^{-1}\left(2InBER_{avg}\right). \tag{1.6}$$

Используя последнюю формулу, мы можем получить соотношение Сигнал-Шум для каждой из голограмм (страниц):

$$SNR_{page} = 20log_{10}\left[\sqrt{2}erfc^{-1}\left(2InBER_{avg}\right)\right]. \tag{1.7}$$

Логарифм отношения правдоподобия на голографическом АБГШ-канале задается уравнением:

$$LLR(y_i) = \frac{P(y_i|x_i=1)}{P(y_i|x_i=0)} = log\frac{\sigma_1}{\sigma_0} - \frac{1}{2}\left(\frac{y_i-\mu_0}{\sigma_0}\right) + \frac{1}{2}\left(\frac{y_i-\mu_1}{\sigma_1}\right), \tag{1.8}$$

где $y_i$ – интенсивность детектированного излучения $i$-го бита, $x_i$ – значение переданного $i$-го бита.

Эти значения используются в качестве апостериорных вероятностей при декодировании помехоустойчивых кодов.

Для достижения, требуемого для ВЗУ ЭВМ уровня вероятности ошибки на бит в диапазоне $10^{-8}$-$10^{-10}$ (в зависимости от назначения данных) в настоящее время в голографической памяти на страничном уровне (рис. 1.4) используются помехоустойчивые низкоплотностные коды [1].

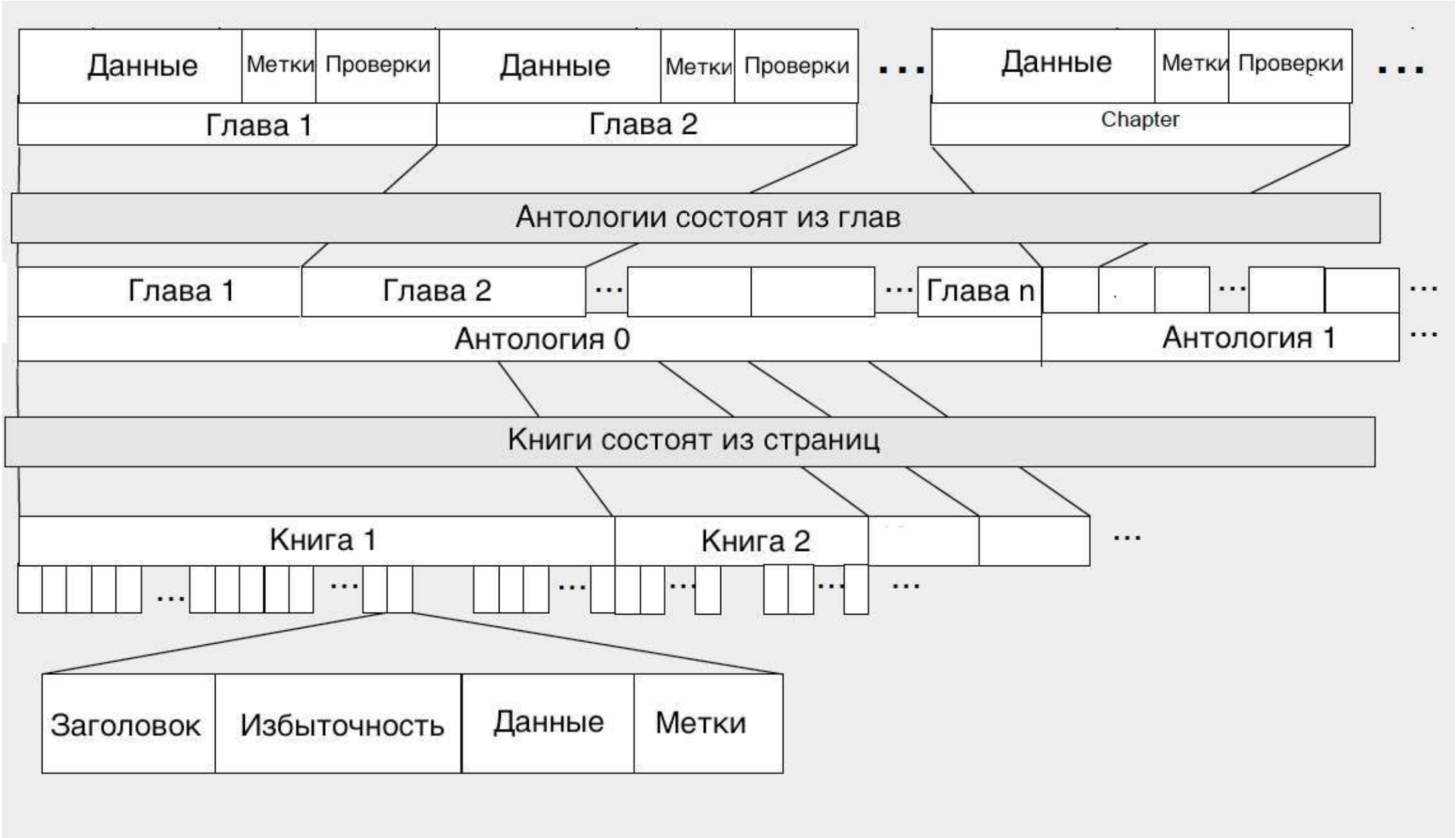

Рис. 1.4 – Организация хранения данных в архивной голографической памяти

На страничном уровне коррекция ошибок выполняется для 2-мерного массива пикселей (голограммы - слоя в апертуре), содержащего битовые значения 0 и 1. Детектор снимает значения с носителя и формирует массив логарифмов правдоподобия, используемых при декодировании низкоплотностных кодов.

Страничный уровень работает в условиях больших шумом, поэтому для коррекции ошибок используется код длиной 32000 со скоростью 0.5, содержащий 32 битный код контроля на четность.

Страничный уровень - один из ключевых уровней, определяющих плотность записи информации в архивной голографической памяти. Все последующие уровни (главы, книги и онтологии), использующие коды Рида-Соломона и проверку на четность, предназначены для обнаружения ошибок и исправления стираний.

На рис. 1.5 представлены результаты декодирования низкоплотностного кода с использованием 40 итераций [1].

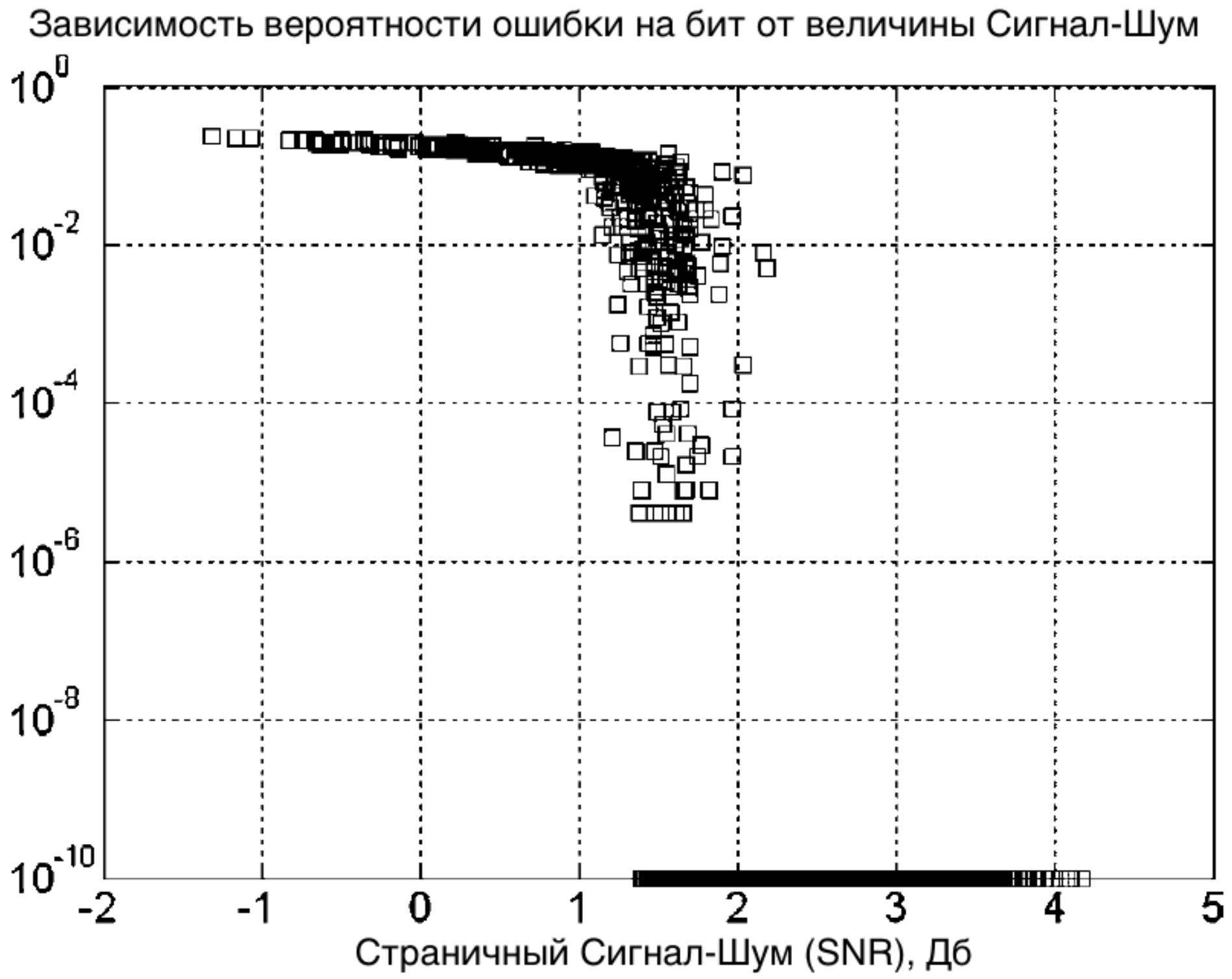

Рис. 1.5 Результаты моделирования низкоплотностного кода со скоростью 0.5, длины N=32000, MS декодер с 40 итерациями [1].

Эффективность коррекции ошибок LDPC-кодом, применяемым в голографической памяти, показана на рис. 1.6.

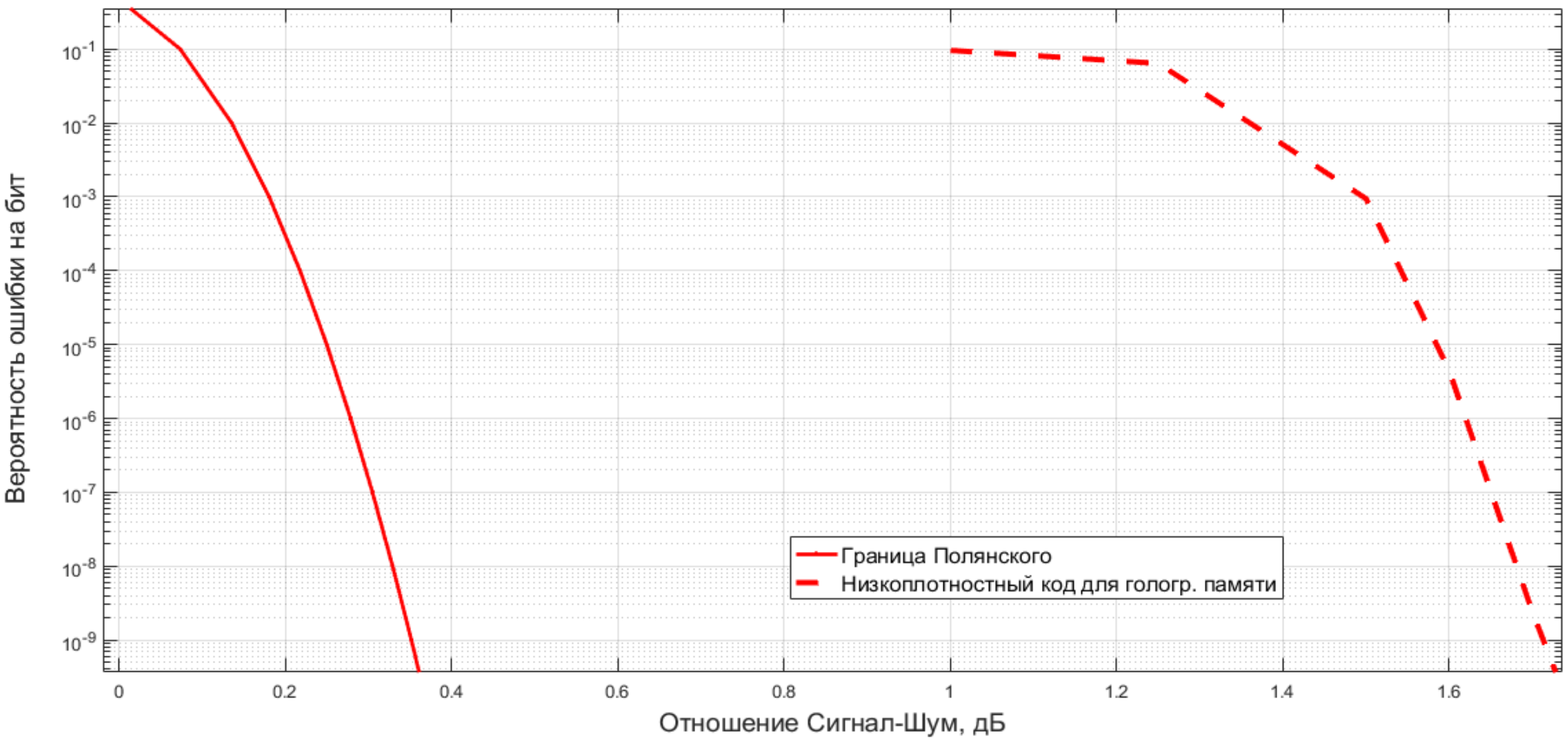

Рис. 1.6 Эффективность коррекции ошибок квазициклическим LDPC-кодом, применяемым в голографической памяти

Рисунок показывает отставание в 1 дБ применяемого в архивной голографической памяти квазициклический LDPC-код от теоретического предела корректирующей способности линейного блочного кода. Дальнейшее улучшение эффективности коррекции ошибок на страничном уровне для повышения надежности считывания данных возможно с построением новых более помехоустойчивых кодов. Эту задачу можно решить, опираясь на новые более совершенные методы построения низкоплотностных кодов.

## 1.2 Низкоплотностные коды для коррекции ошибок в архивной голографической памяти

### 1.2.1 Коды с низкоплотностной проверочной матрицей (LDPC)

$(n, k)$ LDPC-код – это блочный линейный код размерностью $k$ и длиной кодового слова $n$, задаваемый проверочной матрицей $H$ размерностью $(n\text{-}k)\cdot n$, имеющей небольшую плотность отличных от нуля символов. По определению проверочной матрицы для любого кодового слова $v$ LDPC-кода справедливо следующее: $v\cdot H^T = 0$. Каждая строка проверочной матрицы $H$ задает уравнение проверки на четность:

$$\sum_{t=0}^{n-1} v_t \cdot h_{i,t} = 0, \qquad (1.9)$$

где $h_{i,t}$ – элемент проверочной матрицы, $i$ – номер строки проверочной матрицы (номер проверочного уравнения), $t$ – номер символа кодового слова.

Ключевым преимуществом низкоплотностных кодов является возможность применения субоптимального алгоритма мягкого декодирования методом распространения доверия (BP, belief-propagation), обладающего значительно большей помехоустойчивостью по сравнению с алгоритмами жесткого декодирования при приемлемой сложности реализации.

Алгоритм BP предусматривает представление LDPC-кода в виде двудольного графа Таннера (пример графа Таннера приведен на рис. 1.7).

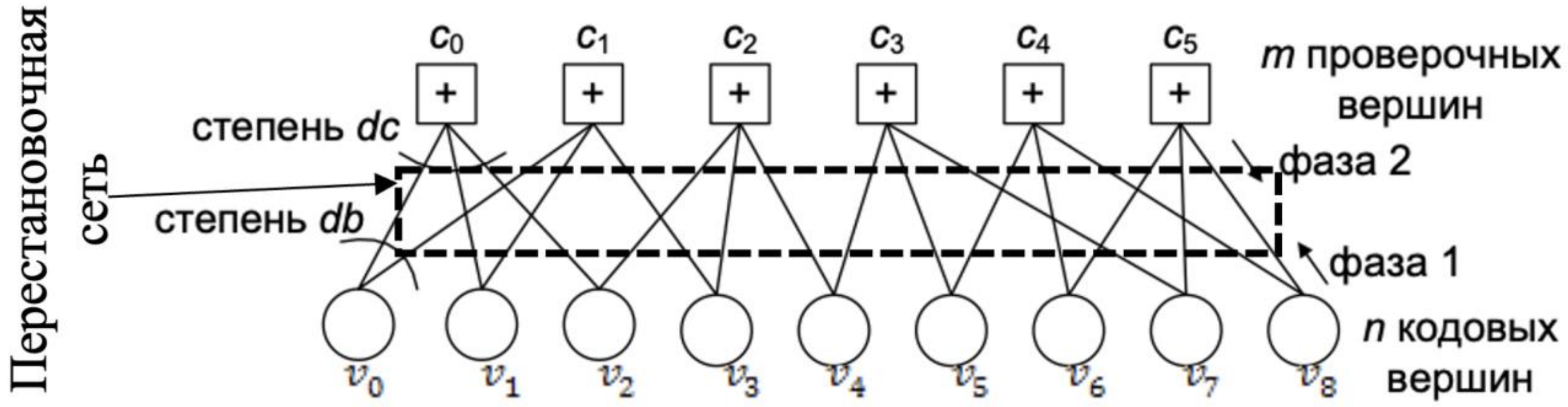

Рис. 1.7. Двудольный граф двоичного регулярного LDPC кода длины 9.

Граф Таннера $G=\{C,V,E\}$ имеет два множества вершин – $C,V$. Одно множество состоит из $m'=N-K$ проверочных вершин $\{c_0, c_1, \dots, c_{m-1}\}$, соответствующих $m'$ строкам матрицы $H$**,** второе - из $n = N$ кодовых вершин $\{v_0, v_1, \dots, v_{n-1}\}$, соответствующих $n$ столбцам матрицы $H$. Кодовая вершина $v_j$ соединяется ребром с проверочной вершиной $c_i$ в том случае, если элемент проверочной матрицы на $j$-столбце и $i$-й строке $H_{i,j} \neq 0$.

Квазициклический регулярный ($J$, $L$) LDPC-код ($J$ – вес столбца матрицы, $L$ – вес строки) задается проверочной матрицей:

$$H = \begin{bmatrix} I(p_{0,0}) & I(p_{0,1}) & \dots & I(p_{0,L-1}) \\ I(p_{1,0}) & I(p_{1,1}) & & I(p_{1,L-1}) \\ \vdots & \vdots & \ddots & \vdots \\ I(p_{J-1,0}) & I(p_{J-1,1}) & \dots & I(p_{J-1,L-1}) \end{bmatrix}, \qquad (1.10)$$

где $0 \le j \le J-1$, $0 \le l \le L-1$ и $I(p_{j,l})$ – подматрица перестановки размера $z \cdot z$ (циркулянт - единичная матрица, циклически сдвинутая вправо на $p_{j,l}$ символов).

Проверочная матрица регулярного кода также может содержать нулевые подматрицы размера $z \cdot z$. Если веса строк (столбцов) проверочной матрицы LDPC-кода принимают различные значения, то такой код называют нерегулярным.

Структура графа Таннера и распределение строчных и столбчатых весов квазициклического LDPC-кода определяется базовой проверочной матрицей (граф Таннера которой называется протографом). Проверочная матрица квазициклического LDPC-кода получается путем расширения базовой матрицы подматрицами циклических перестановок. Использование циркулянт позволяет получать компактное представление квазициклического LDPC-кода и осуществлять параллельное кодирование и декодирование с глубиной параллелизма равной размеру циркулянта. Эти особенности обуславливают широкое применение квазициклических LDPC-кодов на практике.

В соответствии с итеративным алгоритмом декодирования BP получение верных значений бит кодового слова осуществляется в результате многократного обмена сообщениями вершинами графа Таннера. Каждая итерация алгоритма содержит две фазы. В фазе 1 обновляются сообщения проверочных вершин на основе анализа сообщений кодовых вершин; в фазе 2 – сообщения кодовых вершин на основе анализа сообщений проверочных вершин.

На эффективность BP-декодирования отрицательно влияет наличие циклов в графе Таннера, образующих треппин-сеты (Trapping set, TS, [4]) или (*a*, *b*)-подграфы (подграфы в графе Таннера, состоящие из *a* символьных узлов, *b* из которых инцидентны проверочным узлам с нечетными степенями). Эти подграфы обуславливают ошибку BP-декодирования [5]. В случае если вектор ошибки изменит значения символьных узлов, инцидентных нечетному числу проверок, то, вследствие неправильного подсчета условных вероятностей на символьных узлах подграфа ошибка не будет скорректирована, даже если она является корректируемой в соответствии с дистантными свойствами кода. На рис. 1.8 изображены два (*a*, *b*)-подграфа. Левый подграф образован пересечением трех циклов длины 8, правый одним циклом длины 8.

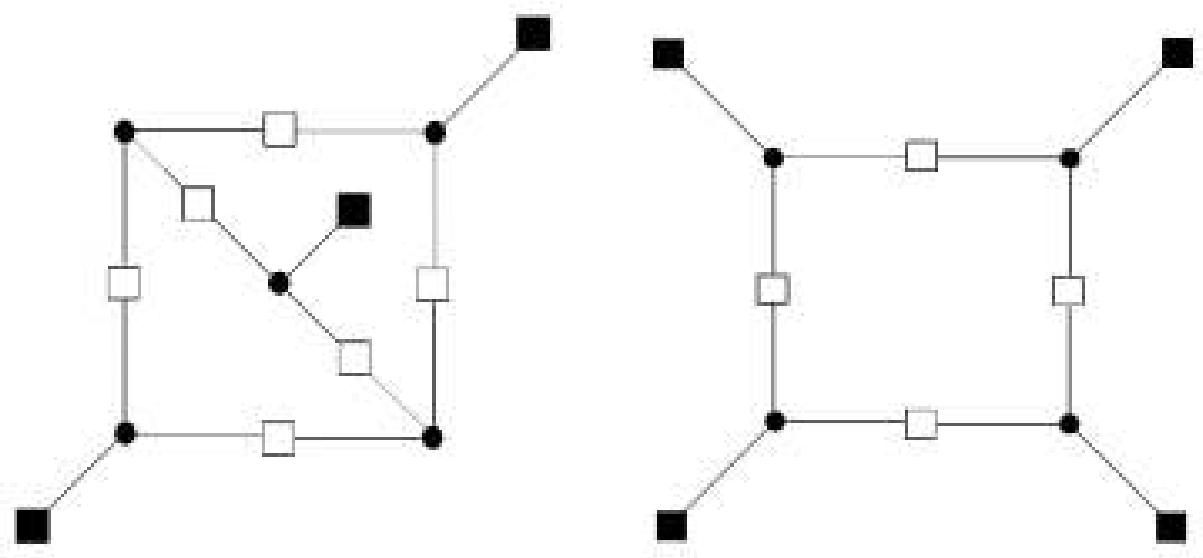

Рисунок 1.8 – Топологическое представление треппин-сетов TS(5, 3) слева, TS(4,4 справа), ●-символьные узлы, □-проверочные узлы с четной степенью, ■ -проверочные узлы с нечетной степенью инцидентности

Важной характеристикой LDPC-кода является приближенный **с**пектр связности (ACE spectrum) графа Таннера, заданного проверочной матрицей кода $H$. Спектр связности представляется в виде вектора:

$$ACE(H) = \left(ACE_{4,0}(VN), ACE_{4,1}(VN), ACE_{4,2}(VN), \ldots, ACE_{i,j}(VN), \ldots, ACE_{k,m}(VN)\right),$$

где $ACE_{i,j}(VN)$ - количество символьных узлов, содержащихся в цикле длины *i* со значение связности *j*. Значение связности (ACE) вычисляется для символьных узлов, содержащихся в подграфе, образованном циклом $v_i \in C$, $ACE(C) = \sum_{v \in C} \left(d(v) - 2\right)$, где $d(v)$- степень инцидентности символьного узла.

Пример вычисления значения связности для циклов длины 8 в подграфах треппин-сетов представлен на рис. 1.9. треппин-сету TS(5,3), более вероятно обуславливающему ошибку, соответствуют циклы длины 8 с меньшими значениями спектра связности (14) по сравнению с треппин-сетом TS(5,5).

Таким образом, код с лучшим спектром связности при одинаковом спектре кода при использовании декодирования методом распространения доверия лучше реализует корректирующие возможности, предоставляемыми весовым спектром. Этому коду не будут мешать псевдокодовые слова, обусловленные TS. Эта особенность обуславливает при построение низкоплотностных кодов необходимость совместной оптимизации дистантных свойств кода (кодового расстояния) и свойств графа (спектров связности).

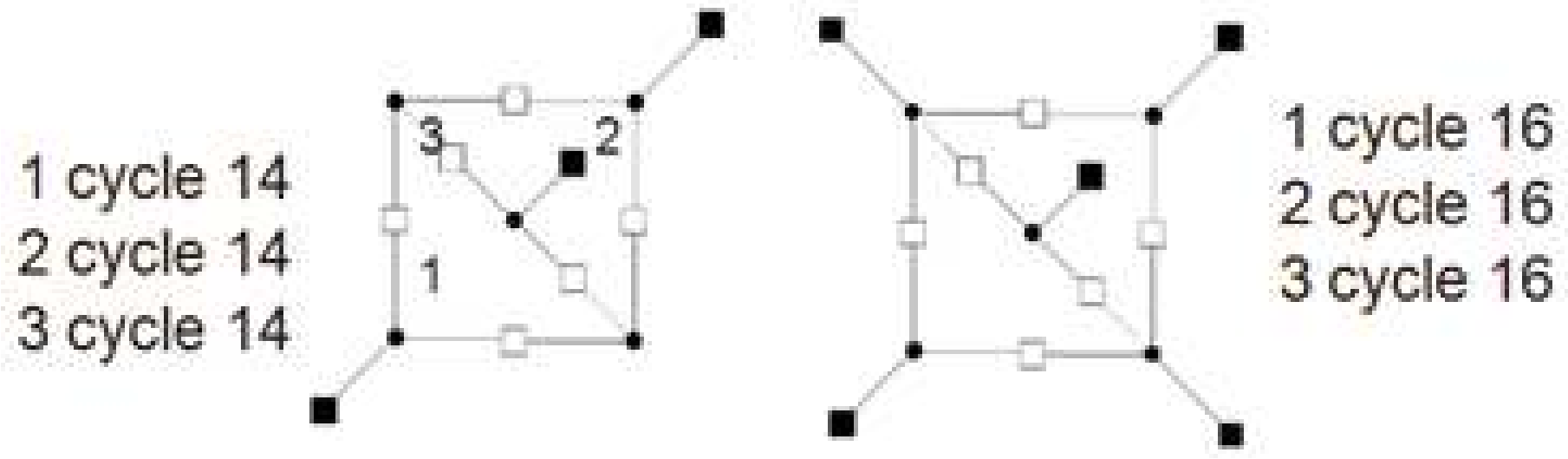

Рис. 1.9 – Пример вычисления ACE для подграфов TS: (5,5), (5,3)

### 1.2.2 Алгоритмы мягкого декодирования низкоплотностных кодов

Требования высокой надежности (низкого значения $P_{BER}$) в сочетание с разумной сложностью и сохранением высокой пропускной способности, предъявляемые к голографической памяти, обуславливают применение методов декодирования с «мягкими метриками». Использование декодера LDPC-кода для исправления ошибок в голографической памяти показано на рис. 1.10.

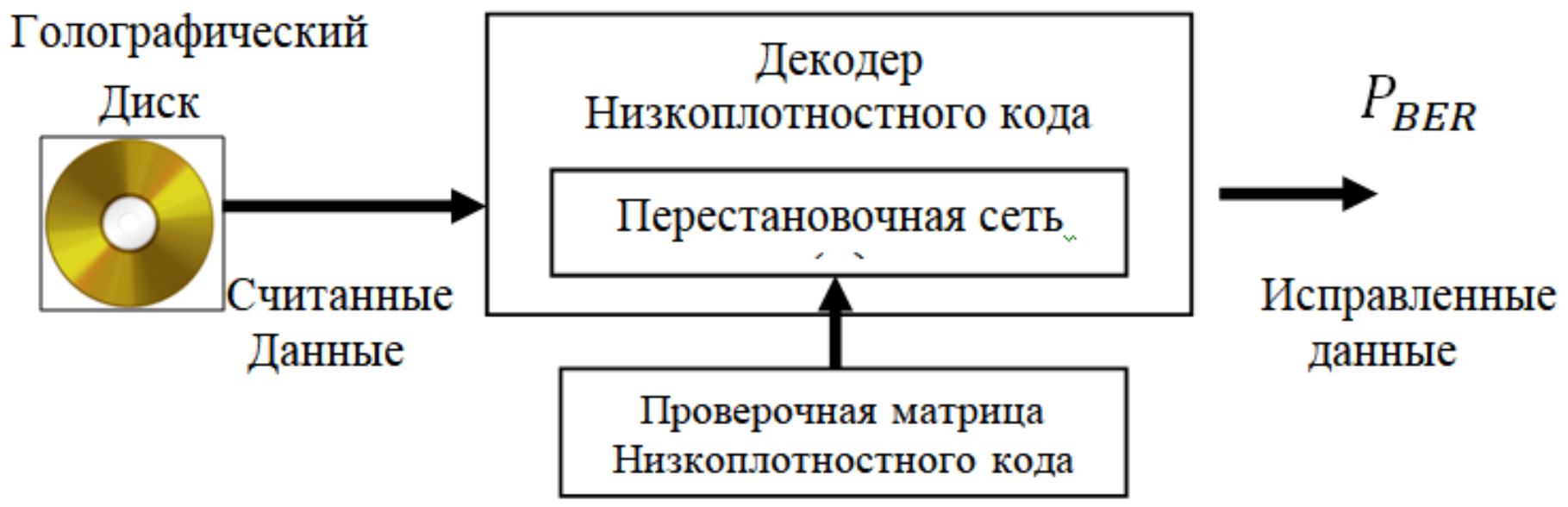

Рисунок 1.10. Включение декодера низкоплотностного кода в накопителе голографической памяти

Повышение надежности голографической памяти осуществляется за счет уменьшения вероятности ошибки на бит $P_{BER}$, в области рабочих значений отношений сигнал-шум.

Сформулируем математическую постановку задачи. Пусть кодовое слово $x$, полученное в результате кодирования линейным кодом $C[n,k]$ над $q-$ арным алфавитом, передано по АБГШ-каналу. Приемник получает зашумленный сигнал $y$. Совместная вероятность символов кодового слова - случайных величин определяется фактор графом, заданным проверочной матрицей $H$.

Совместную плотность вероятностей фактор графа без циклов можно разбить (факторизовать) на множители (product rule):

$$f(X_1,X_2,X_3)=f_1(X_1,X_2)f_2(X_1,X_3) \tag{1.11}$$

в соответствие с заданной структурой фактор графа, рис. 1.11.

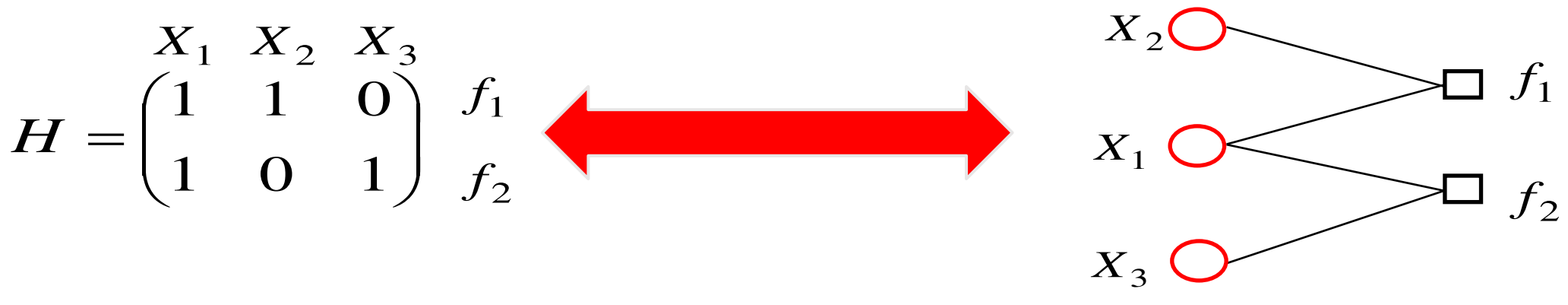

Рис. 1.11 Проверочная матрица и соответствующей ей фактор граф

Рассмотрим байесовый подход теории вероятностей применительно к вероятностным моделям на графах (probabalistical graphical models).

По теореме Байеса апостериорная вероятность кодового слова $x$:

$$P(x|y) = \frac{P(y|x)P(x)}{P(y)} \quad (1.12)$$

В случае АБГШ-канала без памяти с дисперсией $\sigma$ передачей $\pm a$ и двоичным алфавитом $q = 2$ условные вероятности $P(y|x) = \prod_{i=1}^{N} P(y_i|x_i)$ примут вид

$$P(y|x_i = 1) = \frac{1}{\sqrt{2\pi\sigma^2}} e^{\left(-\frac{(y_i-a)^2}{2\sigma^2}\right)},\ P(y|x_i = 0) = \frac{1}{\sqrt{2\pi\sigma^2}} e^{\left(-\frac{(y_i+a)^2}{2\sigma^2}\right)}. \quad (1.13)$$

Тогда отношения правдоподобия, определяемые апостериорными вероятностями для двоичной фазовой манипуляции (BPSK), имеют вид:

$$\frac{P(y_i|x_i = 1)}{P(y_i|x_i = 0)} = e^{\left(\frac{2ay_i}{\sigma^2}\right)}. \quad (1.14)$$

Априорное распределение кодовых слов считаем равномерным, т.е. передачу каждого из кодовых слов считаем случайным и равновероятным событием. Знаменатель $P(y) -$ нормализующая константа выписывается в виде:

$$P(y) = \sum_x P(y|x)P(x). \quad (1.15)$$

По апостериорным вероятностям символов, полученных из канала, и известной совместной вероятностью распределения символов можно сформулировать и эффективно решить две задачи декодирования.

1. Задача полного декодирования (Block-wise ML decoding) – определение наиболее вероятного переданного кодового слова $\dot{x}\,(y) = arg \max_{x \in C} P_{x|y}(x|y)$.
2. Задача побитового декодирования (Symbol-wise ML decoding) – для каждого переданного $q -$ арного символа кодового слова определить наиболее вероятное его значение $\dot{x_i}\,(y) = arg \max_{x_i \in X} P_{X_i|y}(x_i|y)$.

Решение задачи побитового декодирования, реализующего минимизацию вероятности ошибки на бит (bit error rate, BER), осуществляется путем маргинализации каждого из символов фактор графа:

$$P(x_i|y) = \sum_{X \backslash x_i} P(x|y). \quad (1.16)$$

Без утраты общности аналогичные рассуждения применимы для комплексного АБГШ-канала.

В общем виде задача является экспоненциально сложной, однако возможность разбить на множители (факторизовать) вероятности внутри суммирования (generalized distributed law, [6]), позволяет значительно упростить вычисление маргиналов.

Например, маргинальная вероятность символа $X_1$

$$P(X_1) = \sum_{X \setminus X_1} f(X_1, X_2, X_3) = \sum_{X_2} \sum_{X_3} f_1(X_1, X_2,) f_2(X_1, X_3)$$

$$= \sum_{X_2} f_1(X_1, X_2) \times \sum_{X_3} f_2(X_1, X_3) \quad (1.17)$$

Такой способ упрощенного вычисления маргиналов при помощи сумм и произведения вероятностей представляет собой алгоритм Sum Product, известный так же, как метод распространения доверия (Belief Propagation).

Нетрудно убедиться, что упрощение маргинализации выполнимо в случае, если символы не участвуют в циклах, т.е. на деревьях такой алгоритм за, конечно, линейное число операций дает точное значение маргиналов. Однако, фактор графы, заданные проверочной матрицей кода, являющиеся деревьями, обладают малой корректирующей способностью. С приближением кода к максимально возможному кодовому расстоянию (МДР-кодам) растет число циклов в фактор графе (корреляций между символами) и сложность его маргинализации (рис. 1.12).

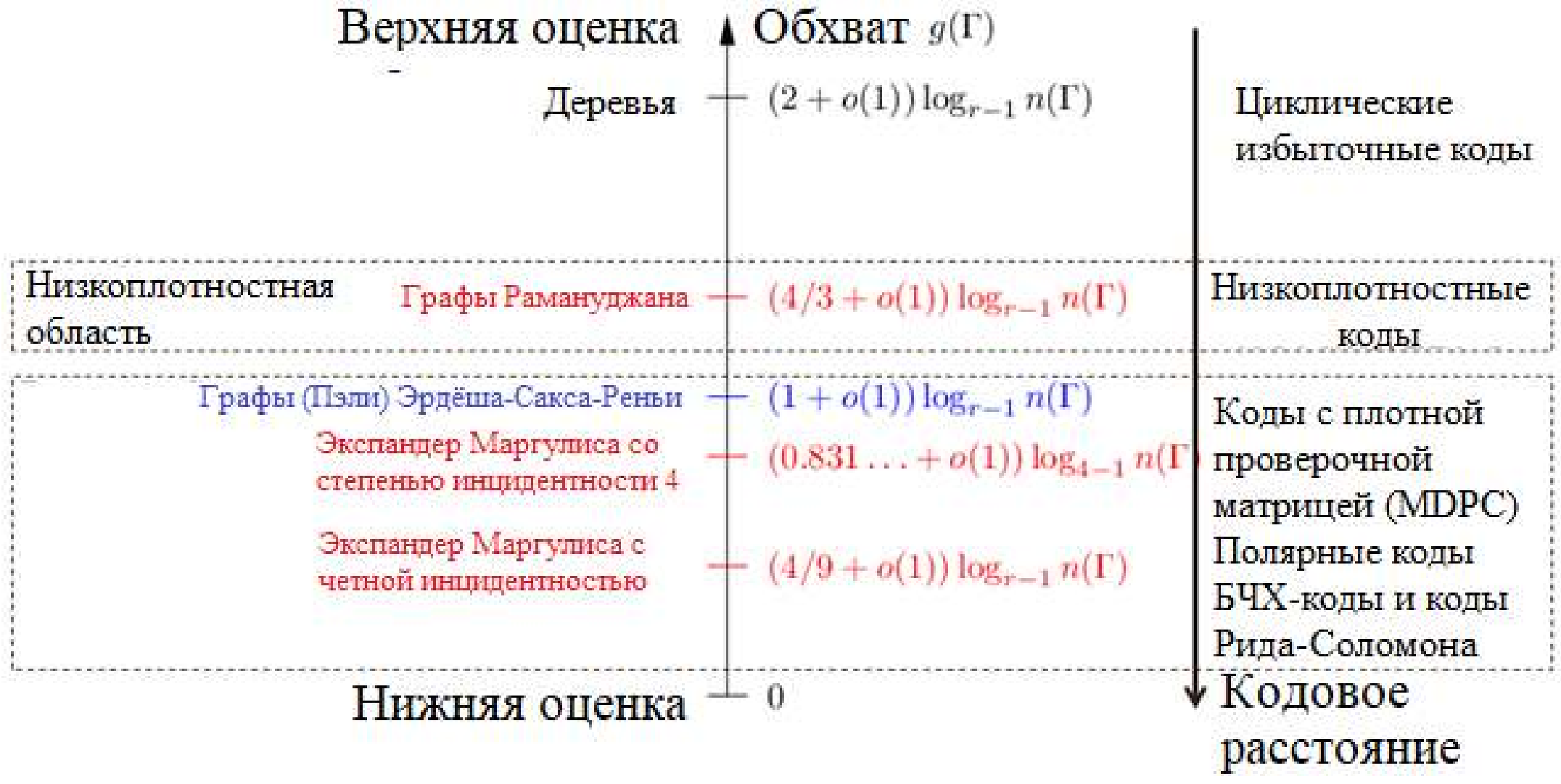

Рис. 1.12 Зависимость обхвата и кодового расстояния от длины кода, [7]

Решение задачи маргинализации для фактор-графов с циклами становится экспоненциально сложным. В этом случае применяются следующие подходы к учету циклов в фактор графе:

- Субоптимальное декодирование без учета циклов (loopy Belief Propagation) – вычисление маргиналов, как если бы циклы в фактор графе отсутствовали. Для эффективного применения этого подхода необходимо строить фактор графы со специальной структурой циклов. В дальнейшем будут рассмотрены подходы, позволяющие эффективно применять этот метод.

- Применение алгоритма для дерева сочленений (junction tree algorithm) - представление символов, участвующих в цикле, в качестве кластеризованного узла с оптимальным расчетом маргиналов в нем. На этом подходе основаны обобщённые (GLDPC) и недвоичные низкоплотностные коды (NB-LDPC), осуществляющие оптимальное декодирования в кластерах проверок и внутри символа соответственно.

- Применение алгоритма разреза графа (cutset condition), осуществляющего разбиение циклов путем назначения определенного значения символьному узлу. Этот алгоритм эффективен для полярных кодов (Polar codes) и низкоплотностных кодов с мультиграфовой структурой факторов (Multi-edge Type LDPC, ME-LDPC). Применение этого метода позволяет игнорировать циклы, инцидентные узлу, участвующему в разрезе графа. В случае полярного кода разбиение циклов осуществляется путем последовательной маргинализации символов нормального фактор графа с использованием априорного знания значений (замороженных) символов. В низкоплотностных кодах с мультиграфовой структурой факторов выделяются специальные символы, которые содержатся в большом числе факторов и в отличие от остальных узлов не обладают апостериорной вероятностью. Эти символы восстанавливаются на основе маргинальных значений всех остальных символов. В случае специальной структуры корреляций символов (в некотором смысле образующих «деревья» - компенсирующие друг друга подграфы с циклами), возникает возможность однозначного восстановления этих символов на основание маргиналов остальных узлов.

По своей сути каждому классу кодов, достигающему пределов его корректирующей способности, соответствует некоторый фактор граф, определяющий метод декодирования.

В данной диссертационной работе мы будем применять комбинацию перечисленных методов с целью увеличения числа корреляций между символами для улучшения корректирующей способности кода и устранения отрицательного влияния циклов.

### 1.2.3 Методы оценки эффективности исправления ошибок низкоплотностными кодами

Помехоустойчивость кодов в области высоких значений сигнал-шум с мягким декодером определяется весовым спектром кода в соответствии с формулой (Union bound), ошибка на блок [8]:

$$P_s \leq \sum_{i=d_{\min}}^{N} \omega_i Q\left(\sqrt{i}\right), \tag{1.18}$$

где $d_{\min}$ -кодовое расстояние, $\omega_i$ -кратность слов минимального веса, $i$ - вес кодового слова, Q - Q-функция $Q(x) = \left(\pi N_0\right)^{-1/2} \int_x^{\infty} e^{-n^2/N_0} dn$.

При высоких отношениях сигнал-шум ошибка на блок приближенно вычисляется:

$$P_s \approx \omega_{d_{\min}} Q\sqrt{d_{\min}}\ , \tag{1.19}$$

$\omega_{d_{\min}}$ - кратность слов минимального веса.

Однако в случае субоптимального декодирования методом распространения доверия (Belief propagation) возникают псевдокодовые слова, обусловленные подграфами треппин-сетов. Псевдокодовые слова относительно большого веса определяют характер кривой помехоустойчивости в области “водопада” (рис. 1.13). Псевдокодовые слова малого веса определяют характер полки кода (рис. 1.14). Задача определения спектра кодовых и псевдокодовых слов имеет экспоненциальную сложность. Однако существуют классы эффективных методов, позволяющих определить поведение водопада и полки с использованием свойств фактор графа.

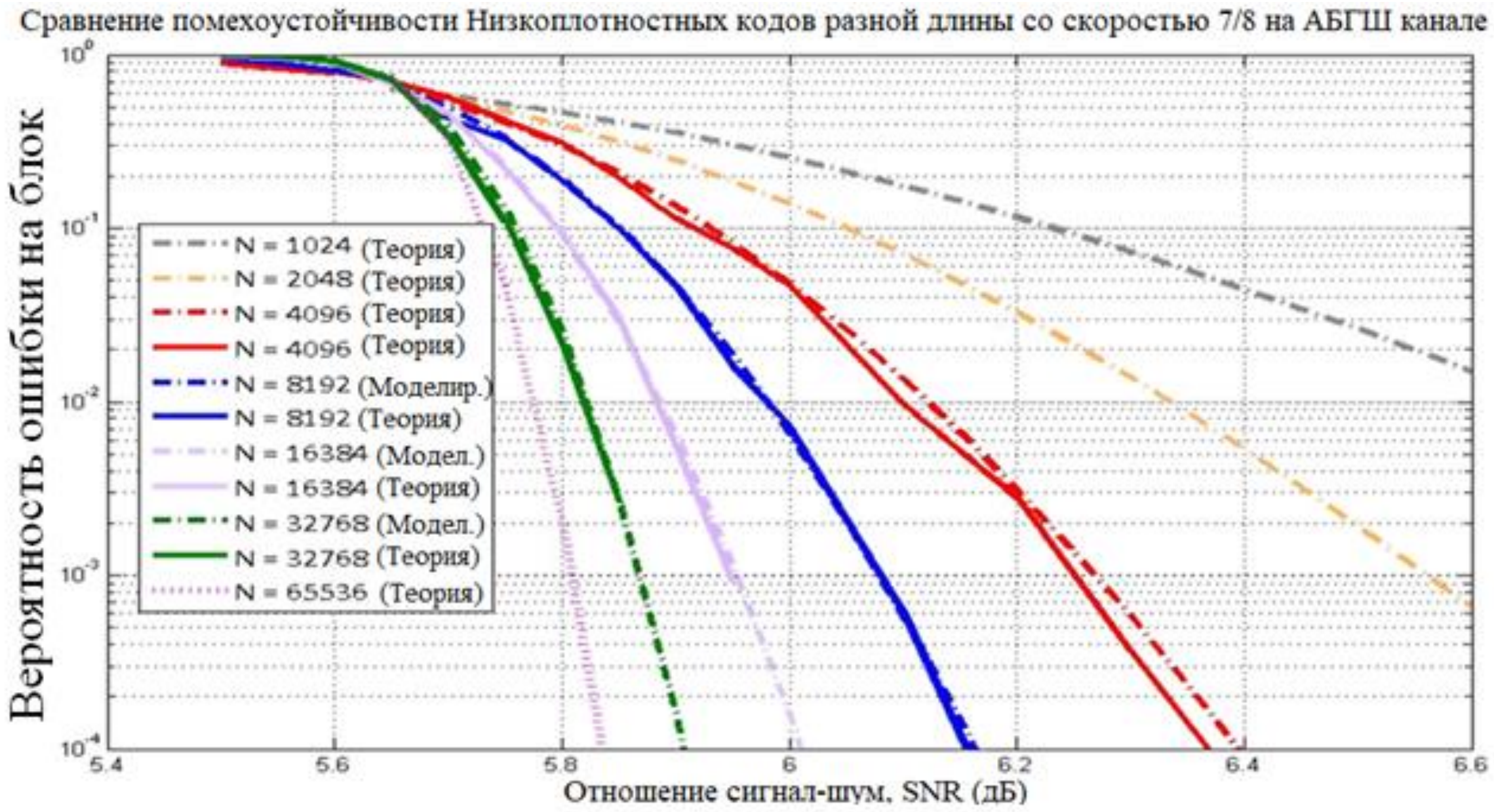

Рис. 1.13- Оценка вероятности ошибки на блок с изменением длины кода, полученная методом Ковариационной Эволюции плотностей (Covariance Evolution), [11]

Для определения полки применяется метод выборки по значимости (Importance Sampling, см. приложение 4). Наиболее эффективным (точным) вариантом его реализации является применение релаксированных симплексов (Фундаментальных конусов, Fundamental Cone) с использованием линейного программирования [9,10]. Для определения водопада используется метод Ковариационной Эволюции, предложенный в работе [11].

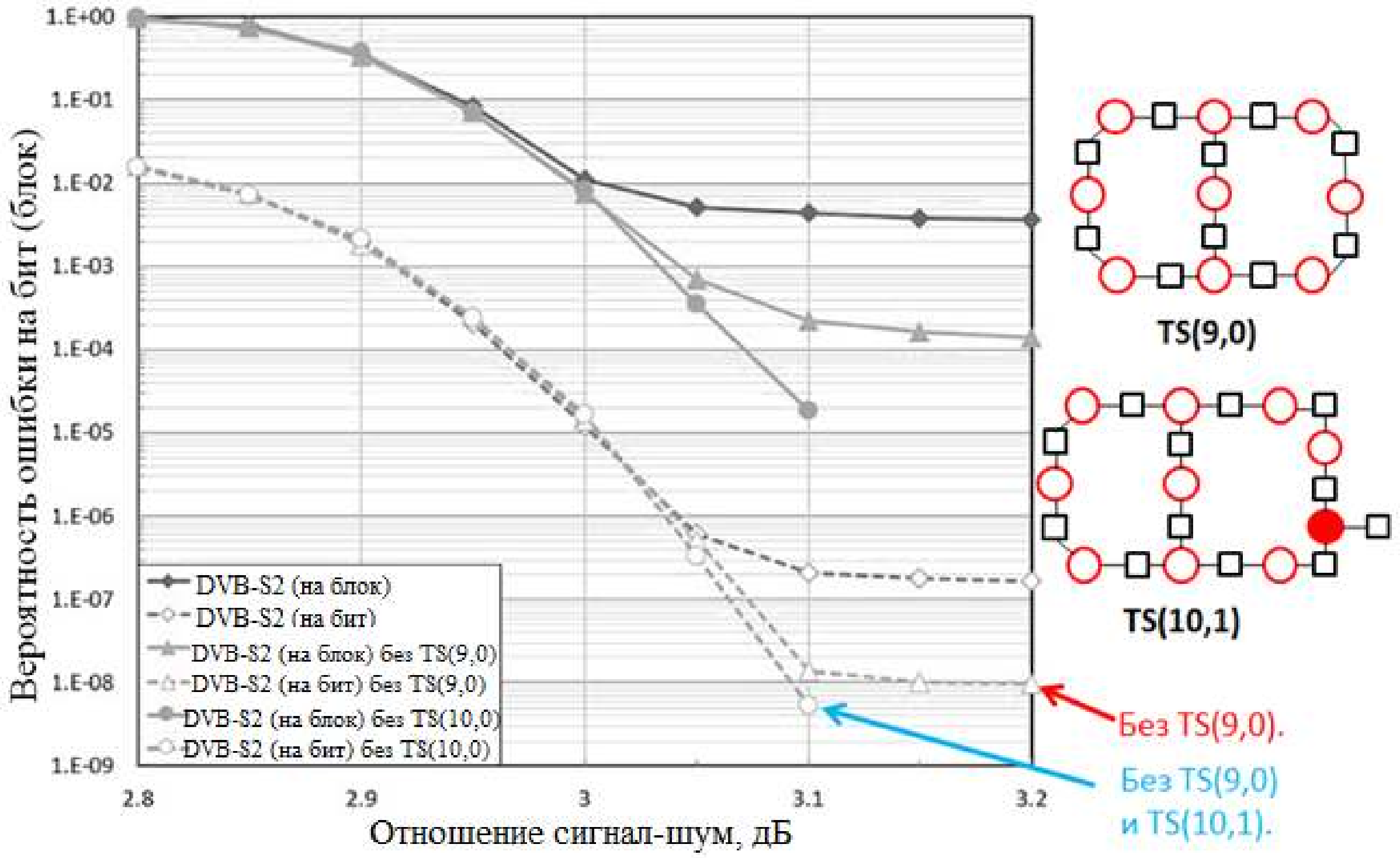

Рис. 1.14 - Вероятность ошибки на бит (блок) для квазициклических низкоплотностных кодов DVB-S2 информационной длины K=43200 и скоростью 2/3 для кодов, полученных модификацией циклических сдвигов, не содержащих TS (9, 0), TS (10,1).

Отметим, что веса псевдокодовых слов зависят от параметров декодера (числа уровней квантования, внутреннего представления коэффициентов правдоподобия при обработке символов и проверок в методе распространения доверия и прочих) и их изменение способно изменить весовой спектр TS и соответственно характер кода в области полки и водопада [12].

В силу субоптимальности метода распространения доверия линейно растущие треппин-сэты обуславливают отклонение «водопада» от порога итеративного декодирования. Влияние их на вероятность ошибки на блок, полученную методом Ковариационной Эволюции плотностей (Covariance Evolution) [11], с увеличением длины кода изображено на рис. 1.13.

Метод Ковариационной Эволюции плотностей работает для ансамблей кодов, чьи короткие циклы распределены по закону Пуассону, [14].

Сублинейные треппин-сеты наряду с кодовым расстоянием определяют помехоустойчивост кодов в области полки, рис. 1.13.

## 1.3 Методы построения низкоплотностных кодов

### 1.3.1 Классификация методов построения низкоплотностных кодов

Процедура построения квазициклических LDPC-кодов предусматривает выполнение 2-х этапов:

1) выбор протографа (базовой матрицы),

2) расширение базовой проверочной матрицы.

Расширение (lifting) базовой проверочной матрицы заключается в замене единичных символов базовой матрицы на циркулянты. На этом этапе определяются значения циклических сдвигов всех циркулянт.

Для реализации этапов построения LDPC-кодов используются комбинаторно-алгебраические методы и методы статистической физики.

К первым относятся методы, основанные на конечных геометриях, блочных дизайнах, числовых сетках, словах малого веса кодов Рида-Соломона, метод Саливана [14-18].

Достоинством этих методов является возможность выбора кода с хорошими дистантными свойствами, простота расширения матрицы циркулянтами, а также быстрое получение высокоскоростных регулярных кодов с обхватом 6 и более. Однако применение их затруднительно для построения кодов произвольной размерности (numerology problem) и требует поиска маски для получения нерегулярных кодов.

Наиболее эффективными при поиске протографов являются методы статистической физики, предложенные Ричардсом, Урбанке и Монтанари. В их число входят «Эволюция плотностей» (Density evolution), предусматривающая оптимизацию асимптотических свойств ансамблей кодов при помощи минимизации порога итеративного декодирования, и ее аппроксимации: Exit-chart, Protograph-Exit chart, Reciprocal channel approximation, Gaussian Approximation. Достоинство этих методов - возможность получения протографа (маски/весового распределения), позволяющего строить нерегулярные коды. Недостатком этого класса методов является игнорирование комбинаторно-алгебраических свойств

Таннер-графа и дистантных свойств кода, в результате полученные коды содержат кодовые и псевдокодовые слова малого веса.

Для расширения базовой матрицы чаще всего применяются следующие методы: улучшенный PEG-алгоритм, Hill-Climbing, Guess-and-Test.

### 1.3.2 Методы поиска протографа

#### 1.3.2.1 Метод эволюции плотностей распределения (Density Evolution)

Метод «Эволюции плотностей», предложенный Ричардсоном и Урбанке [19, 20], в предположении отсутствия циклов в графе, позволяет оценить значение «порога протографа» (protograph threshold) - соотношения сигнал-шум (например, величину дисперсии шума), при котором вероятность битовой ошибки не превосходит заданную величину.

Процедура поиска протографа заключается в минимизации порога с ограничениями на структуру циклов в графе, поскольку на последующем этапе нам потребуется расширить этот протограф таким образом, чтобы циклы, возникающие из-за использования циклических матриц перестановок и конечной длины кода, не помешали достичь предполагаемую порогом корректирующую способность.

Рассмотрим подробнее этот метод.

Одним из ключевых параметров, характеризующих ансамбль низкоплотностных кодов, является распределение весов в столбцах и строках проверочной матрицы. Это распределение принято описывать при помощи многочленов $\lambda(x) = \sum \lambda_i x^{i-1}$, $\rho(x) = \sum \rho_i x^{i-1}$, $\sum_i \lambda_i = 1, \sum_i p_i = 1$, где $\lambda_i$ ($\rho_i$) это доля ребер, соединенных с символьным (проверочным) узлом, со степенью вершины $i$.

Например, данное распределение

$$\lambda(x) = \frac{4}{24}x^5 + \frac{4}{24}x^3 + \frac{5}{24}x^2 + \frac{11}{24}x, \quad \rho(x) = \frac{1}{12}x^7 + \frac{5}{12}x^6 + \frac{4}{12}x^5 + \frac{2}{12}x^4 \qquad (1.20)$$

описывает ансамбль кодов, у которого:

- 11/24 доли проверок по столбцам имеют вес 2, 5/24 - вес 3, 4/24 - вес 4, 4/24 - вес 6;

- 2/12 доли проверок по строкам имеют вес 5, 4/12 вес 6, 5/12 вес 7, 1/12 вес 8.

Тогда средний вес символьного и проверочного узлов вычисляется по формулам:

$$l_{avg} = \frac{1}{\sum_i (\lambda_i / i)}, r_{avg} = \frac{1}{\sum_i (p_i / i)}. \quad (1.21)$$

где $l_{avg}$- средний вес символьного узла, $r_{avg}$ – средний вес проверочного узла.

В случае отсутствия линейно-зависимых строк скорость кода определяется

$$rate(\lambda, \rho) = 1 - \frac{l_{avg}}{r_{avg}}. \quad (1.22)$$

При устремлении длины кода к бесконечности и отсутствия автоморфизмов в графе (циклической и квазициклической структуры) граф представляет собой дерево. Для квазициклических кодов оценки обхватов графа были получены Таннером, Вонтобелем и Фоссорье в работах [21-23].

В случае графа - дерева символьный узел с вероятностью $\lambda_i$ имеет $i$ инцидентных к нему узлов, т.е. один родительский узел и $i$-1 «дочерних» (узлов потомков). Каждый из дочерних узлов в Таннер-графе с вероятностью $p_i$ имеет $i-1$ соседних дочерних к нему узлов. Проверочный узел $c$ инцидентен символьному узлу $v$ в качестве родительского узла и имеет $i-1$ символьных узлов $v_s$, $s \in i,...,i-1$ в качестве «дочерних» узлов.

Пусть $\gamma_s$ логарифм правдоподобия, представляющий собой оценку правдоподобия символьного узла $v_s$. Вычисляя маргинальные вероятности методом распространения доверия, узел $c$ получает сообщения с оценкой $\gamma_s$ от узлов потомков. Узел $c$, получая сообщения, осуществляет оценку инцидентных ему узлов $m_{cv} = 2\tanh^{-1}\left[\prod_{s=1}^{i-1} \tanh(\gamma_s)\right]$, и затем отправляет эту оценку $m_{cv}$ в качестве нового сообщения родительскому узлу $v$.

Независимые случайные величины $\gamma_s$ заданы функцией плотности вероятности $f$:

$$P(a < \gamma_s < b) = \int_a^b f(x)dx, \qquad (1.23)$$

Функция плотности вероятностей $f_{cv}$ случайной величины $m_{cv}$ задана уравнением:

$$f_{cv} = \sum_{i=2}^{i_{\max}} p_i f^{(i-1)\otimes}, \qquad (1.24\ )$$

где $f^{(i-1)\otimes}$ свертка вероятностей.

Тогда символьный узел веса $i$ получает сообщения с оценками логарифмов правдоподобия $\delta_s$ от инцидентных к нему проверочных узлов. В силу статистической независимости этих оценок $\delta$ представляет собой сумму случайных величин:

$$\delta = \sum_{s=1}^{i-1} \delta_s - \delta_0, \qquad (1.25)$$

где $\delta_0$ начальная оценка, полученная из канала.

Метод эволюции плотностей предполагает, что оценки заданы той же функцией плотности вероятностей $f$. Тогда для оценки $\delta$ функция плотности вероятностей $f_{cv}$ задана уравнением

$$f_{cv} = f \times f \times \ldots . \times f \times f_0 = f^{(i-1)\times} \times f_0, \qquad (1.26)$$

$\times$-операция свертки плотности вероятностей.

Функция плотности вероятностей определяется случайными величинами чьи значения связаны с распределением весов $\lambda$:

$$f_{vc} = \left(\sum_{i=2}^{i_{max}} \lambda_i f^{(i-1)\times}\right) \times f_0 \,. \qquad (1.27)$$

Вероятность ошибки вычисляется по формуле:

$$P_e = \int_{-\infty}^{0} f_{vc}(x)dx. \qquad (1.28)$$

**Алгоритм:**

Пусть $f_0$ функция плотности вероятностей, заданная логарифмами правдоподобия при некотором значение сигнала-шум, заданном дисперсией ошибки $\sigma_+$; $l_\lambda$ максимальный вес символьного узла and $l_p$ максимальный вес проверочного узла, $\varepsilon$ - требуемая вероятность ошибки на бит, $i$ - номер итерации в алгоритме распространения доверия.

Проинициализируем $f_{cv} = f_0$.

В цикле по $i$ вычисляем функции плотностей вероятностей:

$$f_{cv} = \sum_{i=2}^{i_{\max}} p_i f^{(i-1)\otimes}, \tag{1.29}$$

$$f_{vc} = \left(\sum_{i=2}^{i_{max}} \lambda_i f^{(i-1)\times}\right) \times f_0 \,. \tag{1.30}$$

Если вероятность ошибки на бит $P_e = \int_{-\infty}^{0} f_{vc}(x)dx < \varepsilon$ не превосходит требуемую, прекращаем работу алгоритма и выдаем $\sigma_+$.

Код, заданный распределением $(\lambda, p)$ называется оптимальным, если его порог итеративного декодирования, заданный дисперсией ошибки ( $\sigma_+$ ), максимален:

$$(\lambda, p) = \arg\max_{\tilde{\lambda}, \tilde{p}} \sigma_+\left(\tilde{\lambda}, \tilde{p}\right).. \tag{1.31}$$

Зависимость порога итеративного декодирования $\sigma$ АБГШ-канала для регулярных кодов с весом столбца $a$, $\lambda(x) = x^a$ и скоростями 0.5, 0.75, 0.9, 0.95 приведены на рис. 1.15, [24]. Уменьшение веса символа приводит к улучшению порога итеративного декодирования, однако значительно затрудняет построение кода вследствие недостаточного кодового расстояния. Это обусловлено свойствами кода или же особенностями структура графа, которые мы рассмотрим в дальнейшем.

По этой причине при построении кодов одним из подходов будет попытка компенсации малого веса символов путем добавления символов большого веса. В Табл. 1.1, показана разница порога итеративного декодирования и предела Шеннона для нерегулярного кода со скоростью ½ с ростом максимального веса символьного узла, [25]. Начиная со значений веса столба более 10, разница в пороге не превышает 0.1 ДБ. Именно этот вес определяет размеры протографа низкоплотностного кода. Число строк в графе квазициклического кода не может быть меньше веса максимального символа $\max deg\lambda(x)$ .

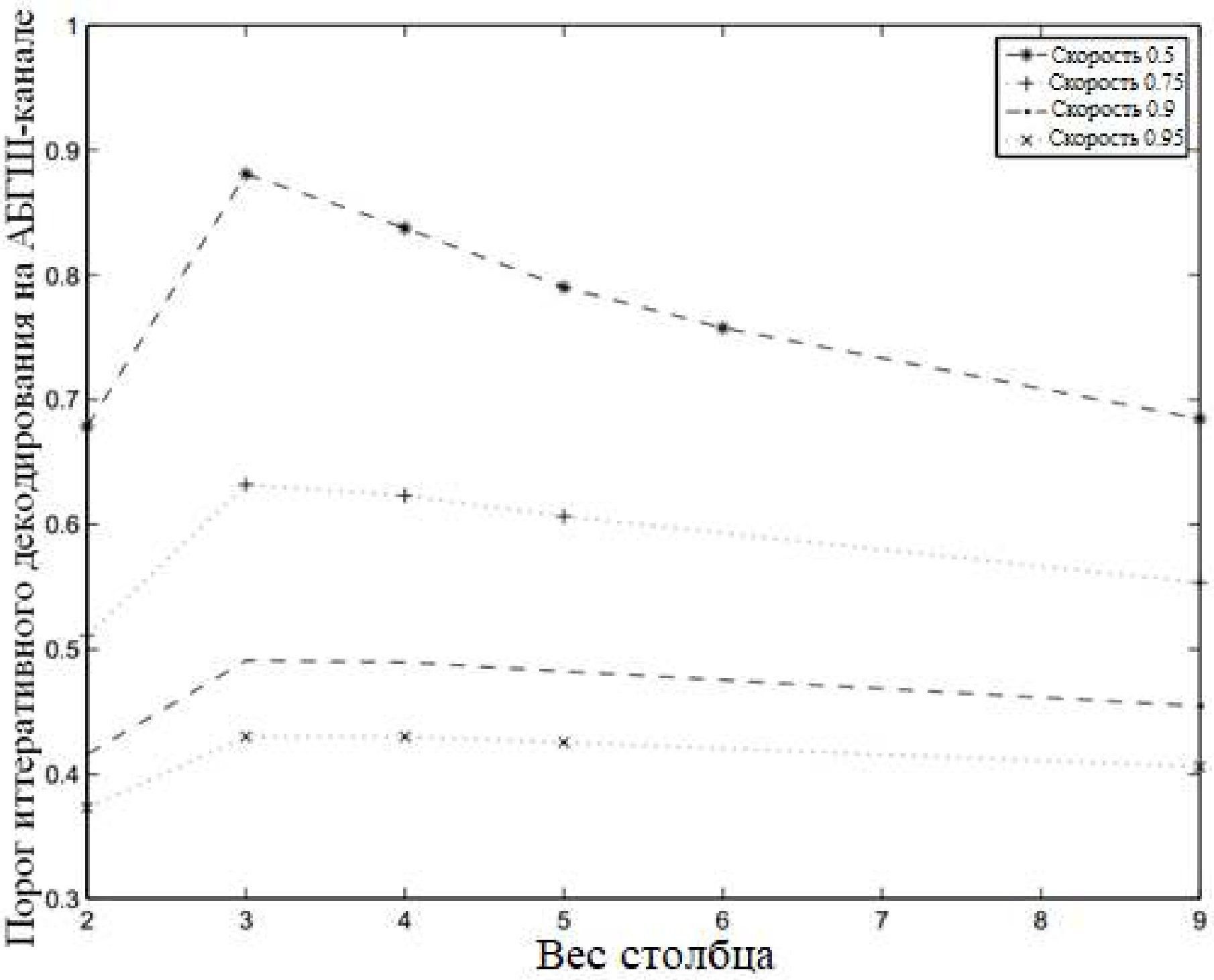

Рис. 1.15 Зависимость порога итеративного декодирования от веса столбца в случае регулярных низкоплотностных кодов различных скоростей для АБГШ-канала

Скорость кода, заданного протографом, содержащим выколотые “символы”, без использования укорочения кода с целью адаптации его параметров длинны и скорости определяется формулой:

$$Rate(H) = \frac{VN-CN}{CN-PN}, . \qquad (1.31)$$

где $VN$ число столбцов в протографе $(VN \geq maxdeg\lambda(x))$, $CN$ число строк в протографе, $PN$- число «выколотых» символов. Последний параметр потребуется в дальнейшем, поскольку он позволяет при увеличении числа итераций значительно улучшить порог итеративного декодирования.

Однако размер такого протографа велик (число строк в нем не может быть меньше веса столбца). При фиксированной длине рассматриваемых нами кодов это приводит к значительному уменьшению циркулянта.

Таблица 1.1 - Влияние распределения весов символьного узла (веса столбца) на порог итеративного декодирования нерегулярных низкоплотностных кодов, [25]

| Максимальный вес столбца, $d_l$ | 100 | | 200 | | 8000 | |
|---|---|---|---|---|---|---|
| | $x$ | $\lambda_x$ | $x$ | $\lambda_x$ | $x$ | $\lambda_x$ |
| | 2 | 0.170031 | 2 | 0.153425 | 2 | 0.096294 |
| | 3 | 0.160460 | 3 | 0.147526 | 3 | 0.095393 |
| | 6 | 0.112837 | 6 | 0.041539 | 6 | 0.033599 |
| | 7 | 0.047489 | 7 | 0.147551 | 7 | 0.091918 |
| | 10 | 0.011481 | 18 | 0.047938 | 15 | 0.031642 |
| | 11 | 0.091537 | 19 | 0.119555 | 20 | 0.086563 |
| | 26 | 0.152978 | 55 | 0.036379 | 50 | 0.093896 |
| | 27 | 0.036131 | 56 | 0.126714 | 100 | 0.018375 |
| | 100 | 0.217056 | 200 | 0.179373 | 150 | 0.086919 |
| | | | | | 400 | 0.089018 |
| | | | | | 900 | 0.057176 |
| | | | | | 2000 | 0.085816 |
| | | | | | 3000 | 0.006163 |
| | | | | | 6000 | 0.003028 |
| | | | | | 8000 | 0.118165 |
| Средний вес столбца, $\lambda_{AVG}$ | 10.9375 | | 12.0000 | | 18.5000 | |
| Среднеквадратическое отклонение, $\sigma$ | 0.97592 | | 0.97704 | | 0.9781869 | |
| Нормализованный сигнал-шум, $SNR_{norm}$ | 0.0247 | | 0.0147 | | 0.00450 | |

Малый размер циркулянта не позволяет расширить такие протографы с приемлемой структурой циклов, а также уменьшает пропускную способность декодера, т.к. максимальный параллелизм определяется размером циркулянта.

Поскольку в аппаратуре алгоритм декодирования реализуется при помощи арифметики с фиксированной точкой, уменьшение числа бит влияет на динамический диапазон, ухудшая порог итеративного декодирования кода. Результаты построения двоичных кодов, опирающиеся на традиционный подход изображенны на рисунке 1.16 и будут рассмотрены в дальнейшем.

Существует другой подход, позволяющий улучшить порог итеративного декодирования без значительного увеличения веса символьного узла в протографе. Суть этого метода заключается в использование протографов со специальной структурой восстановления символов после стирания (Multi-edge Type LDPC codes). Такие протографы требуют использования специальной версии эволюции плотностей, предложенной Ричардсоном-Урбанке [26] и развитое МакЭллисом-Торпом-Дивсаларом [27-28].

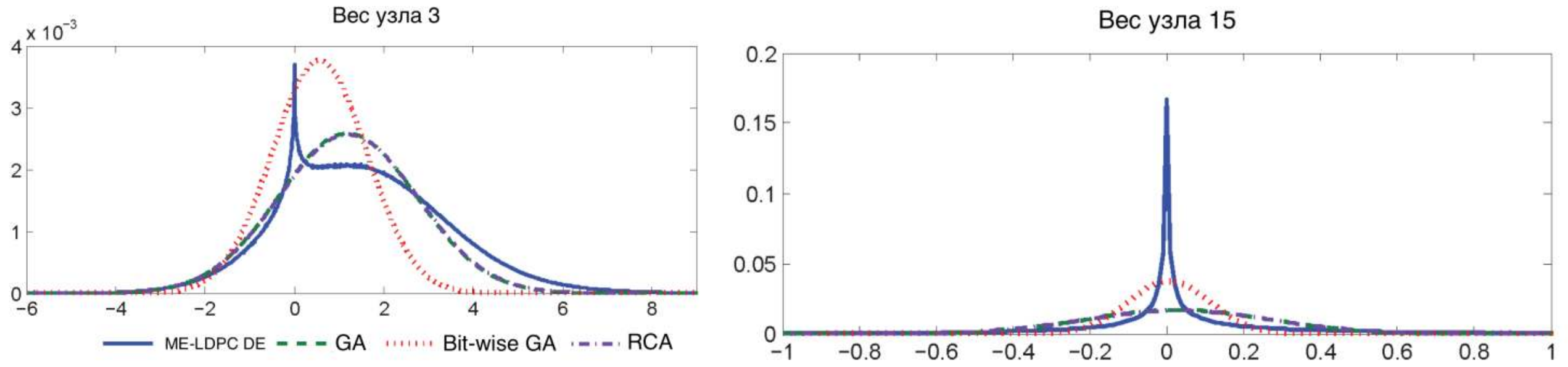

Рисунок 1.16 - Значение функции плотности вероятности переданной от символьного узла к проверочному для протографа со скоростью ½, для АБГШ-канала с дисперсией 0.85 на второй итерации декодирования методом распространения доверия вычисленной при помощи методов: Эволюции плотностей (ME-LDPC DE), Гауссовой Аппроксимации (GA, Gaussian Approximation), обобщенной Гауссовой Аппроксимации (Bit-wise GA) и Взаимного канала (RCA, Reciprocal Channel Approximation)

1.3.2.2 Методы, использующие аппроксимацию эволюции функции плотности вероятностей

В силу сложности вычисления интеграла на практике применяются аппроксимации метода Эволюции плотностей: Гауссова аппроксимация (Gaussian Approximation, GA, [29]), Гауссова аппроксимация на бит (Bit-wise Gaussian approximation, Bit-wise GA [30], аппроксимация через взаимные каналы (Reciprocal-channel approximation, RCA [31]). Однако в случае Multi-edge type LDPC все эти методы демонстрируют крайне неточное и неустойчивое поведение, особенно в случае ME-LDPC кодов с низкой скоростью, содержащих символьные узлы малого веса, [32]. На Рис. 1.17 показана функция плотности вероятностей для символов веса 15 и 3. Уже для веса 3 наблюдается существенное отклонение функции плотности вероятности.

Последнее обстоятельство делает эти методы малопригодными для использования в случае оптимизации протографа ME-LDPC кодов.

С целью решения этой проблемы будет применяться метод аппроксимации вероятностей через взаимную информацию (Mutual Information) и битовый вариант взаимной информации (Bit-wise Mutual Information, Generalize Mutual Information). Взаимная информация особенно хорошо аппроксимирует результаты декодирования недвоичных кодов с большим алфавитом $q > 64$, для которых характерно оптимальное декодирования символьной информации из модуляционного созвездия [33]. Битовая взаимная информация наилучшим образом аппроксимирует двоичные коды декодируемые без турбо петли на созвездие, [34].

Для оптимизации протографов мы будем использовать графики Protograph Exit-Chart (Generalize Exit-Chart) для согласования параметров кода при заданном числе итераций, [35]. За счет работы метода на фиксированном протографе точность метода составляет 0.01 дБ в случае использования аппроксимации функции взаимной информации при помощи таблицы и менее 0.05 дБ при помощи аппроксимации полиномом [36]. Метод был предложен Тен Бринком [37] и обобщен Ливой [38].

Оценка порога итеративного декодирования при помощи графиков эволюции априорной и апостериорной взаимной информации случайных величин и факторов для заданного распределения весов символов и проверок низкоплотностного кода (EXIT-charts) также как в методе Эволюции плотностей опирается на отсутствие циклов в Таннер-графе низкоплотностного кода.

Вычисление порога итеративного декодирования осуществлятся путем построения графиков эволюции взаимной информации в процессе работы алгоритма распространения доверия (belief propagation) двух компонентов низкоплотностного кода: символов и проверок $I_{E,V}$ и $I_{E,C}$. Пример Эволюции взаимной информации этих компонентов приведен на графике рисунка 1.17.

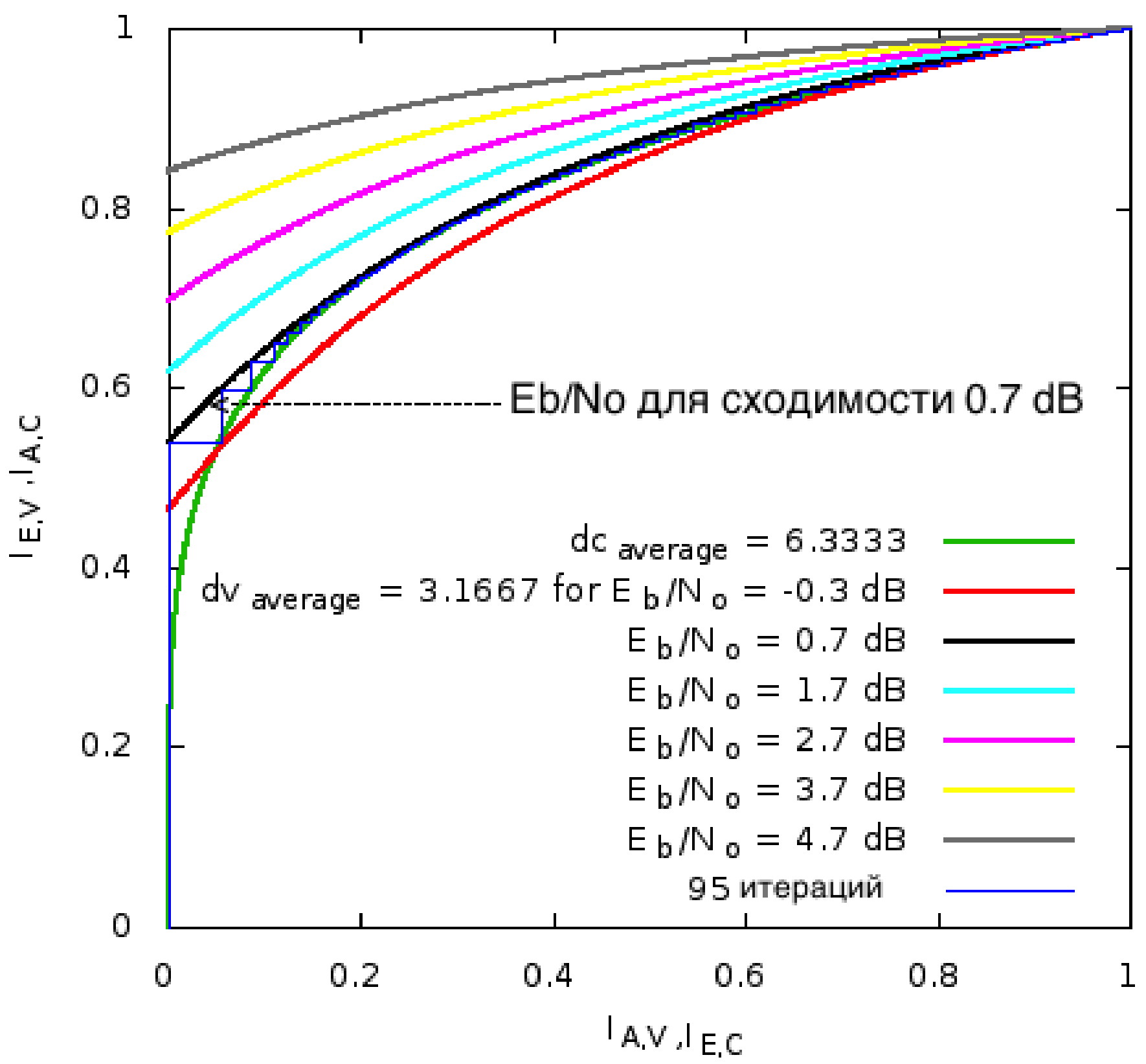

Рисунок 1.17 – График взаимной информации проверочного символа и символьного узла с ростом числа итераций алгоритма декодирования

Один компонент изображен с его вводом на горизонтальной оси и его выходом на вертикальной оси. Другой компонент строится с его вводом на вертикальной оси и его выходом на горизонтальной оси. Для успешного декодирования эти графики должны лежать максимально близко друг к другу и пересекаться в 1. Порог декодирования — это величина дисперсии АБГШ, при которой эти кривые пересекаются в 1.

Апостериорная информация (внешняя) $I_{E,V}$ символьных узлов вычисляется в соответствие с уравнениями:

$$I_{E,V} = J(\sigma) = J\left(\sqrt{(d_v - 1)\sigma_A^2 + \sigma_{ch}^2}\right), . \qquad (1.32)$$

$$J(\sigma) = 1 - \int_{-\infty}^{\infty} \frac{1}{\sqrt{2\pi\sigma}} e^{-(l-\frac{\sigma^2}{2})^2/2\sigma^2} \log(1 + e^{-l})dl, \qquad (1.33)$$

где $J(\sigma)$ – взаимная информация.

Априорная информация $I_{A,V}$ символьных узлов вычисляется в соотвествие с уравнениями:

$$I_{A,V} = J(\sigma_A), \tag{1.34}$$

$$I_{E,V} = J(\sigma) = J\left(\sqrt{(d_v - 1)\left[J^{-1}(I_{A,V})\right]^2 + \sigma_{ch}^2}\right). \tag{1.35}$$

Апостериорная информация (внешняя) $I_{E,C}$ проверочных символов (факторов) вычисляется в соотвествие с уравнениями:

$$I_{E,C} = 1 - J\left(\sqrt{(d_c - 1)\left[J^{-1}(1 - I_{A,C})\right]^2}\right). \tag{1.36}$$

Веса задаются в соответствие с полиномом распределения проверок $\lambda(z) = \sum_{d=1}^{d_v} \lambda_d z^{d-1}$ and $\rho(z) = \sum_{d=1}^{d_c} \rho_d z^{d-1}$, где $\lambda_d$ доля проверок веса $d$ в столбце проверочной матрицы, и $\rho_d$ доля проверок веса $d$ в строке.

Для нерегулярных кодов

$$I_{E,V} = \sum_{d=1}^{d_v} \lambda_d I_{E,V}(d, I_{A,V}), \tag{1.37}$$

$$I_{E,C} = \sum_{d=1}^{d_v} \rho_d I_{E,C}(d, I_{A,C}). \tag{1.38}$$

Функция взаимной информации $J(\cdot)$ приближается полиномом либо просмотровой таблицей с некоторой заданной точностью.

Путем вычисления функций $I_{E,V}(d, I_{A,V})$ и $I_{E,C}(d, I_{A,C})$ осуществляется итеративный процесс эволюции взаимной информации начиная с 0:

$$I_{A,V}^{(0)} = 0, \tag{1.39}$$

$$I_{A,V}^{(n+1)} = I_{E,C}(d, I_{E,V}(d, I_{A,V}^{(n)})). \tag{1.40}$$

Процесс останавливается, когда априорная информация $I_{A,V}^{(n)}$ на некоторой иттерации получает значение близкое к 1. Номер итерации показывает нам какое число иттераций требуется для соответсвующего порога итеративного декодирования.

Задача оптимизации протографа заключается в выборе протографе и максимизации дисперсии шума (уменьшения величины Сигнал-Шум $\frac{E_b}{N_0}$) для соответствующего канала с требуемой величиной ошибки на бит в результате выполнения заданного числа итераций декодирования. При этом аналогично методу Эволюции плотностей осуществляется маргинализация на основе знания априорной функции распределения, заданной фактор графом.

Примеры наилучших среди известных протографов изображены на рисунках 1.18-1.22.

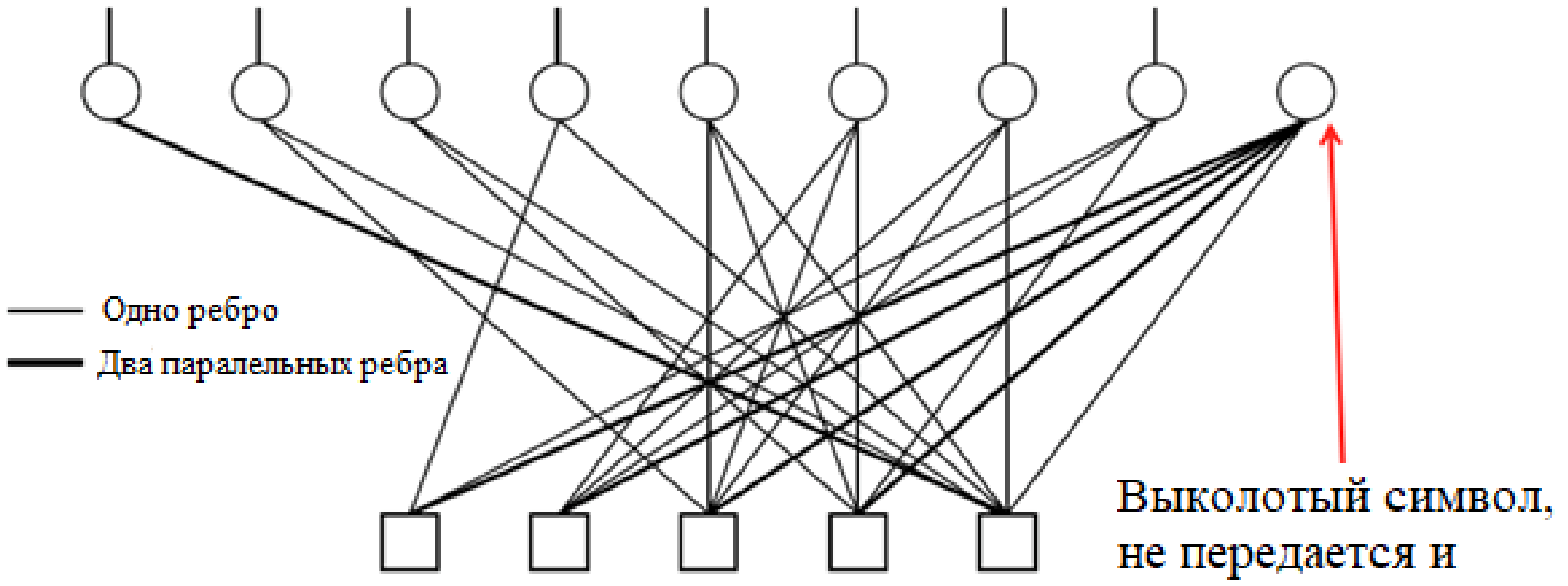

$$\text{Protograph}=\begin{pmatrix} 0 & 0 & 0 & 1 & 0 & 0 & 0 & 1 & 2 \\ 0 & 0 & 0 & 0 & 0 & 1 & 1 & 1 & 2 \\ 0 & 1 & 0 & 0 & 1 & 1 & 1 & 0 & 2 \\ 0 & 0 & 1 & 0 & 1 & 1 & 0 & 1 & 2 \\ 2 & 1 & 1 & 1 & 1 & 0 & 1 & 0 & 1 \end{pmatrix}$$

Параллельные ребра представляют сумму двух циркулянтов $I^{14}+I^{0}$

Рисунок 1.18 - Протограф Торпа, [27]

$$H=\begin{bmatrix} 2 & 0 & 1 & 1 & 0 & 1 & 0 & 1 & 1 & 1 & 1 & 0 & 1 & 0 & 0 & 0 \\ 4 & 2 & 0 & 1 & 0 & 1 & 0 & 1 & 1 & 0 & 0 & 0 & 1 & 0 & 0 & 0 \\ 3 & 0 & 0 & 0 & 0 & 0 & 2 & 0 & 0 & 0 & 1 & 0 & 0 & 1 & 0 & 0 \\ 3 & 2 & 1 & 0 & 2 & 0 & 0 & 0 & 0 & 0 & 0 & 0 & 0 & 1 & 0 & 0 \\ 3 & 0 & 0 & 0 & 0 & 0 & 0 & 0 & 0 & 1 & 0 & 1 & 0 & 0 & 0 & 1 \\ 3 & 0 & 0 & 1 & 0 & 1 & 0 & 1 & 0 & 0 & 0 & 0 & 0 & 0 & 0 & 1 \\ 3 & 0 & 0 & 0 & 1 & 0 & 1 & 0 & 0 & 0 & 0 & 1 & 0 & 0 & 1 & 0 \\ 3 & 4 & 1 & 0 & 0 & 0 & 0 & 0 & 1 & 0 & 0 & 0 & 0 & 0 & 1 & 0 \end{bmatrix}$$

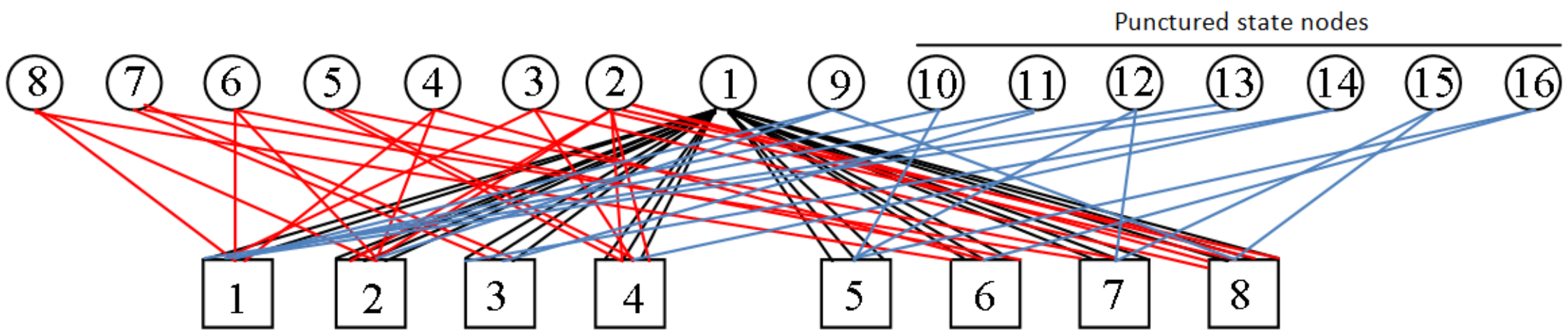

Рисунок 1.19 - Протограф $E^2RC$ –коды 10-16 символы выколотые, [41]

$$\begin{matrix} 1 & 2 & 3 & 4 & 5 & & \dots & & & & 11 \end{matrix}$$

$$H_{4/5} = \begin{pmatrix} 1 & 0 & 0 & 0 & 0 & 0 & 0 & 0 & 0 & 0 & 2 \\ 0 & 1 & 1 & 1 & 3 & 1 & 3 & 1 & 3 & 1 & 3 \\ 0 & 1 & 2 & 2 & 1 & 3 & 1 & 3 & 1 & 3 & 1 \end{pmatrix}$$

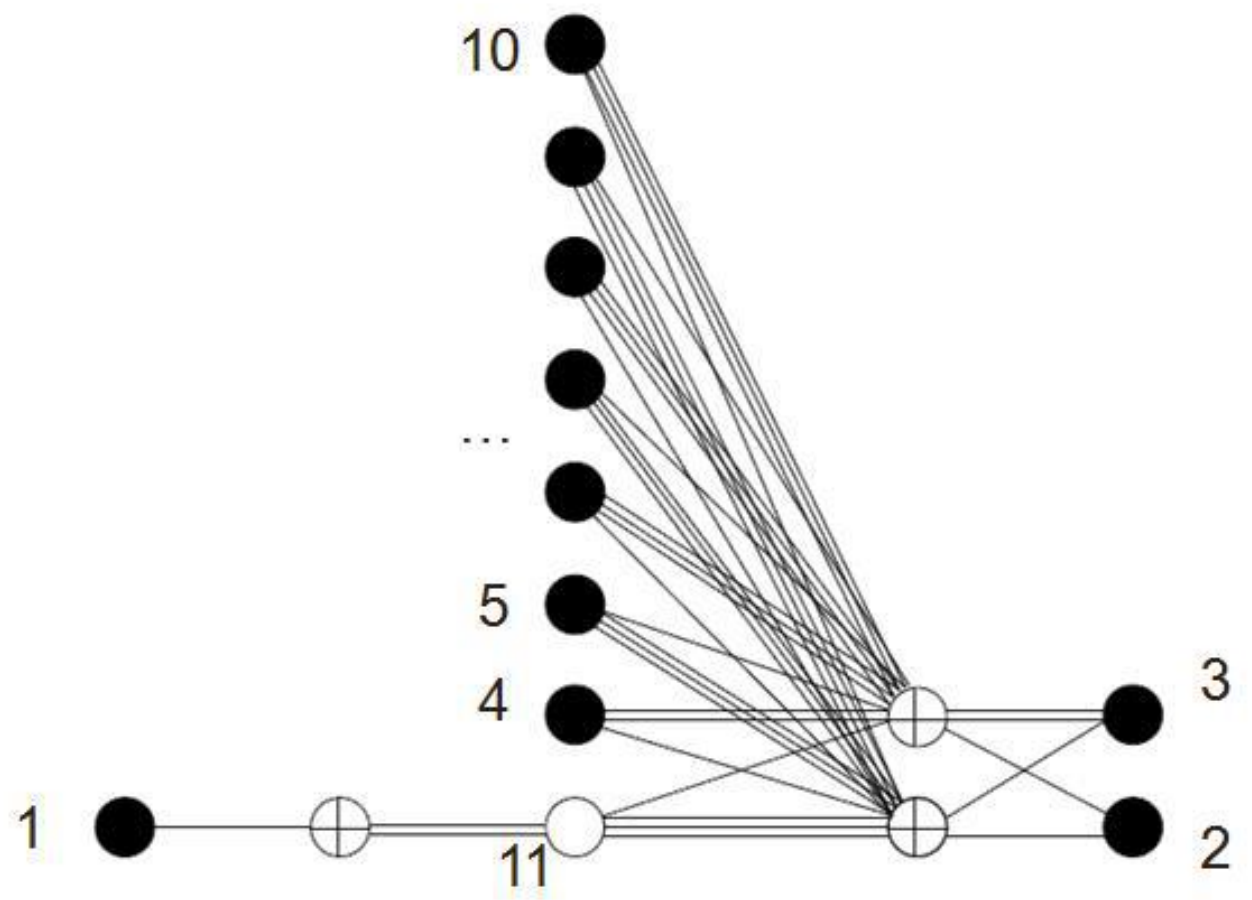

Рисунок 1.20 Протограф ансамбля AR4JA-кодов Дивсалара [39, 40]

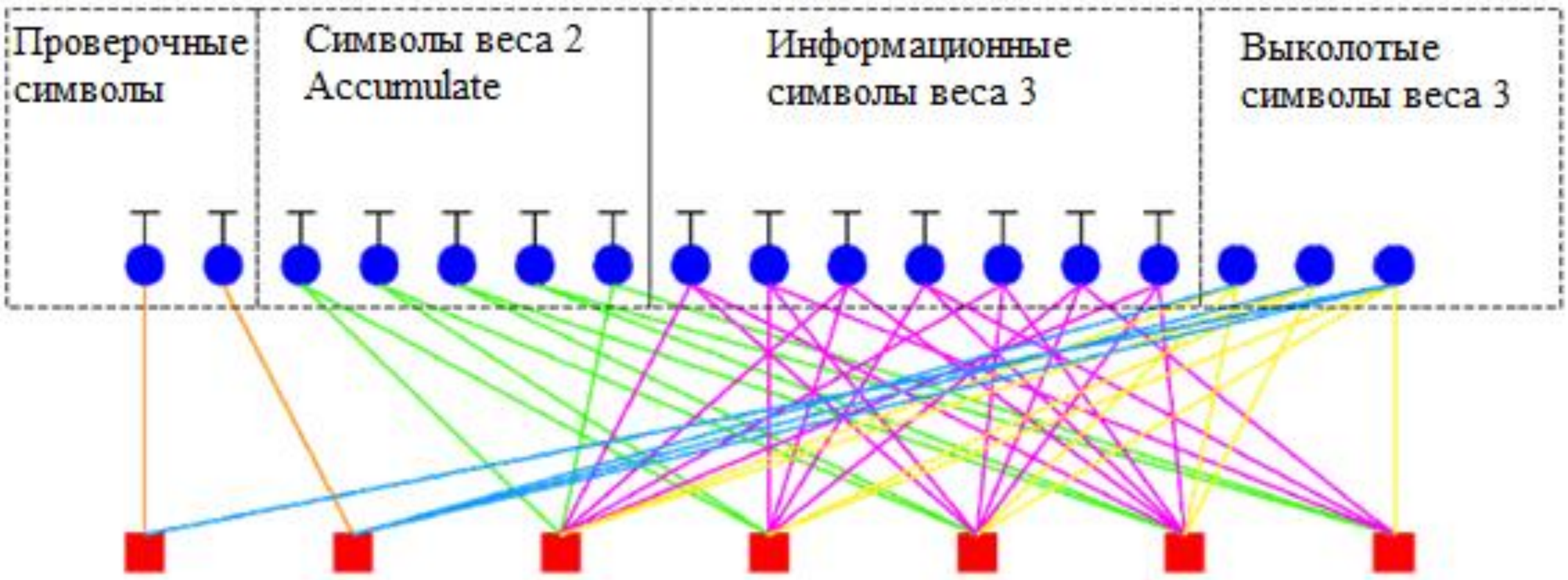

$$H = \begin{pmatrix} 1 & 0 & 0 & 0 & 0 & 0 & 0 & 0 & 0 & 0 & 0 & 0 & 0 & 0 & 1 & 0 & 1 \\ 0 & 1 & 0 & 0 & 0 & 0 & 0 & 0 & 0 & 0 & 0 & 0 & 0 & 0 & 0 & 1 & 0 \\ 0 & 0 & 1 & 0 & 0 & 0 & 1 & 1 & 0 & 1 & 0 & 1 & 0 & 1 & 0 & 1 & 0 \\ 0 & 0 & 1 & 1 & 0 & 0 & 0 & 0 & 1 & 1 & 1 & 0 & 1 & 0 & 0 & 0 & 0 \\ 0 & 0 & 0 & 1 & 1 & 0 & 0 & 1 & 1 & 0 & 0 & 1 & 1 & 1 & 1 & 0 & 1 \\ 0 & 0 & 0 & 0 & 1 & 1 & 0 & 1 & 0 & 1 & 1 & 1 & 0 & 1 & 1 & 1 & 0 \\ 0 & 0 & 0 & 0 & 0 & 1 & 1 & 0 & 1 & 0 & 1 & 0 & 1 & 0 & 0 & 0 & 1 \end{pmatrix}$$

Рисунок 1.21 - Протограф Ричардсона, [26]

Протограф, построенный предложенным в диссертации методом, с порогом итеративного декодирования 1.29 дБ изображен на рис. 1.22. Порог итеративного декодирования наилучшего из протографов такой скорости AR4JA протограф 1.414, рис. 1.22. Построенный нами ансамбль кодов, заданный данным протографом, находится в 0.23 дБ от предела Шэннона.

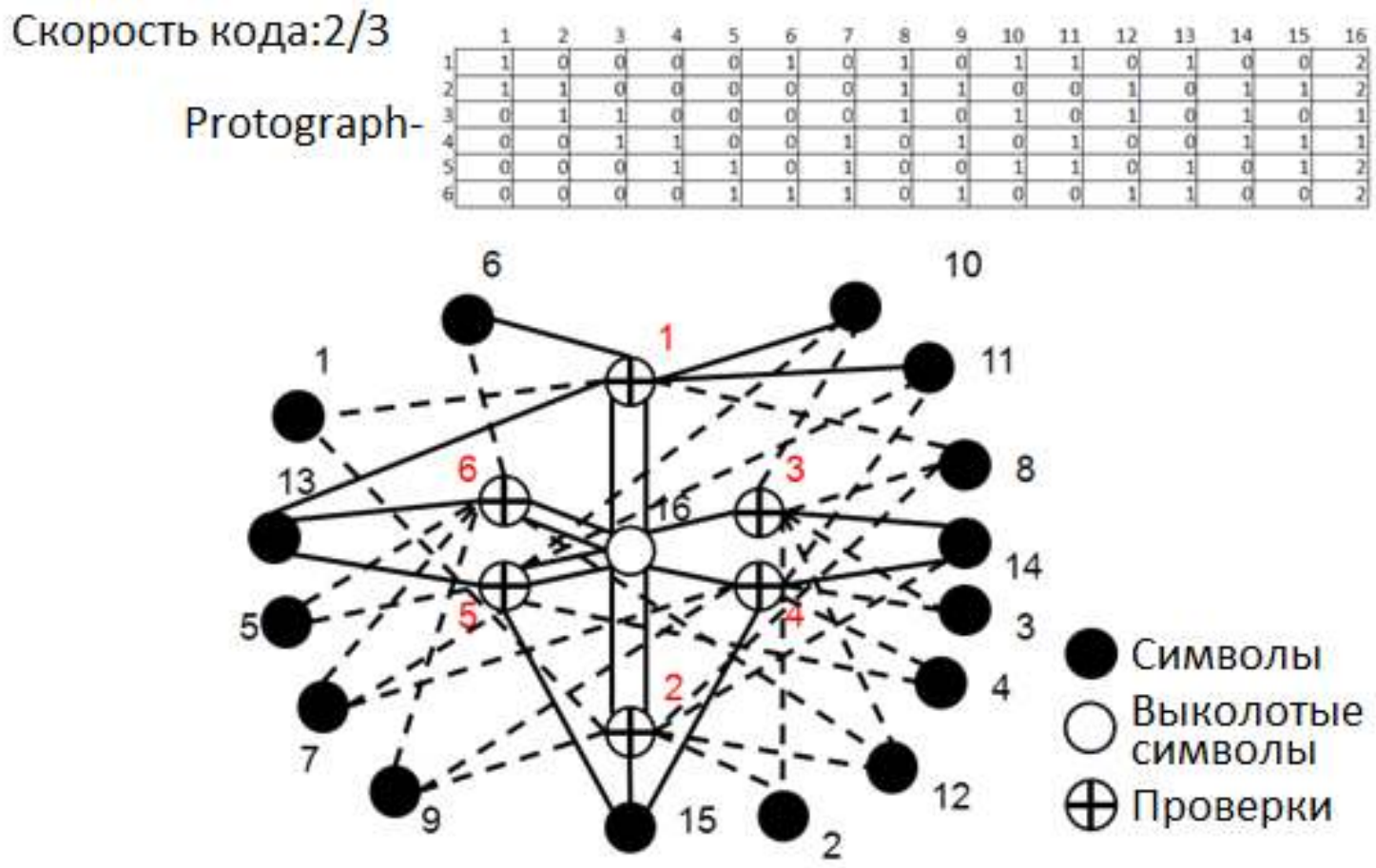

Рисунок 1.22 - Протограф и базовая матрица LDPC кода

Размерность базовой матрицы и размер циркулянта определяются исходя из имеющихся аппаратных ресурсов (вентилей и мультиплексоров) и требований к кривой помехоустойчивости итогового кода (оптимизация по «водопаду» или «полке»). Чем меньше размер циркулянта, тем легче оптимизировать кривую помехоустойчивости, но сложнее аппаратная реализация кода.

### 1.3.3 Методы расширения базовой проверочной матрицы

Расширение (lifting) базовой проверочной матрицы (протографа) заключается в замене единичных символов базовой матрицы на циркулянты размера $z \times z$. На этом этапе определяются значения циклических сдвигов всех циркулянт.

Для расширения базовой матрицы чаще всего применяются следующие методы: улучшенный PEG-алгоритм, Hill-Climbing, Guess-and-Test.

PEG-алгоритм (Progressive Edge-Growth), предложенный в [46], предусматривает последовательное добавление ребер в Таннер граф низкоплотностного кода на основе распределения весов символьных и проверочных узлов. Идея алгоритма заключается в последовательном добавлением проверок в матрицу низкоплотностного кода столбец за столбцом, начиная со столбца малого веса. Каждая из проверок добавляется таким образом, чтобы максимизировать обхват в Таннер-графе низкоплотностного кода. При этом осуществляется локальная оптимизация кратчайших циклов для каждого символьного узла.

Недостатком метода является отсутствие оптимизации дистантных свойств протографа и спектра связности.

Улучшенный PEG-алгоритм, предложенный в [47], позволяет избегать не детектируемые циклы длины 8 и 10.

В работе [45] исследователями из лаборатории Mitsubishi (Mitsubishi Electric Research Laboratories) был предложен метод лифтинга на основе алгоритма восхождения на вершину (Hill-Climbing). Hill-Climbing алгоритм аналогично методу отжига пытается прийти к глобальному оптимуму, итеративно изменяя значения циркулянта Таннер-графа и пересчитывая функции цели (температура, число циклов).

Алгоритм поиска восхождением к вершине находит такую комбинацию параметров, при которой значение обхвата максимально. Поиск начинается из случайной точки. Перечень новых состояний, доступных из текущего, строится путём изменения каждого из параметров по очереди. То есть каждое следующее состояние отличается от предыдущего только в одном параметре. Следующим состоянием из списка доступных выбирается так, что обхват максимален. Алгоритм останавливается, когда нет таких соседних состояний, которые имеют значение целевой функции большее, чем у текущего узла.

Этот алгоритм позволяет находить графы с максимальным возможным обхватом и минимальным размером циркулянта. Он обладает существенно большей вычислительной сложностью, чем PEG. Кроме того, максимизируя обхват, он не оптимизирует спектр связности.

Hill-Climbing алгоритм применялся для построения низкоплотностных кодов, предназначенных для оптической связи и до появления улучшенного PEG-алгоритма, считался наилучшим из известных алгоритмов с точки зрение способности разрывать циклы в графе.

Guess-and-Test использует случайные броски значений сдвигов циркулянта и проверяет полученный граф на обхват в соответствии с уравнением:

$$\sum_{k=0}^{m-1} p_{j_{k+1},l_{k+1}} - p_{j_k,l_k} = 0 \bmod z\,, \qquad (1.41)$$

где $2 \le m \le g/2$, $p_{j_k,l_k}$ - значения сдвигов циркулянт, входящих в цикл. Guess-and-Test метод требует большого числа испытаний и не гарантирует хороших дистантных свойств кода и хорошей связности графа. Это самый простой метод построения графа, однако, он требует большого числа испытаний и не гарантирует хороших дистантных свойств кода и связности графа.

Коротко говоря, процедура построения квазициклических LDPC-кодов предусматривает выполнение 2-х этапов:

1) выбор протографа (базовой матрицы);

2) расширение базовой проверочной матрицы.

Расширение базовой проверочной матрицы заключается в замене единичных символов базовой матрицы на циркулянты размера $z$. На этом этапе определяются значения циклических сдвигов всех циркулянт. Для реализации этапов построения LDPC-кодов используются комбинаторно-алгебраические методы и методы статистической физики. Сравнение методов представлено в табл. 1.2.

Таблица 1.2 – Классификация методов построения низкоплотностных кодов

| Классы методов построения кодов | Методы | Недостатки методов |
|---|---|---|
| Комбинаторно-алгебраические методы: | конечные геометрии, блочные дизайны, числовые сетки, слова малого веса кодов Рида-Соломона, метод Саливана (O’Sullivan2006) | Строит коды с фиксированными параметрами (длина, скорость). Затруднено построение нерегулярных кодов. |
| Методы статистической физики | Density Evolution (Эволюция плотностей), Covariance Evolution (Эволюция ковариации), PEXIT-chart | Игнорирование комбинаторно-алгебраических свойств Таннер-графа и дистантных свойств кода |

Недостатки перечисленных методов для построения низкоплотностных кодов средней длины (32000 бит), которые используются в голографической памяти, делают актуальной задачу построения кодов с повышенной корректирующей способностью для голографической памяти.

### 1.4 Выводы

В связи с интенсивным ростом объема информации, хранимой в электронном виде, в настоящее время актуальными являются вопросы развития систем голографического архивного хранения данных. Достоинствами голографического метода хранения информации является: высокая плотность информации на квадратный дюйм; низкое потребление энергии, произвольный доступ к любому из секторов носителя; низкая стоимость носителей информации.

Особенностью канала считывания информации голографического носителя является группирование ошибок и высокий уровень вероятности ошибки на бит $5 \cdot 10^{-2}$. Для достижения требуемого для архивной памяти уровня вероятности ошибки на бит $10^{-8} - 10^{-10}$ в настоящее время в голографической памяти используются помехоустойчивые низкоплотностные коды.

Высокая помехоустойчивость и относительная простота реализации декодеров обусловили применение квазициклических LDPC-кодов в архивной голографической памяти. Квазициклические LDPC-коды применяются на нижнем страничном уровне трехуровневой системы помехоустойчивого кодирования. Страничный уровень работает в условиях больших шумом, поэтому для коррекции используется LDPC-код длиной 32768 со скоростью 0.5, содержащий 32 битный код контроля на четность. Страничный уровень один из ключевых уровней, определяющих плотность записи архивной голографической памяти. Все последующие уровни, использующие коды Рида-Соломона и проверку на четность, предназначены для обнаружения ошибок и исправления стираний.

Эффективность коррекции ошибок LDPC-кодом, применяемым в архивной голографической памяти, показывает отставание в 1 дБ от предельного значения, заданного границей Полянского.

Дальнейшее улучшение помехоустойчивости квазициклического LDPC-кода на страничном уровне требует улучшение методов построения квазициклических LDPC-кодов.

Процедура построения квазициклических LDPC-кодов предусматривает выполнение 2-х этапов: 1) выбор протографа (базовой матрицы), 2) расширение базовой проверочной матрицы. Расширение (lifting) базовой проверочной матрицы заключается в замене единичных символов базовой матрицы на циркулянты.

На практике для расширения базовой матрицы чаще всего применяются следующие методы: улучшенный PEG-алгоритм, Hill-Climbing, Guess-and-Test. Однако они направлены на улучшение обхвата Таннер-графа LDPC-кода, простейшей и самой грубой характеристики спектра связности LDPC-кода, без улучшения его дистантных свойств. Недостатки перечисленных методов для построения низкоплотностных кодов средней длины (32000 бит), которые используются в голографической памяти, делают актуальной задачу построения кодов с повышенной корректирующей способностью для голографической памяти.

При построении низкоплотностных кодов необходимо выполнять совместную оптимизацию дистантных свойств кода (кодового расстояния) и свойств графа (спектров связности).

Недостатки перечисленных методов для построения низкоплотностных кодов средней длины (32000 бит), которые используются в голографической памяти, делают актуальной задачу построения кодов с повышенной корректирующей способностью для голографической памяти.

# 2 СОЗДАНИЕ МЕТОДА ПОСТРОЕНИЯ НИЗКОПЛОТНОСТНЫХ КОДОВ ДЛЯ НАКОПИТЕЛЕЙ АРХИВНОЙ ГОЛОГРАФИЧЕСКОЙ ПАМЯТИ

## 2.1 Метод построения квазициклических низкоплотностных кодов

В диссертации предлагается метод построения низкоплотностных кодов, состоящий из двух фаз: построения и расширения протографа. Особенностью метода является комбинирование жадного алгоритма запрещенных коэффициентов и стохастического алгоритма отжига. Это позволяет улучшить дистантные свойства кодов и их спектры связности для фильтрации кодов кандидатов и обеспечивает повышение надежности считывания информации в голографической памяти.

Входными данными метода построения низкоплотностного кода являются: требуемая длина кода $\boldsymbol{N}$, минимальный размер циркулянта $\boldsymbol{z_{min}}$ , $\boldsymbol{z}$ – текущий размер циркулянта, параметры алгоритма декодирования, вероятность ошибки на бит на выходе декодера $\boldsymbol{P_{BER}}$ при требуемом отношении сигнал-шум **SNR,** размерность начальной базовой проверочной матрицы (протографа) $\boldsymbol{J} \times \boldsymbol{L}$; битовая маска M размером $\boldsymbol{J} \times \boldsymbol{L}$**,** определяющая подлежащие инвертированию значения базовой матрицы; число кодов кандидатов card и число итераций $\boldsymbol{iteration}$ процедуры построения низкоплотностного кода.

Выходные данные метода построения низкоплотностных кодов: массив кодов кандидатов $\{C'\}$. В результате работы метода определяется множество аппаратно-ориентированных кодов кандидатов с длиной $\boldsymbol{N} = \boldsymbol{L} \times \boldsymbol{z}$ и скоростью $R = 1 - dv_{avg} / dc_{avg}$, где $dv_{avg}$, $dc_{avg}$ средний вес столбца (число единиц в столбце) и строки в проверочной матрице, соответственно.

Процедура поиска протографа предусматривает выполнение следующих этапов:

1. На основе вычисленного распределения весов строк и столбцов при помощи метода Эволюции плотностей (Density Evolution) осуществляется инициализация базовой матрицы. Метод рассматривался в разделе 1.3.2. Выполняется инициализация множества кодов кандидатов пустым множеством $\{C'\} = \emptyset$.

Фаза оптимизации протографа:

2. Случайным образом в позициях битовой маски $M_{j,l} = rand(0,1)$ инвертируются значения базовой матрицы, и вычисляется порог итеративного декодирования $Threshold = Eb/No(\sigma)\colon I_{A,V}^{(n)} \approx 1, \sigma = \frac{1}{\sqrt{4R}} * 10^{-\frac{Eb/No}{20}}$, получаемый при помощи метода статистической физики PEXIT-chart. Метод рассматривался в разделе 1.3.2 и будет приведен кратко далее по тексту.

3. Если в величину $P_{BER}$ основной вклад будет вносить кодовое расстояние треппин-сета минимального веса, для его оценки применяется обобщение неравенства Буля (Union Bound), $P_{TS}$, приведенное далее в методе выборки по значимости (Importance Sampling). Если в величину $P_{BER}$ основной вклад будет вносить кодовое расстояние треппин-сета TS(a,0), ( $d_{\min} = a$), то верхняя оценка вероятности ошибочного декодирования $P_{UB}$:

$$P_{UB} \approx \frac{\omega_{d_{min}K}}{N} Q\left(\sqrt{\frac{d_{min}K}{N} 2E_b/N_0}\right), \tag{2.1}$$

где $d_{\min}$ -кодовое расстояние, $\omega_{d\ \min}$ - кратность слов минимального веса, $Q(x)$ - Q-функция $Q(x) = (\pi N_0)^{-1/2} \int_x^{\infty} e^{-n^2/N_0} dn$, $K$- информационная длина кода, $N$ – кодовая длина кода, $E_b/N_0$ –отношение сигнал-шум, $E_b/N_0 = SNR - 10log_{10}(R)$, R – скорость кода.

4. Шаг проверки полученной вероятности ошибочного декодирования

4.1 Если $P_{UB} \leq P_{BER}$& $Treshold \leq SNR$ **,** переход к п.5.

4.2 Если ($P_{UB} > P_{BER}$) & $Treshold > SNR$ & ($z > z_{min}$), увеличивается размер протографа $\boldsymbol{J} \times \boldsymbol{L}$ , уменьшается значение z, переход к п.2.

4.3 Если ($P_{UB} > P_{BER}$) & $Treshold > SNR$ & ($z < z_{min}$), останов.

Фаза оптимизации протографа необходима для получения базовой матрицы, удовлетворяющей требуемой вероятности ошибки при заданном отношение сигнал-шум SNR.

Фаза расширения протографа:

5. Расширяется протограф жадным методом запрещённых коэффициентов, накапливается требуемое число кодов кандидатов в множестве {C'}.
6. Вычисляется уточненное кодовое расстояние кодов кандидатов, применяя метод оценки кодового расстояния, подробно рассмотренный в следующей главе. В множестве {C'} остаются коды с достаточным кодовым расстоянием $P_{UB} \leq P_{BER}$:

$$P_{UB} \approx \sum_{i=d_{min}}^{N} \omega_i/KQ\left(\sqrt{\frac{iK}{N} 2E_b/N_0}\right), \tag{2.2}$$

где $\omega_i$ - кратность слов веса $i$, $i$ -вес кодового слова, $Q(x)$ - Q-функция, $K$ - информационная длина кода, $N$ – кодовая длина кода, $E_b/N_0$- отношение сигнал-шум.

7. Для оставшихся в множестве $\{C'\}$ кодах осуществляется поиск пересечения коротких циклов, которые потенциально содержат малое число невыполненных проверок – треппин сетов. Поиск треппин-сетов осуществляется методом выборки по значимости Коула (Cole's Importance Sampling). Вероятность ошибки на бит в символьных узлах треппин-сета вычисляется по формуле:

$$P_{TS_{BER}} = Q\left(\frac{2m_\lambda + 2\sum_{j=1}^{iter}\frac{m_{\lambda^{ext}}^{(j)} + m_\lambda}{\mu_{max}^{j}}}{\sqrt{\left(1+\sum_{j=1}^{iter}\frac{1}{\mu_{max}^{j}}\right)m_\lambda + \sum_{j=1}^{iter}\frac{m_{\lambda^{ext}}^{(j)}}{\mu_{max}^{j}}}}\right), \quad (2.3)$$

где $\mu_{max}^{j}$ - спектральный параметр роста логарифмов коэффициэнтов правдоподобия на $j$ итерации алгоритма распространения доверия, показывает, насколько логарифмы правдоподобия $\lambda$ растут быстрее в треппин-сете по отношению к оставшемуся графу; $Q(x)$ - $Q-$функция; $m_\lambda$ начальное значение логарифмов правдоподобия в символьных узлах, полученных из канала, $m_{\lambda^{ext}}^{(j)}$ - сообщение в методе распространения доверия на $j$ итерации, вычисляемое при помощи Эволюции плотностей или ее аппроксимаций, например $m_{\lambda^{ext}}^{(j)} = \phi^{-1}\left(1 - \left[1-\phi\left(m_\lambda + (d_v-1)m_{\lambda^{ext}}^{(j-1)}\right)\right]^{d_c-1}\right)$, $\lambda(\lambda^{ext})$ - логарифмы правдоподобия в символьных узлах из канала (внешние логарифмы правдоподобия из декодера на $j-1$ итерации, extrinsic information), $d_c$ -вес проверочного узла, $d_v$ -вес символьного узла, $\phi$ - функция проверочного узла в Эволюции плотностей или ее приближении (Reciprocal Channel Approximation, RCA Gaussian Approximation). Например, для аппроксимации Гауссианами получим:

$$\phi(x) = \begin{cases} \exp\left(-0.4527x^{0.86} + 0.0218\right), \text{ для } 0 \le x < 10 \\ \sqrt{\frac{\pi}{x}}\exp(-0.25x)\left(1-\frac{10}{7x}\right), \text{для } x \ge 10 \end{cases}. \quad (2.4)$$

*8.* Для каждого кода кандидата $c \in \{C'\}$ вычисляется значение штрафа $Penalte$ от $Treshold$ порога итеративного декодирования (PEXIT chart). Штраф обусловлен конечной длиной LDPC-кода и рассчитывается по формуле: $Penalte = Treshold - P_{watterfall}$,

$$P_{waterfall}(N,\sigma) \cong Q\left(\frac{\sqrt{N}\left(C(\sigma)-C(\sigma^*)-\beta N^{-\frac{2}{3}}\right)}{\alpha}\right)+O\left(N^{-\frac{1}{3}}\right), \quad (2.5)$$

где $N$ - длина кода, $\alpha$ - масштабирующий коэффициент, $\beta$ - коэффициент сдвига, $\sigma^*$ -порог итеративного декодирования (среднеквадратичное отклонение), $\sigma$ - среднеквадратичное отклонение в АБГШ-канале, $C(\sigma)$ - пропускная способность канала, соответствующая мощности шума $\sigma$, $Q(x)$ - $Q$ −функция. Коэффициэнты $\alpha$ и $\beta$ вычисляются путем решения системы дифференциальных уравнений в методе Эволюции Ковариации (Covariance Evolution). По соотношению штрафа *Penalte* и порога $Treshold$ упорядочиваются коды кандидаты согласно выражению, $C \subset \{C'\}$:

$$c_1 \prec c_2 : c_1(Threshold + Penalte)_{P_{BER}} < c_2(Threshold + Penalte)_{P_{BER}}, c_1, c_2 \in C. \quad (2.6)$$

В случае, одинаковых штрафов, упорядочивание происходит по дополнительным параметрам, сложность аппаратной имплементации. При полной эквивалентности один из кодов выбирается случайно.

9. При помощи метода имитации отжига модифицируются коды кандидаты из множества $\{C'\}$ для улучшения спектра связности (при помощи ACE spectrum или EMD) графа Таннера, [8].

10. Если число итераций процедуры не превысило требуемую величину и число кодов кандидатов меньше заданной величины $|\{C'\}| < \text{card}$: осуществляется переход к шагу 2, иначе завершается работа алгоритма.

Введенная фаза расширения протографа позволяет получить проверочные матрицы квазициклических низкоплотностных кодов, ориентированные на аппаратную реализации перестановочной сети при помощи управляемых сдвигов.

Шаги 5 и 9 направлены на получение уточнённой оценки порога итеративного декодирования с учетом особенностей расширения протографа и реализуются оригинальными алгоритмами А и Б, соответственно.

Для нахождения порога итеративного декодирования $Treshold$ (этап 2) перебираются возможные значения порога с их последующей оценкой при помощи метода эволюции взаимной информации (PEXIT-chart) [35].

Метод PEXIT-chart предусматривает обмен взаимной информации между символьными и проверочными вершинами графа Таннера: $I_{E,V}$ и $I_{E,C}$.

Апостериорная информация (внешняя) $I_{E,V}$ символьных узлов вычисляется в соответствие с уравнениями:

$$I_{E,V} = J(\sigma) = J\left(\sqrt{(d_v - 1)\sigma_A^2 + \sigma_{ch}^2}\right), \quad (2.7)$$

$$J(\sigma) = 1 - \int_{-\infty}^{\infty} \frac{1}{\sqrt{2\pi}\sigma} e^{-(l-\frac{\sigma^2}{2})^2/2\sigma^2} \log(1 + e^{-l})dl, \quad (2.8)$$

где $J(\sigma)$ – функция взаимной информации, $\sigma_A^2$ и $\sigma_{ch}^2$ -априорная дисперсия и дисперсия шума.

Априорная информация $I_{A,V}$ символьных узлов вычисляется в соответствие с уравнением:

$$I_{A,V} = J(\sigma_A), I_{E,V} = J(\sigma) = J\left(\sqrt{(d_v - 1)\left[J^{-1}(I_{A,V})\right]^2 + \sigma_{ch}^2}\right) \quad (2.9)$$

Апостериорная информация (внешняя) $I_{E,C}$ проверочных узлов (факторов) вычисляется в соответствие с уравнением:

$$I_{E,C} = 1 - J\left(\sqrt{(d_c - 1)\left[J^{-1}(1 - I_{A,C})\right]^2}\right). \quad (2.10)$$

Веса символьных и проверочных узлов графа Таннера задаются в виде полиномов распределения весов $\lambda(z) = \sum_{d=1}^{d_v} \lambda_d z^{d-1}$ и $\rho(z) = \sum_{d=1}^{d_c} \rho_d z^{d-1}$, где $\lambda_d$ доля проверок веса $d$ в столбце проверочной матрицы, и $\rho_d$ доля проверок веса $d$ в строке.

Для нерегулярных кодов

$$I_{E,V} = \sum_{d=1}^{d_v} \lambda_d I_{E,V}(d, I_{A,V}), I_{E,C} = \sum_{d=1}^{d_v} \rho_d I_{E,C}(d, I_{A,C}). \quad (2.11)$$

Вычисление функции взаимной информации $J(\cdot)$ реализуется путем вычисления значений аппроксимирующего полинома либо путем выбора из таблицы предварительно вычисленных значений функции (LUT).

Путем вычисления функций $I_{E,V}(d, I_{A,V})$ и $I_{E,C}(d, I_{A,C})$ выполняется итеративный процесс эволюции взаимной информации начиная с нулевого значения $I_{A,V}^{(0)}$:

$$I_{A,V}^{(0)} = 0, I_{A,V}^{(n+1)} = I_{E,C}(d, I_{E,V}(d, I_{A,V}^{(n)})). \qquad (2.12)$$

Процесс останавливается, когда априорная информация $I_{A,V}^{(n)}$ на некоторой итерации принимает значение близкое к 1. Номер итерации показывает нам какое число итераций требуется для соответствующего порога итеративного декодирования.

Метод "Эволюции плотностей" (Density Evolution) используется на 2-м этапе процедуры построения LDPC-кода следующим образом.

Пусть $f_0$ - функция плотности вероятностей, заданная логарифмами правдоподобия, $\sigma_+$ - дисперсия ошибки, соответствующая отношению сигнал/шум; $I_\lambda$ - максимальный вес символьного узла, $I_p$ - максимальный вес проверочного узла, $\varepsilon$ - требуемая вероятность ошибки на бит, $i_{max}$ - число итераций в алгоритме распространения доверия.

Проинициализируем $f_{c\to v} = f_0$. В цикле по $i$ вычисляем $f^{(i-1)\otimes}$ свертку функции плотностей вероятностей:

$$f_{c\to v} = \sum_{i=2}^{i_{max}} p_i f^{(i-1)\otimes}, f_{v\to c} = \left(\sum_{i=2}^{i_{max}} \lambda_i f^{(i-1)\otimes}\right) \times f_0. \qquad (2.13)$$

Если вероятность ошибки на бит:

$$P_e = \int_{-\infty}^{0} f_{vc}(x)dx < \varepsilon \qquad (2.14)$$

не превосходит требуемую, тогда прекращаем работу алгоритма и выдаем $\sigma_+$.

Этапы 2 и 3 обеспечивают модификацию протографа, если вероятность ошибочного декодирования, обусловленная дистантными свойствами, выше заданной $P_{UB}$.

На этапе 3 для оценки кодового расстояния протографа применяется верхняя оценка кодового расстояния квазициклических кодов Маккея-Вонтобеля-Смарандаши-Сигеля-Батлера по протографу (см. раздел 3.1.2).

На 6-м этапе поиск и взвешивание треппин-сетов может выполняться при помощи решения задачи линейного программирования, [92-93]:

$$\begin{gathered} \min \lambda^T x\,, \\ H \otimes x = 0 \end{gathered} \tag{2.15}$$

где $H-$ проверочная матрица, $x-$ кодовое слово LDPC-кода, $\lambda-$ вектор логарифмов правдоподобия символов кодового слова, полученный в результате применения метода Монте-Карло, $H \otimes x = 0$ – условия целочисленной программы Фелдмана, (Fundamental Cone, [53]). Пример решающей области представлен на Рис. 2.1.

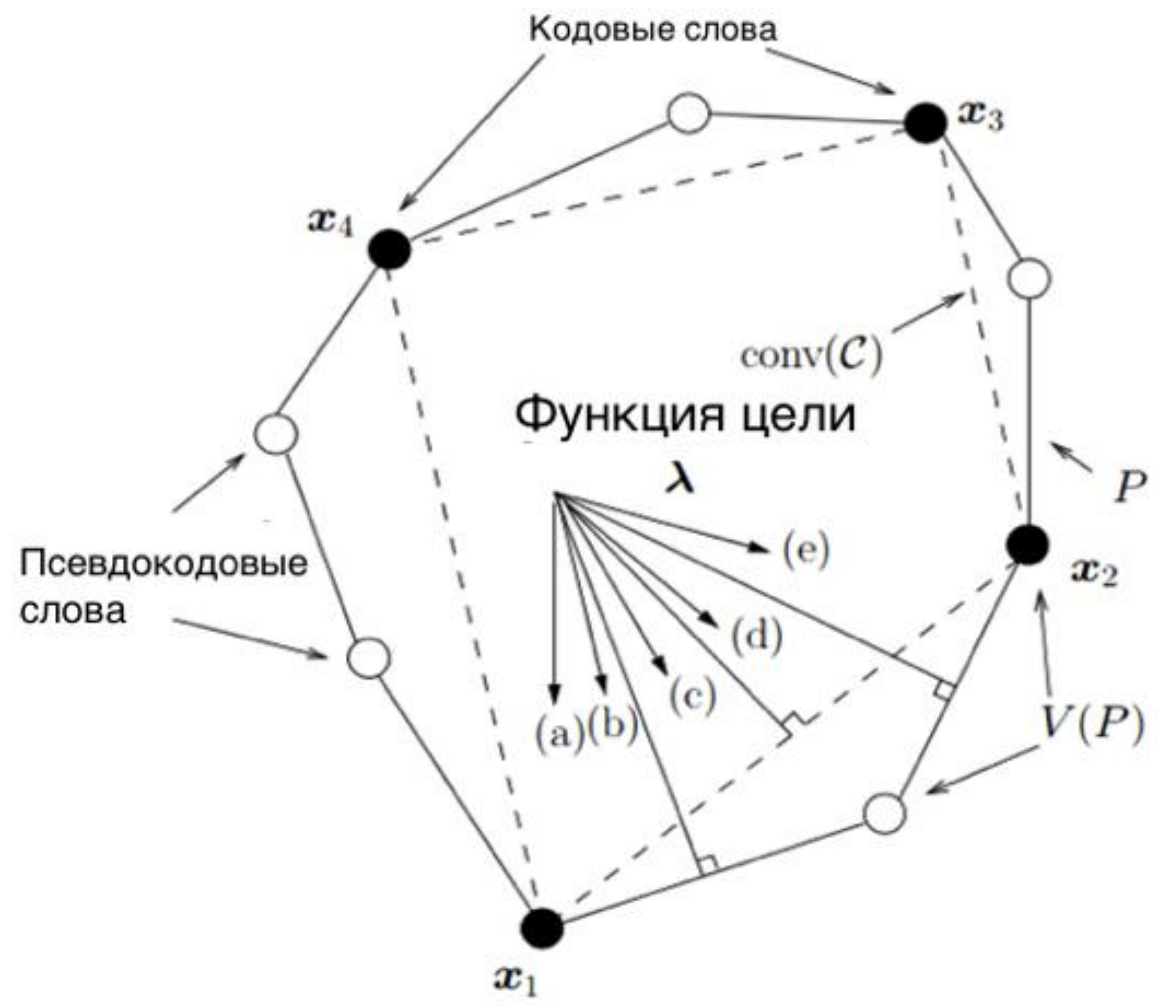

Рисунок 2.1 Пример решающей области для поиска треппин-сета

Одним из способов поиска треппин-сетов по проверочной матрице LDPC-код $H^{m \times n}$ с Таннер-графом $G = (V \cup C, E)$ осуществляется путем решения целочисленной программы с $n + 2m$ переменных и m+1 условиями:

$$min \sum_{v_i \in V} x_i : \sum_{c_j \in C} I_j^{odd} = b, \tag{2.16}$$

$$2\alpha_j + I_j^{odd} - \sum_{(v_i, c_j) \in E} x_i = 0,\ \forall j \in \{1,2, \dots, m\}, \tag{2.17}$$

$$x_i, I_j^{odd} \in \{0,1\}, \alpha_k \in \left\{0, \dots, \left\lfloor \frac{\rho_{max}}{2} \right\rfloor\right\}, \tag{2.18}$$

$$\forall k \in \{1,2,\dots,m\}, \forall i \in \{1,2,\dots,n\}, \forall j \in \{1,2,\dots,m\}, \tag{2.19}$$

где $b_{\max}$ - задает максимальный параметр нечетной степени инцидентности $TS(a,b): b < b_{\max}$ , $\rho_{\max}$ наибольшая степень инцидентности проверки в TS-подграфе (вес строки), $I_j^{odd}$- фиктивная переменная равная 1, если символьный узел инцидентен нечетному число проверок и 0 в противном случае, вспомогательная переменная $\alpha_j$ для проверки условия степени инцидентности фиктивной переменной. Такая программа является применением метода линейного программирования для поиска треппин-сетов.

Пример 1. Проиллюстрируем решение этой задачи на примере для кода (4,1) с проверочной матрицей

$$H = \begin{pmatrix} 0 & 1 & 0 & 1 \\ 0 & 0 & 1 & 1 \\ 1 & 1 & 1 & 1 \end{pmatrix}.$$

По определению проверочной матрицы выполняется

$$H \otimes x = \begin{pmatrix} 0 & 1 & 0 & 1 \\ 0 & 0 & 1 & 1 \\ 1 & 1 & 1 & 1 \end{pmatrix} \otimes x = \begin{pmatrix} x_2 \oplus x_4 \\ x_3 \oplus x_4 \\ x_1 \oplus x_2 \oplus x_3 \oplus x_4 \end{pmatrix} = 0.$$

Рассматривая каждую из строк матрицы $H$, мы будем строить политоп - набор неравенств, состоящий из нечетного числа символьных вершин Таннер-графа.

Для построения политопа используется следующая формула [92-93]:

$$\sum_{t \in V} x'_t - \sum_{t \in h_i / V} x'_t \leq |V| - 1, \tag{2.20}$$

где $x'_t$ -символ кодового слова, входящий в политоп, $V$ - подмножество из нечетного числа символов, соответствующих еденицам строки проверочной матрицы $h_i$.

Координата символа (вершины) в проверке, содержащейся в текущем подмножестве проверочной строки, равна +1, не содержащейся в текущем подмножестве проверочной строки равна -1. Для символов, отсутствующих в текущей строке $h_i$, значение равно 0.

Для первой строки проверочной матрицы, задающей уравнение $x_2 \oplus x_4 = 0$, политоп задается двумя неравенствами. Одно из которых соответствует $V=\{x'_2\}$, второе - $V=\{x'_4\}$:

$$\begin{pmatrix} +1 & -1 \\ -1 & +1 \end{pmatrix}\begin{pmatrix} x'_2 \\ x'_4 \end{pmatrix} \le \begin{pmatrix} 1-1 \\ 1-1 \end{pmatrix} \begin{matrix} V=\{x'_2\} \\ V=\{x'_4\} \end{matrix}.$$

Для второй строки, задающей уравнение $x_2 \oplus x_4 = 0$, политоп также задается двумя неравенствами. Одно из которых соответствует $V=\{x'_3\}$, второе - $V=\{x'_4\}$:

$$\begin{pmatrix} +1 & -1 \\ -1 & +1 \end{pmatrix}\begin{pmatrix} x'_3 \\ x'_4 \end{pmatrix} \le \begin{pmatrix} 1-1 \\ 1-1 \end{pmatrix} \begin{matrix} V=\{x'_3\} \\ V=\{x'_4\} \end{matrix}.$$

И третьей строке соответсвует политоп, состоящий из 8 неравенств, которые соответствуют значениям $|V|$, равным 1 или 3:

$$\begin{pmatrix} +1 & -1 & -1 & -1 \\ -1 & +1 & -1 & -1 \\ -1 & -1 & +1 & -1 \\ -1 & -1 & -1 & +1 \\ +1 & +1 & +1 & -1 \\ +1 & +1 & -1 & +1 \\ +1 & -1 & +1 & +1 \\ -1 & +1 & +1 & +1 \end{pmatrix}\begin{pmatrix} x'_1 \\ x'_2 \\ x'_3 \\ x'_4 \end{pmatrix} \le \begin{pmatrix} 1-1 \\ 1-1 \\ 1-1 \\ 1-1 \\ 3-1 \\ 3-1 \\ 3-1 \\ 3-1 \end{pmatrix} \begin{matrix} V=\{x'_1\} \\ V=\{x'_2\} \\ V=\{x'_3\} \\ V=\{x'_4\} \\ V=\{x'_1,x'_2,x'_3\} \\ V=\{x'_1,x'_2,x'_4\} \\ V=\{x'_1,x'_3,x'_4\} \\ V=\{x'_2,x'_3,x'_4\}, \end{matrix}$$

В результате решения системы неравенств линейной модели получим один треппин-сет TS(3,1), состоящий из трех символов $x_2, x_3, x_4$, веса 3.000549, см. Рис 2.2.

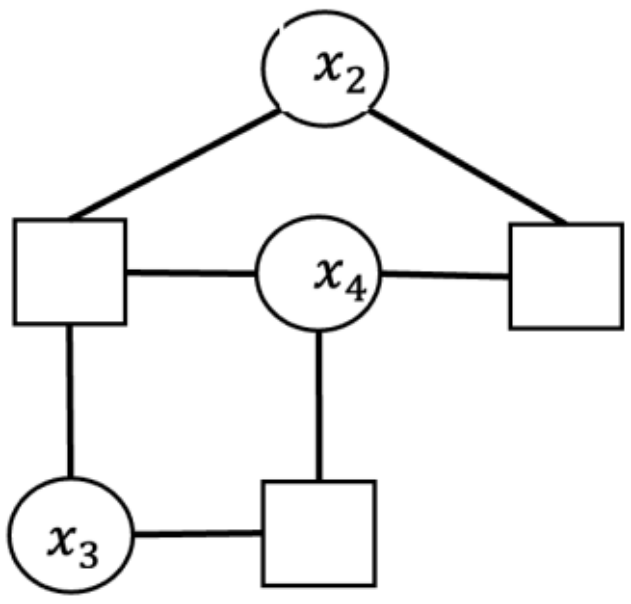

Рисунок 2.2- «Трэппинг-сет» TS (3,1) веса $\omega_1$ =3.000549

Пример 2. Ниже в таблице приведены псевдокодовые слова, принадлежащие 17 классам подграфов треппин-сетов, с весом псевдокодового слова $\omega \le 21$ для квазициклического низкоплотностного кода (Приложение 3) с базовой матрицей 4 на 36 и циркулянтом 278 в случае его декодирования шестью итерациями методом распространения доверия.

Таблица 2.1– Таблица треппин-сетов для квазициклического низкоплотностного кода 4х36

| TS($a$, $b$) | Вес псевдокодового слова, $\omega(p)$ | Положение символьных узлов в графе, образующих подграф Треппин-сэта |
|---|---|---|
| (6, 6) | [10.713756] | 671 835 1245 1652 1732 4744 |
| (6, 6) | [11.861630] | 1096 1228 1715 4727 6537 9033 |
| (6, 6) | [12.367398] | 1989 4241 6901 7319 8341 8830 |
| (7, 6) | [12.858173] | 1114 1391 1698 2140 2272 2679 3501 |
| (8, 8) | [14.304589] | 1228 1505 1812 1976 2386 3337 4836 8357 |
| (6, 6) | [14.496057] | 204 2462 3721 5496 6358 7229 |
| (7, 6) | [15.401395] | 2118 2488 2724 3013 3059 3614 3643 |
| (8, 8) | [15.503993] | 423 749 879 2102 4772 6545 7488 9175 |
| (10, 10) | [16.226536] | 296 706 835 1193 1411 2724 5573 7980 8593 9378 |
| (8, 8) | [16.284406] | 725 1782 4171 4820 6364 6404 6681 7214 |
| (8, 8) | [16.485658] | 255 3870 5839 6323 6578 6896 7045 9140 |
| (6, 6) | [16.749574] | 3390 4860 5193 5554 5561 5925 |
| (9, 8) | [17.552285] | 557 1112 2122 2352 2698 6058 6525 9192 9269 |
| (8, 8) | [17.898337] | 1439 3446 4005 5283 6028 6361 7750 7791 |
| (8, 8) | [18.179891] | 2016 2142 2906 3195 4625 4989 8982 9453 |
| (8, 8) | [18.241773] | 1871 2148 2455 5189 7507 8694 9459 9592 |
| (11, 10) | [20.129338] | 301 1167 1911 1947 3067 4272 4300 6682 7426 8336 9295 |

На седьмом этапе, после нахождения и взвешивания треппин-сетов вероятность перехода в соответствующие псевдокодовые слова вычисляется по формуле:

$$P_{TS} = Z * \sum_i Q\left(\frac{\sqrt{a_i \otimes \omega_i(p)}}{\sigma}\right), \qquad (2.20)$$

где $Q -$ функция, $Z$ -размер циркулянта (кратность псевдокодовых слов для квазициклического кода), $\omega_i -$ вес -го псевдокодового слова, образованного треппин-сетом TS $(a_i, b)$ , $\sigma$ - дисперсия ошибки, $p$ – политоп, заданный проверочной матрицей кода.

Используя полученные оценки вероятностей ошибки для циклов, образующих треппин-сеты, для кода кандидата с наибольшей нижней оценкой псевдокодового веса, полученного путем спектрального просеивания кодов кандидатов [64, 67, 102], применим их для оценки $Penalte$ кодов кандидатов.

Ниже на Рис. 2.3 представлена оценка полки для кода из примера 2.

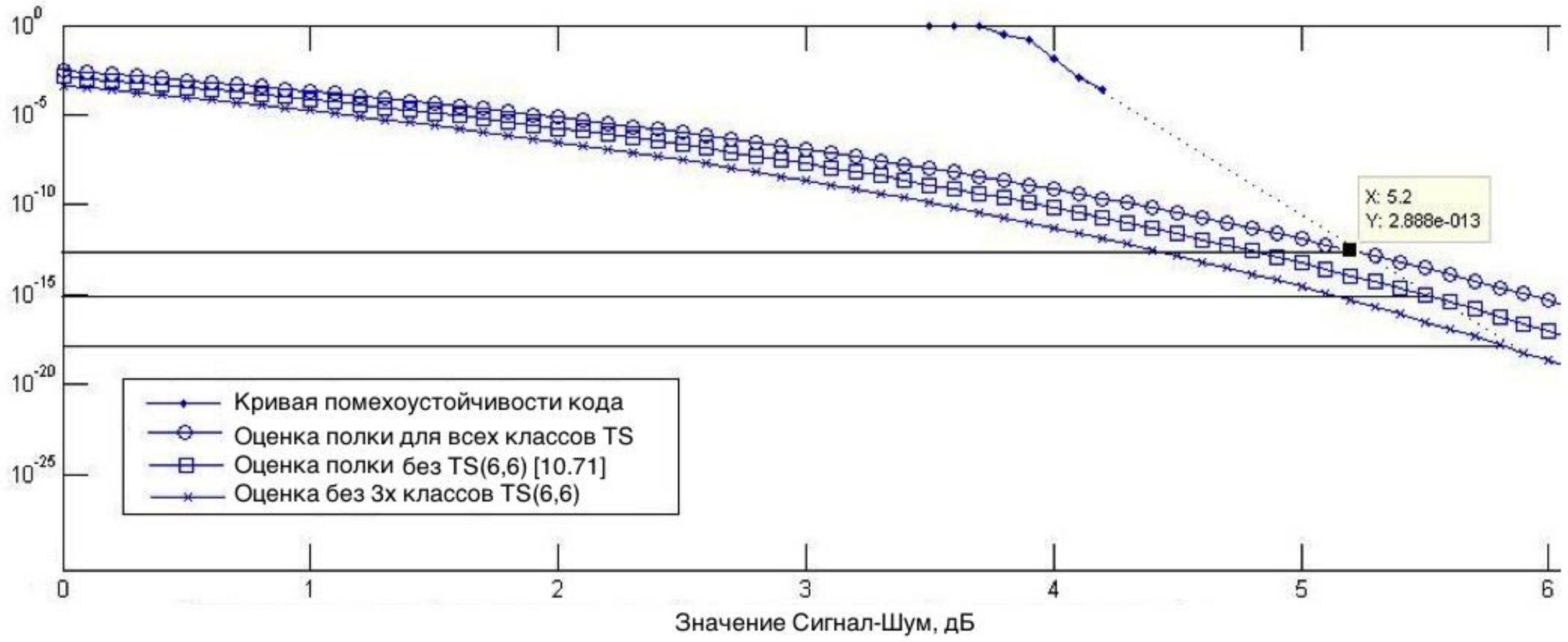

Рисунок 2.3 - Оценка помехоустойчивости кода. Зависимость вероятности ошибки на блок (FER) от значения Сигнал-Шум (EB/No) для квазициклического кода с проверочной матрицей размера 4x36 с циркулянтом 278

Кривая 1 соответствует вероятности ошибки на блок $2.8 \cdot 10^{-13}$, (вероятности ошибки на бит $2.8 \cdot 10^{-17}$ ) при равной вероятности ошибки каждого из битов для длины кода 10000, в случае, если параметры декодера (scale, offset, scale of LLR quantizer mapper, bitwise of message) обуславливают возникновение всех псевдокодовых слов, приведенных в Таблице 2.1.

Кривая 2 соответствует вероятности ошибки на блок $10^{-15}$ (на бит $10^{-19}$) в случае, если параметры декодера исключают появление 278 псевдокодовых слов треппин-сета TS(6,6), $\omega = 10.713756$. Кривой номер 3 соответствует вероятность ошибки на блок $10^{-18}$ (на бит $10^{-22}$ ) в случае, если параметры декодера исключают появление 278 псевдокодовых слов для каждого из трех классов треппин-сетов (6, 6) с весом псевдокодовых слов: $\omega = [10.71375,$ 11.861630, 12.367398]. Все кривые построены при условии, что соответствующие параметры декодера не повлияют на вероятность появления кодовых слов в рассматриваемых областях вероятностей (не ухудшат кривую помехоустойчивости в областях $P_{FER} < 10^{-13}, 10^{-15}$, $10^{-18}$, соответственно).

Проилюстрируем на примере процесс построения проверочной матрицы ME-LDPC кода длиной 10000 и скоростью 0.875 (2 последних символа выколоты), для которого необходимо обеспечить уровень ошибки на бит порядка $10^{-10}$ и порог меньше 5 дБ при 40 итерациях layered BP.

В качестве начальной проверочной матрицы используем матрицу размера 42x4, $H_0$ приведенную на стр. 75.

Оценив порог итеративного декодирования (Этап 2) и кодовое расстояние (Этап 3), получим 10.5 и 1 соответственно. Данного кодового расстояния и порога недостаточно для выполнения поставленных требований.

Добавим еще одну строку проверочной матрицы (Этап 2). Начнем перебирать все возможные положения проверок в этой строке. Получим матрицу протографа $H_1$ размера 42x5 (2 последних символа выколоты) с порогом итеративного декодирования 7.47 и кодовым расстоянием 2.

Продолжим процесс добавления строк проверочной матрицы и жадного поиска оптимального положения проверок в них. Получим матрицу протографа размера 42x6 $H_2$ с кодовым расстоянием 5, порогом итеративного декодирования 3.81.

В итоге получилась матрица протографа размера 42x7, $H_3$ с кодовым расстоянием 11, порогом итеративного декодирования 3.095, удовлетворяющая заданным требованиям.

Не всегда модификация матриц приводит к улучшению дистантных свойств кода. Например, представленная ниже матрица $H_{3_b}$ протографа имеет порог итеративного декодирования 3.076 и потенциально может быть расширена с меньшим числом коротких циклов, однако имеет худшую верхнюю оценку кодового расстояния, равную 10.

На Рис. 2.4 представлены оценки вероятности ошибки на бит для циркулянта размера 780, соответствующего длине кода $N = 780 \times 42 = 32760$, приведенных выше протографов $H_0, H_1, H_2, H_3$.

Теперь расширим протографа размера 42 на 7, циркулянтом размера 30 и продемонстрируем работу уточненной оценки кодового расстояния Маккея-Вонтобеля-Смарандаши-Сигеля-Батлера в расширенном графе ([59-62], для квазициклической проверочной матрицы теорема 7, [60]). Отметим, что коды не достигают верхней границы кодового расстояния, Рис. 2.5.

$H_0 =$

| | | | | | | | | | | | | | | | | | | | | | | | | | | | | | | | | | | | | | | | | | |
|---|---|---|---|---|---|---|---|---|---|---|---|---|---|---|---|---|---|---|---|---|---|---|---|---|---|---|---|---|---|---|---|---|---|---|---|---|---|---|---|---|---|
|1|0|1|1|1|0|0|0|1|1|0|0|1|1|1|1|1|0|1|0|0|0|1|0|0|0|0|0|1|1|1|1|0|0|1|0|0|0|0|0|1|1|
|0|0|0|0|1|1|0|0|0|1|1|1|0|0|0|0|0|1|0|0|0|0|0|0|0|1|1|1|0|0|0|1|1|0|0|0|0|1|1|1|1|1|
|1|0|1|1|0|0|0|1|0|1|0|0|1|0|1|1|0|1|0|1|1|1|0|1|0|1|1|0|1|0|1|0|1|1|0|0|1|0|0|0|1|1|
|1|1|0|1|1|0|1|1|0|0|0|1|0|1|0|0|1|1|1|1|0|1|1|1|1|0|0|1|0|0|0|1|0|1|1|0|0|0|0|0|1|0|

$H_1 =$

| | | | | | | | | | | | | | | | | | | | | | | | | | | | | | | | | | | | | | | | | | |
|---|---|---|---|---|---|---|---|---|---|---|---|---|---|---|---|---|---|---|---|---|---|---|---|---|---|---|---|---|---|---|---|---|---|---|---|---|---|---|---|---|---|
|1|0|1|1|1|0|0|0|1|1|0|0|1|1|1|1|1|0|1|0|0|0|1|0|0|0|0|0|1|1|1|1|0|0|1|0|0|0|0|0|1|1|
|0|0|0|0|1|1|0|0|0|1|1|1|0|0|0|0|0|1|0|0|0|0|0|0|0|1|1|1|0|0|0|1|1|0|0|0|0|1|1|1|1|1|
|1|0|1|1|0|0|0|1|0|1|0|0|1|0|1|1|0|1|0|1|1|1|0|1|0|1|1|0|1|0|1|0|1|1|0|0|1|0|0|0|1|1|
|1|1|0|1|1|0|1|1|0|0|0|1|0|1|0|0|1|1|1|1|0|1|1|1|1|0|0|1|0|0|0|1|0|1|1|0|0|0|0|0|1|0|
|0|1|1|0|0|1|1|1|1|0|1|1|1|1|0|0|0|0|1|1|1|0|0|1|1|1|1|0|1|1|0|0|0|0|1|1|1|1|0|0|0|1|

$H_2 =$

| | | | | | | | | | | | | | | | | | | | | | | | | | | | | | | | | | | | | | | | | | |
|---|---|---|---|---|---|---|---|---|---|---|---|---|---|---|---|---|---|---|---|---|---|---|---|---|---|---|---|---|---|---|---|---|---|---|---|---|---|---|---|---|---|
|1|0|1|1|1|0|0|0|1|1|0|0|1|1|1|1|1|0|1|0|0|0|1|0|0|0|0|0|1|1|1|1|0|0|1|0|0|0|0|0|1|1|
|0|0|0|0|1|1|0|0|0|1|1|1|0|0|0|0|0|1|0|0|0|0|0|0|0|1|1|1|0|0|0|1|1|0|0|0|0|1|1|1|1|1|
|1|0|1|1|0|0|0|1|0|1|0|0|1|0|1|1|0|1|0|1|1|1|0|1|0|1|1|0|1|0|1|0|1|1|0|0|1|0|0|0|1|1|
|1|1|0|1|1|0|1|1|0|0|0|1|0|1|0|0|1|1|1|1|0|1|1|1|1|0|0|1|0|0|0|1|0|1|1|0|0|0|0|0|1|0|
|0|1|1|0|0|1|1|1|1|0|1|1|1|1|0|0|0|0|1|1|1|0|0|1|1|1|1|0|1|1|0|0|0|0|1|1|1|1|0|0|0|1|
|0|1|0|0|0|1|1|0|1|0|1|0|0|0|1|1|1|0|0|0|1|1|1|0|1|0|0|1|0|1|1|0|0|0|0|1|0|0|0|1|1|0|

$H_3$=

| | | | | | | | | | | | | | | | | | | | | | | | | | | | | | | | | | | | | | | | | | |
|---|---|---|---|---|---|---|---|---|---|---|---|---|---|---|---|---|---|---|---|---|---|---|---|---|---|---|---|---|---|---|---|---|---|---|---|---|---|---|---|---|---|
|1|0|1|1|1|0|0|0|1|1|0|0|1|1|1|1|1|0|1|0|0|0|1|0|0|0|0|0|1|1|1|1|0|0|1|0|0|0|0|0|1|1|
|0|0|0|0|1|1|0|0|0|1|1|1|0|0|0|0|0|1|0|0|0|0|0|0|0|1|1|1|0|0|0|1|1|0|0|0|0|1|1|1|1|1|
|1|0|1|1|0|0|0|1|0|1|0|0|1|0|1|1|0|1|0|1|1|1|0|1|0|1|1|0|1|0|1|0|1|1|0|0|1|0|0|0|1|1|
|1|1|0|1|1|0|1|1|0|0|0|1|0|1|0|0|1|1|1|1|0|1|1|1|1|0|0|1|0|0|0|1|0|1|1|0|0|0|0|0|1|0|
|0|1|1|0|0|1|1|1|1|0|1|1|1|1|0|0|0|0|1|1|1|0|0|1|1|1|1|0|1|1|0|0|0|0|1|1|1|1|0|0|0|1|
|0|1|0|0|0|1|1|0|1|0|1|0|0|0|1|1|1|0|0|0|1|1|1|0|1|0|0|1|0|1|1|0|0|0|0|1|0|0|0|1|1|0|
|0|0|0|0|1|0|0|0|0|0|0|0|0|0|0|0|0|0|0|0|0|0|0|0|0|0|0|0|0|0|0|0|0|0|0|0|1|0|1|0|1|1|

$H_{3_b}$=

| | | | | | | | | | | | | | | | | | | | | | | | | | | | | | | | | | | | | | | | | | |
|---|---|---|---|---|---|---|---|---|---|---|---|---|---|---|---|---|---|---|---|---|---|---|---|---|---|---|---|---|---|---|---|---|---|---|---|---|---|---|---|---|---|
|1|0|1|1|1|0|0|0|1|1|0|0|1|1|1|1|1|0|1|0|0|0|1|0|0|0|0|0|1|1|1|1|0|0|1|0|0|0|0|0|1|1|
|0|0|0|0|1|1|0|0|0|1|1|1|0|0|0|0|0|1|0|0|0|0|0|0|0|1|1|1|0|0|0|1|1|0|0|0|0|1|1|1|1|1|
|1|0|1|1|0|0|0|1|0|1|0|0|1|0|1|1|0|1|0|1|1|1|0|1|0|1|1|0|1|0|1|0|1|1|0|0|1|0|0|0|1|1|
|1|1|0|1|1|0|1|1|0|0|0|1|0|1|0|0|1|1|1|1|0|1|1|1|1|0|0|1|0|0|0|1|0|1|1|0|0|0|0|0|1|0|
|0|1|1|0|0|1|1|1|1|0|1|1|1|1|0|0|0|0|1|1|1|0|0|1|1|1|1|0|1|1|0|0|0|0|1|1|1|1|0|0|0|1|
|0|1|0|0|0|1|1|0|1|0|1|0|0|0|1|1|1|0|0|0|1|1|1|0|1|0|0|1|0|1|1|0|0|0|0|1|0|0|0|1|1|0|
|0|0|0|0|1|0|0|0|0|0|0|0|0|0|0|0|0|0|0|0|0|0|0|0|0|0|0|0|0|0|0|0|0|0|0|0|0|0|1|0|1|1|

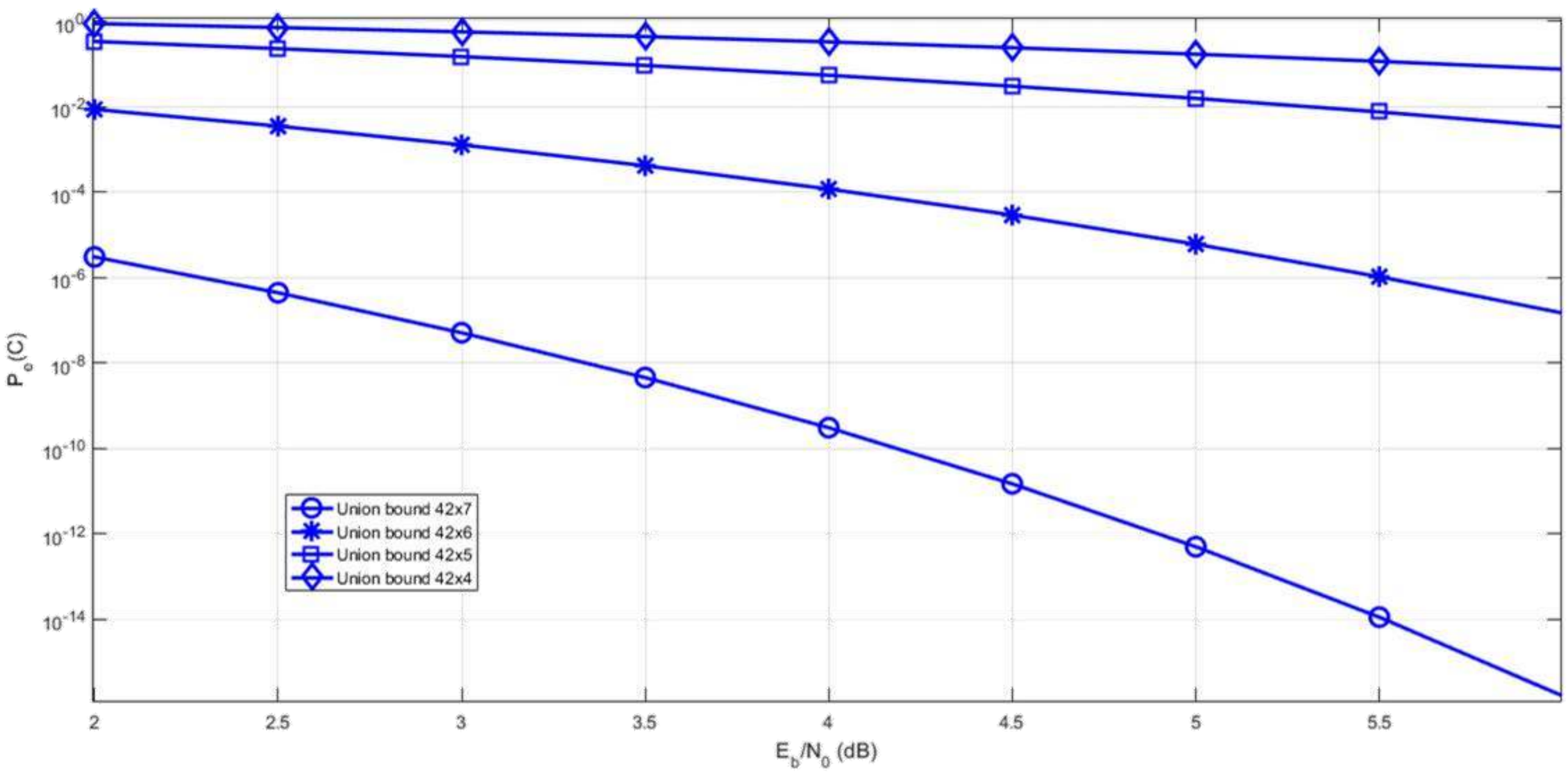

Рисунок 2.4 Оценка вероятности ошибки на бит для протографов $H_0, H_1, H_2, H_3$.

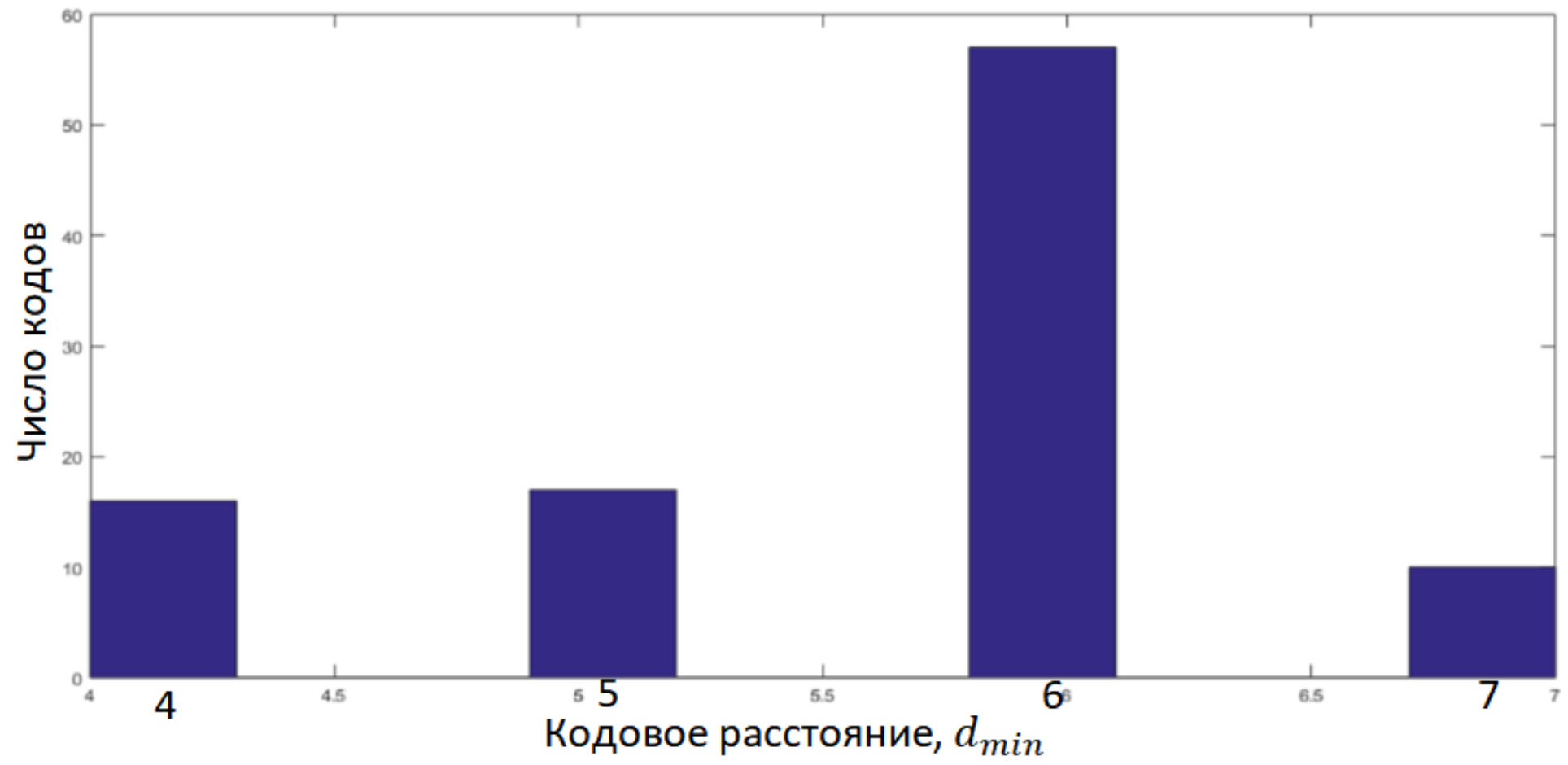

Рисунок 2.5 - Распределение кодового расстояния среди 100 расширенных квазициклических низкоплотностных кодов

В приложение 3, представлены примеры кодов кандидатов с кодовым расстоянием 4, 5, 6, 7. Ниже на рис. 2.6-2.9 представлены спектры их связности (ACE Spectrum), вычисленные по формуле $ACE(C) = \sum_{v \in C} (d(v) - 2)$, [44], где $d(v)$ - степень инцидентности символьного узла $v$, входящего в циклы длины $i$, $n_i$ число таких символьных узлов. Ось абсцисс соответствует значению связности в подграфе (ACE), ось ординат числу символьных узлов (столбцов):

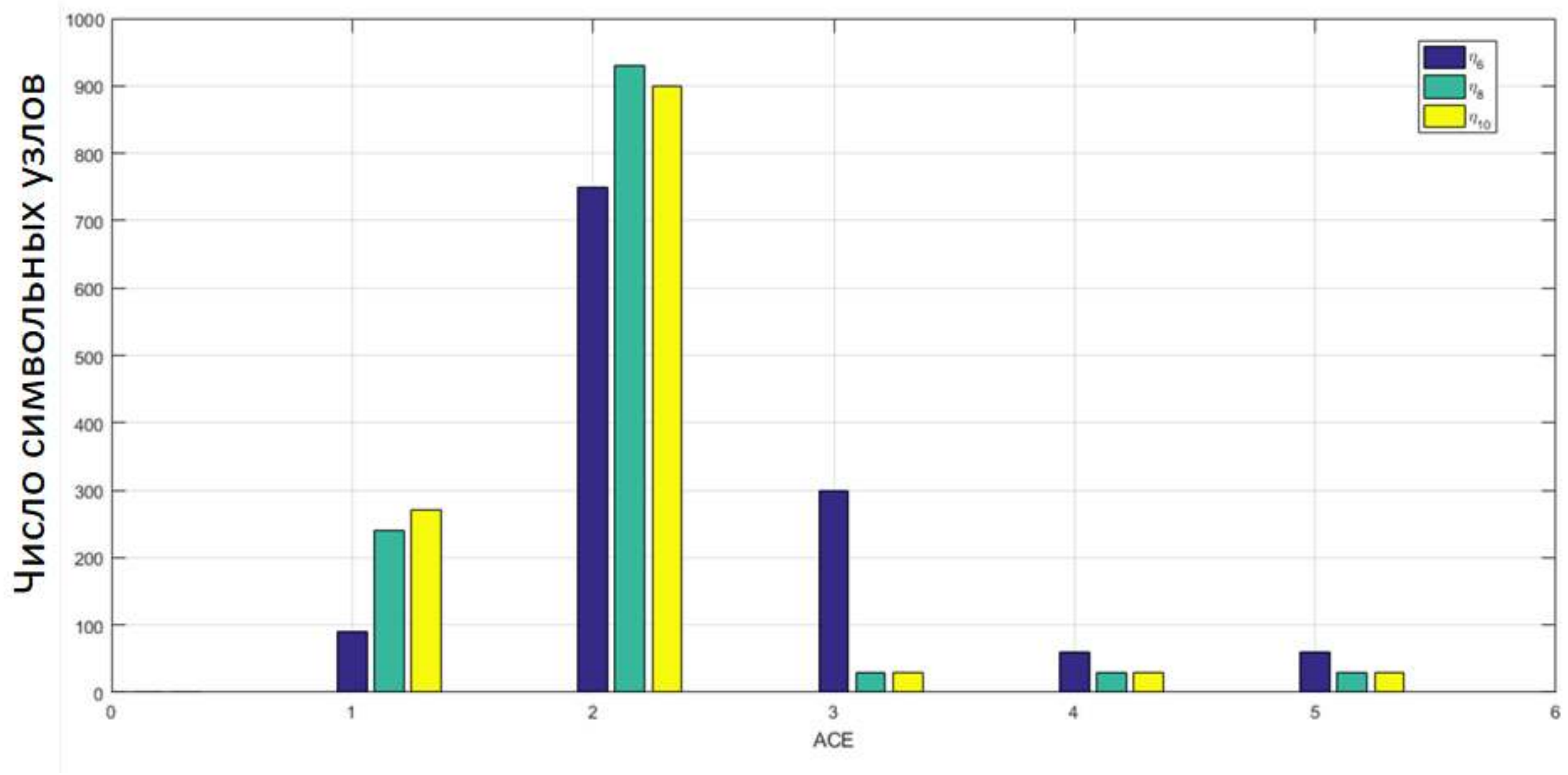

Рисунок 2.6 - Спектр связности $(\eta_6, \eta_8, \eta_{10})$, расширенного протографа 42 на 7 с кодовым расстоянием 4.

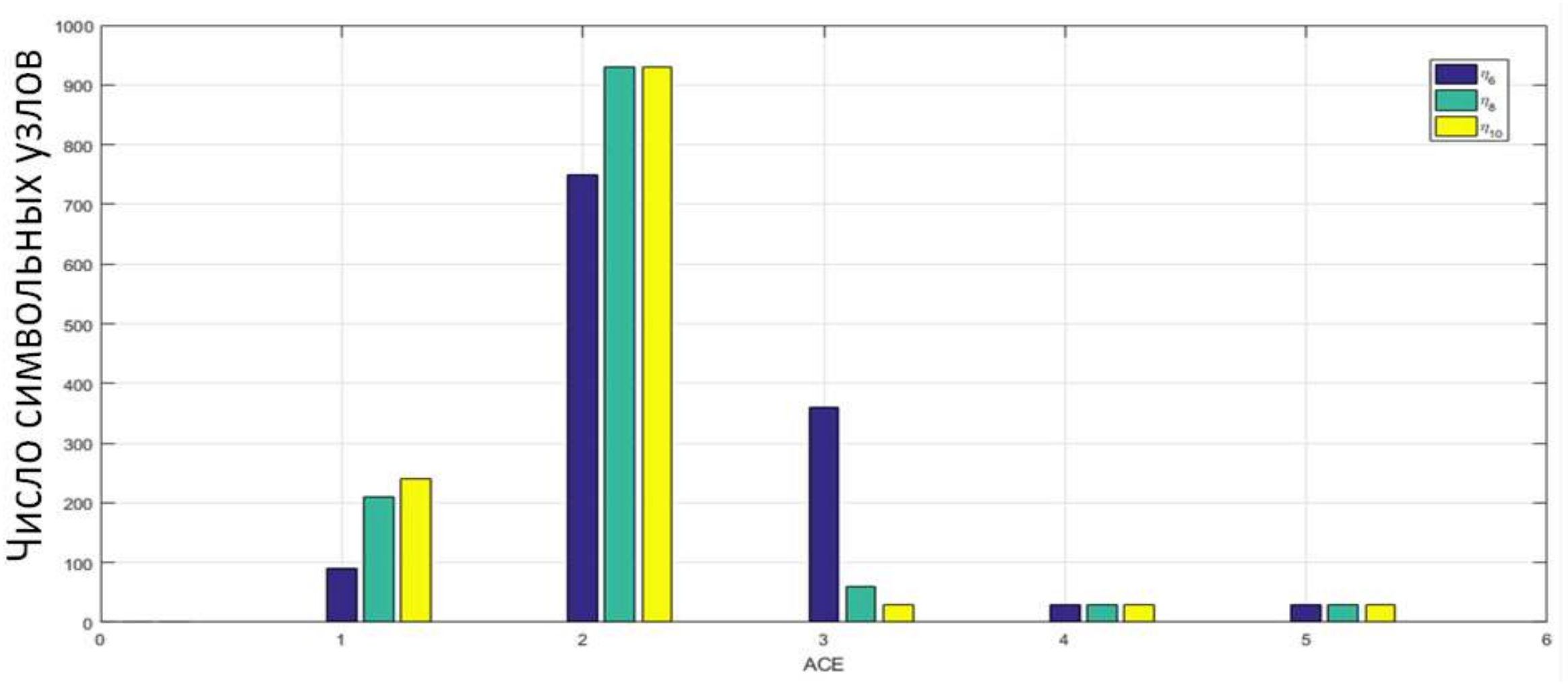

Рисунок 2.7 - Спектр связности $(\eta_6, \eta_8, \eta_{10})$, расширенного протографа 42 на 7 с кодовым расстоянием 5

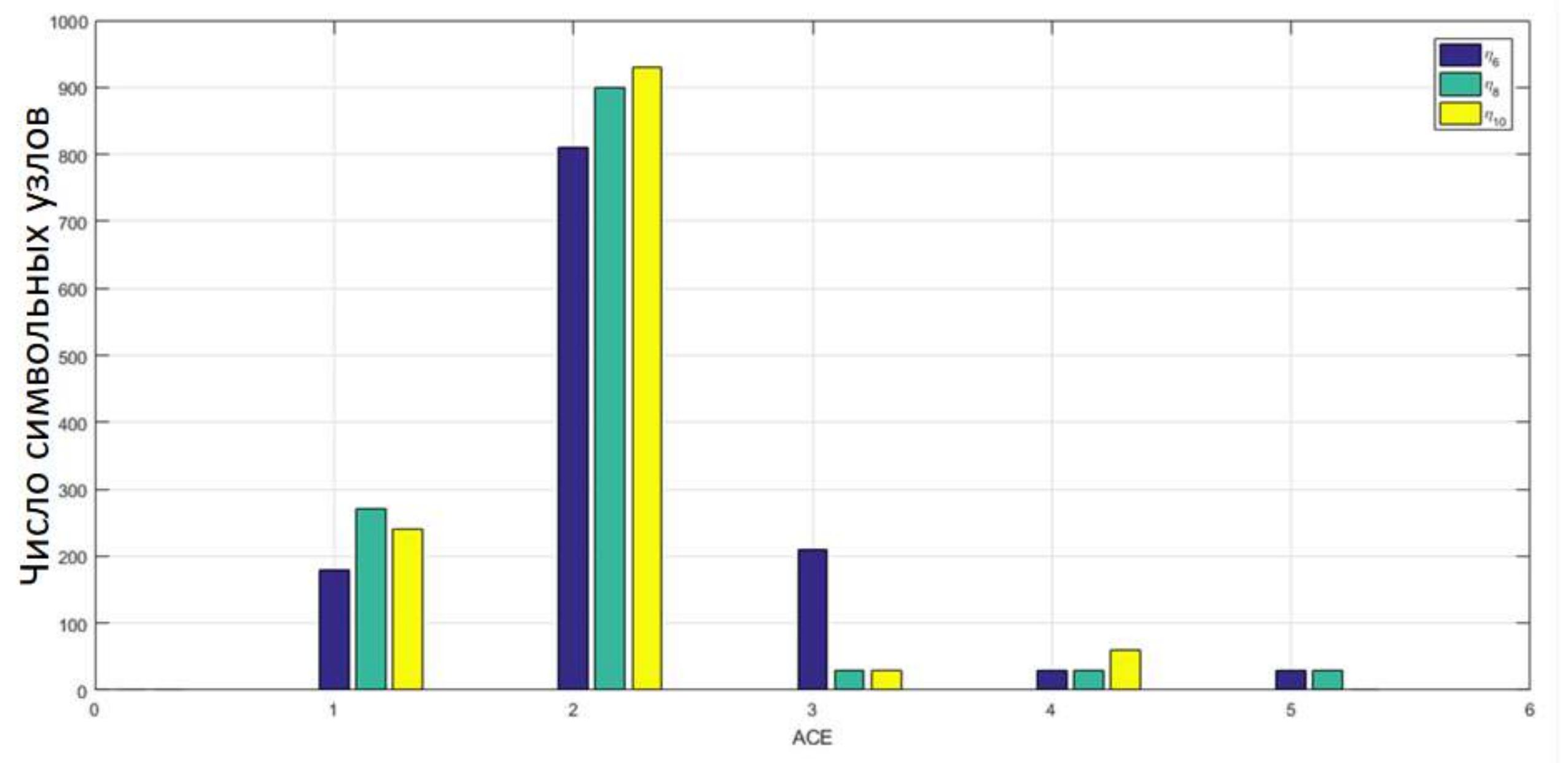

Рисунок 2.8 - Спектр связности $(\eta_6, \eta_8, \eta_{10})$, расширенного протографа 42 на 7 с кодовым расстоянием 6

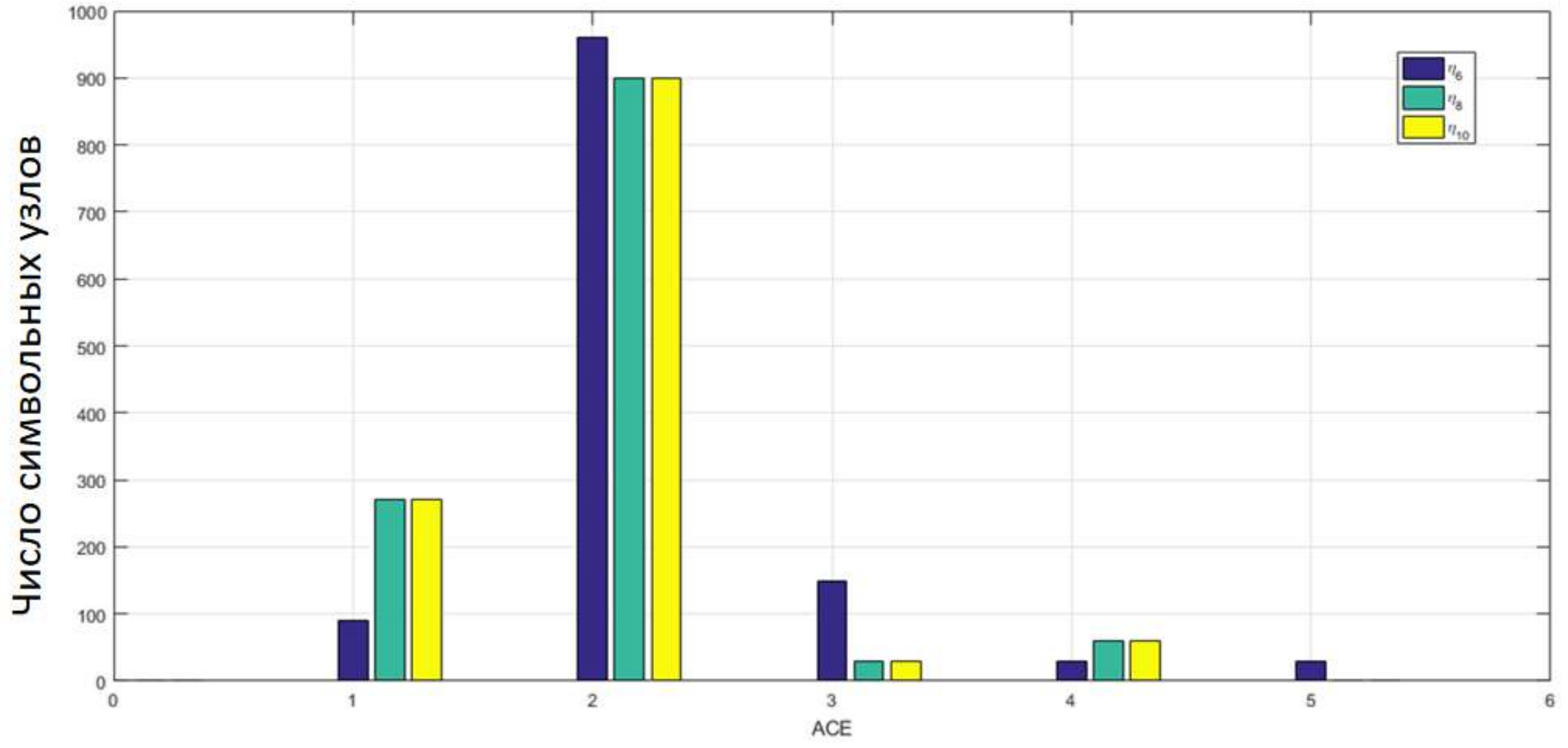

Рисунок 2.9 - Спектр связности $(\eta_6, \eta_8, \eta_{10})$, расширенного протографа 42 на 7 с кодовым расстоянием 7.

В результате работы предложенной процедуры были получены ряд протографов кодов с порогами итеративного декодирования на уровне $10^{-15}$ (ошибка на бит) превосходящими известные ранее:

1) протограф кодов со скоростью 4/5, с максимальным весом столбца 19, с порогом итеративного декодирования $E_B/N_0 = 2.3345$, что лучше, чем порог итеративного декодирования протографа AR4A-кодов, $E_B/N_0$ =2.396, см. Рис. 2.10, [49];

2) протограф кодов со скоростью 2/3, с максимальным весом столбца 10, с порогом итеративного декодирования $E_B/N_0 = 1.292$, что превосходит порог итеративного декодирования протографа AR4JA-кодов, $E_B/N_0 = 1.4$, см. Рис. 2.12, [39,103];

3) протограф кодов со скоростью ½, с максимальным весом столбца 20, с порогом итеративного декодирования $E_B/N_0$ =0.3994, что значительно меньше порога итеративного декодирования протограф AR4A-кода, $E_B/N_0$ =0.56, см. Рис. 2.11, [49].

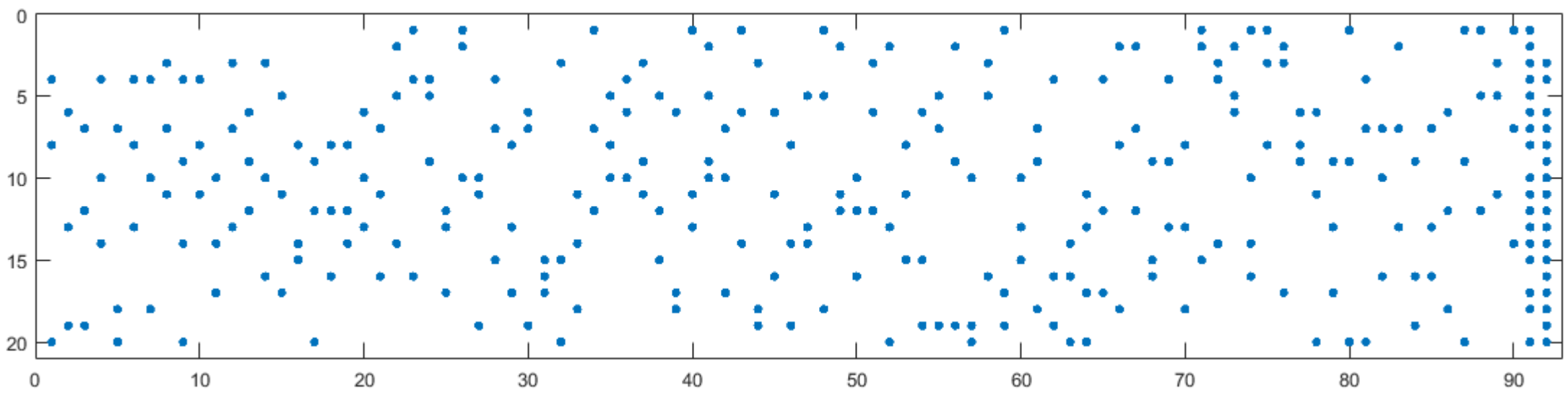

Рисунок 2.10 - Графическое представление базовой матрицы (протографа) кодов со скоростью 4/5, точки обозначают единицы в базовой матрице

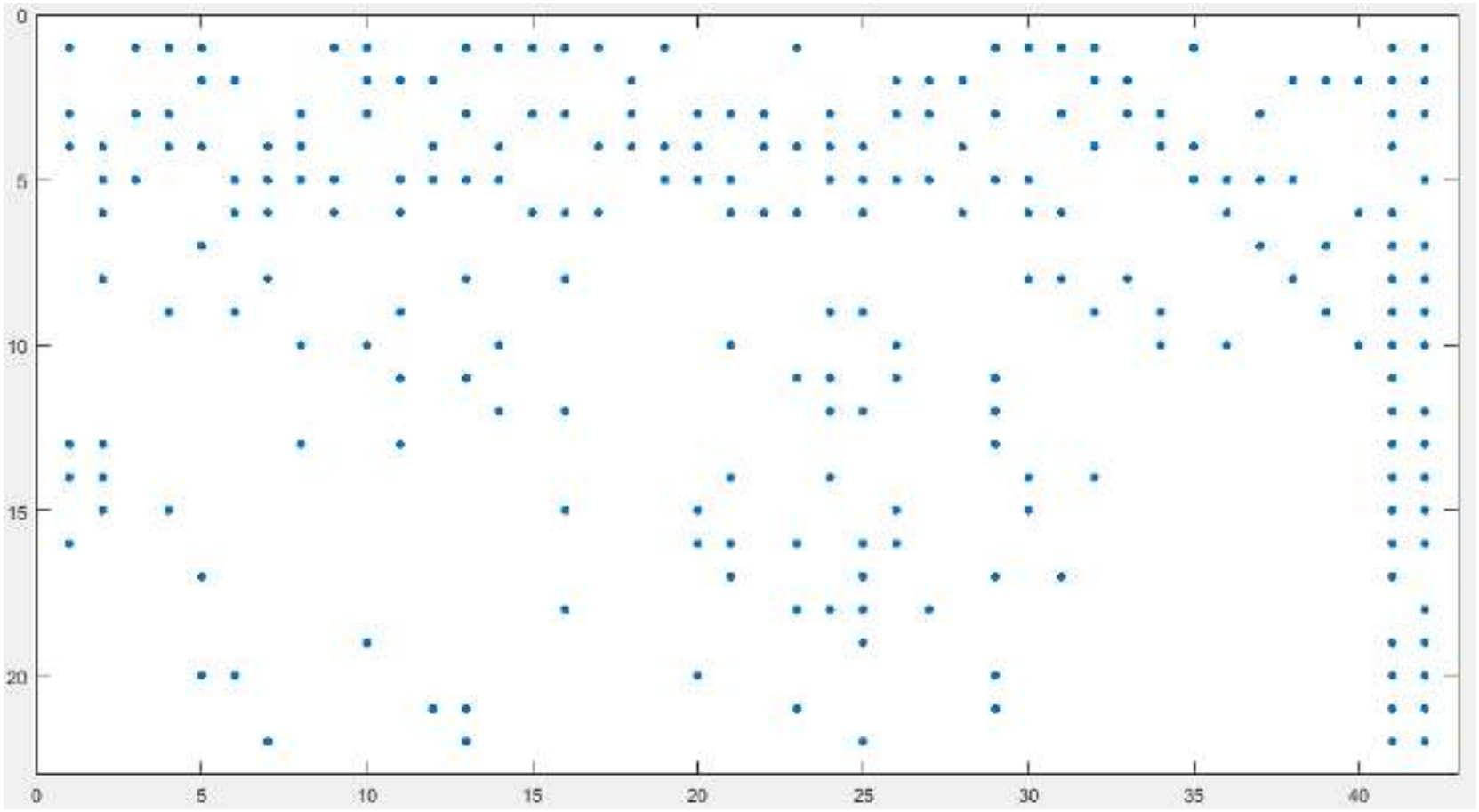

Рисунок 2.11 - Рис. Графическое представление базовой матрицы кодов со скоростью 1/2, точки обозначают единицы в базовой матрице

Скорость кода:2/3

Protograph-

| | 1 | 2 | 3 | 4 | 5 | 6 | 7 | 8 | 9 | 10 | 11 | 12 | 13 | 14 | 15 | 16 |
|---|---|---|---|---|---|---|---|---|---|---|---|---|---|---|---|---|
| 1 | 1 | 0 | 0 | 0 | 0 | 1 | 0 | 1 | 0 | 1 | 1 | 0 | 1 | 0 | 0 | 2 |
| 2 | 1 | 1 | 0 | 0 | 0 | 0 | 0 | 1 | 1 | 0 | 0 | 1 | 0 | 1 | 1 | 2 |
| 3 | 0 | 1 | 1 | 0 | 0 | 0 | 0 | 1 | 0 | 1 | 0 | 1 | 0 | 1 | 0 | 1 |
| 4 | 0 | 0 | 1 | 1 | 0 | 0 | 1 | 0 | 1 | 0 | 1 | 0 | 0 | 1 | 1 | 1 |
| 5 | 0 | 0 | 0 | 1 | 1 | 0 | 1 | 0 | 0 | 1 | 1 | 0 | 1 | 0 | 1 | 2 |
| 6 | 0 | 0 | 0 | 0 | 1 | 1 | 1 | 0 | 1 | 0 | 0 | 1 | 1 | 0 | 0 | 2 |

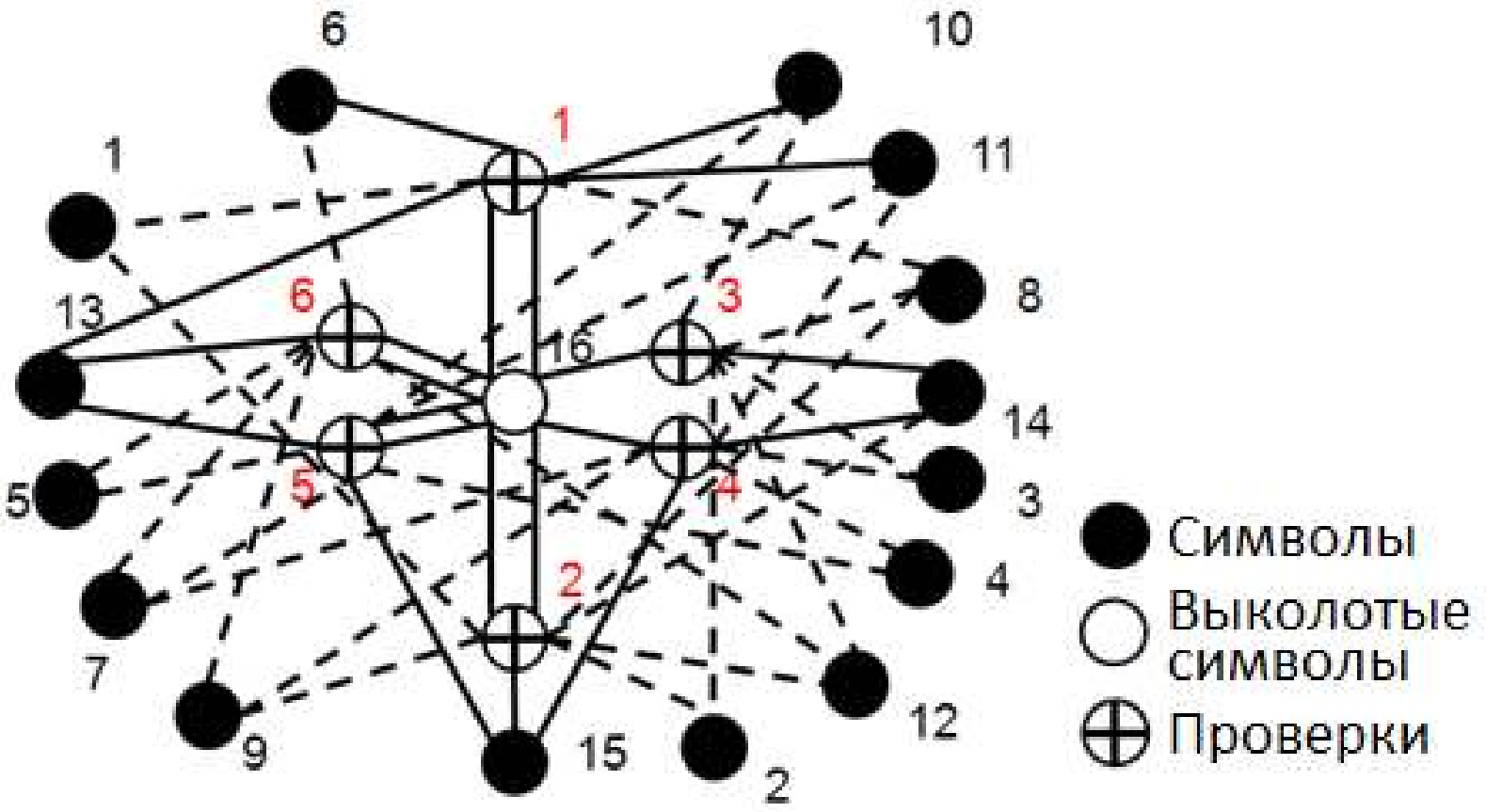

Рисунок 2.12 - Полученный протограф кодов со скоростью кода 2/3, [103]: а) базовая матрица, б) протограф соответствующей базовой матрице

## 2.2 Алгоритмы для фазы расширения предложенного метода

В диссертации предложенно использовать на 5 и 9 этапах метода построения низкоплотностных кодов жадный алгоритм с запрещенными коэффициентами (Алгоритм А) и алгоритм имитации отжига (Алгоритм Б). Их совместное использование позволяет улучшить дистантные свойства кодов и их спектры связности.

Часто используемый Guess-and-Test метод требует большого числа испытаний для достижения требуемого обхвата графа. Причина этой особенности метода, заключается в использование случайных бросков значений сдвигов циркулянта, с последующей проверкой полученного графа на обхват в соответствие с уравнением (2), [22].

Ниже приводится формализация жадного алгоритма запрещенных коэффициентов (Алгоритм А), [109, 143]. На вход Алгоритма А подаются: требуемый обхват графа $g$ , базовая матрица в форме списка проверочных уравнений $h_{j,l}$ (единиц в базовой проверочной матрице, заданных номерами строки $j$ и столбца $l$). В алгоритме используется массив флагов запрещенных циклических сдвигов циркулянтов $noshift$ (циклических сдвигов, формирующих циклы длиной меньше g) размерности $J \times L \times z$ .

1. Инициализируется массив флагов запрещенных циклических сдвигов циркулянт: $noshift_{J\times L\times z} = 0$ ,$0 \le j \le J-1$ , $0 \le l \le L-1,$ $0 \le s \le z-1$.

2. В цикле по $0 \le j \le J-1$ , $0 \le l \le L-1$, для всех ненулевых элементов базовой проверочной матрицы:

2.1. Задается случайный сдвиг циркулянта $p_{j,l} = rand(0, z-1)$ с нулевым значением флага запрещенных циклических сдвигов.

2.2. Выполняется модификация массива флагов запрещенных сдвигов для элементов базовой матрицы с неопределенным до сих пор сдвигом циркулянта. Для всех потенциально возможных циклов глубиной до $g$ , в которых участвует элемент базовой матрицы $h_{j,l}$ , проверяется выполнение условия наличия циклов, формула 4 в [22, 104]. Если условия выполняются, то соответствующий флаг запрещенного циклического сдвига устанавливается в единицу $noshift_{j,l,z} = 0$.

2.3. Если все флаги запрещенных сдвигов равны 1, запускается работа алгоритма заново. В случае превышения заданного числа запусков работа алгоритма завершается с фиксацией неудачи построения квазициклического кода.

Алгоритм запрещенных коэффициентов демонстрирует высокую вероятность (примерно на порядок большую, чем PEG) успешного построения LDPC-кода кандидата, присущую методу Hill-Climbing при существенно меньшей вычислительной сложности, сравнимой со сложностью метода Guess-and-Test, Рис. 2.13.

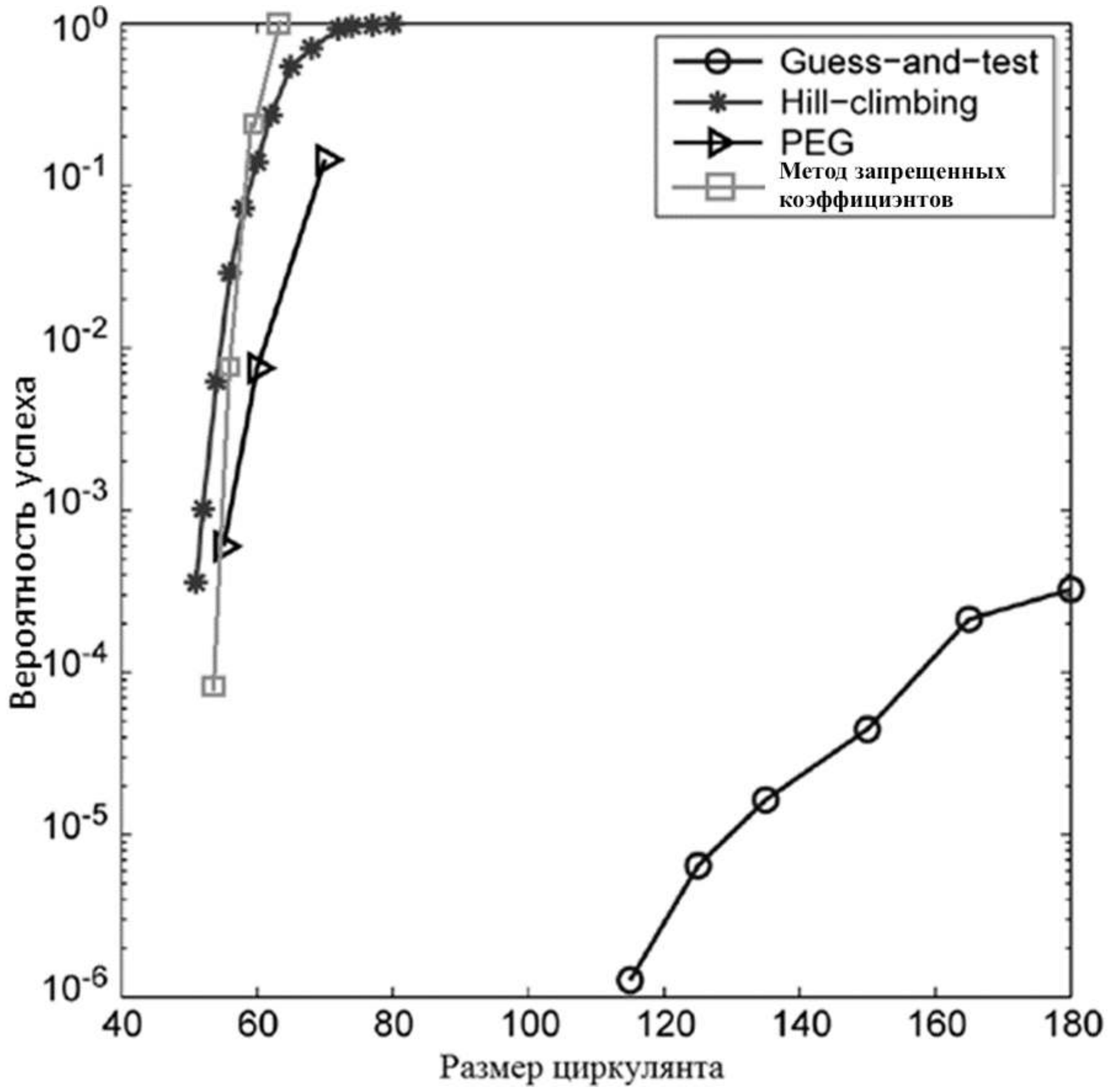

Рисунок 2.13. - Сравнение вероятности успешного построения квазициклического регулярного кода обхватом 8 с весом строки 12 и весом столбца 3 методами расширения базовой матрицы: Guess-and-Test, Hill-Climbing, запрещенных коэффициентов, [109,143] и PEG

На 9 этапе метода построения квазициклических низкоплотностных кодов предусматривается использование алгоритма имитации отжига (Simulated Annealing), [101,111,122,123].

Алгоритм Б (имитации отжига) представлен ниже:

1. Инициируем счетчик шагов алгоритма Nstep=0;
2. Выбираем случайно не нулевой элемент в базовой матрице $h_{j,l}$;
3. Перечисляем все циклы, проходящие через этот циркулянт $h_{j,l}$, используемые в формуле расчета штрафной функции $Penalte$;
4. Вычисляется число циклов Θ для всевозможных циклических сдвигов $a_{j,l} \in \{0 \dots L-1\}$ циркулянта $a_{j,l}$, при помощи формулы 4 в [22, 104];

5. Выбирается случайно одно из значений $a_{j,l}$ с вероятностью, зависящей от числа циклов и температуры Temp, (величины, зависящая от числа циклов в графе и количество шагов алгоритма Б). Если Nstep=0, то Temp=$max_{a_{j,l}}\Theta$. Вероятность выбора сдвига циркулянта $a_{j,l}$, задана функцией:

$$P\left(a_{jl}\right)=w(a_{jl})/\sum_{i=0}^{L-1}w(i), \qquad (2.21)$$

где функция плотности вероятности $w\left(a_{jl}\right)$ возрастает с уменьшением числа циклов $\Theta$ и падает с уменьшением температуры:

$$w\left(a_{jl}\right)=e^{\frac{-\Theta\left(a_{jl}\right)}{Temp}}; \qquad (2.22)$$

6. Инкрементируем переменную Nstep, вычисляем новое значение температуры:

$$Temp=\eta\,\frac{\Phi}{Nstep^{2}}, \qquad (2.23)$$

где $\eta$ - константа, $\Phi$-общее число циклов в расширенной матрице $H$ .

7. Останавливаем работу алгоритма, если через заданное число шагов число циклов в графе $\Phi$ не изменилось, иначе перезапускаем алгоритм.

На выходе Алгоритмы А и Б выдают проверочные матрицы *H* квазициклических кодов. Алгоритм А быстро вычисляет начальное приближение расширенной проверочной матрицы. Оригинальность Алгоритма Б в предлагаемом методе построения квазициклических низкоплотностных кодов заключается в пропуске значительного числа локальных минимумов числа треппин-сетов, что позволяет получить квазициклические коды с дистантными свойствами и спектром связности, недостижимые предложенными ранее методами, а также на порядок более высокую вероятность успешного расширения протографов.

С помощью алгоритма имитации отжига удалось уменьшить минимальные значения циркулянтов в базовой матрице регулярного кода (см. Табл. 2.2-2.3) (уменьшение минимального значения требуемого циркулянта позволяет улучшить помехоустойчивость кода).

Таблица 2.2 - Минимальное значение размера циркулянта полученное методами лифтинга для регулярного кода с числом строк $m$=3, числом столбцов $L$ с обхватом 10

| Число столбцов, $L$ | Предложенный Метод (Simulated Annealing), [101] | Hill Climbing, [45] | Улучшенный PEG, [47] | Нижняя оценка, [48] |
|---|---|---|---|---|
| 4 | 37 | 39 | 37 | 37 |
| 5 | 61 | 63 | 61 | 61 |
| 6 | 91 | 103 | 91 | 91 |
| 7 | 155 | 160 | 155 | 127 |
| 8 | 215 | 233 | 227 | 168 |
| 9 | 304 | 329 | 323 | 217 |
| 10 | 412 | 439 | 429 | 271 |
| 11 | 545 | 577 | 571 | 331 |
| 12 | 709 | 758 | - | - |

Таблица 2.3 - Минимальное значение размера циркулянта полученное методами лифтинга для регулярного кода с числом строк $m$=3, числом столбцов $L$ с обхватом 12

| Число столбцов | Предложенный метод (Simulated Annealing), [101] | Улучшенный PEG, [47] | Таблица V, [18] |
|---|---|---|---|
| 4 | 73 | 73 | 97 |
| 5 | 160 | 163 | 239 |
| 6 | 320 | 369 | 479 |
| 7 | 614 | 679 | 881 |
| 8 | 1060 | 1291 | 1493 |
| 9 | 1745 | 1963 | 2087 |
| 10 | 2734 | - | - |
| 11 | 4083 | - | - |
| 12 | 5964 | - | - |

Предложенный метод расширения базовой матрицы обладает на порядок большей вероятностью построения квазициклического кода с требуемым обхватом графа Таннера (см. рис. 2.14).

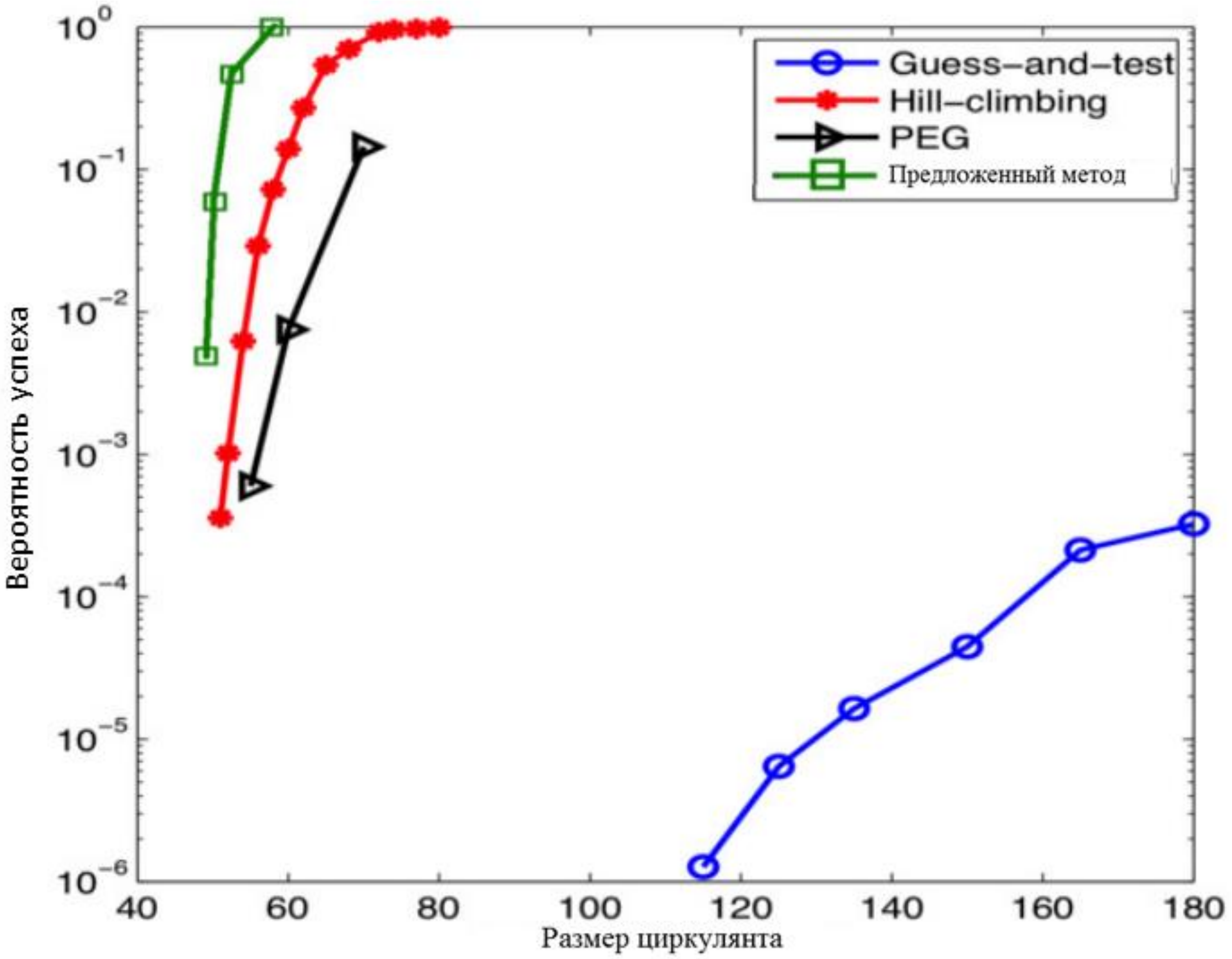

Рисунок 2.14 - Сравнение вероятности успешного построения квазициклического регулярного кода обхватом 8 с весом строки 12 и весом столбца 3 методами расширения базовой матрицы: Guess-and-test, Hill-climbing, PEG и алгоритмом имитации отжига (Simulated Annealing), [101]

Вычислительная сложность расчета кодового расстояния известными методами на 6 этапе растет экспоненциально от информационной длины кода $O(2^k)$, что может быть реализовано только на суперЭВМ. В связи с этим в диссертации (в разделе 3) разработан более быстродействующий метод вычисления кодового расстояния, использующий верхние и нижние оценки кодового расстояния и позволяющий по ним ранжировать коды кандидаты.

## 2.3 Методы адаптации по длине квазициклических низкоплотностных кодов

В диссертации был предложен метод адаптации кода по длине «Floor scale modular lifting» с целью повышения корректирующей способности кода для компенсации старения носителя. Метод обобщает ранее известный метод (Floor lifting) и позволяет применять коды с лучшим спектром связности, [100, 102, 104, 124,125, 129, 133, 147].

Для метода адаптации кода по длине «Floor lifting», значения сдвигов циркулянт вычисляются по следующей формуле, [94]:

$$E_{ij}(H_k) = \begin{cases} -1, & E_{ij}(H_0) = -1, \\ \left\lfloor \frac{L_k}{L_0} \times E_{ij}(H_0) \right\rfloor & \end{cases}, \tag{2.24}$$

где $E_{ij}(H_0)$ -значение сдвига циркулянта в проверочной матрице кода $E_{ij}(H_0)$ максимальной длины, $L_0$ - размер циркулянта кода максимальной длинны, $L_k$ - требуемый размер циркулянта, $k$ – число различных размеров циркулянт поддерживаемых методом.

Для метода адаптации кода по длине «Modular lifting», значения сдвигов циркулянт, вычисляются по следующей формуле, [94]:

$$E_{ij}(H_k) = \begin{cases} -1, & E_{ij}(H_0) = -1, \\ E_{ij}(H_0) \bmod L_k & \end{cases}. \tag{2.25}$$

Ключевой идеей предложенного нами метода была максимизация числа допустимых сдвигов циркулянта для получения заданного обхвата. Это позволяет эффективно использовать полученную свободу выбора циркулянта, при использовании полного перебора циклов, метода запрещенных коэффициентов, метода имитации отжига и других методов.

В соответствии с предложенным нами методом адаптации кода по длине «Floor Scale modular lifting", значения сдвига циркулянт вычисляются по следующей формуле, [100, 102]:

$$E_{ij}(H_k) = \begin{cases} -1, & E_{ij}(H_0) = -1, \\ \left\lfloor \frac{L_k}{L_0} ((r \times E_{ij}(H_0)) \bmod L_0) \right\rfloor & \end{cases}, \tag{2.26}$$

где $r$ - коэффициент отображения циклического сдвига.

В работе [102] дано доказательство того, что вероятность коротких циклов в «Modular lifting» методе адаптации кода по длине меньше, чем в «Floor lifting» методе адаптации кода по длине.

Теорема 1. Вероятность разрыва циклов длины 4 в проверочной матрице низкоплотностного квазициклического кода с циркулянтом размера $q$ в методе адаптации по длине «Modular lifting» $P_{mod}$:

$$P_{mod} = (1 - p_{mod})^{y} = 1 - yp_{mod} + O(q^{-2}), q \to \infty, \tag{2.27}$$

где $p_{mod}$ - вероятность образования цикла, $p_{mod} \triangleq 1/(2q-1)$ [102, лемма 3], $y$-число цепочек сдвигов циркулянтов, не образующих циклов длины 4.

Вероятность разрыва циклов длины 4 в проверочной матрице низкоплотностного квазициклического кода с циркулянтом размера $q$ для метода «Floor lifting»:

$$P_{fl} = \left(1 - p_{fl}\right)^{y} = 1 - yp_{fl} + O(q^{-2}), q \to \infty, \tag{2.28}$$

где $p_{fl}$ - вероятность образования цикла, по лемме 2 [102] $p_{fl} \triangleq 5/\left(4(2q-1)\right)$, $y$ - число цепочек сдвигов циркулянтов, не образующих циклов длины 4.

Вероятность разрыва циклов длины 4 в проверочной матрице низкоплотностного квазициклического кода с циркулянтом размера $q$ для метода адаптации по длине «Floor Scale Modular lifting»:

$$P_{fsml}(N_r) = \begin{cases} 1 - O(q^{-N_r}) \\ 1, y < N_r, q \to \infty, \end{cases} \tag{2.29}$$

где $y$ - число цепочек сдвигов циркулянтов, не образующих циклов длины 4, $N_r$-число масштабирующих значений.

Для доказательства теоремы 1 необходима следующая лемма.

Лемма 1. Биноминальное тождество

$$\sum_{k=0}^{n} (-1)^{k-1} \binom{n}{k} g(k) = 0 \tag{2.30}$$

выполняется для полинома $g(k)$ степени меньше $n$.

Доказательство:

Любой полином $g(k)$ степени меньше $t$ может быть представлен в форме

$$g(k) = \sum_{l=0}^{l} c_l (k)_l, \tag{2.31}$$

где $(k)_l = k(k-1)\dots(k-l+1)$ и $c_l$ - некоторая константа.

Для любого $l < n$ выполняется

$$\sum_{k=0}^{n} (-1)^{k} \binom{n}{k} (k)_l = \sum_{k=l}^{n} (-1)^{k} \binom{n}{k} (k)_l = \sum_{k=l}^{n} (-1)^{k} \frac{n!k!}{k!(n-k)!(k-l)!} \tag{2.32}$$

$$=\frac{n!}{(k-l)!}\sum_{k=l}^{n}(-1)^k\frac{(n-l)!}{(n-k)!(k-l)!}=\frac{n!}{(k-l)!}\sum_{k=l}^{n}(-1)^k\binom{n-l}{n-k}=0.\ \#$$

Доказательство теоремы 1, [102]:

Вероятность $P_k$ того, что первые $k$ строк не содержат циклов, равна $P_k = \left(1-kp_{fl}\right)^y$ . Используя принцип включения-исключения, можно получить вероятность разрыва циклов в случае использования «Floor Scale Modular lifting» $P_{fsml}$,

$$P_{fsml} = \sum_{k=1}^{N_r}(-1)^{k-1}\binom{N_r}{k}\left(1-kp_{fl}\right)^y. \tag{2.33}$$

Применив Лемму 1 и асимптотическое тождество

$$p_{fl} \triangleq \frac{5}{\left(4(2q-1)\right)} = O\left(\frac{1}{q}\right), q \to \infty \tag{2.34}$$

к выражению (2.33) вероятности $P_{fsml}$ получим

$$P_{fsml}(N_r) = \begin{cases} 1 - O(q^{-N_r}) \\ 1, y < N_r, q \to \infty \end{cases}. \tag{2.35}$$

#

Предложенный метод адаптации кода по длине уменьшает число кратчайших циклов в графе Таннера, позволяя улучшить помехоустойчивость, Табл. 2.4.

Таблица 2.4 - Число циклов и их длина для различных методов адаптации длины для кода IEEE 802.16 (WiMAX) со скоростью 0.5.

| Предложенный метод | | | Метод Samsung (Floor lifting), $r=1$ | |
|---|---|---|---|---|
| $L_k$, циркулянты | $r$ | обхват / число циклов | $L_k$, циркулянты | обхват / число циклов |
| **24** | **95** | **6 / 13** | **24** | **6 / 20** |
| 28 | 1 | 4 / 1 | 28 | 4 / 1 |
| 32 | 1 | 6 / 11 | 32 | 6 / 11 |
| **36** | **95** | **6 / 7** | **36** | **6 / 13** |
| 40 | 1 | 6 / 7 | 40 | 6 / 7 |
| **44** | **95** | **6 / 5** | **44** | **6 / 10** |
| 48 | 1 | 6 / 7 | 48 | 6 / 7 |
| 52 | 1 | 6 / 6 | 52 | 6 / 6 |
| 56 | 1 | 6 / 5 | 56 | 6 / 5 |
| 60 | 1 | 6 / 6 | 60 | 6 / 6 |
| **64** | **34** | **6 / 5** | **64** | **6 / 9** |
| **68** | **53** | **6 / 4** | **68** | **6 / 8** |
| **72** | **11** | **6 / 6** | **72** | **6 / 9** |
| **76** | **91** | **6 / 4** | **76** | **6 / 5** |
| **80** | **2** | **6 / 5** | **80** | **6 / 7** |
| **84** | **11** | **6 / 3** | **84** | **6 / 8** |
| **88** | **41** | **6 / 3** | **88** | **6 / 6** |
| **92** | **13** | **6 / 4** | **92** | **6 / 8** |
| 96 | 1 | 6 / 5 | 96 | 6 / 5 |

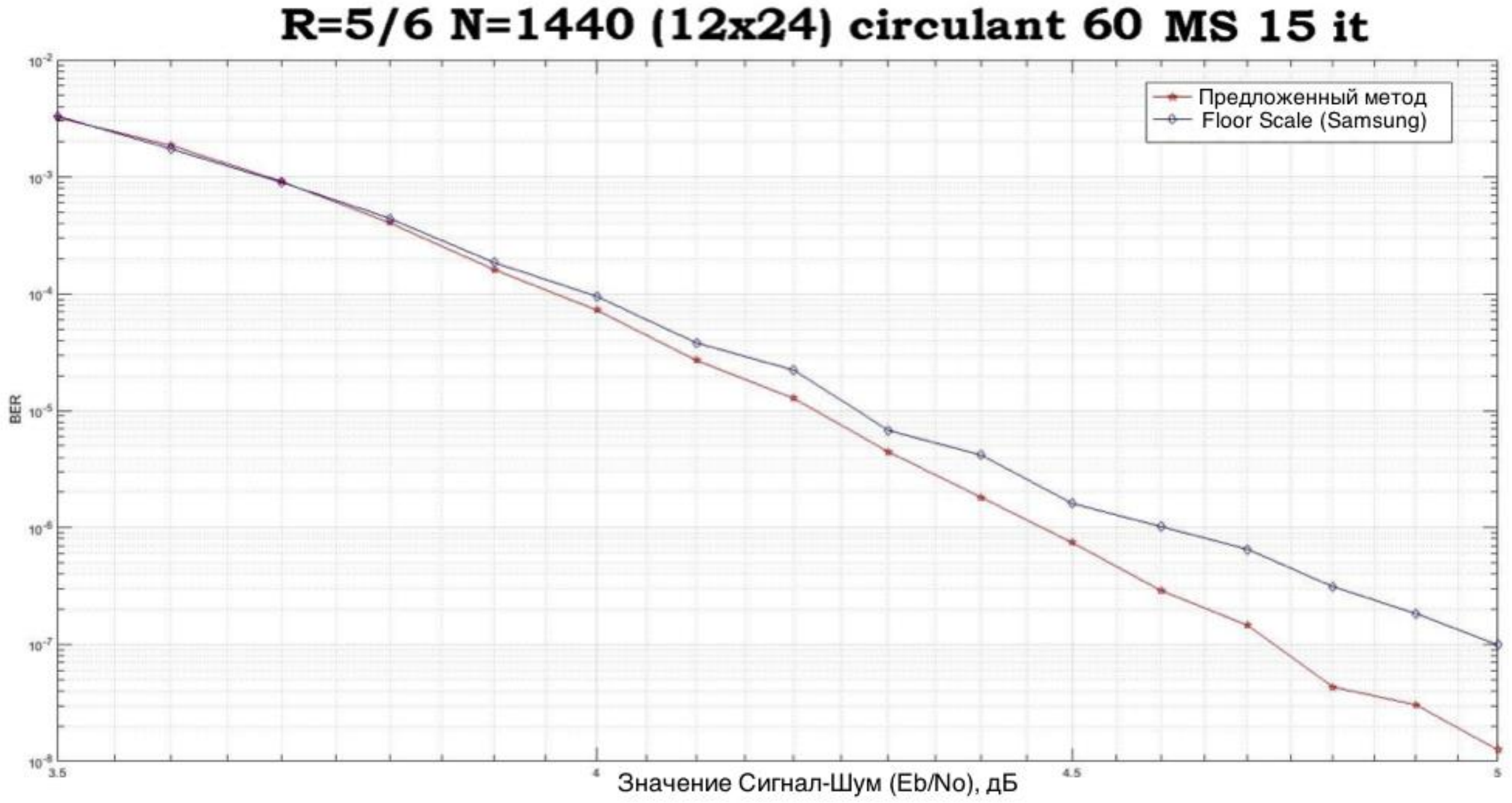

Рисунок 2.15 - Сравнение помехоустойчивости кода (вероятность ошибки на бит) со скоростью 5/6 используя 15 итераций алгоритма min-sum

Применяя предложенный метод адаптации по длине, удалось построить код с помехоустойчивость превосходящей коды А, Б на коротких длинах, где особенно наблюдается вклад коротких циклов, Рис. 2.16-2.17 [Код А, 51], [Код Б, 52].

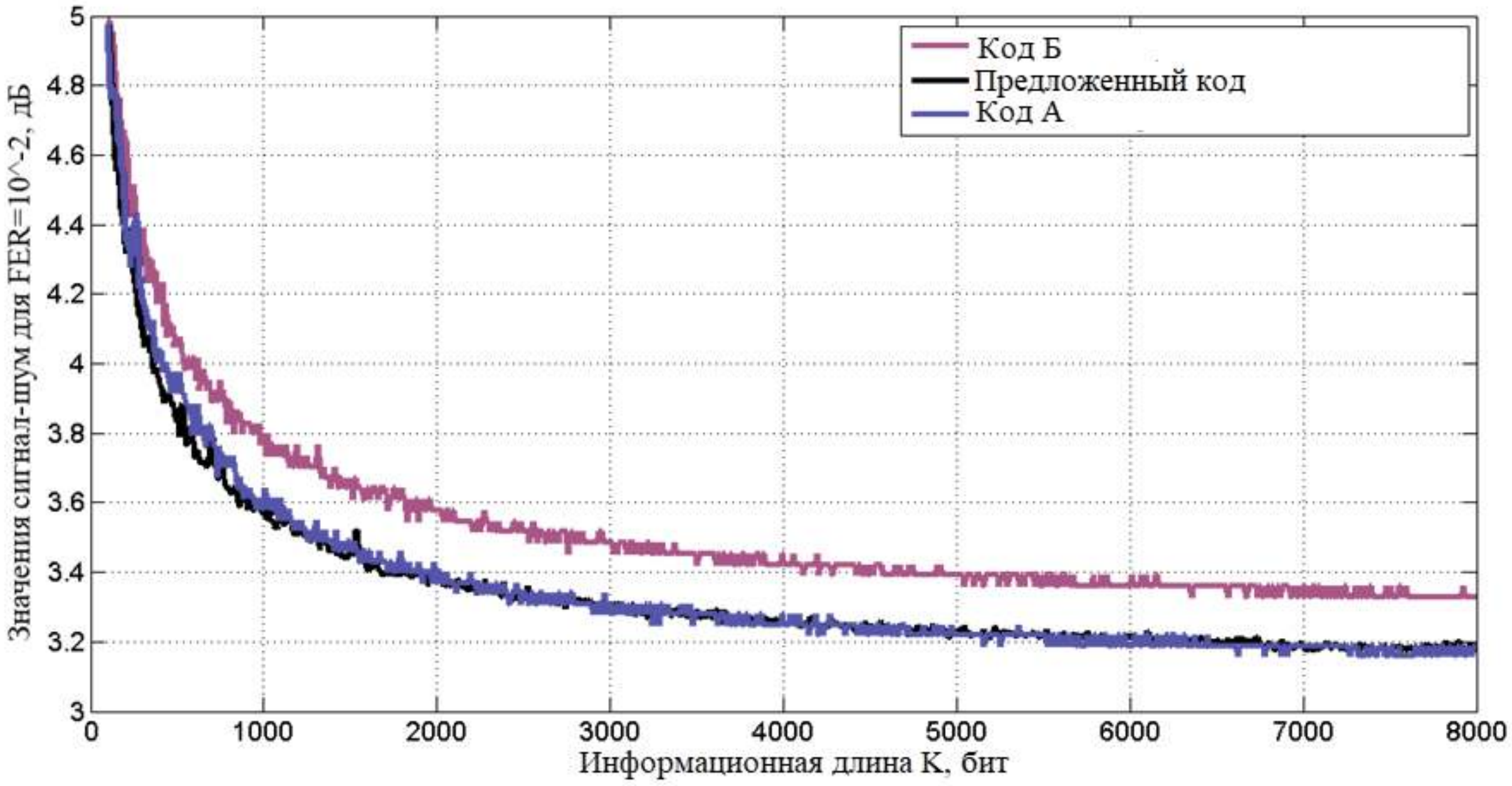

Рисунок 2.16 - Сравнение помехоустойчивости кодов различной информационной длины с уровнем блочной ошибки $10^{-2}$, используя 15 иттераций алгоритма min-sum, с шагом K=8 бит и со скоростью кода 0.67

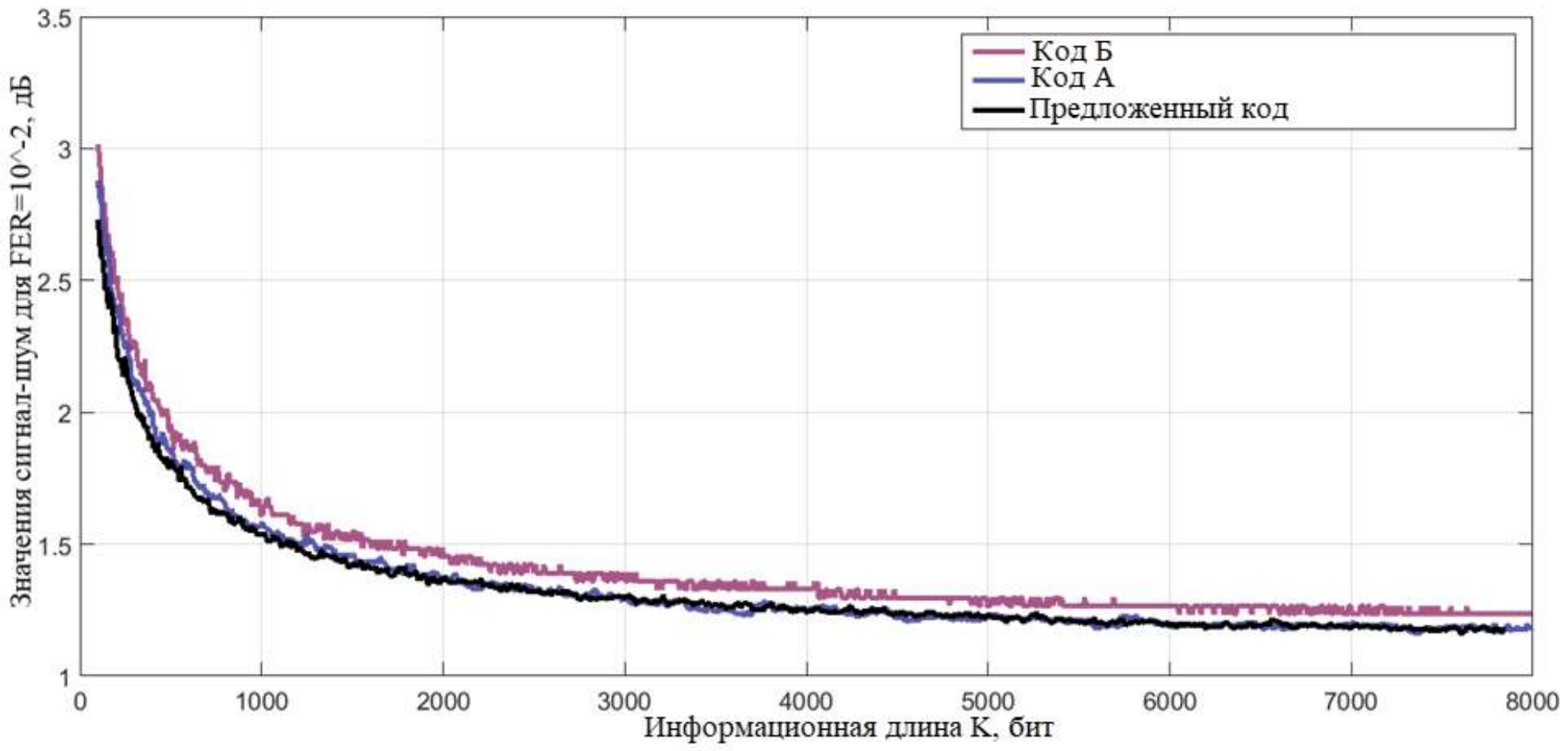

Рисунок 2.17 - Сравнение помехоустойчивости кодов различной информационной длины с уровнем блочной ошибки $10^{-2}$, используя 15 итераций алгоритма min-sum, с шагом K=8 бит и со скоростью кода 0.5

## 2.4 Выводы

Созданный метод построения низкоплотностных кодов, отличающийся использованием в фазе расширения протографа жадного метода запрещенных коэффициентов и метода имитации отжига обеспечивает дистантные свойства кодов и их спектры связности для фильтрации кодов кандидатов. Что позволяет позволяет улучшить дистантные свойства кодов (в первую очередь их кодовые расстояния) и их спектры связности для фильтрации кодов кандидатов.

Жадный алгоритм запрещенных коэффициентов быстро вычисляет начальное приближение расширенной проверочной матрицы. Оригинальность алгоритма имитации отжига в предлагаемом методе построения квазициклических низкоплотностных кодов заключается в пропуске значительного числа локальных минимумов числа треппин-сетов, что позволяет получить квазициклические коды с дистантными свойствами и спектром связности, не достижимые предложенными ранее методами, а также на порядок более высокую вероятность успешного расширения протографов.

Предложенный метод позволил увеличить вероятность успешного расширения базовой матрицы на порядок и уменьшить минимальный размер циркулянта на 10%.

Предложенный метод построение низкоплотностных кодов был защищен международными патентами: WO/2018/093286, EP3533145, US20190273511, CN110024294, WO/2017/105270.

Предложен метод адаптации по длине квазициклических двоичных низкоплотностных кодов, отличительной особенность которого является расширение множества допустимых циклических сдвигов в кольце циклических матриц перестановок. Метод позволил улучшить спектры связности, обеспечив тем самым дополнительный энергетическим выигрыш в 0.1 дБ. На метод адаптации длины были получены международные патенты WO/2018/030909, US 10,931,310 B2, EP3529899.

Вычислительная сложность расчета кодового расстояния известными методами на 6 этапе метода построения низкоплотнстных кодов растет экспоненциально от информационной длины кода $O(2^k)$, что может быть реализовано только на суперЭВМ. В связи с этим в диссертации разработан более быстродействующий метод вычисления кодового расстояния, использующий верхние и нижние оценки кодового расстояния и позволяющий по ним ранжировать коды кандидаты.

# 3 В РАМКАХ МЕТОДА ПОСТРОЕНИЯ НИЗКОПЛОТНОСТНЫХ КОДОВ СОЗДАНЫ ЧАСТНЫЙ МЕТОД ОЦЕНКИ КОДОВОГО РАССТОЯНИЯ И АППАРАТНО-ОРИЕНТИРОВАННЫЙ АЛГОРИТМ ОЦЕНКИ КОДОВОГО РАССТОЯНИЯ С ИСПОЛЬЗОВАНИЕМ ГЕОМЕТРИИ ЧИСЕЛ

## 3.1 Методы определения кодового расстояния помехоустойчивых блочных кодов

Можно выделить три основных класса методов оценки кодового расстояния:

- Классические переборные алгебраические методы. Одним из наиболее известных представителей этого класса является метод оценки кодового расстояния по информационным совокупностям Брауэра-Циммермана;
- Импульсные методы Берроу-Фоссорье и их модификации;
- Генетические методы.

Переборные алгебраические методы в процессе работы вычисляют оценки верхних и нижних границ кодового расстояния. Они позволяют осуществить верхнюю оценку времени, необходимого для получения оценки кодового расстояния. В случае применения полного перебора гарантируют точность оценки кодового расстояния. Именно эти свойства обуславливают их применение в качестве эталонных методов при построении кодов.

Импульсные методы эффективно работают на высокоскоростных итеративно декодируемых кодах, однако не гарантируют точность верхних и нижних границ кодового расстояния.

Генетические методы применяются к кодам, для которых применение импульсных и алгебраических методов не эффективно.

### 3.1.1 Метод Брауэра-Циммермана

Чаще всего на практике для оценки кодового расстояния используют метод Брауэра-Циммермана, представленный в работе [53] вместе с таблицей лучших по кодовому расстоянию известных линейных блочных кодов. Дальнейшее развитие метода предложено в работе [54, 55]. Применение этого метода позволило построить множество новых наилучших линейных блочных кодов [56]. Реализация метода Брауэра-Циммермана была осуществлена в алгебраическом пакете MAGMA. Подробное описание реализации этого метода в пакете алгебраическом пакете MAGMA было приведено в работе [57].

Дадим описание этого метода.

**Определение 3.1** Информационная совокупность. Пусть $C=[n,k,d]_q$ - линейный блочный код длины $n$ и размерности $k$ определенный над полем $F_q$. Подмножество $\mathrm{I}\subseteq\{1,2,...,n\}$ размера $|\mathrm{I}|=k$ является информационной совокупностью, если соответствующие столбцы порождающей матрицы $G$ кода $C$ линейно независимы.

Используя систематическую матрицу для кодирования векторов сообщений $i\in F_q^k$, нетрудно убедиться в истинности следующей леммы, лежащей в основе метода Брауэра-Циммермана.

**Лемма 3.1** Пусть имеется систематическая матрица $G=(I\,|\,A)$ и вектор сообщения $i\in F_q^k$. Тогда для кодового слова $c=iG=(i,iA)$ выполняется неравенство $wgt(c)\geq wgt(i)$, где $wgt(c)$ - Хэммингов вес кодового слова $c$.

В соответствии с методом Брауэра-Циммермана оценка кодового расстояния линейного блочного кода осуществляется путем перебора информационных сообщений, соответствующих информационной совокупности, с увеличением их веса и одновременным уточнением нижней и верхней границы веса кодового слова.

**Алгоритм 3.1** оценки кодового расстояния по одной информационной совокупности:

Вход: Порождающая матрица $G = (I \mid A)$;

Выход: Кодовое расстояние $d_{\min}$.

---

1. Инициализация переменных: $d_{lb} = 1$; $d_{ub} = n - k + 1$; $\omega = 1$;
2. Выполнять в цикле пока: $\omega \le k$ и $d_{lb} < d_{ub}$
3. Перебирать кодовые слова, $d_{ub} = \min\left(d_{ub}, \min\left\{wgt(iG) : i \in F_q^k \mid wgt(i) = \omega\right\}\right)$;
4. $d_{lb} = \omega + 1$;
5. $\omega = \omega + 1$;
6. Конец цикла
7. Вернуть $d_{\min} = d_{ub}$.

---

В этом алгоритме слова сообщений $i \in F_q^k, wgt(i) = \omega$ перечисляются в порядке возрастания веса Хэмминга $\omega$. Из Леммы 3.1 следует, что $wgt(c) \ge wgt(i) > \omega$, это определяет нижнюю оценку $d_{lb}$. В процессе работы алгоритма 3.1 осуществляется уточнение верхней и нижней оценок кодового расстояния. Процесс перебора кодовых слов останавливается, если все кодовые слова, удовлетворяющие условию $d_{lb} \le iG \le d_{ub}$, были перечислены.

Если кодовое расстояние больше $k+1$, то будут перечислены все $q^k$ слов. В этом случае единственный способ уменьшить перебор — найти способ улучшить верхнюю оценку.

Для этого можно воспользоваться следующим свойством линейного блочного кода: для систематического кодирования можно использовать любую порождающую матрицу, соответствующую какой-то информационной совокупности. Это позволяет ввести другие порождающие матрицы в алгоритм 3.1.

Пусть у нас имеется $\mathrm{I}_1, \mathrm{I}_2, \ldots, \mathrm{I}_m$ множество попарно различных информационных совокупностей кода $C$, соответствующих порождающим матрицам $G_j$. Запишем алгоритм оценки кодового расстояния по множеству непересекающихся информационных совокупностей.

**Алгоритм 3.2** оценки кодового расстояния по множеству непересекающихся информационных совокупностей:

---

Вход: Порождающие матрицы $G_j, j = 1...m$;

Выход: Кодовое расстояние $d_{\min}$.

---

1. Инициализация переменных: $d_{lb} = 1;\ d_{ub} = n - k + 1;\ \omega = 1;$
2. Выполнять в цикле пока: $\omega \le k$ и $d_{lb} < d_{ub}$
3. Выполнять в цикле по $j = 1$ до $m$
4. Перебирать кодовые слова

$d_{ub} = \min\left(d_{ub}, \min\left\{wgt\left(iG_j\right) : i \in F_q^k \mid wgt(i) = \omega\right\}\right);$

5. Конец цикла
6. $d_{lb} = m(\omega + 1);$
7. $\omega = \omega + 1;$
8. Конец цикла
9. Вернуть $d_{\min} = d_{ub}$.

---

Закодировав порождающими матрицами $G_j$ все сообщения $i$ веса $wgt(i) \le \omega$, соответствующие информационным совокупностям $\mathrm{I}_j$, легко убедиться, что оставшиеся кодовые слова различаются не более чем в $\omega + 1$ ненулевых элементах в позициях $\mathrm{I}_j$. Так как все информационные совокупности различны, получим нижнюю оценку на вес кодовых слов $d_{lb} = m(\omega + 1)$.

В сравнение с алгоритмом оценки кодового расстояния по одной информационной совокупности алгоритм 3.2 увеличивает число перебираемых кодовых слов веса $\omega$ в $m$ раз. Однако нижняя оценка минимального веса растет также в $m$ раз быстрее, что позволяет значительно снизить максимальное значение веса векторов сообщений.

В лучшем случае для кода $C=[n,k]_q$ у нас будет $\lfloor n/k \rfloor$ информационных совокупностей. Однако для произвольного кода с порождающей матрицей $G_1=(I\,|\,A_1)$, соответствующей первой информационной совокупности, ранг матрицы $A_1$ может быть меньше $k$, из чего следует, что информационной совокупности, не пересекающейся с $I_1$ не существует. В этой ситуации, мы можем получить частичную информационную совокупность $\mathrm{I}_2'$ размера $r_2=|\mathrm{I}_2'|=rank\ A_1$. Для того чтобы получить информационную совокупность $\mathrm{I}_2'\subseteq \mathrm{I}_2$, $k-r_2$ элементов $\mathrm{I}_1$ должны быть включены в $\mathrm{I}_2$. Оставшиеся кодовые слова содержат не более $\omega+1$ ненулевых символов в позициях $\mathrm{I}_1$ и $\mathrm{I}_2$.

Так как информационные совокупности пересекаются, и некоторые из позиций вычисляются дважды, нижняя оценка равна $2(\omega+1)-|\mathrm{I}_1\cap \mathrm{I}_2|$.

Эта оценка может быть обобщена на произвольное число информационных совокупностей $(\mathrm{I}_1,\mathrm{I}_2,\ldots,\mathrm{I}_m)$. Чтобы вычислить ее, введем относительной ранг $r_j$ информационной совокупности $\mathrm{I}_j$:

$$r_j=k-\left|\mathrm{I}_j\cap\bigcup_{l=1}^{j-1}\mathrm{I}_l\right|. \qquad (3.1)$$

Т.е. $r_j$ равны числу позиций в информационной совокупности $\mathrm{I}_j$, которые не содержаться ни в одной $\mathrm{I}_l$ такой что $l<j$. В случае $m=2$, рассмотренном ранее, порождающая матрица $G_j$ соответствует $(\omega+1)-(k-r_j)$ словам веса $\omega$. Рассмотренное улучшение алгоритма [58] позволяет использовать дополнительные порождающие матрицы, ускоряя рост нижней оценки:

Алгоритм 3.3 оценки кодового расстояния по множеству пересекающихся информационных совокупностей:

---

Вход: Порождающие матрицы $G_j, j=1...m$;

Выход: Кодовое расстояние $d_{\min}$.

---

1. Инициализация переменных: $d_{lb}=1;\ d_{ub}=n-k+1;\ \omega=1;$
2. Выполнять в цикле пока: $\omega\le k$ и $d_{lb}<d_{ub}$

3. Выполнять в цикле по $j=1$ до $m$

4. Перебирать кодовые слова

$$d_{ub}=\min\left(d_{ub},\min\left\{wgt\left(iG_j\right):i\in F_q^k \mid wgt(i)=\omega\right\}\right);$$

5. Конец цикла

6. $d_{lb}=\sum_{j=1}^{m}\max\left(0,(\omega+1)-\left(k-r_j\right)\right);$

7. $\omega=\omega+1;$

8. Конец цикла

9. Вернуть $d_{\min}=d_{ub}.$

---

В Магме применяется следующий метод вычисления информационных совокупностей. Первая порождающая матрица имеет вид $G_1=\left(I\mid A_1\right)$. Методом Гаусса из матрицы $G_1$ получается подматрица $A_1$ вида

$$A_1=\begin{pmatrix} I_{r_2} & B_{12} \\ 0 & 0 \end{pmatrix}, \tag{3.2}$$

Вторая порождающая матрица примет вид

$$G_2=\begin{pmatrix} B_{11} & 0 & I_{r_2} & B_{12} \\ B_{21} & I_{k-r_2} & 0 & 0 \end{pmatrix}, \tag{3.3}$$

где $r_2$ ранг $A_1$. Следующая порождающая матрица $G_3$ вычисляется методом Гаусса из подматрицы $B_{12}$ в $G_2$ аналогично (3). Процесс прекращается, если ранг следующей подматрицы стал равен 0, или информационные совокупности покрывают все позиции, т.е. $\bigcup_j \mathrm{I}_j=\{1,2,\ldots,n\}$. Для последовательности информационных совокупностей, полученных этим методом, последовательность рангов $r_j$ является не возрастающей и даже иногда принимает значение $r_j-1$.

Порождающие матрицы с небольшими значениями ранга вносят небольшой вклад в рост нижней оценки, поэтому они могут быть пропущены при вычислении. Определив заданную последовательностью $(r_1,r_2,\ldots,r_m)$ относительного ранга, можно вычислить последовательность нижних оценок $d_{lb}$ и вес $\omega_0$, при котором нижняя оценка пересечет верхнюю. Все порождающие матрицы с $r_j < k - \omega_0$ могут быть пропущены.

Кроме этого, используя последовательности нижних оценок, можно вычислить число кодовых слов, подлежащих перебору и определить количество времени, требуемое для оценки.

Время вычисления минимально, когда пересечения между совокупностями минимальны. Наилучшим является случай, когда имеется $m = \lceil n/k \rceil$ информационных совокупностей относительного ранга $r_1 = r_2 = \ldots = r_{m-1} = k$ и $r_m = n - (m-1)k$ . Если на первом шагу не удается получить хорошую последовательность рангов и достаточное число совокупностей, можно применить случайную перестановку символов кода.

Ключевым недостатком метода Брауэра-Циммермана является экспоненциальный рост вычислительной сложность относительно $k$. Например, для LDPC кода с информационной длиной $K = 1000$ метод требует проверки около $2^{200}$ кодовых слов. Это не позволяет его использовать при построении кодов для архивной голографической памяти.

### 3.1.2 Верхняя граница кодового расстояния методом Маккея-Вонтабеля-Смарандаши

Еще один алгебраический метод оценки кодового расстояния квазициклических низкоплотностных кодов был предложен Маккеем [59], Вонтабелем и Смарандаши [60]. В работе [59] для низкоплотностных кодов, расширенных циклической матрицей перестановок с весом циркулянта 1 (без пересечения между циркулянтами), с весом столбца $j$. Маккеем была получена верхняя оценка кодового расстояния $d_{\min} \le (j+1)!$. Там же было установлено, что величина определителя протографа до расширения (лифтинга) характеризует верхнюю оценку кодового расстояния.

Вонтобель и Смарандаши в работе [60] обобщили предложенную Маккеем идею верхней оценки кодового расстояния квазициклического кода и получили ее уточнение, опираясь на конкретные значения сдвигов в расширенной матрице.

Дадим краткое описание этого метода.

Перманентом матрицы B размером $m\, x\, m$ называется величина, вычисляемая по формуле [60]:

$$perm(B) = \sum_{\sigma} \prod_{j\epsilon[m]} b_{j,\sigma(j)}$$

где $\sigma$ берется по всем $m!$ возможным перестановкам элементам множества $[m]$.

Пусть $[\mathbf{W}]$ это множество значений сдвига циклической матрицы перестановки, принимающей значения из отрезка $[0...W-1]$. Обозначим сдвиги циркулянтов в расширенной матрице при помощи возведения в степень, $x$ будет соответсвовать сдвигу 0, $x^2$ - соответствовать сдвигу 1 и так далее, $x^W$ - соответствует сдвигу циклической матрицы перестановки $W-1$.

Теорема 7 из работы [60]. Пусть низкоплотностный квазициклический код, заданный проверочный матрицей $H(x)$ имеет протограф $\boldsymbol{wt}\big(C(H(x))\big)$, тогда верхняя оценка кодового расстояния вычисляется по формуле:

$$d_{\min}\left(C(H)\right) \le \min_{\substack{S\subseteq[W]\\ |S|=C+1}}{}^{+} \sum_{i\in S} wt\left(perm\left(H_{S\setminus i}(x)\right)\right), \tag{3.4}$$

где оператор $min^{+}()$ находит минимальное среди возможных положительных значений аргументов.

Теорема 8 из работы [60]. Верхняя оценка кодового расстояния квазициклического низкоплотностного кода на основе расширенной матрицы, $A = wt(H(x))$:

$$d_{\min}\left(C(H)\right) \le \min_{\substack{S\subseteq W\\ |S|=C+1}}{}^{+} \sum_{i\in S} perm\left(A_{S\setminus i}\right). \tag{3.5}$$

Рассмотрим пример. Пусть у нас имеется протограф, задающий квазициклический низкоплотностный регулярный код с весом строки 2 и весом столбца 4:

$$wt(C(H(x))) = \begin{pmatrix} 1 & 1 & 1 & 1 \\ 1 & 1 & 1 & 1 \end{pmatrix}.$$

Тогда в соответствие с теоремой 7 верхняя оценка кодового расстояния этого кода равна 6:

$$d_{min} \leq min^{+}\{(1+1)+(1+1)+(1+1)\} = 6.$$

Пусть у нас есть два регулярных кода $C(H_1)$, $C(H_2)$ с весом строки 4 и весом столбца 2 с проверочными матрицами, записанными в полиномиальной форме:

$$H_1(x) = \begin{pmatrix} x & x^2 & x^4 & x^8 \\ x^5 & x^6 & x^3 & x^7 \end{pmatrix}, H_2(x) = \begin{pmatrix} x & x^2 & x^4 & x^8 \\ x^6 & x^5 & x^3 & x^9 \end{pmatrix}.$$

Тогда в соответствие с теоремой 8 верхняя оценка кодового расстояния кода с проверочной матрицей $H_1$ равна 4:

$$d_{min}(C(H_1)) \leq min^{+}\left\{\begin{matrix} wt(x^4+x^9)+ wt(x^5+x^{10})+wt(x^7+x^7) \\ wt(x^9+x^{14})+ wt(x^8+x^{13})+wt(x^7+x^7) \\ wt(x^{11}+x^{11})+ wt(x^8+x^{13})+wt(x^4+x^9) \\ wt(x^{11}+x^{11})+ wt(x^9+x^{14})+wt(x^5+x^{10}) \end{matrix}\right\}$$

$$= min^{+}\{4,4,4,4\} = 4.$$

Кодовое расстояние кода, заданного проверочной матрицей $H_2$ равно 6, что соответствует верхней оценке протографа этого кода

$$d_{min}(C(H_2)) \leq min^{+}\left\{\begin{matrix} wt(x^5+x^9)+ wt(x^4+x^{10})+wt(x^6+x^8) \\ wt(x^{11}+x^{13})+ wt(x^{10}+x^{14})+wt(x^6+x^8) \\ wt(x^{13}+x^{11})+ wt(x^{10}+x^{14})+wt(x^4+x^{10}) \\ wt(x^{13}+x^{11})+ wt(x^{11}+x^{13})+wt(x^5+x^9) \end{matrix}\right\}$$

$$= min^{+}\{6,6,6,6\} = 6.$$

Баталёр и Сегель ускорили вычисление перманента [61] и обобщили верхнюю оценку на класс Multi-edge (Multigraph) низкоплотностных кодов [62].

Верхняя оценка кодового расстояния методом Вонтабеля-Смарандаши является вычислительно сложной задачей, сложность ее растет с размером протографа. Например, для протографа размера 30 на 60 нужно вычислить 1.1826e+17 перманентов размера 30x30. Нами был реализован параллельный алгоритм вычисления этой оценки.

Верхняя оценка Вонтабеля-Смарандаши не является точной (плотной) и требует дальнейшего уточнения.

Верхняя оценка Вонтабеля-Смарандаши будет использоваться в предлагаемых методах определения кодового расстояния с использованием геометрических решеток в качестве начальной длины искомого вектора в решетке, что в некоторых случаях позволяет уменьшить пространство перебора.

### 3.1.3 Нижняя оценка кодового расстояния Таннера

Первые оценки кодового расстояния в зависимости от обхвата графа были получены в работе Таннера [63].

В работе доказана нижняя оценка:

$$d_{\min} \geq 1 + \frac{d_{VN}\left[\left(d_{VN}-1\right)^{\lfloor (g-2)/4 \rfloor} - 1\right]}{d_{VN}-2}, \tag{3.6}$$

где $g$ - обхват графа, $d_{VN}$ - вес символа (столбца) в проверочной матрице.

Например, для Таннер-графа с обхватом 6 нижняя оценка кодового расстояния $d_{\min} \geq d_{VN} + 1$.

Одним из чрезвычайно эффективных средств отсева кодов-кандидатов (см. шаг 6 процедуры построения квазициклических низкоплотностных кодов 2.1) является улучшенная нижняя оценка кодового расстояния Таннера, опирающаяся на быстрый спектральный анализ графа [64]. Использование нижней оценки кодового расстояния позволяет остановить поиск кратчайшего вектора в решетке.

Пусть у нас имеется проверочная матрица кода с $r$ проверками и $n$ символами. Матрица смежности такого кода представляет собой матрицу размера $(r+n)\times(r+n)$:

$$A = \begin{pmatrix} 0 & H \\ H^T & 0 \end{pmatrix}. \tag{3.7}$$

Вычислим собственные числа матрицы $\mu_1 > \mu_2 > \dots > \mu_{r+n}$. Они представляют собой спектральную оценку Таннер-графа кода.

Тогда нижняя оценка кодового расстояния матрицы вычисляется по формулам

$$d \geq \frac{n(2m-\mu_2)}{(mj-\mu_2)},\ d \geq \frac{2n(2m+j-2-\mu_2)}{j(mj-\mu_2)}, \tag{3.8}$$

где $m$-вес символа (вес столбца) и $j$-вес проверки (вес строки).

Спектральная оценка также может выступать эвристикой в процессе построения кодов. Код, отношение второго собственного значения матрицы смежности которого к первому меньше ( $\min \frac{\mu_2}{\mu_1}$ ), является лучшим. Второе собственное значение характеризует алгебраическую связность графа, первое собственное значение характеризует сепарабельность графа на два подмножества. Отношение $\min \frac{\mu_2}{\mu_1}$ характеризует графы экспандеры [65, 66]. Обобщение нижней оценки кодового расстояния Таннера на случай нерегулярных кодов предложено в работе [67].

## 3.2 Решетки и их свойства

В предложенном нами в разделе 2 методе построения низкоплотностных кодов определение кодового расстояния осуществляется при помощи поиска кратчайшего вектора в решетке. Приведем необходимые сведения из геометрии чисел.

**Определение 3.2.** Решетка - дискретная абелева подгруппа, заданная в пространстве $R^n$.

Пусть базис $B = \{b_1, ..., b_n\}$ задан линейно-независимыми векторами в $R^m$. Тогда под решеткой будем понимать множество целочисленных линейных комбинаций этих векторов:

$$L(b_1, ..., b_n) = \{\sum_{i=1}^{n} x_i b_i : (x_1, ..., x_n) \in Z^n\}, \tag{3.9}$$

где $m$ и $n$, размерность и ранг решетки соответственно, $m \geq n$.

**Определение 3.3.** Задача поиска короткого вектора ( $\Delta$ -short vector problem, $SVP_\Delta(m)$ ): Пусть дана $m$-мерная решетка $L(B)$, ранга $n$ и вещественное $\Delta > 1$. Найти нетривиальный вектор в $\Delta$ -раз больший кратчайшего вектора в решетке $\bar{b} \in L : \|\bar{b}\| \leq \Delta \cdot \lambda_1(L)$.

При $\Delta = 1$, решается задача поиска кратчайшего вектора в решетке, при $\Delta > 1$ - короткого вектора.

**Определение 3.4.** Задача поиска короткого базиса решетки ($\Delta$-short basis problem, $SBP_{\Delta}(m)$): Пусть дан базис полной решетки $B$ и вещественное $\Delta > 1$. Найти базис $B' = \{b_1^{'}, b_2^{'}, ..., b_m^{'}\} : L(B) = L(B'), \prod_{i=1}^{m} \|b_i^{'}\| \leq \Delta \cdot \prod_{i=1}^{m} \|b_i^{\perp}\|$.

Подробнее методы геометрии чисел: алгоритм поиска короткого базиса решетки (блочный метод Коркина-Золотарева) и алгоритм поиска кратчайшего вектора рассмотрены в работах [27].

**Определение 3.5.** Базис $B = \{b_1, b_2, \ldots, b_m\}$ решётки $L \subset R^m$ приведён по длине, если в результате ортогонализации решетки методом Грамма-Шмидта выполняется следующее неравенство:

$$|\mu_{i,j}| \leq \frac{1}{2}, 1 \leq j < i \leq m,$$

где $\mu_{i,j}$ - коэффициенты Грамма-Шмидта.

**Определение 3.6.** Упорядоченный по длине базис $B = \{b_1, b_2, \ldots, b_m\}$ решётки $L \subset R^m$ приведён блочным методом Коркина- Золотарева с размером блока $\beta \in [2, m]$ и точностью $\delta \in \left(\frac{1}{2}; 1\right]$, если:

- базис $B$ приведён по длине;

- $\delta^2 \cdot \|b_i^{\perp}\|^2 \leq \lambda_1^2(L_i), i = 1, \ldots, m,$ где $\lambda_1(L_i)$ -длина кратчайшего вектора в решётке $L_i$, образованной ортогональным дополнением векторного пространства с базисом $b_i, \ldots, b_{\min(i+\beta-1, m)}$.

**Определение 3.7.** Базис решетки приведен по Ленстра-Ленстра-Ловасу (LLL-алгоритмом), если он приведен блочным методом Коркина-Золотарева с размером блока $\beta = 2$ и точностью ортогонализации $\delta \in \left(\frac{1}{2}; 1\right)$.

Пусть базис решетки имеет вид:

$A = (a_1, a_2, \ldots, a_n) \in R^{m \times n}$, где $a_i$ - вектора столбцы.

**Определение 3.8**. Матрица $A$ размера $n \times n$ ортогональна, тогда и только тогда, когда $AA^T = I$.

**Определение 3.9**. QR-разложением называют представление матрицы $A=(a_1,a_2,\ldots,a_n)$ размера $m\times n$ в виде произведения $Q$ -унитарной и $R$ -верхней треугольной матрицы размерами $m\times n$ и $n\times n$, соответственно:

$$A=(a_1,a_2,\ldots,a_n)=Q^T\times R=(q_1,q_2,\ldots,q_n)\begin{pmatrix} r_{1,1} & r_{1,2} & \cdots & r_{1,n} \\ 0 & r_{2,2} & \cdots & r_{2,n} \\ \vdots & \ddots & \ddots & \vdots \\ 0 & \cdots & 0 & r_{n,n} \end{pmatrix}. \qquad (3.10)$$

Если $A$ состоит из вещественных чисел, то $Q$ - является ортогональной матрицей.

Ортогональные матрицы обладаю рядом следующих свойств необходимых нам в дальнейшем:

**Свойство 3.10.** Произведение двух ортогональных матриц, есть ортогональная матрица.

**Свойство 3.11.** $A^{-1}=A^T$, $A^TA=I$.

**Свойство 3.12.** Ортогональное преобразование сохраняет скалярное произведение векторов, т.е. $\forall x,y\in R^n: y^Tx\overset{def}{=}(x,y)=(Ax,Ay)$, в частности, сохраняется норма вектора: $\|Ay\|=A\|y\|$.

3.2.1 Метод определения кодового расстояния с использованием геометрических решеток

Для определения кодового расстояния необходимо найти в коде слова малого веса.

Предлагаемый метод для решения этой задачи предусматривает вложение кода в решетку, масштабирование $n-k$ подпространств базиса решетки $B_c$ некоторой константой $N$ такой, чтобы расстояние Хэмминга между кодовыми словами отобразилось в евклидово расстояния между векторами кодовых слов решетки. Легко убедится, что полученная решетка – вырожденная, ранга $rank(B_c)=k$.

Осуществив поиск короткого базиса решетки методом Коркина-Золотарева, получим некоторый набор векторов малого веса с нормой $r_{\max}$.

Вектор решетки с наименьшей нормой соответствует предполагаемому кратчайшему вектору. Однако пока мы не осуществим перебор всех линейных комбинаций методом Каннана-Финке-Поста и не убедимся, что полученная норма вектора является наименьшей возможной в решетке, данное кодовое слово будем считать словом малого веса.

Количество отличных от нуля компонент кратчайшего вектора дает искомое кодовое расстояние. Оценим масштабирующий коэффициент $N$ необходимый для корректной работы метода:

**Теорема 3.4.** Базис решетки $L(B_c)$ несистематического кода, $\delta$ - приведённый по Ленстра-Ленстра-Ловасу, в качестве приведенных коротких базисных векторов решетки будет давать слова малого веса несистематического кода, если константа:

$$N \geq \left(\frac{4}{4\delta-1}\right)^{n/2} \times 2\sqrt{(m+1)(n+1)} \times r_{\max} \times M^m, \tag{3.11}$$

где $M = \max\{\|A_0\|, \|A_1\|, \ldots, \|A_{n-1}\|, \|d\|\}$, $A_i$ – коэффициенты эквивалентного линейного диофантова уравнения $A \times x = d$.

Доказательство теоремы приведено в работе [82]. Оценим $N$ в случае приведения базиса блочным методом Коркина-Золотарева с размером блока $\beta$.

**Теорема 3.5.** Базис решетки $L(B_c)$ несистематического кода, приведённый блочным методом Коркина-Золотарева с размером блока $\beta$, в качестве приведенных коротких базисных векторов решетки будет давать слова малого веса несистематического кода, если константа:

$$N \geq \gamma_\beta^{\frac{n-1}{\beta-1}} \times 2\sqrt{(m+1)(n+1)} \times r_{\max} \times M^m, \tag{3.12}$$

где $\gamma_\beta$ -константа Эрмита.

**Доказательство**:

Первая часть доказательства берется из доказательства Теоремы 3.4, [82]. Длина векторов, образующих избыточности в исходном коде, ограничена сверху величиной $\|V_i\| \leq 2\sqrt{(n+1)(m+1)} \times r_{\max} \times M^m, 0 \leq i \leq n-k$.

Для доказательства оставшейся части нам необходимо оценить сверху длину векторов избыточности решетки после приведения ее базиса блочным методом Коркина-Золотарева с размером блока $\beta$, $\bar{b}_i \in L' : 0 \le i \le n-k$. В работе [83] Шнор доказал, что длина приведенных векторов ограничена сверху величиной $\|b_i\|^2 \le \gamma_\beta^{2\frac{n-1}{\beta-1}} \dfrac{i+3}{4} \lambda_i^2$, $\lambda_i$ - $i$-соответствующий минимум решетки. Подстановкой получим искомую оценку.

В случае систематической порождающей матрицы кода, легко убедиться, что константа $N=1$.

Предлагаемый метод поиска кодовых слов минимального веса требует построения решетки на основе порождающей матрицы систематического кода $G \in R^{k\times n}, B_c \in R^{(n+k)\times(k+n)}$:

$$B_c = \begin{pmatrix} G^T & qI_n \\ I_k & 0 \end{pmatrix}, \tag{3.13}$$

В случае несистематического кода:

$$B_c = \begin{pmatrix} N\cdot G' & N\cdot qI_n \\ I_k & 0 \end{pmatrix}, \tag{3.14}$$

где $G'$ - порождающая матрица несистематического кода, $q$ - размерность алфавита, $I_k$ - единичная матрица размера $k\times k$.

После построения решетки в ней ищется вектор $\|v\|_\infty = 1,\ v \in L,\ \ rank(L) = n$ с минимальным числом отличных от нуля координатных компонент. Переход от метрики Евклида к метрике Чебышева дает нам эквивалентную формулировку задачи поиска слов минимального веса в случае решетки, а именно поиск среди точек решетки, образованных различными $2^n$ шарами с центрами $\left(\pm\frac{d}{2},\dots,\pm\frac{d}{2}\right)$ и радиусами $\frac{d}{2}\sqrt{n}$, [84]. Кодовые слова двоичного $q=2$, $-1 \equiv 1 \bmod 2$ и троичного кодов $q=3$, $-1 \equiv 2 \bmod 3$ будут представлены векторами решетки, с координатными компонентами, принимающими значения из множества $\{-1,0,1\}$.

В соответствии с предлагаемым методом определения кодового расстояния с использованием геометрических решеток выполняется следующая последовательность действий.

1. *Вложение кода в решетку.* Для этого базис решетки $B_c$ масштабируется некоторой константой $N$ такой, чтобы в результате приведения базиса $B_c^T \in R^{(n+k)\times(k+n)}$ решетки получился базис размера $n \times n$:

$$B_c^T = \begin{pmatrix} N \cdot G & I_k \\ N \cdot q \cdot I_n & 0 \end{pmatrix}, \quad (3.15)$$

где $G \in F_q^{k \times n}$ - порождающая матрица кода, $I_k$ - единичная матрица, $B_c^T$ - транспонированный базис решетки.

В случае систематического кода $G = (I_k \mid P), P \in F_q^{k \times (n-k)}$, масштабирующий коэффициент $N$ равен 1.

2. *Приведение базиса решетки.* Для приведения решетки используется блочный метод Коркина-Золотарева с размером блока $\beta$. После удаления линейно зависимых строк и столбцов получается некоторый короткий базис решетки $\bar{b}_0, \bar{b}_1, ..., \bar{b}_{n-m} \in R^{n+1}$, $rank(L(b_0, b_1, ..., b_{n-m})) = n - m + 1$.

3. *Ортогонализация базиса решетки $B_c$ методами QR-разложения* с целью получения ортогонального базиса решетки $\bar{b}_0^{\perp}, \bar{b}_1^{\perp}, ..., \bar{b}_{n-m}^{\perp}$ и коэффициентов Грама-Шмидта.

Ортогонализация базиса по Граму-Шмидту выполняется следующим образом:

$$b_1^{\perp} = b_1,\ b^{\perp}{}_i = b_i - \sum_{j=1}^{i-1} \mu_{i,j} b_j^{\perp}, i = 2, 3, ..., n, \quad (3.16)$$

где $\mu_{i,j}$ -коэффициенты Грама-Шмидта вычисляемые по формуле

$$\mu_{i,j} = \frac{\langle b_i, b_j^{\perp} \rangle}{\langle b_j^{\perp}, b_j^{\perp} \rangle} = \frac{\langle b_i, b_j^{\perp} \rangle}{\left\| b_j^{\perp} \right\|^2}. \quad (3.17)$$

Коэффициенты Грама-Шмидта образуют верхнюю треугольную матрицу с $n \times (n-1)/2$ ненулевыми элементами.

4. *Поиск кратчайшего вектора в решетке*. После чего в решетке ищется вектор с минимальным числом отличных от нуля координатных компонент. Количество отличных от нуля компонент найденного вектора равняется искомому кодовому расстоянию.

Задача поиска кратчайшего вектора $x$ в решетке сводится к целочисленному решению системы неравенств:

$$\begin{cases} x_n^2 \left\| b_n^{\perp} \right\|^2 \le A^2, \\ (x_{n-1} + \mu_{n,n-1} x_n) \left\| b_{n-1}^{\perp} \right\|^2 \le A^2 - x_n^2 \left\| b_n^{\perp} \right\|^2 \\ \dots \\ (x_1 + \sum_{i=2}^{n} x_i \mu_{i,j})^2 \left\| b_1^{\perp} \right\|^2 \le A^2 - \sum_{j=2}^{n} l_j \end{cases}, \qquad (3.18)$$

где $A$ -верхняя оценка кодового расстояния (начальная норма $A = (n+1) \times r_{\max}^2$), $x_i$ - координатная компонента искомого вектора $l_j = (x_j + \sum_{i=j+1}^{n} x_i \mu_{i,j})^2 \left\| b_j^{\perp} \right\|^2$ -частичная сумма.

Для этой цели используется метод Каннана-Финке-Поста (КФП).

Метод КФП представляет собой вариант метода ветвей и границ и заключается в переборе линейных комбинаций базисных векторов решётки, дающих вектор с нормой, ограниченной сверху оценкой $A$ , которая может уменьшаться в процессе поиска.

Для ускорения ортогонализации базиса решетки (этап 3 метода оценки кодового расстояния), предложено применять вместо модифицированного метода Грамма-Шмидта, параллельные методы QR-разложения матриц: блочный метод Хаусхолдера при использовании многоядерных процессоров и метод поворота Гивенса при использовании видеокарт.

Метод Хаусхолдера представляет собой линейное преобразование векторного пространства, которое описывает его отображение относительно гиперплоскости, которая проходит через начало координат. Этот метод эффективен в случае невысокого уровня параллелизма и размещения данных в кэш-памяти. Его целесообразно применять при использовании многоядерных процессоров общего назначения с мощным устройством управления, оснащённых кэш памятью большого объема, характерных для X86 архитектур.

Метод ортогонализации поворотом Гивенса, осуществляет обнуление координатных компонент. Метод легко распараллеливаем и особенно эффективен в случае низкоплотностных матриц. Эти свойства обуславливают его чрезвычайную эффективность в случае реализации вычислений на видеокартах.

В диссертации выполнено сравнение быстродействия последовательного модифицированного метода Грама-Шмидта, преобразования Хаусхолдера, выполняемых на многоядерных процессорах общего назначения (Intel Math Kernel Library i5-7700K), и вращения Гивенса, реализуемого на видеоускорителе NVidia GeForce 1070. Методы QR-разложения приведены в [138], реализация в Приложение 4.

Метод Гивенса, исполняемый на видеоускорителе, обеспечивает 895-кратное ускорение работы алгоритма ортогонализации в сравнение с модифицированным последовательным методом Грама-Шмидта и 5.5-кратное ускорение в сравнение с многоядерной реализацией (8-потоков) метода Хаусхолдера на размерах ортогонализуемого базиса порядка 5500, Табл. 3.1.

Таблица 3.1 - Сравнение времени работы алгоритмов ортогонализации базиса решеток, в секундах

| Размер базиса решетки | Поворот Гивенса | Преобразования Хаусхолдера | Метод Грама-Шмидта |
|---|---|---|---|
| 110 | 0.0027 | 0.0002 | 0.0029 |
| 160 | 0.0031 | 0.0003 | 0.0093 |
| 210 | 0.0041 | 0.0006 | 0.0211 |
| 260 | 0.0043 | 0.0008 | 0.0392 |
| 310 | 0.0046 | 0.0012 | 0.0662 |
| 360 | 0.0048 | 0.0014 | 0.0819 |
| 410 | 0.0051 | 0.0021 | 0.1355 |
| 460 | 0.0056 | 0.0025 | 0.1918 |
| 510 | 0.0065 | 0.0037 | 0.2515 |
| 660 | 0.0075 | 0.0063 | 0.5039 |
| 710 | 0.0084 | 0.0090 | 0.6283 |
| 760 | 0.0093 | 0.0097 | 0.7925 |
| 810 | 0.0107 | 0.0112 | 0.9362 |
| 860 | 0.0136 | 0.0129 | 1.1160 |
| 910 | 0.0222 | 0.0142 | 1.1367 |
| 960 | 0.0123 | 0.0170 | 1.5428 |
| 1000 | 0.015 | 0.02 | 1.7644 |
| 1500 | 0.02 | 0.06 | 5.9053 |
| 2000 | 0.04 | 0.13 | 14.4604 |
| 2500 | 0.06 | 0.22 | 28.1351 |
| 3000 | 0.10 | 0.50 | 48.7143 |
| 3500 | 0.15 | 0.57 | 78.0097 |
| 4000 | 0.17 | 0.80 | 119.3055 |
| 4500 | 0.23 | 1.15 | 165.1248 |
| 5000 | 0.28 | 1.47 | 227.5358 |
| 5500 | 0.35 | 1.90 | 313.0146 |

В случае размеров базиса около 100 многоядерная реализация (8-потоков) метода Хаусхолдера обеспечивает 10 кратное устроение работы алгоритма, метод Гивенса на видеокарте не дает ускорение. Причиной отсутствия ускорения являются затраты времени на копирование базиса для ортогонализации из ОЗУ ЭВМ в ОЗУ видеокарты.

Для 32000-мерной решетки, соответствующей низкоплотностному коду страничного уровня голографической памяти, работающему в условиях больших шумом, справедливы следующие оценки: по сравнению с последовательным методом Грамма-Шмидта преобразование Хаусхолдера на многоядерных процессорах демонстрирует прирост скорости ортогонализации на два порядка, метод вращения Гивенса, исполняемый на процессорах видеокарт, обеспечивает дополнительный прирост еще на порядок.

Поиск кратчайшего вектора в решетке (этап 4) является самой вычислительно сложной частью предложенного метода оценки минимального кодового расстояния. Для реализации поиска используется метод Каннана-Финке-Поста (КФП). Рассмотрим его подробнее.

Решение задачи поиска кратчайшего вектора в решетке по методу КФП заключается в полном переборе всех линейных комбинаций векторов базиса решетки $\|x\|^2 = \left\|\sum_{i=1}^{m} x_i b_i\right\|^2 \le A^2, x_i \in Z$, где $A$ - норма искомого кратчайшего вектора. В качестве нормы берётся верхняя оценка длины кратчайшего вектора, $A = \sqrt{\gamma_m}\det(L)^{\frac{1}{m}}$, где $\gamma_m$ -константа Эрмита, в тех случаях, когда наименьший из векторов в базисе решетки превосходит оценку, $\|b_1\| > \sqrt{\gamma_m}\det(L)^{\frac{1}{m}}$. По этой причине предварительное приведение базиса решетки позволяет уменьшить пространство перебора $x_i$ . В случае оценки кодового расстояния в квазициклических кодах, мы можем брать гораздо более плотную верхнюю оценку Вонтобеля-Смарандаши, $A = d_{min}(C(H))$, (см. 3.1.2).

С целью уменьшения пространства перебора распишем базис решетки через ортогональные вектора, для простоты изложения используя метод ортогонализации Грамма-Шмидта. Получим $b_i = \sum_{j=1}^{i} \mu_{i,j} b_j^{\perp}, 2 \le i \le m, 1 \le j < i \le m$, где $\mu_{i,j}$ - коэффициенты Грама-Шмидта. В этом случае поиск кратчайшего вектора сводиться к решению системы неравенств:

$$\begin{cases} x_m^2 \left\|b_m^{\perp}\right\|^2 \le A^2, \\ (x_{m-1} + \mu_{m,m-1} x_m)\left\|b_{m-1}^{\perp}\right\|^2 \le A^2 - x_m^2\left\|b_m^{\perp}\right\|^2 \\ \dots \\ (x_1 + \sum_{i=2}^{m} x_i \mu_{i,j})^2 \left\|b_1^{\perp}\right\|^2 \le A^2 - \sum_{j=2}^{m} l_j \end{cases}, \text{ где } l_j = (x_j + \sum_{i=j+1}^{m} x_i \mu_{i,j})^2 \left\|b_j^{\perp}\right\|^2 \quad (3.19)$$

и выбору одного из целочисленных векторов, у которого норма скалярного произведения с базисом решетки минимальна.

Формализуя задачу, получим обход дерева от корня к листу, в каждой из вершин которого решается соответствующее линейное уравнение. Из корня этого дерева выходит $2\cdot\left\lceil\frac{A}{\left\|b_m^{\perp}\right\|}\right\rceil=2\cdot\left\lceil\frac{\sqrt{\gamma_m}\det(L)^{\frac{1}{m}}}{\left\|b_m^{\perp}\right\|}\right\rceil$ ветвей или $2\cdot\left\lceil\frac{\left\|b_1\right\|}{\left\|b_m^{\perp}\right\|}\right\rceil$ ветвей в случае предварительного приведения базиса решетки. В силу симметричности дерева (по свойствам нормы) для получения искомого кратчайшего вектора нам необходимо перебрать только половину его вершин, см. рис. 3.1. В результате полного обхода дерева от корня к листу, мы будем получать предполагаемый кратчайший вектор $x$ с нормой меньше либо равной искомой. Если норма полученного вектора будет меньше заданной ранее, целесообразно обновить ее с целью уменьшения пространства перебора. Остановка алгоритма осуществляется, когда завершен обход вершин дерева, или когда мы получили вектор с достаточной для нас нормой в случае поиска короткого вектора.

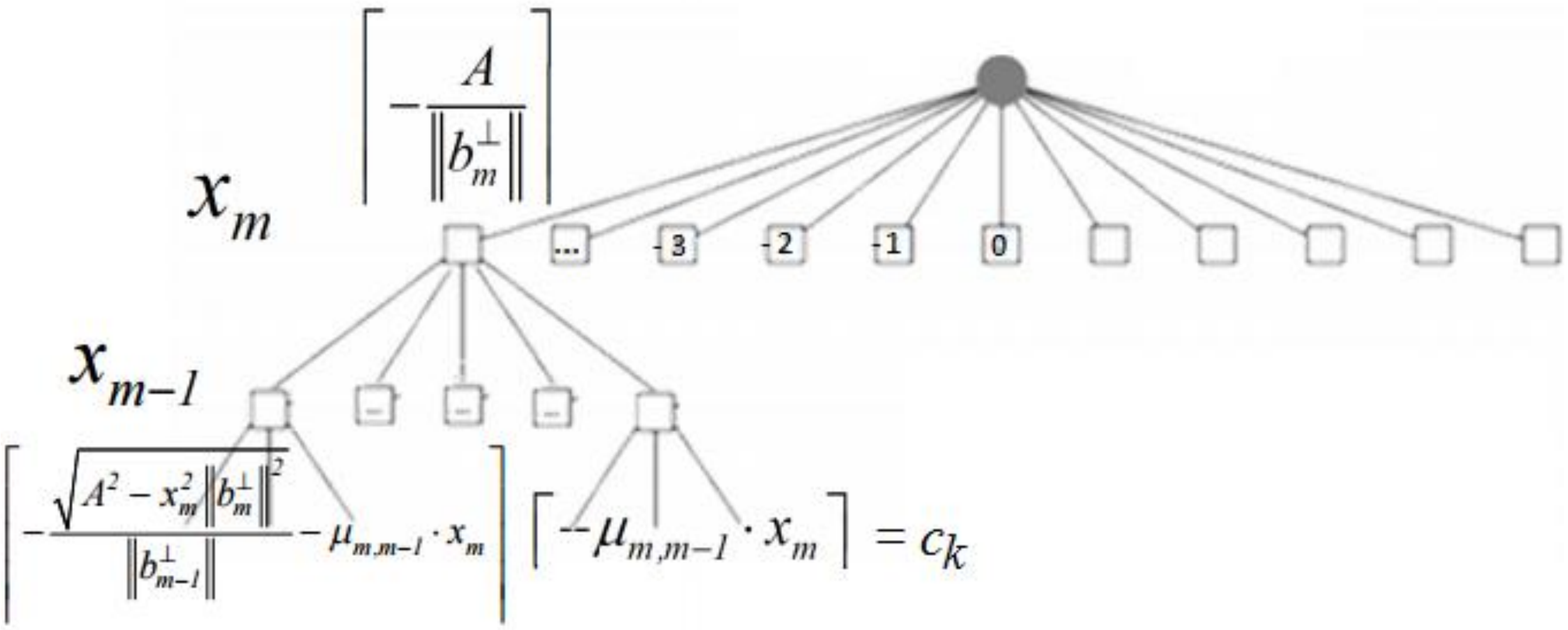

Рисунок 3.1 - Дерево перебора в методе Каннана-Финке-Поста

Из корня дерева выходит $N$ ветвей, каждая из которых соответствует конкретному решению первого неравенства. При этом существует Гауссова эвристика, позволяющая с достаточно высокой точностью оценить число веток на каждом из уровней дерева перебора, $H_k=\frac{1}{2}\times\frac{V_k(A_k)}{\prod_{i=m+1-k}^{m}\left\|b_i^{\perp}\right\|}$, [78]. Пусть $\alpha_m$ — произвольное решение этого неравенства. Ветвь дерева поиска, соответствующую данному решению, будем называть $\alpha_m$-ветвью. Следующая вершина $\alpha_m$-ветви помечается переменной $x_{m-1}$. Данная вершина является корнем дерева поиска для системы неравенств, полученной подстановкой $x_m = \alpha_m$. Блок схема метода КФП приведена на рисунке 3.2.

Описанный метод КФП представлен ниже в виде алгоритма 3.6, реализованного с использованием потоковой модели NPTL. Каждый из потоков осуществляет вычисление своей $\alpha_m$-ветви, исходящей из корня дерева, [96, 138].

**Алгоритм 3.6** поиска кратчайшего вектора в решетке (Каннана-Финке-Поста) $L(B), B = \{b_1, b_2, \dots b_m\}$:

**Вход:** $\|b_1^\perp\|, \|b_2^\perp\|, \dots, \|b_m^\perp\|$, $\mu_{i,j} : 1 \le j < i \le m$, $A$;

**Выход:** Вектор $x = (x_1, x_2, ..., x_m) \in Z^m : \left\|\sum_{i=1}^m x_i b_i\right\| = \lambda_1(L(B))$;

Инициализация переменных:

$x^m = (0)^m, x_1 = 1, c^m = (0)^m, l^{m+1} = (0)^{m+1}, y^m = (0)^m, \Delta x = x, \Delta^2 x = (-1)^m, \Delta^2 x(1) = 1,\ i = 1.$

Повторять в цикле:

1. $l_i = l_{i+1} + (x_i - c_i)^2 \times \|b_i^\perp\|^2$

2.**Если** $l_i \le A^2$ и $i = 1$ то

возвращаем $x^m$, обновляем значение нормы кратчайшего вектора $A = \|x^m\|$.

Запускаем программу вновь с нормой $A = A - 1$.

3.**Если** $l_i \le A^2$ и $i > 1$

$i = i - 1$ (Опускаемся на размерность ниже по дереву), $c_i = -\sum_{j=i+1}^{m} x_j \mu_{j,i}$,

$c_i = -c_i$

$x_i = round(c_k)$,

$\Delta x_i = 0.$

**Если** $c_i < x_i$ то $\Delta^2 x_i = 1;$

**Иначе** $\Delta^2 x_i = -1;$

**Иначе Если** $l_i > A^2$ и $i = m$ то

**Иначе** $i = i + 1,\ \Delta^2 x_i = -\Delta^2 x_i,\ \Delta x_i = -\Delta x_i + \Delta^2 x_i,\ x_i = x_i + \Delta x_i$.

4. Не существует линейной комбинации, дающей вектор в решетке короче ранее найденого. Останавливаем программу, возвращаем вектор $x^m$ с нормой $A = A+1$. Если ранее вектор не был найден, значит кратчайшим вектором решетки является самый короткий из векторов исходного базиса решетки.

Отметим, что зигзагообразный обход узлов предложенный Шнором-Охнером (Schnorr-Euchner zig-zag path) значительно ускоряет работу метода Каннана-Финке-Поста [79]. Однако в силу неоднородности нагрузки каждого вычислительного ядра требует более сложной декомпозиции решетки (дерева перебора).

Для повышения быстродействия процедуры оценки минимального кодового расстояния в диссертации предлагается параллельная процедура поиска кратчайшего вектора в решетке, основанного на КФП.

Сущность параллельного поиска заключается в предварительном вычислении значений координатных компонент дерева перебора, разбиение дерева и параллельный перебор на многоядерных процессорах общего назначения или процессорах видеокарт.

Для выполнения паралельного поиска кратчайших векторов мы будем хранить верхнюю треугольную матрицу Грамма-Шмидта, состоящую из $m\times(m-1)/2$ ненулевых элементов битовой точности $p$ для каждого из исполнительных элементов (ядер). Таким, образом каждое из ядер потребует $p\times m\times(m-1)/2$ бит памяти. Поскольку точность битовой представления коэффициэнтов будет влиять на сложность каждого из ядер, один из подходов заключается в аппостериорном оценки точности на основе моделирования процесса приведения базиса решетки, предложенного в работе [81].

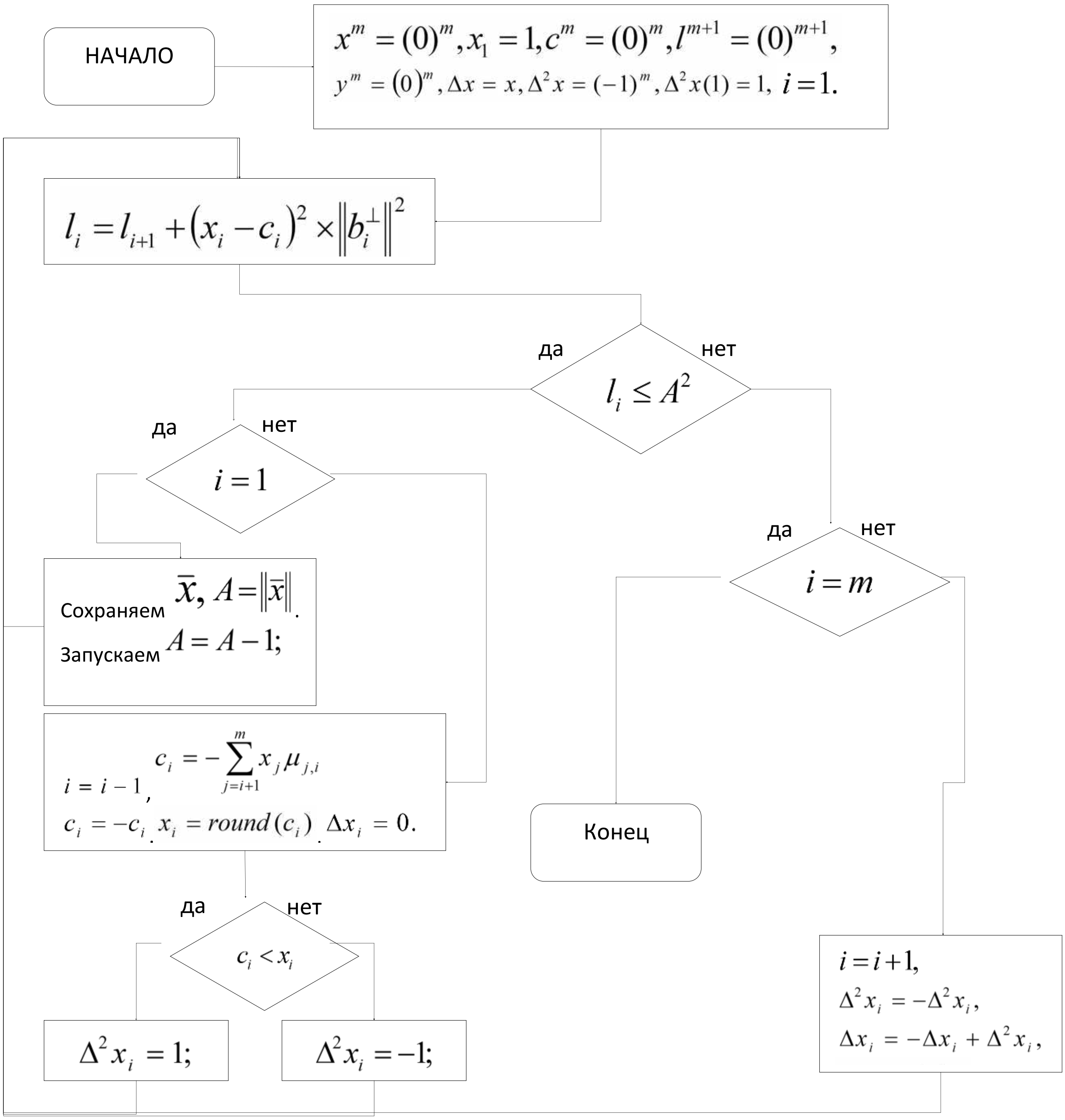

Рисунок 3.2 - Блок схема метода Каннана-Финке-Поста

Дерево перебора для неприведенного базиса решетки крайне несбалансированно, эффективность работы алгоритма будет зависить от качества декомпозиции дерева перебора на поддеревья. При этом нами предлагается использовать гетерогенную вычислительную среду: центральный процессор будет выступать устройством управления - выполнять декомпозицию дерева на составляющие, выделяя ту часть, которую он будет обрабатывать самостоятельно и ту, которую будет выполнять сопроцессор в соответсвие с Гауссовой эвристикой. Еще одним важным фактором, влияющим на эффективность поиска кратчайшего вектора, является частота обновления значения нормы предполагаемого кратчайшего вектора. В случае слишком частого обновления задержка на обновление может превысить совокупное ускорение работы алгоритма за счет отсечения веток новой нормой. Таким образом эффективность работы алгоритма зависит от качества декомпозиции и частоты обновления нормы.

Параллельный поиск обеспечивает уменьшение времени оценки кодового расстояния примерно на порядок. Например, для оценки кодового расстояния тернарного кода [64, 97] реализованному в пакете MAGMA 2.2-09 алгоритму Брауэра-Циммермана потребуется порядка 134 дней, для аналогичной оценки предложенным методом потребуется около 15 дней. Для верхней оценки кодового расстояния CCSDS AR4JA-кода со скоростью 4/5 пакету Magma (без знания автоморфизмов) потребуется 344 дня, с использованием предложенного метода на это требуется около 27 дней.

Поиск кратчайшего вектора в решетке является самой вычислительно сложной частью предложенного в [82] метода оценки минимального кодового расстояния. Для повышения быстродействия процедуры оценки минимального кодового расстояния его целесообразно реализовывать аппаратно. См. раздел 4 диссертации.

Рассмотрим работу предложенного метода определения кодового расстояния на примере тернарного совершенного кода Голея (11,6) с порождающей матрицей:

$$G = \begin{pmatrix} 2 & 2 & 1 & 2 & 0 & 1 & 0 & 0 & 0 & 0 & 0 \\ 0 & 2 & 2 & 1 & 2 & 0 & 1 & 0 & 0 & 0 & 0 \\ 0 & 0 & 2 & 2 & 1 & 2 & 0 & 1 & 0 & 0 & 0 \\ 0 & 0 & 0 & 2 & 2 & 1 & 2 & 0 & 1 & 0 & 0 \\ 0 & 0 & 0 & 0 & 2 & 2 & 1 & 2 & 0 & 1 & 0 \\ 0 & 0 & 0 & 0 & 0 & 2 & 2 & 1 & 2 & 0 & 1 \end{pmatrix}.$$

Решетка для поиска минимального расстояния в случае приведения блочным методом Коркина-Золотарева с размером блока $\beta = 2$ и $N$=6 примет вид:

$$
B_c^T = \begin{pmatrix}
12 & 12 & 6 & 12 & 0 & 6 & 0 & 0 & 0 & 0 & 0 & 1 & 0 & 0 & 0 & 0 & 0 \\
0 & 12 & 12 & 6 & 12 & 0 & 6 & 0 & 0 & 0 & 0 & 0 & 1 & 0 & 0 & 0 & 0 \\
0 & 0 & 12 & 12 & 6 & 12 & 0 & 6 & 0 & 0 & 0 & 0 & 0 & 1 & 0 & 0 & 0 \\
0 & 0 & 0 & 12 & 12 & 6 & 12 & 0 & 6 & 0 & 0 & 0 & 0 & 0 & 1 & 0 & 0 \\
0 & 0 & 0 & 0 & 12 & 12 & 6 & 12 & 0 & 6 & 0 & 0 & 0 & 0 & 0 & 1 & 0 \\
0 & 0 & 0 & 0 & 0 & 12 & 12 & 6 & 12 & 0 & 6 & 0 & 0 & 0 & 0 & 0 & 1 \\
18 & 0 & 0 & 0 & 0 & 0 & 0 & 0 & 0 & 0 & 0 & 0 & 0 & 0 & 0 & 0 & 0 \\
0 & 18 & 0 & 0 & 0 & 0 & 0 & 0 & 0 & 0 & 0 & 0 & 0 & 0 & 0 & 0 & 0 \\
0 & 0 & 18 & 0 & 0 & 0 & 0 & 0 & 0 & 0 & 0 & 0 & 0 & 0 & 0 & 0 & 0 \\
0 & 0 & 0 & 18 & 0 & 0 & 0 & 0 & 0 & 0 & 0 & 0 & 0 & 0 & 0 & 0 & 0 \\
0 & 0 & 0 & 0 & 18 & 0 & 0 & 0 & 0 & 0 & 0 & 0 & 0 & 0 & 0 & 0 & 0 \\
0 & 0 & 0 & 0 & 0 & 18 & 0 & 0 & 0 & 0 & 0 & 0 & 0 & 0 & 0 & 0 & 0 \\
0 & 0 & 0 & 0 & 0 & 0 & 18 & 0 & 0 & 0 & 0 & 0 & 0 & 0 & 0 & 0 & 0 \\
0 & 0 & 0 & 0 & 0 & 0 & 0 & 18 & 0 & 0 & 0 & 0 & 0 & 0 & 0 & 0 & 0 \\
0 & 0 & 0 & 0 & 0 & 0 & 0 & 0 & 18 & 0 & 0 & 0 & 0 & 0 & 0 & 0 & 0 \\
0 & 0 & 0 & 0 & 0 & 0 & 0 & 0 & 0 & 18 & 0 & 0 & 0 & 0 & 0 & 0 & 0 \\
0 & 0 & 0 & 0 & 0 & 0 & 0 & 0 & 0 & 0 & 18 & 0 & 0 & 0 & 0 & 0 & 0
\end{pmatrix}.
$$

После приведения базиса решетки получим:

$$
B_c = \begin{pmatrix}
0 & 0 & 0 & 0 & 0 & 0 & 0 & 0 & 0 & 0 & 0 & 0 & -3 & 0 & 0 & 0 & 0 \\
0 & 0 & 0 & 0 & 0 & 0 & 0 & 0 & 0 & 0 & 0 & 0 & 0 & -3 & 0 & 0 & 0 \\
0 & 0 & 0 & 0 & 0 & 0 & 0 & 0 & 0 & 0 & 0 & 0 & 0 & 0 & 0 & 0 & -3 \\
0 & 0 & 0 & 0 & 0 & 0 & 0 & 0 & 0 & 0 & 0 & -3 & 0 & 0 & 0 & 0 & 0 \\
0 & 0 & 0 & 0 & 0 & 0 & 0 & 0 & 0 & 0 & 0 & 0 & 0 & 0 & -3 & 0 & 0 \\
0 & 0 & 0 & 0 & 0 & 0 & 0 & 0 & 0 & 0 & 0 & 0 & 0 & 0 & 0 & -3 & 0 \\
6 & 6 & 6 & 0 & 0 & 0 & 6 & 0 & 0 & 6 & 0 & -1 & 0 & 1 & 0 & 1 & 0 \\
6 & 0 & 0 & -6 & 0 & -6 & 0 & 0 & -6 & 6 & 0 & -1 & 1 & 1 & -1 & 1 & 0 \\
0 & 6 & 6 & 6 & 0 & 0 & 0 & 6 & 0 & 0 & 6 & 0 & -1 & 0 & 1 & 0 & 1 \\
6 & 0 & 0 & 0 & 6 & 0 & 0 & 6 & 0 & 6 & 6 & -1 & 1 & 1 & 1 & 1 & 1 \\
6 & 6 & 0 & -6 & -6 & 0 & 0 & -6 & 0 & 0 & 0 & -1 & 0 & -1 & 0 & 0 & 0 \\
0 & -6 & 0 & 6 & 0 & 6 & -6 & 0 & 0 & 0 & 6 & 0 & 1 & -1 & 1 & 0 & 1 \\
0 & -6 & 0 & 0 & 0 & 0 & 0 & 6 & 6 & -6 & 6 & 0 & 1 & -1 & -1 & -1 & 1 \\
-6 & 0 & -6 & -6 & -6 & 0 & 0 & 0 & -6 & 0 & 0 & 1 & -1 & 0 & -1 & 0 & 0 \\
0 & 0 & -6 & 0 & -6 & -6 & -6 & 0 & 0 & 0 & -6 & 0 & 0 & 1 & -1 & 0 & -1 \\
0 & 0 & 0 & -6 & 0 & 0 & -6 & 0 & -6 & -6 & -6 & 0 & 0 & 0 & 1 & -1 & -1 \\
0 & 0 & -6 & 0 & 0 & -6 & 0 & -6 & -6 & -6 & 0 & 0 & 0 & 1 & -1 & -1 & 0
\end{pmatrix}.
$$

Отбрасываем первые 6 строк и последние 6 столбцов, как вспомогательные координатные компоненты (подпространство векторов избыточности кода), и отмасштабируем значения координатных компонент в решетке. Получим базис решетки, дающий некоторое слово малого веса, которое может оказаться словом наименьшего веса:

$$B_c = \begin{pmatrix} 1 & 1 & 1 & 0 & 0 & 0 & 1 & 0 & 0 & 1 & 0 \\ 1 & 0 & 0 & -1 & 0 & -1 & 0 & 0 & -1 & 1 & 0 \\ 0 & 1 & 1 & 1 & 0 & 0 & 0 & 1 & 0 & 0 & 1 \\ 1 & 0 & 0 & 0 & 1 & 0 & 0 & 1 & 0 & 1 & 1 \\ 1 & 1 & 0 & -1 & -1 & 0 & 0 & -1 & 0 & 0 & 0 \\ 0 & -1 & 0 & 1 & 0 & 1 & -1 & 0 & 0 & 0 & 1 \\ 0 & -1 & 0 & 0 & 0 & 0 & 0 & 1 & 1 & -1 & 1 \\ -1 & 0 & -1 & -1 & -1 & 0 & 0 & 0 & -1 & 0 & 0 \\ 0 & 0 & -1 & 0 & -1 & -1 & -1 & 0 & 0 & 0 & -1 \\ 0 & 0 & 0 & -1 & 0 & 0 & -1 & 0 & -1 & -1 & -1 \\ 0 & 0 & -1 & 0 & 0 & -1 & 0 & -1 & -1 & -1 & 0 \end{pmatrix}.$$

В качестве базиса решетки мы получили одиннадцать векторов, соответствующих 11 кодовым словам веса 5. Вектору решетки в последней строке $x = (0,0,-1,0,0,-1,0,-1,-1,-1,0)$ соответствует кодовое слово $(0,0,2,0,0,2,0,2,2,2,0)$ веса 5. Для того чтобы убедиться, что не существуют слова меньшего веса, нам необходимо выполнить полный перебор среди всех возможных линейных комбинаций векторов решетки, отличающихся своими координатными компонентами. Запуск алгоритма Каннна-Финке-Поста показывает, что таковые вектора в решетке отсутствуют, $d_{\min} = 5$.

Корректность предложенного метода и программного комплекса, его осуществляющего были проверены на более чем 30 различных кодах. Использовались Полярные коды, коды Рида-Маллера, коды БЧХ, код Голея, Турбо-коды, LDPC-коды [85]. Например, для оценки кодового расстояния эквивалентной блочной матрицы Турбо-кода [156, 48, 13] при однопоточной реализации метода на ЭВМ (Phenom x4-965/ 8 Gb DDR3) потребовалась 21 секунда, тогда как оценка кодового расстояния в GAP 4.7.8 (Guava 3.12, Sonata 2.6) требует более 648000 секунд. Однопоточная реализация предложенного метода доступна на сайте [86], так же приведена в Приложение 4.

Предложенный метод поиска короткого вектора в решетке занял первое место на международном конкурсе Технического Университета г. Дармштадт и оставался лучшим результатом на протяжении полутора лет, Приложение 5.

Применение параллельного метода ортогонализации, параллельного поиска кратчайшего вектора и оценок кодового расстояния (отсекающие ветки в дереве перебора) обеспечивает ускорение определения кодового расстояния низкоплотностного кода величиной от 6 раз до 7 порядков, Табл. 3.2.

Таблица 3.2 - Время поиска кодового слова минимального веса, сек

| Параметры кода, $[n, k, d]$ / Метод оценки | [384,192, 15] | [1008,504, $12 \le d \le 20$] | [2016, 1008, $12 \le d \le 62$] | [32000, 16000, $14 \le d \le 82$] |
|---|---|---|---|---|
| Брауэр-Циммерман | 59717 | $10^{13}$ | $10^{66}$ | - |
| Предложенный метод с параллельным КФП | 9432 | $4 \times 10^6$ | $10^7$ | $6 \times 10^8$ |

Дальнейшее ускорение поиска кодового слова минимального веса возможно с использованием аппаратного акселератора, реализующего поиск кратчайшего вектора в подрешетке (решетке). При этом поиск слова минимального веса сводится к многократному поиску кратчайшего вектора в подрешетке.

### 3.2.2 Вероятностный метод определения кодового расстояния с использованием геометрических решеток

Поиск кодового расстояния может быть дополнительно значительно ускорен с использованием предложенного в диссертации вероятностного метода.

Суть метода заключается в переборе всевозможных целых точек решетки $L$, $x \in \{L \cap S \cap P\}$, где $S$ – шар радиуса, равного верхней оценки кодового расстояния $A$, $P$- подмножество точек, наиболее вероятно содержащее кратчайший вектор. Этот метод позволяет уменьшить пространство перебора путем выбора подходящего базиса решетки и области $P$ [87].

Задача вероятностного поиска заключается в выборе области разбиения и набора базисов решетки, позволяющих достичь минимальных значений математического ожидания $Exp$ и дисперсии $Var$ мощности множества необходимого числа перебираемых точек в области $P$ до получения кратчайшего вектора.

В диссертации предлагается следующим образом определить область $P$ :

$$P = \left\{ \sum_{i=1}^{n} x_i b_i^{\perp} : \frac{t_i}{2} < x_i \le \frac{t_i + 1}{2} \right\}, t \in \mathrm{N}^n . \quad (3.18)$$

Для этой области:

$$Exp = \sum_{i=1}^{n} \left( \frac{t_i^2}{4} + \frac{t_i}{4} + \frac{1}{12} \right) \left\| b_i^{\perp} \right\|^2, \; Var = \sum_{i=1}^{n} \left( \frac{t_i^2}{48} + \frac{t_i}{48} + \frac{1}{180} \right) \left\| b_i^{\perp} \right\|^4 . \quad (3.19)$$

Выбор подходящего набора базисов решетки выполняется путем перестановок базиса решеток. Поскольку число, таких перестановок велико, в диссертации для этой цели используется метод Гивенса.

Ниже приводим формальное описание метода вероятностного поиска, [88].

Вход: порождающая матрица, вложенная в решетку $L$ , $d_{\max}$ - верхняя оценка кодового расстояния, $\gamma$ -точность поиска кратчайшего вектора.

1. Приведем решетку блочным метод Коркина-Золотарева с размером блока $\beta$;
2. После удаления линейно зависимых строк и столбцов получим некоторый короткий базис решетки $\overline{b}_0, \overline{b}_1, ..., \overline{b}_{n-m} \in R^{n+1}$ , $rank(L(b_0, b_1, ..., b_{n-m})) = n - m + 1$ . В случае, если среди получившихся коротких векторов, будет вектор нормы меньше изначальной оценки, примем длину полученного вектора, соответствующего некоторому слову малого веса за новую оценку с целью уменьшения дерева перебора;
3. Ортогонализуем полученный базис методами QR-разложения, получим ортогональное дополнение пространства решетки $\overline{b}_i^{\perp}$ и коэффициенты Грама-Шмидта $\mu_{ij}$ (верхняя треугольная матрица);
4. Повторять в цикле пока $\left\| b_1 \right\| > \gamma \cdot \lambda_1(L(B))$ :

В цикле по $x$ от 1 до $2^u$

$v = b_m$ , $\upsilon = \mu_{m,1...m}$

В цикле по $j = m-1$ до 1

$y = \lceil \upsilon_j - 0.5 \rceil$

**Если** $x = 1 \bmod 2$ то

**Если** $\upsilon_j - y \le 0$ то $y = y - 1$

**Иначе** $y = y + 1$.

$x = \lfloor x/2 \rfloor, v = v - yb_j, \upsilon = \upsilon - y \cdot \mu_{j,1...m}$.

**Если** $\|v\|^2 \le 0.99\|b_1\|^2$ то выйти из цикла (break).

5. Если вероятность успеха поиска вектора меньше требуемой, $\text{Success Probability}(B,u,\gamma)$, то остановить работу алгоритма (требуется другая область поиска кратчайшего вектора).

Если $x = 2^u$, то прекратить работу алгоритма (“Короткий вектор не найден”).

6. Выполним поиск кратчайшего вектора $\bar{v} \in Z^{n+1} : -d_{\max} \le v_i \le d_{\max}, i = 0...n$ ) в подрешетке с начальной нормой $A' = (n+1) \times d_{\max}^2$.

Представенный метод предусматривает использование следующих функций.

Функция **Success Probability**

**Вход:** Ортогонализованный базис решетки $B = \left(\|b_1^\perp\|^2, \|b_2^\perp\|^2, \ldots, \|b_m^\perp\|^2\right)$, $u$ - размерность подпространства поиска, $\gamma$ - точность поиска кратчайшего вектора

**Выход:** число измерений $t_{\max} = \max\left\{t \in Z \mid \Pr\left[\min\left\{\|v\|^2 \le \gamma\|b_1^\perp\|^2 \mid v \in V_t\right\}\right] \ge 1/2\right\} \cup \{-\infty\}$

**Success Probability** $(B,u,\gamma)$: $x = \pi\left(\|b_1^\perp\|^2, \ldots, \|b_m^\perp\|^2\right)$, $\pi$ - перестановка

Вернуть $\max\{\text{LogSuccessProbBound}(x,k,u,\gamma) \mid k = 1,\ldots,m-u\}$.

Функция **LogSuccessProbBound** $(x,k,u,\gamma)$:

Если ExpLength $(x,k,u,1) \le \gamma\|b_1^\perp\|^2$, то вернуть $-1$

Иначе если ExpLength $(x,k,u,0) \ge \gamma\|b_1^\perp\|^2$, то вернуть $-\infty$

$q_\gamma =$ RegularFalsi(ExpLength $(x,k,u,q) = \gamma\|b_1^\perp\|^2, q \in [0,1]$)

Вернуть $\left\lfloor \frac{k(k-1)}{4}\log_2(q_\gamma) - 1 \right\rfloor$.

Функция **ExpLength**$(x,k,u,\gamma)$: Вернуть $\frac{1}{12}\sum_{i=1}^{k-1} q^{k-i}x_i + \frac{1}{12}\sum_{i=k}^{m-u-1} x_i + \frac{1}{3}\sum_{i=m-u}^{m-1} x_i + x_m$.

Применение вероятностного метода обеспечивает до 60 порядков ускорения определения кодового расстояния низкоплотностного кода с **приемлемой вероятностью ошибки,** достаточной для просеивания кодов по критерию кодового расстояния, Табл. 3.3, [104, 145, 146]. Вероятностный метод обеспечивает отсев множества кодов кандидатов со слабыми дистантными свойствами.

Таблица 3.3 - Время поиска кодового слова минимального веса, сек

| Параметры кода, $[n, k, d]$ / Метод оценки | [384,192, 15] | [1008,504, $12 \le d \le 20$] | [2016, 1008, $12 \le d \le 62$] | [32000, 16000, $14 \le d \le 82$] |
|---|---|---|---|---|
| Брауэр-Циммерман | 59717 | $10^{13}$ | $10^{66}$ | - |
| Предложенный метод с параллельным КФП | 9432 | $4 \times 10^{6}$ | $10^{7}$ | $6\times10^{8}$ |
| Предложенный вероятностный метод | - | $10^{6}$ | $1.3 \times 10^{6}$ | $3 \times 10^{6}$ |

Опираясь на предложенные методы построения кодов, позволяющие улучшить их свойства: кодовое расстояние и спектр связности, нами был построен код (КОД Б) со скоростью $R = \frac{1}{3}$ и информационной длиной от 80 до 1600 на основе протографа из стандарта 5G (КОД А), [104].

В полученом коде по сравнению с кодом 5G удалось увеличить кодовое расстояние для циркулянта 8 с 20 до 23, для циркулянта 32 с 31 до 44, [104].

На рис. 3.3 приведены результаты исследования эффективности построенного кода, с использование квадратурно фазовой модуляции (QPSK), для 50 итераций итеративного декодера с мягкими метриками Sum-Product на АБГШ-канале. Улучшение свойств кода и графа обеспечило энергетический выигрыш от 0.1 до 0.4 дБ в сравнение с кодом из стандарта 5G.

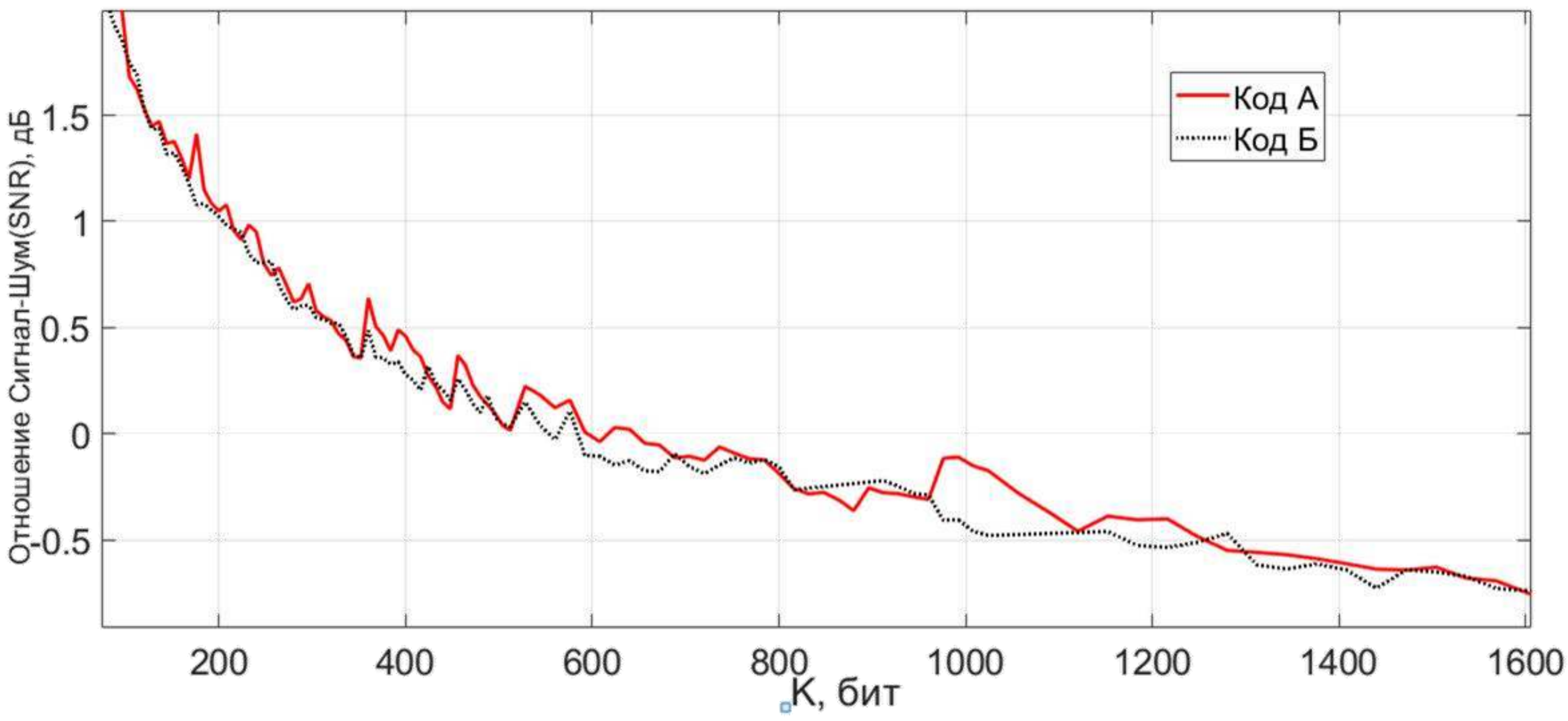

Рисунок 3.3– Требуемое отношение Сигнал-Шум для кода информационной длины $K$, скорости $R = 1/3$, для уровня блочной ошибки $< 10^{-5}$, с использование квадратурно фазовой модуляции (QPSK), 50 итераций итеративного декодера с мягкими метриками Sum-Product на АБГШ-канале.

### 3.3 Выводы

В диссертации предложены методы оценки кодового расстояния, опирающиеся на решение задачи Геометрии чисел – задачи поиска кратчайшего вектор. Для этого код вкладывается в решетку при помощи техники Каннана (Kannan's embedding technique). Кратчайшему вектору в решетке соответствует слово минимального веса. Решение задачи поиска кратчайшего вектора, требует выполнения ортогонализация базиса решетки и поиска кратчайшего вектора в сфере, ограниченной радиусом равным Хэммингову весу кодового слова.

Чаще всего для оценки кодового расстояния применяют метод Брауэра-Зиммермана. Ключевой особенностью метода Брауэра-Зиммермана является перебор кодовых слов на основе информационных совокупностей в сочетании с одновременным вычислением верхних и нижних границ кодового расстояния. Метод Брауэра-Зиммермана применим к произвольным линейным блочным кодам, позволяет получить верхнюю оценку временных затрат. К сожалению, его производительность недостаточна для оценки кодового расстояния низкоплотностных кодов средней длины.

Были предложены параллельные методы ортогонализации базиса решетки, обеспечившие 895-кратное ускорение на видеоускорителе (NVidia GeForce 1070), 162-кратное ускорение на 8-поточном процессоре (Intel i5-7700K) по сравнению с последовательным модифицированным методом Грамма-Шмидта. Для фиксированной области перебора был предложен параллельный метод поиск кратчайшего вектора в решетке. Применение этого метода обеспечило ускорения оценки кодового расстояния от 3 до 10 раз.

Параллельный метод обеспечил пятикратное ускорение (для 8-поточной модели исполнения). Вероятностный метод обеспечил ускорение от 5 до 60 порядков в случае поиска верхних и нижних оценок слов малого веса с погрешностью достаточной для построения кодов с уровнем ошибки на бит от $10^{-8}$ до $10^{-10}$.

Дальнейшее ускорение поиска кодового слова минимального веса возможно с использованием специализированного устройства, реализующего поиск кратчайшего вектора в подрешетке (решетке). При этом поиск слова минимального веса сводится к многократному поиску кратчайшего вектора в подрешетке.

# 4 РАЗРАБОТКА СПЕЦИАЛИЗИРОВАННОГО УСТРОЙСТВА ОСУЩЕСТВЛЯЮЩЕГО ПОИСК КОДОВОГО РАССТОЯНИЯ В ПОДРЕШЕТКЕ М-РАЗМЕРНОСТИ ДЛЯ ПОСТРОЕНИЯ НИЗКОПЛОТНОСТНЫХ КОДОВ И ЭКСПЕРИМЕНТАЛЬНАЯ ОЦЕНКА НАДЕЖНОСТИ СЧИТЫВАНИЯ ДАННЫХ ИЗ ГОЛОГРАФИЧЕСКОЙ

Наибольшие вычислительные затраты предложенного метода оценки кодового расстояния линейного кода занимает поиск кратчайшего вектора в решетке, заключающийся в решении системы КФП. Именно он ограничивает применение анализа дистантных свойств кода при построении низкоплотностных кодов для голографической памяти. Поэтому для реализации поиска кратчайшего вектора в коде разработано специализированное устройство, реализованное на ПЛИС.

Со схемотехнической точки зрения поиск кратчайшего вектора сводится к итерационно-вычислительной процедуре определения значений координатных компонент дерева перебора и использовании этих компонент для вычисления векторов решетки на основе множества операций умножения, сложения 32-битовых чисел, сдвига, их сравнения и перезаписи в процессе перебора элементов решетки.

Построение низкоплотностного кода для архивной голографической памяти осуществлялось с использованием следующих аппаратных средств, [99, 122]: хост-компьютера и специализированного устройства поиска кратчайшего вектора в решетке. Функции, выполняемые хост-компьютером показаны на рис. 4.1.

Наиболее сложные вычисления выполняет специализированное устройство поиска кратчайшего вектора в решетке, реализующее представленный ниже алгоритм.

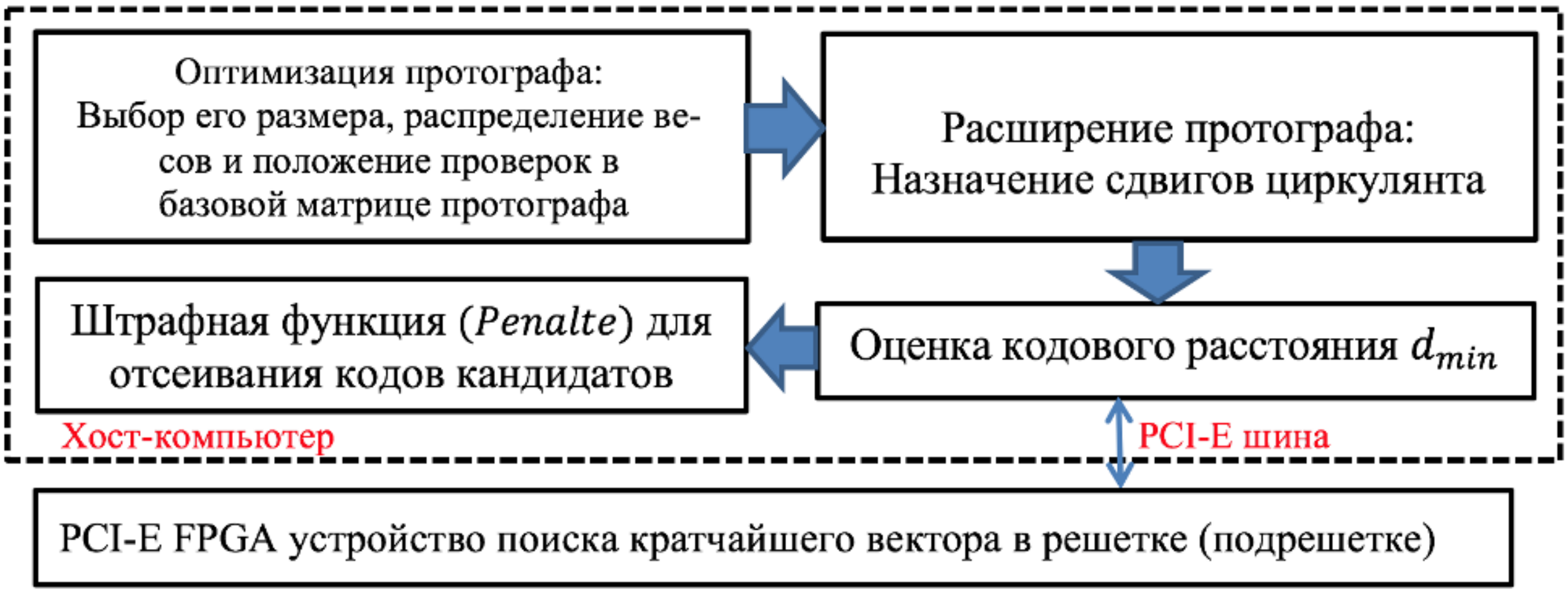

Рисунок 4.1. Аппаратные средства построения LDPC-кодов

## 4.1 Аппаратно-ориентированный алгоритм поиска кратчайшего вектора в решетке

На рисунке 4.2 представлен алгоритм поиска кратчайшего вектора в решетке $B_c^T$ , удобный для аппаратной реализации. Алгоритм реализует метод Каннана-Финке-Поста.

Ключевой особенностью алгоритма, обеспечивающего его высокое быстродействие, является параллельное выполнение всех умножений.

Оценка нормы $A$ кратчайшего вектора следует из определения 5 $A = \sqrt{\gamma_m}\det(L)^{\frac{1}{m}}$ . При ее вычислении используют приближение константы Эрмита, $\gamma_m \le \frac{2}{\pi}\Gamma(2+\frac{m}{2})^{\frac{2}{m}}$, где $\Gamma$ - Гамма-функция [89].

Каждый базисный вектор решетки может быть выражен через ортогональные вектора следующим образом $b_i = \sum_{j=1}^{i}\mu_{i,j}b_j^{\perp}, 2 \le i \le m, 1 \le j < i \le m$ . Координаты кратчайшего вектора решетки $x_i$ удовлетворяют следующей системе неравенств [90, 91]:

$$\begin{cases} x_m^2 \left\|b_m^{\perp}\right\|^2 \le A^2, \\ (x_{m-1} + \mu_{m,m-1}x_m)\left\|b_{m-1}^{\perp}\right\|^2 \le A^2 - x_m^2\left\|b_m^{\perp}\right\|^2 \\ \dots \\ (x_1 + \sum_{i=2}^{m} x_i\mu_{i,j})^2 \left\|b_1^{\perp}\right\|^2 \le A^2 - \sum_{j=2}^{m} l_j \end{cases}, \qquad (4.1)$$

где $$l_j = (x_j + \sum_{i=j+1}^{m} x_i \mu_{i,j})^2 \left\|b_j^{\perp}\right\|^2 .$$

Представленный алгоритм находит решение этой системы в виде вектора ($x_1$, $x_2$,…,$x_m$). Особенностью алгоритма является одновременное (параллельное) выполнение всех мультипликативных операций (умножений), требующих наибольших временных затрат при аппаратной реализации (см. блок 1 на рис. 4.2). Умножения используются при расчете частичных сумм

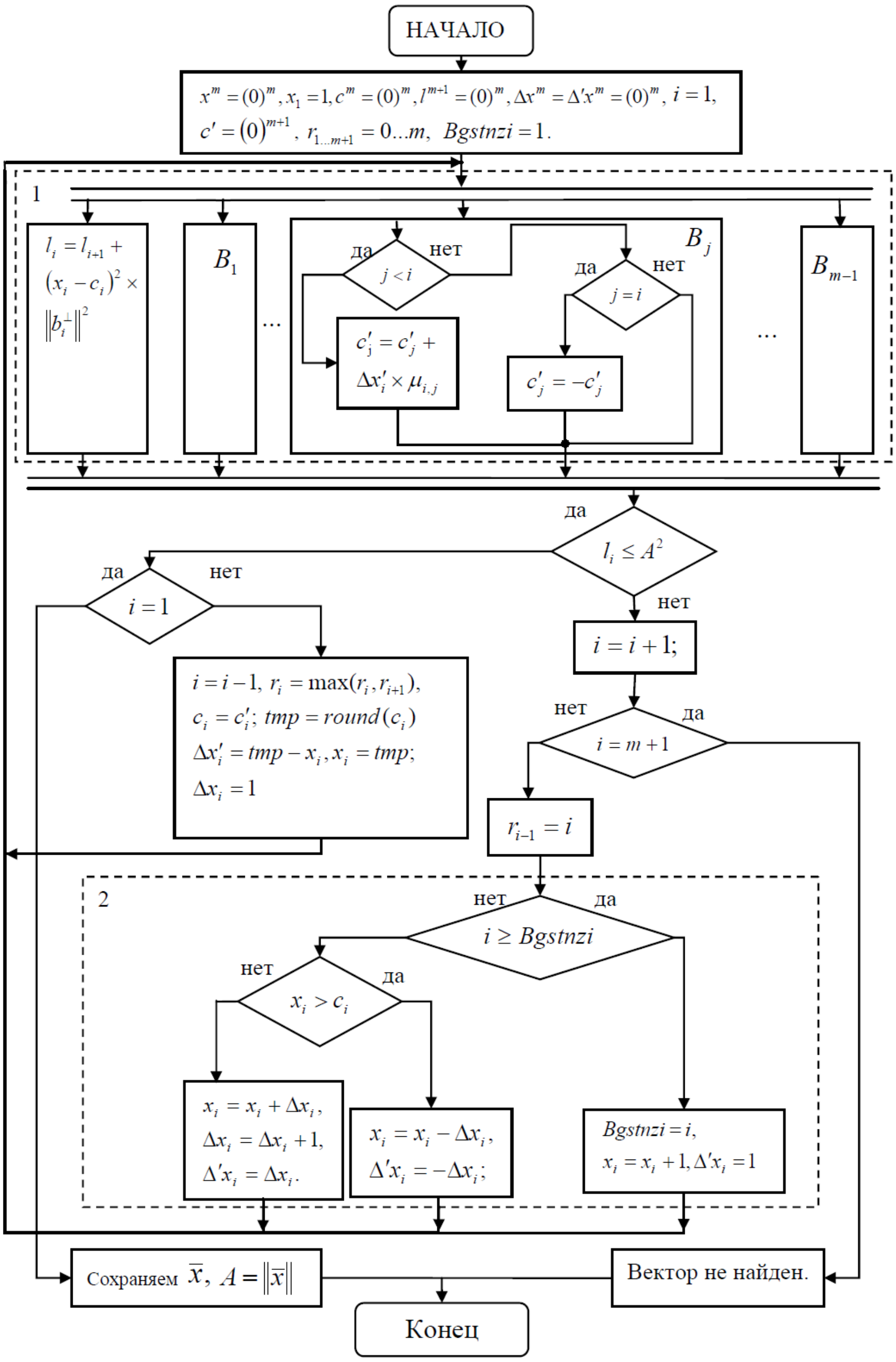

Рисунок 4.2 - Аппаратно-ориентированный алгоритм поиска кратчайшего вектора в решетке

$$l_i = l_{i+1} + (x_i - c_i)^2 \times \left\| b_i^{\perp} \right\|^2 \qquad (4.2)$$

и величин изменения $c_j$ во время перебора координатных компонент:

$$c'_j = c'_j + \Delta x'_i \times \mu_{i,j} \qquad (4.3)$$

Перебор координатных компонент вектора ($x_1$, $x_2$,…, $x_m$) осуществляется при помощи зигзагообразного обхода Шнора (блок 2 на рис. 4.2), [95-96] .

## 4.2 Структурно-функциональная организация специализированного устройства поиска кратчайшего вектора в решетке (подрешетке)

Структурная схема специализированного устройства поиска кратчайшего вектора, реализующая предложенный алгоритм, приведена на рисунке 4.3.

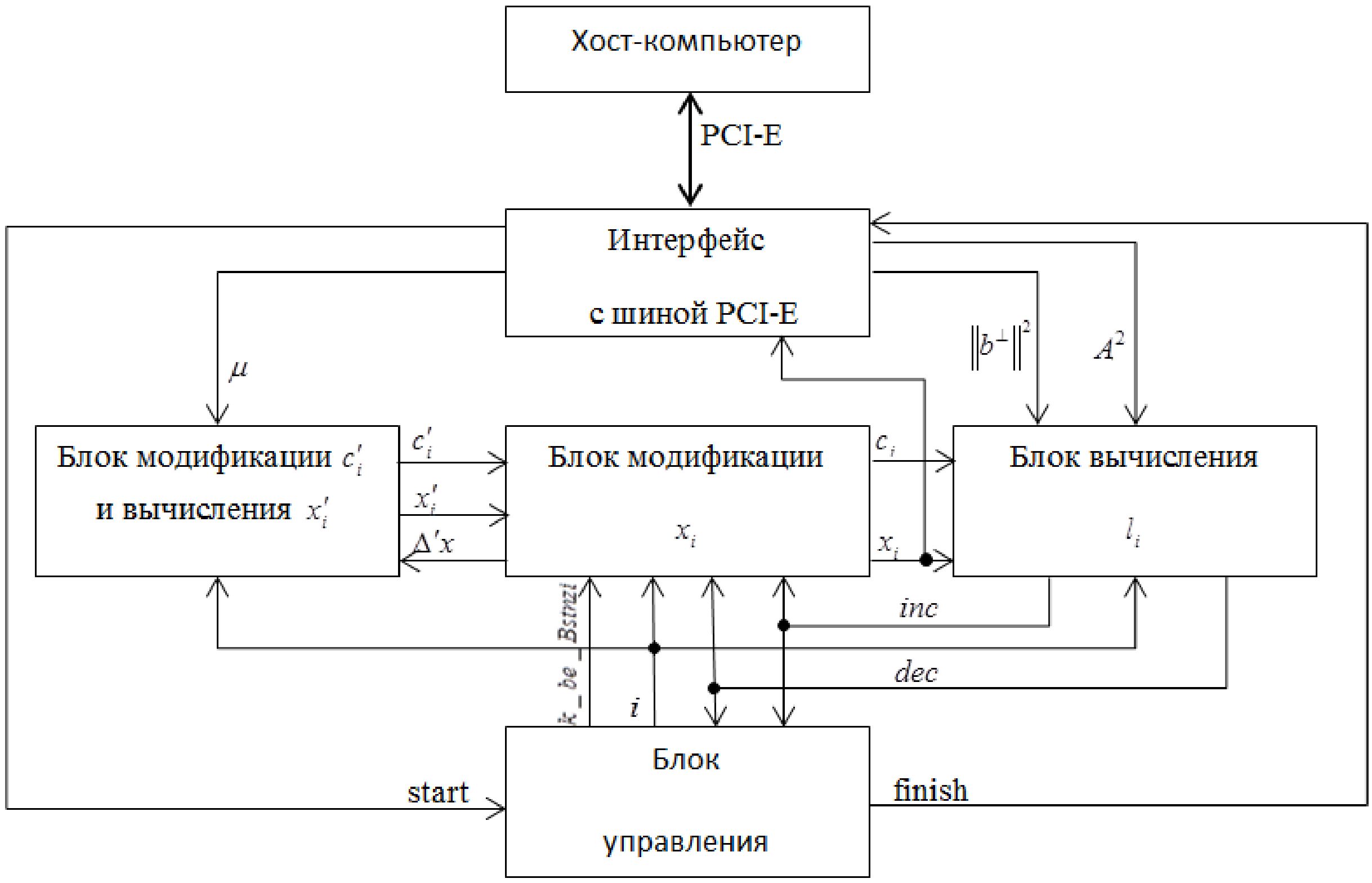

Рисунок 4.3. Специализированное устройства для оценки кодового расстояния

Устройство содержит: интерфейс с шиной PCI-E хост – компьютера, блок модификации $c'_i$ и $x'_i$ , блок модификации $x_i$ , блок вычисления $l_i$ и блок управления.

Быстродействие устройства в основном определяется временем умножения многоразрядных чисел. В предлагаемом специализированном устройстве удалось совместить во времени вычисления $c'_j = c'_j + \Delta x'_i \times \mu_{i,j}$ и $l_i = l_{i+1} + (x_i - c_i)^2 \times \|b_i^{\perp}\|^2$, требующие умножений. Это позволило повысить быстродействие устройства не менее чем в 1,5 раза.

Интерфейсный блок предназначен для загрузки в разработанное устройство: оценки квадрата нормы искомого вектора $A$, коэффициентов Грама-Шмидта $\mu_{i,j}$ и квадратов норм ортогонализованного базиса решетки $\|b_i^{\perp}\|^2$, предварительно вычисленных в хост-компьютере. Также интерфейсный блок передает в хост компьютер, найденный кратчайший вектор. Для оценки кодового расстояния кодов длины $N$, применяется блочный метод Коркина-Золотарева с размером блока $\beta \le 512$.

Блок вычисления частичных сумм $l_i$ (см. рис. 4.4) содержит входной регистр, умножитель, квадратор, два сумматора, вычитатель, компаратор и три стековых регистра. Блок вычисляет значение $l_i$, сравнивает его с оценкой нормы $A$ и в зависимости от результата сравнения формирует один из управляющих сигналов $inc$ или $dec$. Сигналы $inc$ и $dec$ соответствуют увеличению ($inc$) или уменьшению ( $dec$ ) индекса текущей координатной компоненты $i$ . Они определяют направления сдвига регистров, хранящих $l_i$, $\|b_i^{\perp}\|^2$, $c_i$, $x_i$, обеспечивая тем самым правильный выбор элементов из последовательности.

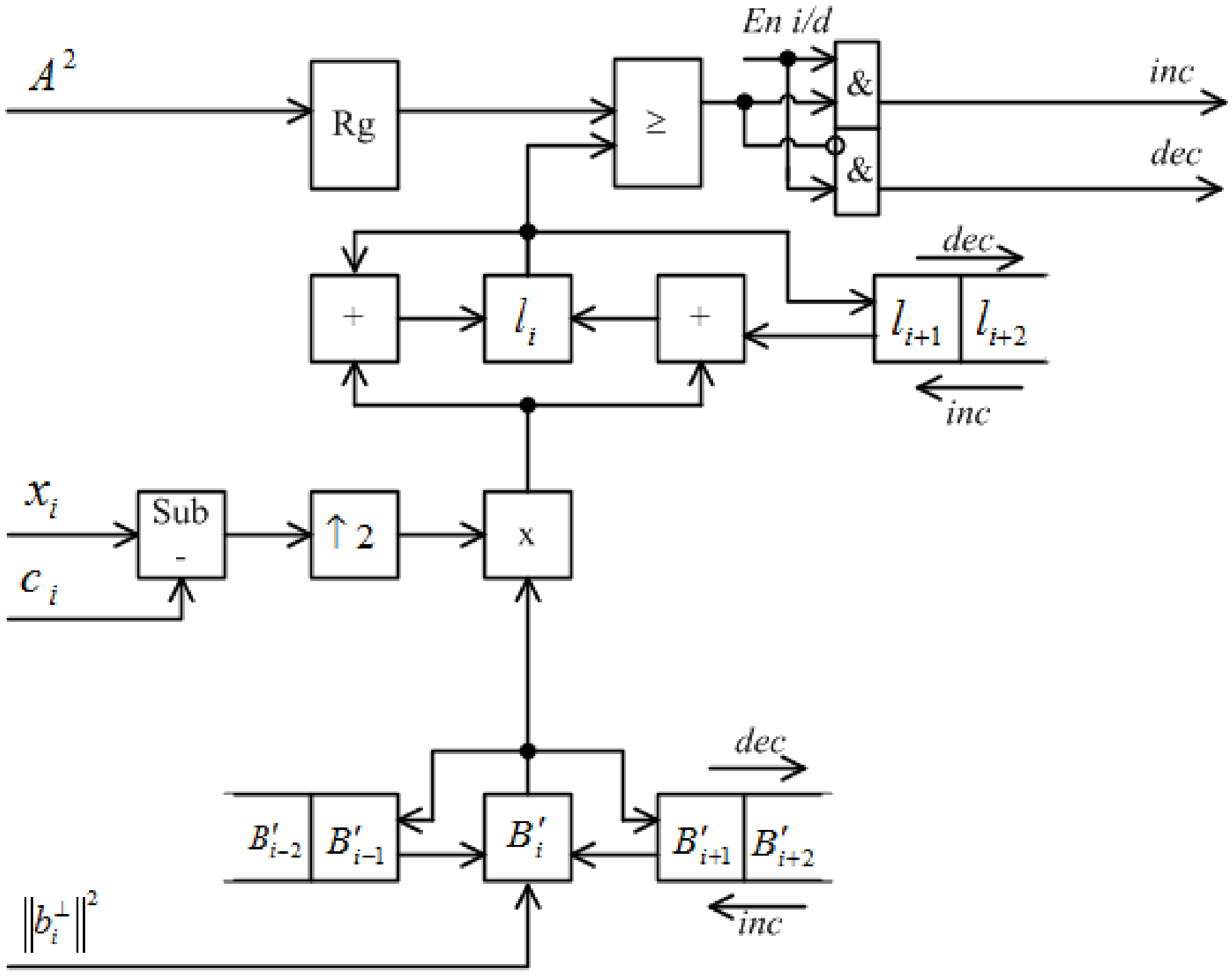

Рисунок 4.4. Структурная схема блок вычисления $l_i$

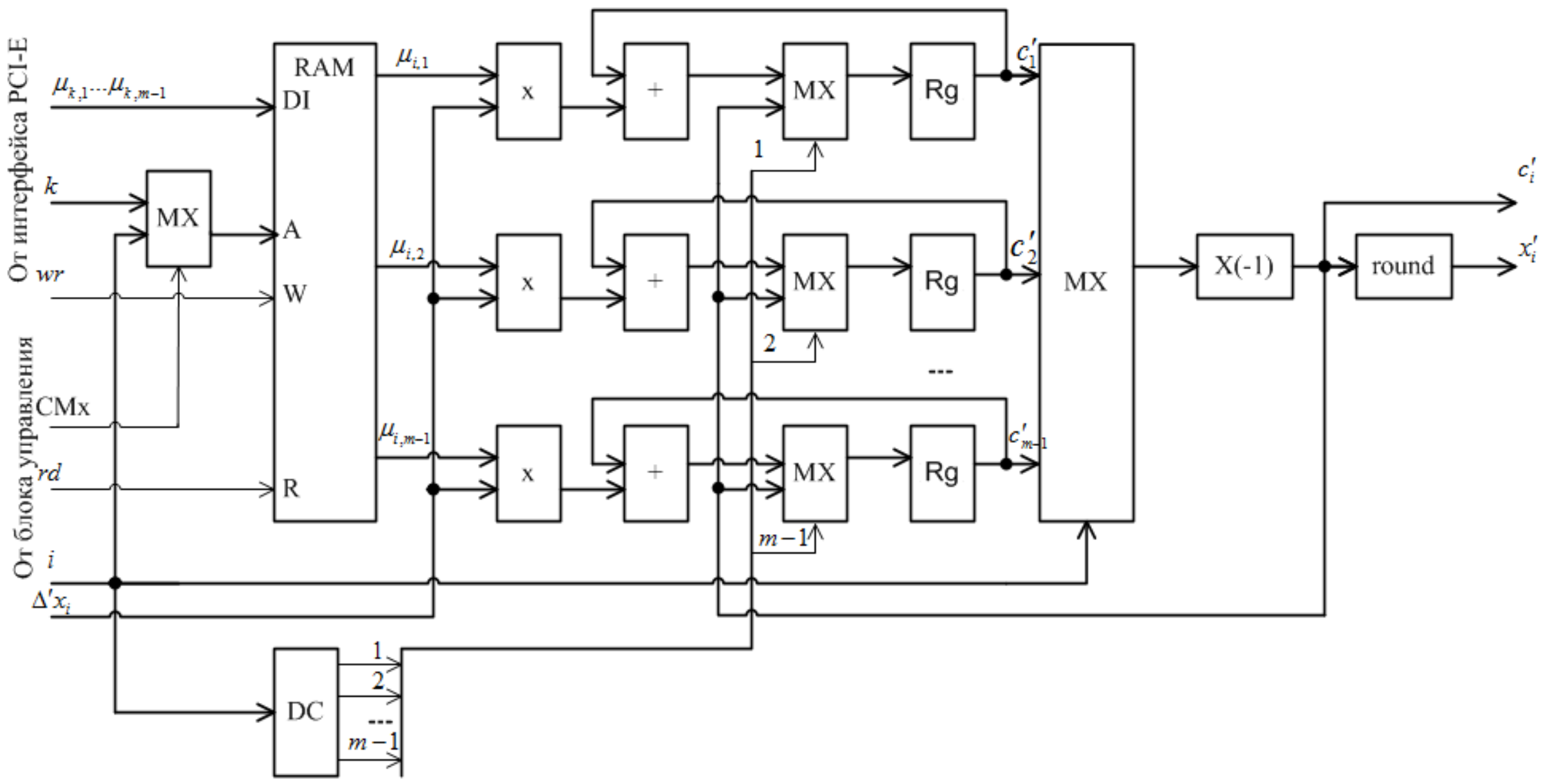

Рисунок 4.5. Структурная схема блока модификации $c_i'$ и вычисления $x_i'$

Блок модификации $x_i$ (рис. 4.6) содержит сумматоры, вычитатель, компаратор, блок изменения знака, мультиплексоры и три стековых регистра. Блок модификации $x_\text{i}$ , осуществляет изменение координатных компонент искомого вектора в процессе зигзагообразный обхода, а также возвращает непосредственно координаты искомого кратчайшего вектора в решетке. В предложенной аппаратной реализации используется 32 бита для коэффициентов и ортогонализованного базиса решетки $\left\|b_i^{\perp}\right\|^2$, а также 8 бит для $c'$ и $\Delta x'$;

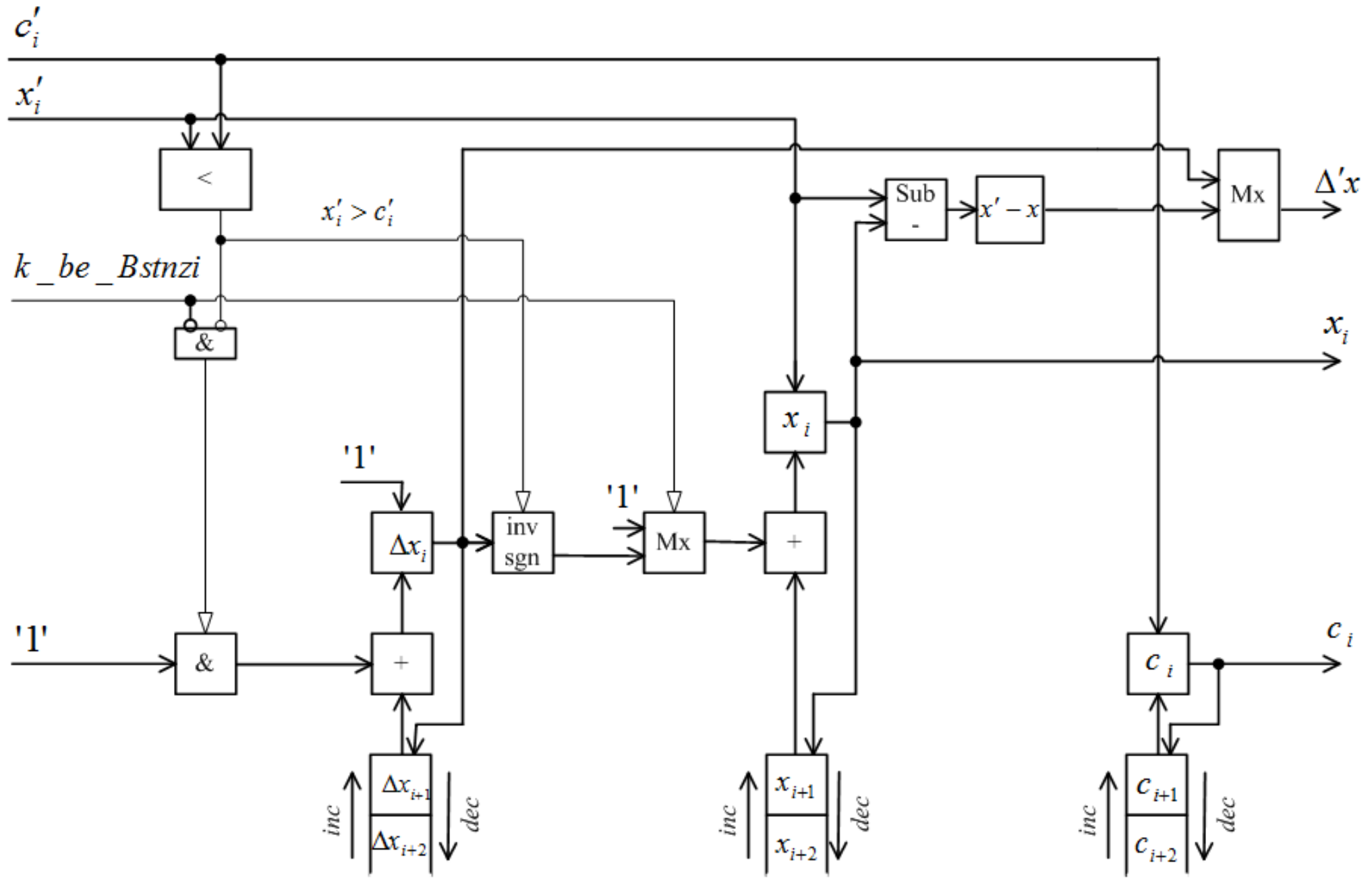

Рисунок 4.6. Структурная схема блок модификации $x_i$

В предложенном устройстве все умножения (требующие наибольших временных затрат) осуществляются параллельно (блок 1 алгоритма на рис. 4.3). Это обеспечивает высокую производительность устройства.

Синтез блоков устройства осуществлялся в САПР Vivado v2016.3 (64-bit) на ПЛИС Virtex UltraScale. В табл. 2 приведены оценки аппаратной сложности операционной части устройства для различных размеров подрешетки m.

Таблица 4.1 - Аппаратной сложность операционной части специализированного устройства для различных размеров подрешетки m.

| m | 64 | 128 | 256 | 512 |
|---|---|---|---|---|
| Количество LUT | 690 | 1301 | 2803 | 6154 |
| Количество FF | 322 | 832 | 1120 | 2242 |
| Количество DSP | 28 | 46 | 92 | 184 |
| Объём памяти, КБ | 8,125 | 32,25 | 128,5 | 513 |

В табл. 4.2 приведено сравнение по быстродействию предлагаемого устройства и программной реализации поиска по методу КФП на AMD Phenom 965/8 Gb DDR2-800. Выигрыш по быстродействию представленного устройства по сравнению с программной реализацией составляет примерно 40 раз для LDPC кодов скоростью ½. Устройство было синтезировано для ПЛИС Virtex 7 Ultrascale (xc7vx485tffg1930-2L) с тактовой частотой 213 МГц.

Для сравнения выполнялась программная реализация поиска кратчайшего слова по методу КФП на ПЭВМ (AMD Phenom 965/8 Gb DDR2-800). В таблице 4.2 приведено сравнение по быстродействию предлагаемого устройства и программной реализации поиска. Исследование проводилось для LDPC кодов со скоростью ½ и длиной $m$ = 64, 128, 256, 512.

Таблица 4.2 – Время поиска слов минимального веса предлагаемым устройством и программной реализацией

| m / Параметры | 64 | 128 | 256 | 512 |
|---|---|---|---|---|
| Сложность аппаратной реализации, вентилях | 286716 | 474462 | 945778 | 1894860 |
| Число циклов работы устройства | $0{,}49\cdot10^9$ | $236\cdot10^9$ | $486\cdot10^9$ | $1005\cdot10^9$ |
| Время $T_1$ поиска слов предлагаемым устройством, с. | 9.84 | 4720 | 9720 | 20100 |
| Время $T_2$ поиска слов программной реализацией, с. | 333.68 | 156011 | 329925 | 681946 |
| $T_2/T_1$ | 33,91 | 33,05 | 33,94 | 33,93 |

## 4.3 Построение низкоплотностных кодов для архивной голографической памяти с повышенной корректирующей способностью

В диссертации с использованием предложенного метода (п. 2.1) был построен квазициклический низкоплотностный код ($n$, $k$) = (32000,16000) для архивной голографической памяти. Рассмотрим процесс построения кода подробнее.

В начале процесса построения кода на фазе оптимизации протографа (п. 2.1, шаги 2-3), были получены протографы, базовые матрицы которых размером 21 на 11 представлены на рис. 4.7-4.9.

| | | | | | | | | | | | | | | | | | | | | |
|---|---|---|---|---|---|---|---|---|---|---|---|---|---|---|---|---|---|---|---|---|
| 0 | 0 | 0 | 0 | 0 | 0 | 0 | 0 | 1 | 1 | 0 | 1 | 0 | 0 | 1 | 1 | 0 | 1 | 0 | 1 | 1 |
| 0 | 0 | 0 | 0 | 0 | 0 | 0 | 1 | 1 | 0 | 0 | 1 | 0 | 1 | 0 | 1 | 0 | 0 | 1 | 1 | 1 |
| 0 | 0 | 0 | 0 | 0 | 1 | 1 | 0 | 0 | 0 | 0 | 0 | 1 | 1 | 0 | 0 | 1 | 0 | 1 | 1 | 1 |
| 0 | 0 | 0 | 0 | 0 | 0 | 0 | 0 | 0 | 1 | 1 | 1 | 0 | 1 | 0 | 0 | 1 | 1 | 0 | 0 | 1 |
| 0 | 0 | 0 | 0 | 0 | 0 | 1 | 1 | 0 | 0 | 0 | 0 | 1 | 0 | 1 | 0 | 1 | 0 | 1 | 0 | 1 |
| 0 | 0 | 0 | 0 | 0 | 1 | 0 | 0 | 0 | 0 | 1 | 0 | 1 | 0 | 1 | 1 | 0 | 1 | 0 | 0 | 1 |
| 1 | 0 | 0 | 0 | 0 | 0 | 0 | 0 | 0 | 0 | 0 | 0 | 1 | 0 | 0 | 1 | 1 | 0 | 0 | 1 | 1 |
| 0 | 0 | 1 | 0 | 0 | 0 | 0 | 0 | 0 | 0 | 0 | 0 | 0 | 1 | 0 | 0 | 1 | 1 | 1 | 0 | 1 |
| 0 | 0 | 0 | 1 | 0 | 0 | 0 | 0 | 0 | 0 | 0 | 1 | 0 | 1 | 0 | 0 | 0 | 0 | 1 | 1 | 1 |
| 0 | 0 | 0 | 0 | 1 | 0 | 0 | 0 | 0 | 0 | 0 | 0 | 0 | 1 | 0 | 1 | 1 | 0 | 0 | 1 | 1 |
| 0 | 1 | 0 | 0 | 0 | 0 | 0 | 0 | 0 | 0 | 0 | 0 | 0 | 1 | 1 | 1 | 0 | 0 | 0 | 0 | 1 |

Рисунок 4.7 – Базовая матрица протографа с кодовым расстоянием 22 и порогом итеративного декодирования при 30 и 200 итерациях $Eb/No=$ 0.8903 и $Eb/No=0.5503$, соответственно.

| | | | | | | | | | | | | | | | | | | | | |
|---|---|---|---|---|---|---|---|---|---|---|---|---|---|---|---|---|---|---|---|---|
| 0 | 0 | 0 | 0 | 0 | 0 | 0 | 0 | 1 | 1 | 0 | 1 | 0 | 0 | 1 | 1 | 0 | 1 | 0 | 1 | 1 |
| 0 | 0 | 0 | 0 | 0 | 0 | 0 | 1 | 1 | 0 | 0 | 1 | 0 | 1 | 0 | 1 | 0 | 0 | 0 | 0 | 1 |
| 0 | 0 | 0 | 0 | 0 | 1 | 1 | 0 | 1 | 0 | 1 | 0 | 1 | 0 | 0 | 0 | 1 | 0 | 1 | 1 | 1 |
| 0 | 0 | 0 | 0 | 0 | 0 | 0 | 0 | 0 | 1 | 1 | 1 | 0 | 1 | 0 | 0 | 0 | 1 | 0 | 0 | 1 |
| 0 | 0 | 0 | 0 | 0 | 0 | 1 | 1 | 0 | 1 | 0 | 0 | 0 | 0 | 1 | 0 | 1 | 0 | 1 | 0 | 1 |
| 0 | 0 | 0 | 1 | 0 | 1 | 0 | 1 | 0 | 0 | 1 | 0 | 1 | 0 | 1 | 0 | 0 | 1 | 0 | 0 | 1 |
| 1 | 0 | 0 | 0 | 1 | 0 | 0 | 0 | 0 | 0 | 0 | 0 | 1 | 0 | 0 | 1 | 1 | 0 | 0 | 1 | 1 |
| 0 | 0 | 1 | 1 | 1 | 1 | 1 | 0 | 0 | 0 | 0 | 0 | 0 | 1 | 0 | 0 | 1 | 1 | 1 | 0 | 1 |
| 0 | 0 | 0 | 1 | 0 | 0 | 0 | 0 | 0 | 0 | 0 | 1 | 0 | 1 | 0 | 0 | 0 | 0 | 1 | 1 | 1 |
| 0 | 0 | 0 | 0 | 1 | 0 | 0 | 0 | 0 | 0 | 0 | 0 | 0 | 1 | 0 | 1 | 1 | 0 | 0 | 0 | 1 |
| 0 | 1 | 0 | 0 | 0 | 0 | 0 | 0 | 0 | 0 | 0 | 0 | 0 | 1 | 1 | 1 | 0 | 0 | 0 | 0 | 1 |

Рисунок 4.8 – Базовая матрица протографа с кодовым расстоянием 24 и порогом итеративного декодирования при 30 и 200 итерациях $Eb/No=$ 1.0403 и $Eb/No=$ 0.8303, соответственно.

| | | | | | | | | | | | | | | | | | | | | |
|---|---|---|---|---|---|---|---|---|---|---|---|---|---|---|---|---|---|---|---|---|
| 0 | 0 | 0 | 0 | 0 | 0 | 0 | 0 | 1 | 1 | 0 | 0 | 0 | 0 | 1 | 1 | 0 | 0 | 0 | 0 | 1 |
| 0 | 0 | 0 | 0 | 0 | 0 | 0 | 1 | 1 | 0 | 0 | 1 | 0 | 0 | 0 | 1 | 0 | 0 | 0 | 0 | 1 |
| 1 | 0 | 1 | 0 | 0 | 1 | 1 | 0 | 1 | 0 | 1 | 0 | 1 | 0 | 0 | 0 | 1 | 0 | 1 | 1 | 1 |
| 0 | 0 | 0 | 0 | 0 | 0 | 0 | 0 | 0 | 1 | 1 | 1 | 0 | 1 | 0 | 0 | 0 | 1 | 0 | 0 | 1 |
| 0 | 0 | 0 | 0 | 0 | 0 | 1 | 1 | 0 | 1 | 0 | 0 | 0 | 0 | 1 | 0 | 1 | 0 | 0 | 0 | 1 |
| 0 | 1 | 1 | 1 | 0 | 1 | 0 | 1 | 0 | 0 | 1 | 0 | 1 | 0 | 1 | 0 | 0 | 1 | 0 | 0 | 1 |
| 1 | 0 | 0 | 0 | 1 | 0 | 0 | 0 | 0 | 0 | 0 | 0 | 1 | 0 | 0 | 1 | 1 | 0 | 0 | 1 | 1 |
| 1 | 1 | 1 | 1 | 1 | 1 | 1 | 0 | 0 | 0 | 0 | 0 | 0 | 1 | 0 | 0 | 0 | 1 | 1 | 0 | 1 |
| 0 | 0 | 0 | 1 | 0 | 0 | 0 | 0 | 0 | 0 | 0 | 1 | 0 | 0 | 0 | 0 | 0 | 0 | 1 | 1 | 1 |
| 0 | 0 | 0 | 0 | 1 | 0 | 0 | 0 | 0 | 0 | 0 | 0 | 0 | 0 | 0 | 1 | 1 | 0 | 0 | 0 | 1 |
| 0 | 1 | 0 | 0 | 0 | 0 | 0 | 0 | 0 | 0 | 0 | 0 | 0 | 1 | 1 | 1 | 0 | 0 | 0 | 0 | 1 |

Рисунок 4.9 – Базовая матрица протографа с кодовым расстоянием 54 и порогом итеративного декодирования при 30 и 200 итерациях $Eb/No=$ 1.1603 и $Eb/No=$ 0.8803, соответственно.

Характеристика треппин-сетов для кода с базовой матрицей, приведенной на рис. 4.8, дана в Таблице 4.3. В таблице приведены символьные узлы, образующие треппин-сеты, и веса псевдокодовых слов, образованных треппин-сетами. Веса вычислены с использованием метода линейного программирования, приведенного во второй главе.

Таблица 4.3 - Треппин-сеты для кода с базовой матрицей, приведенной на рис. 4.8

| TS($a$, $b$) | Вес псевдокодового слова, $\omega(p)$ | Столбцы (символьные узлы), участвующие в псевдокодовом слове треппин-сета |
|---|---|---|
| **(7, 7)** | **[8.781202]** | **9139 13040 14556 15304 17468 19075 28301** |
| **(9, 4)** | **[9.001648]** | **7616 9946 10184 10866 12151 13025 14541 15289 30384** |
| **(8, 8)** | **[9.855998]** | **10662 11380 12254 13770 16730 17487 21945 24782** |
| **(9, 9)** | **[9.861800]** | **7616 9285 11490 13025 14541 15289 16792 28447 29099** |
| **(8, 8)** | **[9.888543]** | **7616 7833 10184 10279 18475 18597 22027 22244** |
| **(10, 6)** | **[10.001831]** | **7616 8355 9400 11605 12172 12479 16028 17474 18380 26196** |
| **(10, 8)** | **[10.001831]** | **7616 9627 10184 10866 11832 13263 14541 15289 29827 30065** |
| **(10, 8)** | **[10.001831]** | **7616 10062 10184 10866 13263 14779 16973 19781 22027 22748** |
| **(8, 8)** | **[10.084940]** | **7616 8201 9246 10184 19060 19182 22149 22612** |
| **(7, 6)** | **[10.213432]** | **12185 12531 14765 15224 16959 17418 18136** |
| **(7, 6)** | **[10.244420]** | **13708 14315 15609 15979 25909 26516 26886** |
| **(7, 6)** | **[10.290990]** | **7616 7833 9341 10184 18597 19060 19277** |
| **(8, 6)** | **[10.719603]** | **9139 11344 13282 14339 16610 17710 18169 18858** |
| **(10, 9)** | **[10.747731]** | **7616 13025 13736 14541 15289 15661 17107 17453 22399 27823** |
| **(10, 10)** | **[10.861930]** | **7616 10062 10184 10866 13263 13698 18126 22027 26980 28977** |
| **(10, 9)** | **[10.938397]** | **7616 9934 10184 10647 12139 13013 14529 19060 24852 27573** |
| **(11, 9)** | **[11.002014]** | **7616 7854 9341 10184 10866 13263 14779 15527 19060 24948 28061** |
| … | … | … |
| **(14, 4)** | **[14.002564]** | **7616 7833 10163 10184 10845 12151 13025 13242 14541 14758 15289 15506 18359 18597** |
| **…** | **…** | **…** |
| **(17, 6)** | **[17.003113]** | **7616 8317 9362 9579 10184 10866 11329 11784 12203 13263 13719 14779 15527 15990 19060 19298 29779** |
| … | … | … |

В таблице 4.3 приведены не все значения треппин-сетов, а только образующие псевдокодовые слова малого веса. Однако вес именно этих псевдоковых слов $\omega(p)$ определяет вероятность блочной ошибки.

Вероятность ошибки в символьном узле треппин-сета вычисляется по следующей формуле:

$$P_{TS} = Q\left(\frac{2m_\lambda + 2\sum_{j=1}^{iter}\frac{m_{\lambda ext}^{(j)} + m_\lambda}{\mu_{max}^{j}}}{\sqrt{\left(1+\sum_{j=1}^{iter}\frac{1}{\mu_{max}^{j}}\right)m_\lambda + \sum_{j=1}^{iter}\frac{m_{\lambda ext}^{(j)}}{\mu_{max}^{j}}}}\right), \quad (4.4)$$

где $\mu_{max}^{j}$ - спектральный параметр роста логарифмов коэффициэнтов правдоподобия на $j$ итерации алгоритма распространения доверия, показывает, насколько логарифмы правдоподобия $\lambda$ растут быстрее в треппин-Сете по отношению к оставшемуся графу; $Q$ - $Q -$функция; $m_{\lambda}$ начальное значение логарифмов правдоподобия в символьных узлах, полученных из канала, $m_{\lambda^{ext}}^{(j)}$ - сообщение в методе распространения доверия на $j$ итерации, вычисляемое при помощи Эволюции плотностей или ее аппроксимаций, например

$$m_{\lambda^{ext}}^{(j)} = \phi^{-1}\left(1 - \left[1 - \phi\left(m_{\lambda} + (d_v - 1)m_{\lambda^{ext}}^{(j-1)}\right)\right]^{d_c - 1}\right), \tag{4.5}$$

$\lambda(\lambda^{ext})$- логарифмы правдоподобия в символьных узлах из канала (внешние логарифмы правдоподобия из декодера на $j-1$ итерации, extrinsic information), $d_c$-вес проверочного узла, $d_v$-вес символьного узла, $\phi$ - функция проверочного узла в Эволюции плотностей или ее приближении (Reciprocal Channel Approximation, RCA Gaussian Approximation), Приложение 4. Например, для аппроксимации Гауссианами получим:

$$\phi(x) = \begin{cases} \exp\left(-0.4527x^{0.86} + 0.0218\right), \text{ для } 0 \le x < 10 \\ \sqrt{\frac{\pi}{x}}\exp(-0.25x)\left(1 - \frac{10}{7x}\right), \text{для } x \ge 10 \end{cases}. \tag{4.6}$$

Применим предложенную линейную модель для оценки вероятности возникновения ошибки в символьном узле треппин-сета $P_{TS}$.

Например, для треппин-сета (17, 6), содержащего символьные узлы: 12185, 12555, 13101, 14158, 14617, 14765, 14789, 15224, 15365, 16083, 16429, 16983, 17529, 18136, 26272, 26990, 27336, при отношении сигнал-шум 1.4 дБ $P_{TS}$ соответственно равны 1.223185e-07, 1.126428e-07, 1.248727e-07, 1.234708e-07, 1.252041e-07, 1.157210e-07, 1.162862e-07, 1.220277e-07, 1.252517e-07, 1.136047e-07, 1.237212e-07, 1.094425e-07, 1.101199e-07, 1.104766e-07, 1.087270e-07, 1.100143e-07, 1.113092e-07. Усредненное значение по всем символам - 1.167771e-07.

Аналогично оценим $P_{TS}$ для треппин-сета (14, 4). 7616: 5.166419e-09; 7833: 4.477989e-09; 10163: 4.619398e-09; 10184: 4.502959e-09; 10845: 5.321935e-09;

12151: 4.635558e-09; 13025: 5.358256e-09; 13242: 5.674400e-09; 14541: 5.729683e-09; 14758: 5.721226e-09; 15289: 5.745688e-09; 15506: 5.164377e-09; 18359: 1.456904e-09; 18597: 1.440904e-09. Среднее: 4.643978e-09

Для треппин-сета (8, 6). 9139: 3.842229e-10; 11344:3.397534e-10; 13282:1.824814e-10; 14339:3.376797e-10; 16610: 3.837266e-10; 17710: 1.805997e-10; 18169: 1.820864e-10; 18858: 2.630548e-10. Среднее: 2.817006e-10.

Таким образом, усредняя по информационным символам кода значения $P_{TS}$ для всех треппин-сетов из Таблицы 4.3, получим оценку вероятности ошибки по множеству треппин-сетов: 1.726665e-09.

При отношении сигнал-шум 1.4 ДБ базовая матрица, приведенная на рис. 4.8, не удовлетворяет условию $P_{UB} < P_{BER}$. Вероятность битовой ошибки по множеству треппин-сетов кода равна $1.7 \cdot 10^{-9}$, что больше требуемых $10^{-12}$.

Характеристика треппин-сетов для кода с базовой матрицей, приведенной на рис. 4.9, дана в Таблице 4.4.

Таблица 4.4 - Треппин-сеты для кода с базовой матрицей, приведенной на рис. 4.9

| TS (a, b) | Вес псевдокодового слова, $\omega(p)$ | Столбцы (символьные узлы), участвующие в псевдокодовом слове треппин-сета |
|---|---|---|
| **(12, 12)** | **[17.297055]** | **4570 5497 7926 7966 8252 9485 10412 18727 18767 19649 27849 28445** |
| **(12, 12)** | **[17.297055]** | **4570 5166 8522 8562 8848 9485 10081 18722 19323 19363 27518 28445** |
| **(12, 12)** | **[17.297055]** | **4721 5648 8077 8117 8403 9636 10563 18277 18878 18918 28000 28596** |
| **(7, 7)** | **[18.891353]** | **6053 7254 7616 8212 18417 19013 19609** |
| **(8, 8)** | **[21.135316]** | **7733 8830 9252 10179 11243 12170 18534 18704** |
| **(7, 7)** | **[21.368022]** | **7896 8492 9077 10000 10596 18355 19293** |
| **(12, 11)** | **[21.598147]** | **7968 8271 8954 13924 15965 16502 16561 18769 19365 19668 19755 25892** |
| **(10, 10)** | **[22.116214]** | **7735 8272 8575 15342 16269 18449 18536 19669 26196 27123** |
| **(12, 11)** | **[22.401390]** | **7968 8027 8867 13924 15634 16502 16561 18741 18769 18828 19365 25892** |
| **(13, 13)** | **[22.953044]** | **7968 8271 8954 13924 15965 16502 16561 18769 19365 19668 19755 25892 26819** |
| **(14, 14)** | **[23.308733]** | **7912 8449 8580 8752 15519 16446 18323 18626 18713 19381 26373 27300 28938 29534** |
| **(14, 14)** | **[23.441228]** | **7735 8151 8272 8575 15342 16269 18449 18536 19548 19669 21323 21919 26196 27123** |
| **(16, 16)** | **[23.914631]** | **4570 5166 7649 8784 8848 8920 9707 10303 18450 18722 18794 19585 22092 22688 29530 30126** |
| **(16, 16)** | **[24.058115]** | **4689 5616 8027 8371 8564 8867 15634 16561 18438 18741 18828 19768 25892 26488 29053 29649** |
| **(11, 11)** | **[24.267965]** | **4574 5501 7766 8256 8580 14405 15460 19381 19653 28938 29534** |
| **(12, 10)** | **[24.552166]** | **7748 7786 8845 9267 10194 10662 11258 12359 13286 18549 18587 18719** |
| **(15, 13)** | **[24.585329]** | **7720 7779 8462 14359 14955 15414 15473 16010 16069 18277 19117 19176 19263 26327 26923** |
| **(14, 14)** | **[24.684611]** | **4728 5324 7807 9006 9078 9112 10929 18608 18880 18952 21323 22250 29688 30284** |
| **(14, 12)** | **[24.855100]** | **7892 7968 8271 8954 12465 13392 15965 16561 18693 19365 19668 19755 25892 26819** |
| **(18, 16)** | **[25.004586]** | **4574 5501 7912 8256 8449 8580 8752 15519 16446 18323 18626 18713 19381 19653 26373 27300 28938 29534** |
| … | … | … |

Линейная модель дает следующие вероятности ошибок $P_{TS}$ для треппин-сета (12, 12) при отношение сигнал-шум 1.4 дБ, 4570: 2.630168e-07; 5166: 2.620616e-07;8522: 4.241899e-07; 8562: 4.536300e-07; 8848: 4.105557e-07; 9485: 2.952741e-07; 10081: 2.923268e-07; 18722: 4.087889e-07; 19323: 4.309396e-07; 19363: 4.571745e-07; 27518: 2.260399e-07; 28445: 2.297366e-07. Средняя вероятность ошибки по подграфу треппин-сета: 3.461445e-07.

Треппин-сет (12, 12). 4570:2.521952e-07; 5497:2.478352e-07;7926: 4.081326e-07; 7966: 4.378850e-07;8252: 3.866187e-07; 9485: 2.852650e-07; 10412: 2.873231e-07; 18727:4.105099e-07;18767: 4.404778e-07; 19649: 3.906458e-07; 27849: 2.182454e-07; 28445:2.208116e-07. Средняя вероятность ошибки по подграфу: 3.321621e-07

Треппин-сет (12, 12). 4721: 2.599513e-07; 5648: 2.644726e-07; 8077: 4.416100e-07; 8117: 4.521678e-07; 8403: 3.983156e-07; 9636: 2.934235e-07; 10563: 2.968188e-07; 18277: 3.961740e-07; 18878: 4.431977e-07; 18918: 4.585798e-07; 28000: 2.308703e-07; 28596: 2.272107e-07; 19365: 3.770099e-10; 21674: 9.448026e-11; 25892: 1.441660e-13. Среднее: 7.656254e-08

Усреднение по информационным символам кода значений $P_{TS}$ для всех треппин-сетов из Таблицы 4.4 дает оценку вероятности ошибки по множеству треппин-сетов, равную 7.656254e-08.

Базовая матрица, приведенная на рис. 4.9, не удовлетворяет условию $P_{UB} < P_{BER}$. При отношении сигнал-шум 1.4 ДБ вероятность битовой ошибки по множеству треппин-сетов кода составляет 7.656254e-08, что больше требуемых $10^{-12}$.

Характеристика треппин-сетов для кода с базовой матрицей, приведенной на рис. 4.10, представлена треппин-сетами (15, 14), (18, 18), (25, 23).

Линейная модель дает следующие вероятности ошибок $P_{TS}$ для треппин-сетов при отношении сигнал-шум 1.4 дБ.

Треппин-сет (15, 14). 10662: 1.069067e-12; 10725: 6.781307e-12; 13059: 1.218794e-12; 13122: 8.135955e-12; 14385: 1.828696e-12; 14459: 1.212976e-12; 14638: 8.077101e-12; 15386: 6.562997e-12; 16656: 1.598787e-12; 16730: 1.159289e-12; 21945: 3.107531e-13; 22008: 7.994288e-13; 22941: 1.286929e-12; 23078: 1.506446e-12; 23194: 2.232073e-12. Среднее: 2.918707e-12.

Треппин-сет (18, 18). 7616: 5.972829e-13;7627: 5.241620e-13;10184: 6.925564e-13;10195: 3.670270e-13;10866: 6.767801e-13;12162: 3.493063e-13;13036: 5.119405e-13;13263: 5.784997e-13;14552: 6.444547e-13;14779: 3.728355e-13;15289: 4.064789e-13;15300: 6.395260e-13;16973: 3.401904e-14;18258: 3.475354e-14;23506: 5.875287e-15;23744: 6.095418e-15;27584: 3.847128e-14;28869: 4.053199e-14. Среднее: 3.622554e-13.

Треппин-сет (25, 23). 7616: 1.029473e-13; 8051: 1.687002e-18; 9306: 4.516580e-14;9773: 6.159254e-15;10184: 6.837219e-14;10619: 1.890528e-19;10866: 3.583253e-14;11063: 6.167990e-16;13025: 9.317471e-14;13207: 8.896956e-16;13460: 2.733849e-16;14541: 1.233733e-13;14860: 1.287711e-15;14976: 7.321990e-17;15289: 1.249857e-13;15608: 6.653103e-16 ;15724: 1.326159e-17;17635: 2.055339e-13;19242: 7.999058e-14;19492: 9.987388e-14;22346: 4.496835e-17;28468: 5.729449e-14;28685: 4.234165e-16 ;30146: 3.778013e-14;31062: 2.134731e-20. Среднее: 4.339094e-14.

Усреднение по информационным символам кода значений $P_{TS}$ для Треппин-сетов (15, 14), (18, 18), (25, 23) дает оценку вероятности ошибки по множеству Треппин-сетов, равную 1. 158302e-013.

Базовая матрица, приведенная на рис. 4.10, удовлетворяет условию $P_{UB} < P_{BER}$ при отношении сигнал-шум 1.4 дБ. Вероятность битовой ошибки по множеству Треппин-сетов кода составляет 1.158302e-013, что меньше требуемых $10^{-12}$.

Полученный протограф (рис. 4.10) является результатом поиска компромисса между дистантными свойствами кода (свойства кода) и их потенциальным спектром связности (свойства графа), выраженного в пороге итеративного декодирования (водопад) и полке.

Полученный протограф (рис. 4.10) был расширен (шаг 5 процедуры построения кода) модификацией метода запрещенных коэффициентов. После чего для полученных кодов-кандидатов выполнялись шаги 5-9.

Для оценки вероятности ошибки на шаге 8 процедуры применялась следующая формула:

$$P_{waterfall}\left(N,\sigma\right) \cong Q\left(\frac{\sqrt{N}\left(C(\sigma)-C(\sigma^{*})-\beta N^{-\frac{2}{3}}\right)}{\alpha}\right)+O\left(N^{-\frac{1}{3}}\right), \tag{4.6}$$

где $N$ - длина кода, $\alpha$ - масштабирующий коэффициент, $\beta$ - коэффициент сдвига, $\sigma^{*}$ -порог итеративного декодирования (среднеквадратичное отклонение), $\sigma$ - среднеквадратичное отклонение в АБГШ-канале, $C(\sigma)$ - пропускная способность канала соответствующая мощности шума $\sigma$ , $Q$- Q-функция.

Для определения штрафной функции на конечной длине применялся метод «Эволюции ковариации» (Covariance Evolution), Приложение 4. В соответствии с этим методом решалась система дифференциальных уравнений, соответствующая декодированию (BP, метод распространения доверия) ансамблей кодов, заданных приведенными выше протографами.

Предварительно выполнялась параметризация для перехода от дискретной к непрерывным величинам (от дискретной кривизны к непрерывной кривизне пространства, от Гомологии к Когомологии):

$$\tau \doteq \frac{\ell}{N}, r_c \doteq \frac{R_c(\tau)}{N}, v_d(\tau) = \frac{V_d(\tau)}{N}, \quad (4.6)$$

где $\ell$ - число итераций peeling декодера, $N$ - длина кода, $V_d(\tau)\big(R_c(\tau)\big)$ - число символьных (проверочных узлов) типа $c_d$ во время $\tau$, позволившая осуществить переход от дискретной модели к непрерывной, представленной следующей системой дифф. уравнений:

$$\frac{\partial \hat{v}_d(\tau)}{\partial \tau} = f(\triangle V_d(\tau)) = E[\triangle V_d(\tau) | \{\hat{v}_d(\tau), \hat{r}_c(\tau)\}_{d\in\mathcal{F}_v, c\in\bar{\mathcal{F}}_c}]$$

$$\frac{\partial \hat{r}_c(\tau)}{\partial \tau} = f(\triangle R_c(\tau)) = E[\triangle R_c(\tau) | \{\hat{v}_d(\tau), \hat{r}_c(\tau)\}_{d\in\mathcal{F}_v, c\in\bar{\mathcal{F}}_c}] \quad (4.7)$$

$$\frac{\partial \delta_{c,c'}(\tau)}{\partial \tau} = Cov[\Delta R_c, \Delta R_{c'} | \mathfrak{E}(\tau)] +$$

$$\sum_{u\in\bar{\mathcal{F}}_c} \delta_{c,u}(\tau) \frac{\partial f(\triangle R_{c'}(\tau))}{\partial r_u}\Big| \mathfrak{E}(\tau) + \sum_{u\in\bar{\mathcal{F}}_c} \delta_{c',u}(\tau) \frac{\partial f(\triangle R_c(\tau))}{\partial r_u}\Big| \mathfrak{E}(\tau) + \quad (4.8)$$

$$\sum_{d\in\mathcal{F}_v} \delta_{c,d}(\tau) \frac{\partial f(\triangle R_{c'}(\tau))}{\partial r_d}\Big| \mathfrak{E}(\tau) + \sum_{d\in\mathcal{F}_v} \delta_{c',u}(\tau) \frac{\partial f(\triangle R_c(\tau))}{\partial r_d}\Big| \mathfrak{E}(\tau),$$

где $\triangle V_d(\tau) = V_d\big(\tau + \frac{1}{N}\big) - V_d(\tau)$, $\triangle R_c(\tau) = R_c\big(\tau + \frac{1}{N}\big) - R_c(\tau)$ и $\mathcal{F}_v(\mathcal{F}_c)$ - множество типов узлов вершин (проверок) в графе Таннера, $\mathfrak{E}(\tau)$-ожидаемое распределение весов во время $\tau$: $\mathfrak{E}(\tau) = \{\hat{v}_d(\tau), \hat{r}_c(\tau)\}_{d\in\mathcal{F}_v, c\in\bar{\mathcal{F}}_c}$.

Размерность пространства системы дифференциальных уравнений равна:

$$\dim(DE) = card(c_d) + \text{max_degree}(R_c) - 1, \quad (4.9)$$

где $card(c_d)$ – число различных весов столбцов в базовой матрице, $\text{max_degree}(R_c)$ – максимальный вес строки в базовой матрице. Например, размерность для базовой матрицы на рис. 4.10, $\dim(DE) = 4 + (11 - 1) = 14$. Поскольку размерность растет линейно с ростом размера базовой матрицы дифференциальные уравнения можно решить численно (квази-аналитически).

Для упрощения оценки штрафной функции АБГШ-канал был заменен на стирающий канал. Возможность замены обусловлена топологическими особенностями (топологическими инвариантами, не допускающими нарушение «локальных свойств» графа, в симметричной фазе высоких шумов - "области водопада" в случае их отображения на другие типы каналов) предложенного метода построения кода, исключающего из политопа кода псевдокодовые слова малого веса. Полученные в результате симплексы (топологические комплексы и их приближения) можно рассматривать в качестве звездной области относительно множества принятых кодовых слов, рис 4.10. Отметим, что модификация кода с целью изменения длины при помощи “укорочения” (shortening) или “выкалывания” (puncturing) может приводить к нарушению топологических свойств. Для сохранения топологических инвариантов автором был предложен метод адаптации длины квазициклических кодов, см. глава 2.3.

Под звёздной областью относительно фиксированной точки $x$ понимается подмножество $D$ Евклидова пространства $R^n$, такое, что отрезок, соединяющий любую точку области $D$ с точкой $x$, целиком принадлежит этой области. Подмножество называется просто звёздной областью, если существует точка, относительно которой это подмножество звёздное.

Кодовые и псевдокодовые слова малого веса, представляющих собой пустоты, нарушают топологию звездных областей. Получение свойств локально выпуклого пространства звездной области, позволяющий осуществлять замену АБГШ-канала на стирающий канал, требует отсутствия таких пустот путем обеспечения достаточных дистантных свойств и спектров связности, обеспечиваемых предложенной процедурой построения кода.

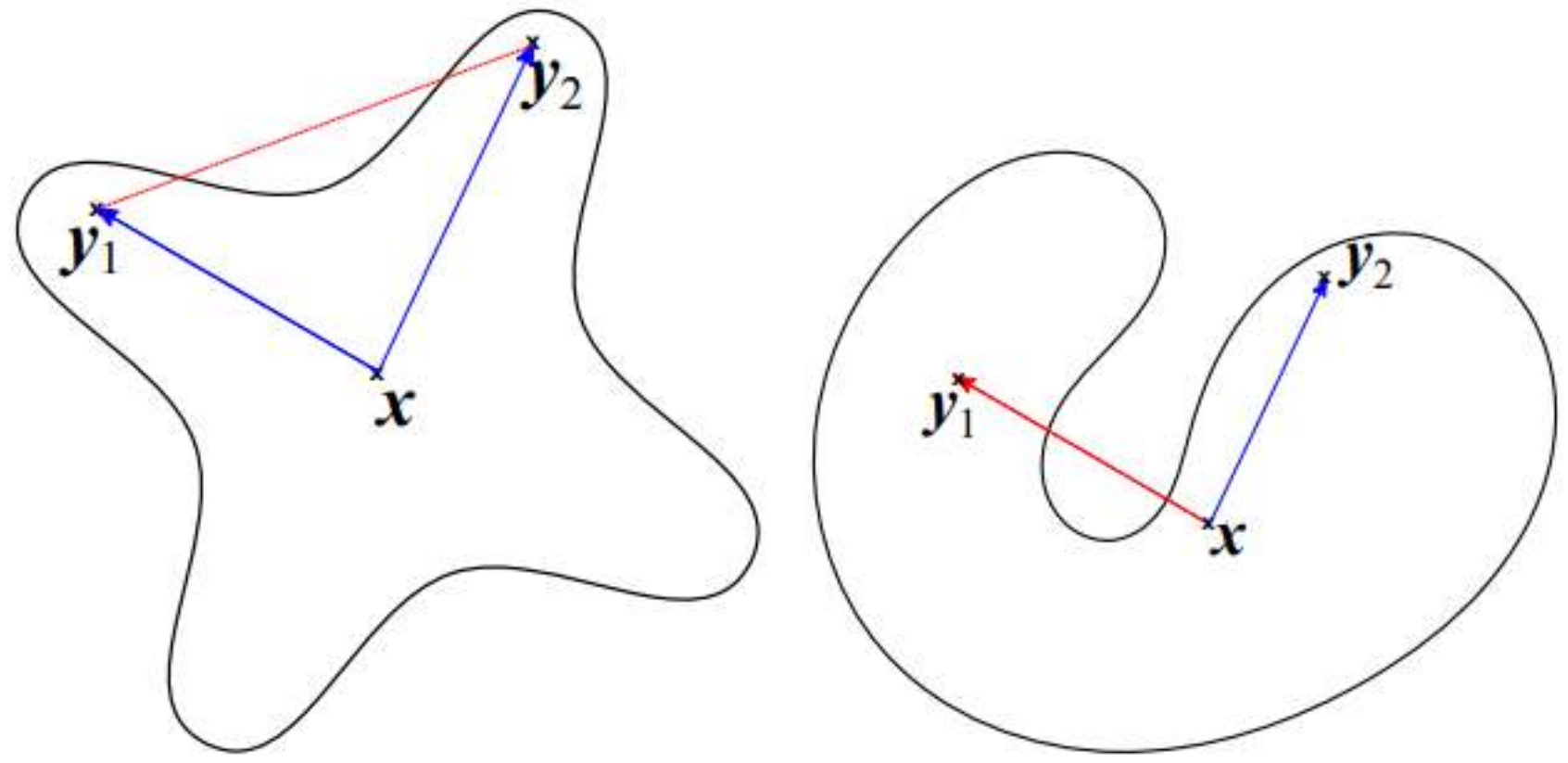

Рисунок 4.10 – (Слева) Звездная область политопа декодирования в случае исключение псевдокодовых слов малого веса на основе предложенной процедуры построения кода, где $x$ -принятое слово из канала, y- кодовые и псевдокодовые слова. (Справа) Область политопа симплекса декодирования в случае наличия псевдокодовых слов малого веса

Результаты численного решения системы дифференциальных уравнений приведены на рис. 4.11, 4.12.

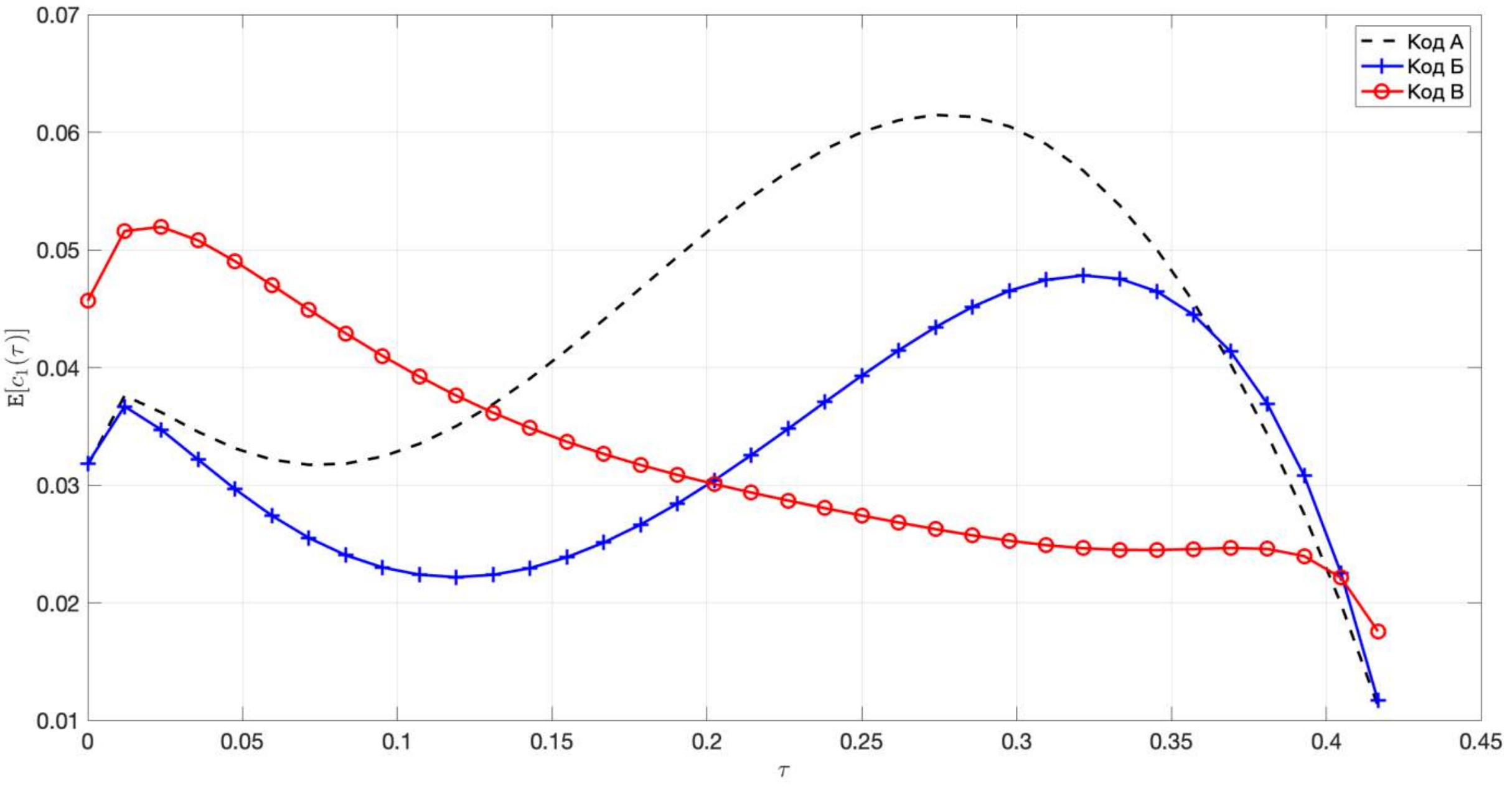

Рисунок 4.11 – Зависимость математического ожидания доли проверочных узлов веса 1 ($E[c_1(\tau)]$) в методе «Эволюции ковариации» от критического времени ($\tau$) для мощности шума $\epsilon = 0.4$ на двоичном симметричном канале связи со стиранием, топологически эквивалентном АБГШ-каналу

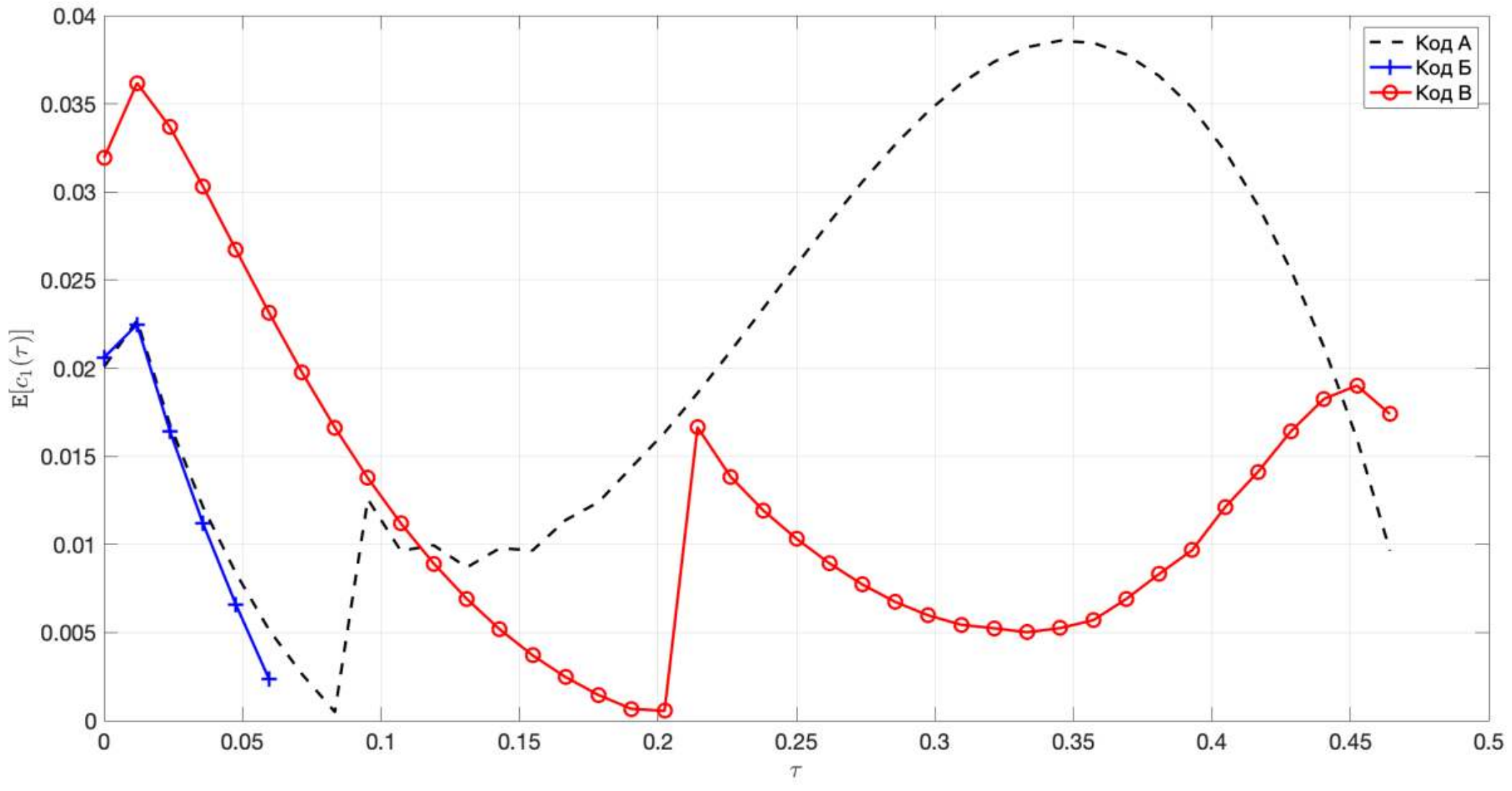

Рисунок 4.12 – Зависимость математического ожидания доли проверочных узлов веса 1 ($E[c_1(\tau)]$) в методе Эволюции ковариации от критического времени ($\tau$) для $\epsilon = 0.45$ на двоичном симметричном канале связи со стиранием, топологически эквивалентном АБГШ-каналу

Полученная зависимость доли проверочных узлов веса 1 позволила вычислить коэффициенты $\alpha, \beta$ для штрафной-функции (водопада) $P_{waterfall}\left(N,\sigma\right)$, пункт 7 процедуры построения.

В результате модификации квазициклических кодов-кандидатов методом имитации отжига (шаг 8 процедуры) были получены 10 двоичных квазициклических кодов-кандидатов, структура циклов которых приведена на рис. 4.13-4.15.

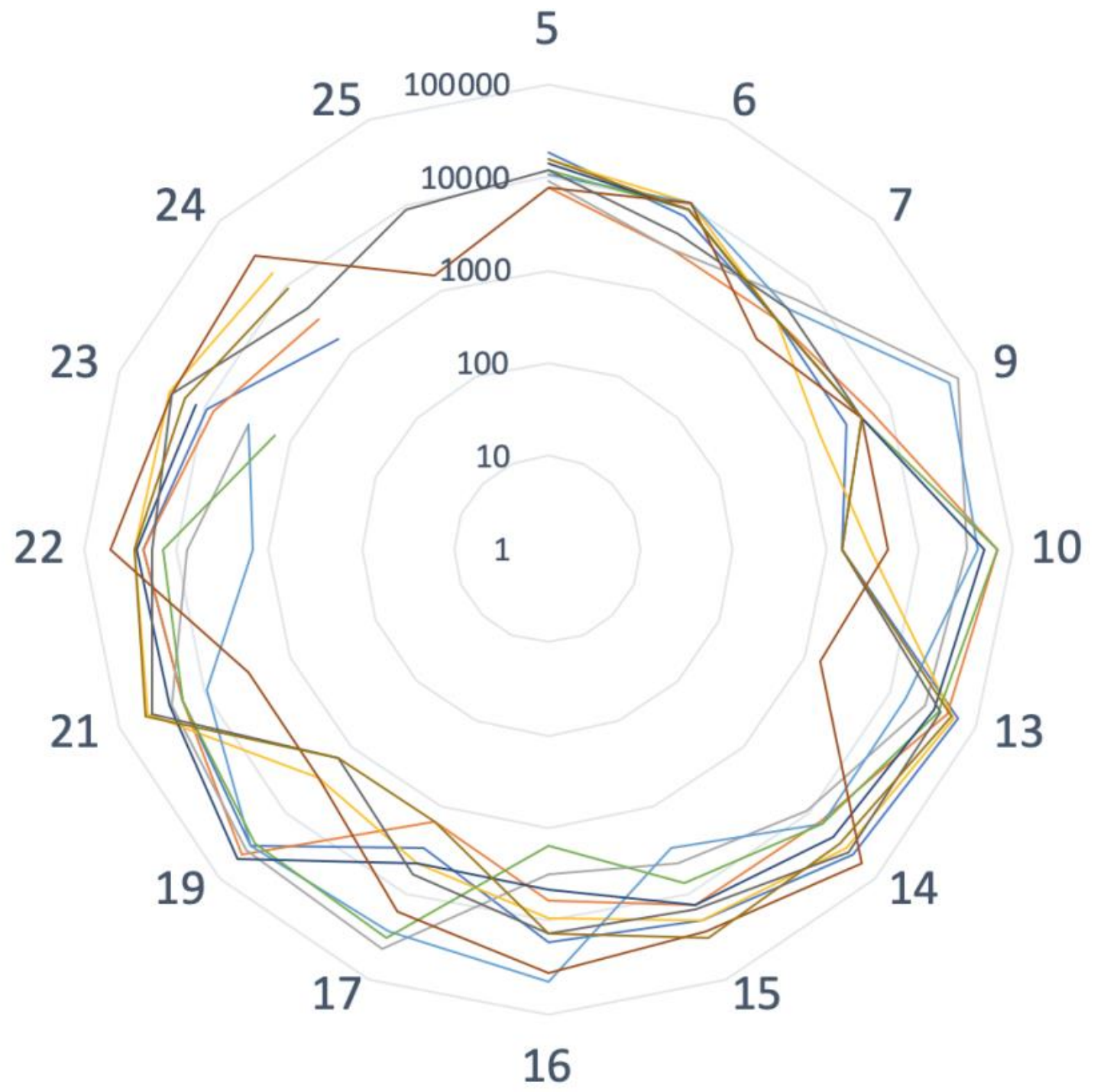

Рисунок 4.13 – Распределение значений EMD циклов длины 10 для 10 лучших кодов кандидатов

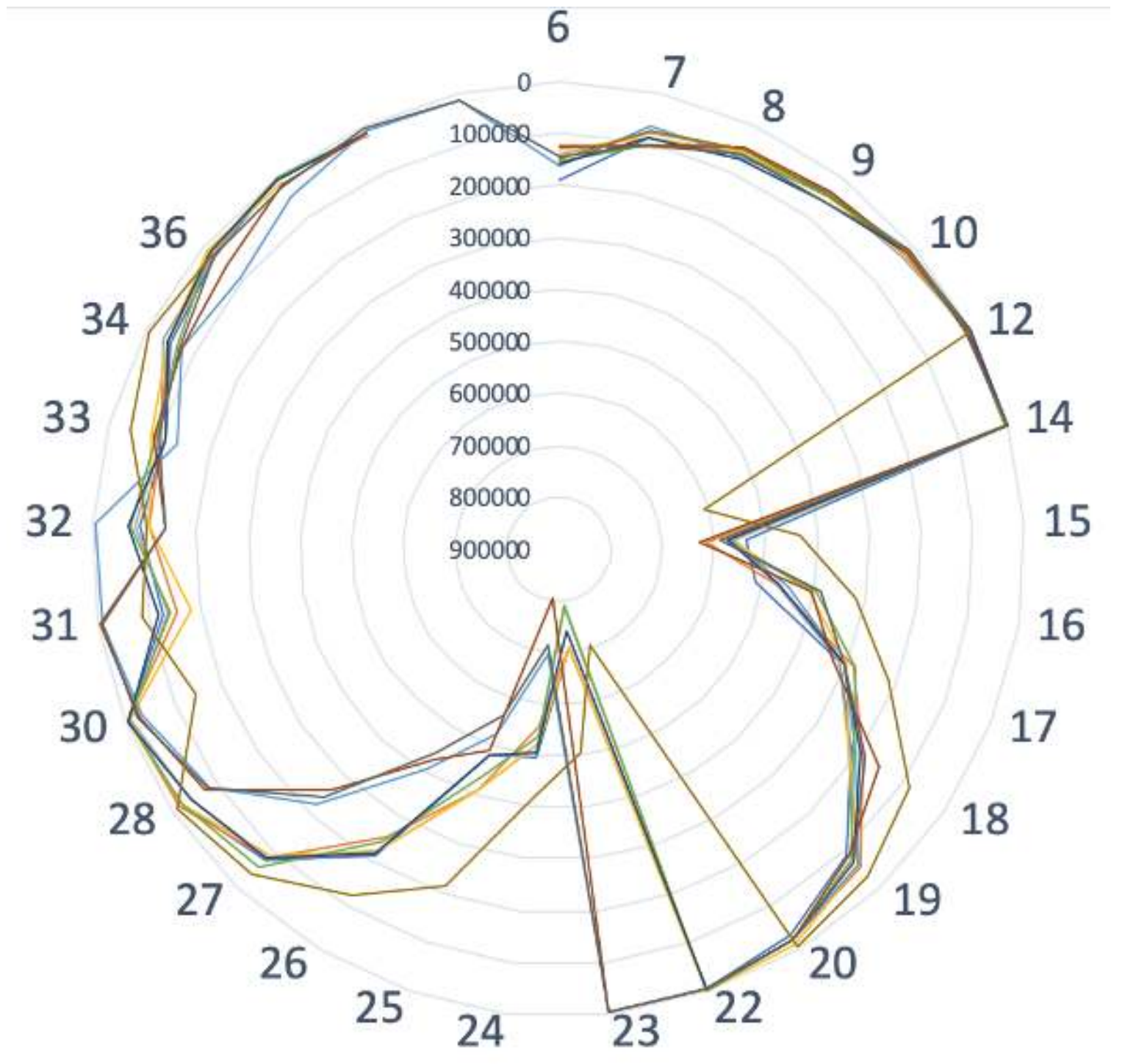

Рисунок 4.14 – Распределение значений EMD циклов длины 12 для 10 лучших кодов кандидатов

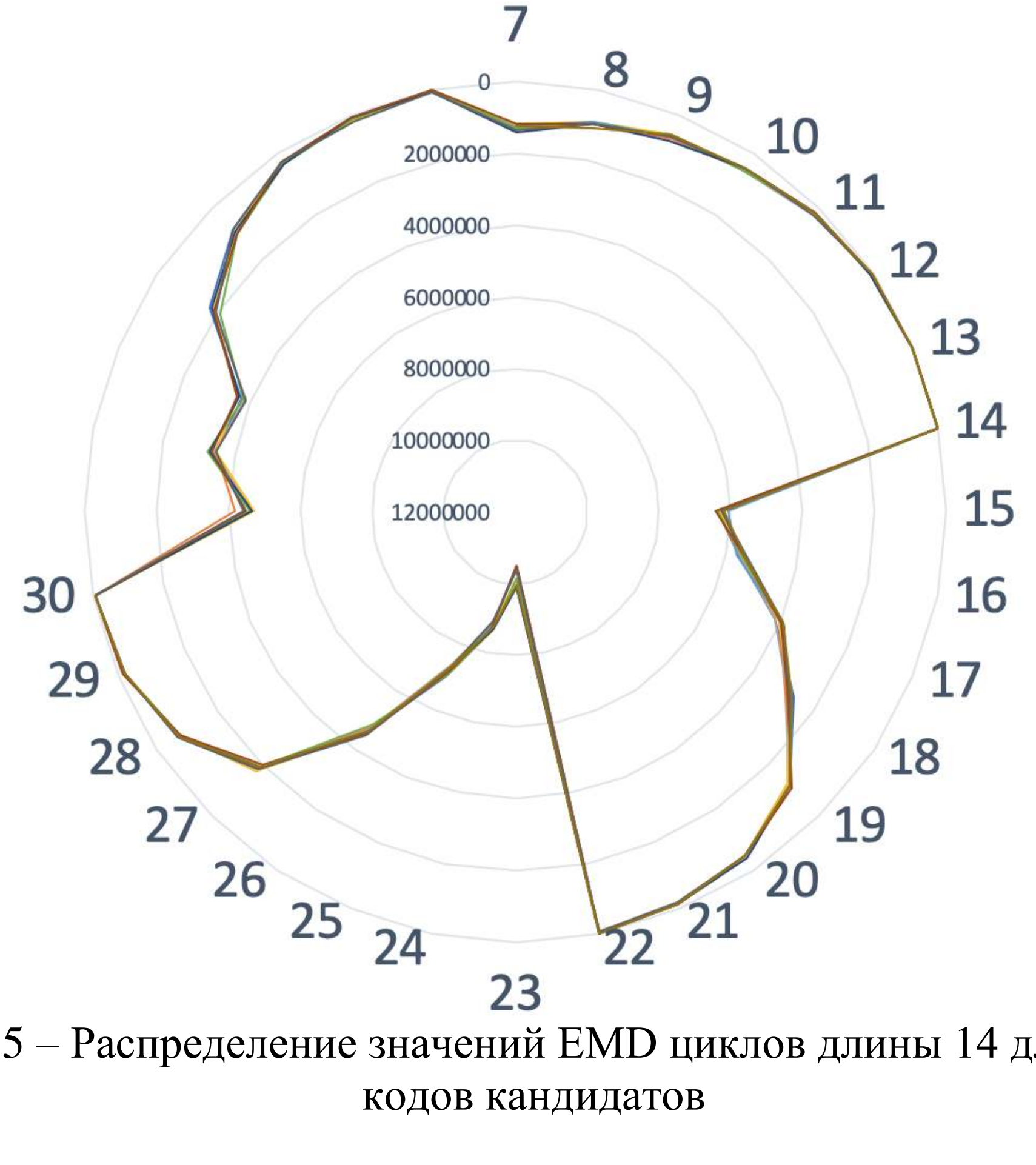

Рисунок 4.15 – Распределение значений EMD циклов длины 14 для 10 лучших кодов кандидатов

На рисунках 14-16 изображен спектр-связности (EMD Spectrum) для расширения фиксированного протографа, выбор кода осуществляется путем максимизации EMD для циклов, образующих треппин-сеты, обуславливающих наибольшую битовую ошибку.

Из 10 полученных квазициклических кодов-кандидатов путем имитационного моделирования был выбран лучший код (КОД А, *R*=0,5 и *N*=32000), обладающий энергетическим выигрышем около 0.15 дБ, позволивший значительно повысить надежность воспроизведения информации, по сравнению с двоичным F-LDPC-кодом (КОД Б), предложенным компанией TrellisWare для голографической архивной памяти, (Рис. 4.16, табл. 4.5), [1].

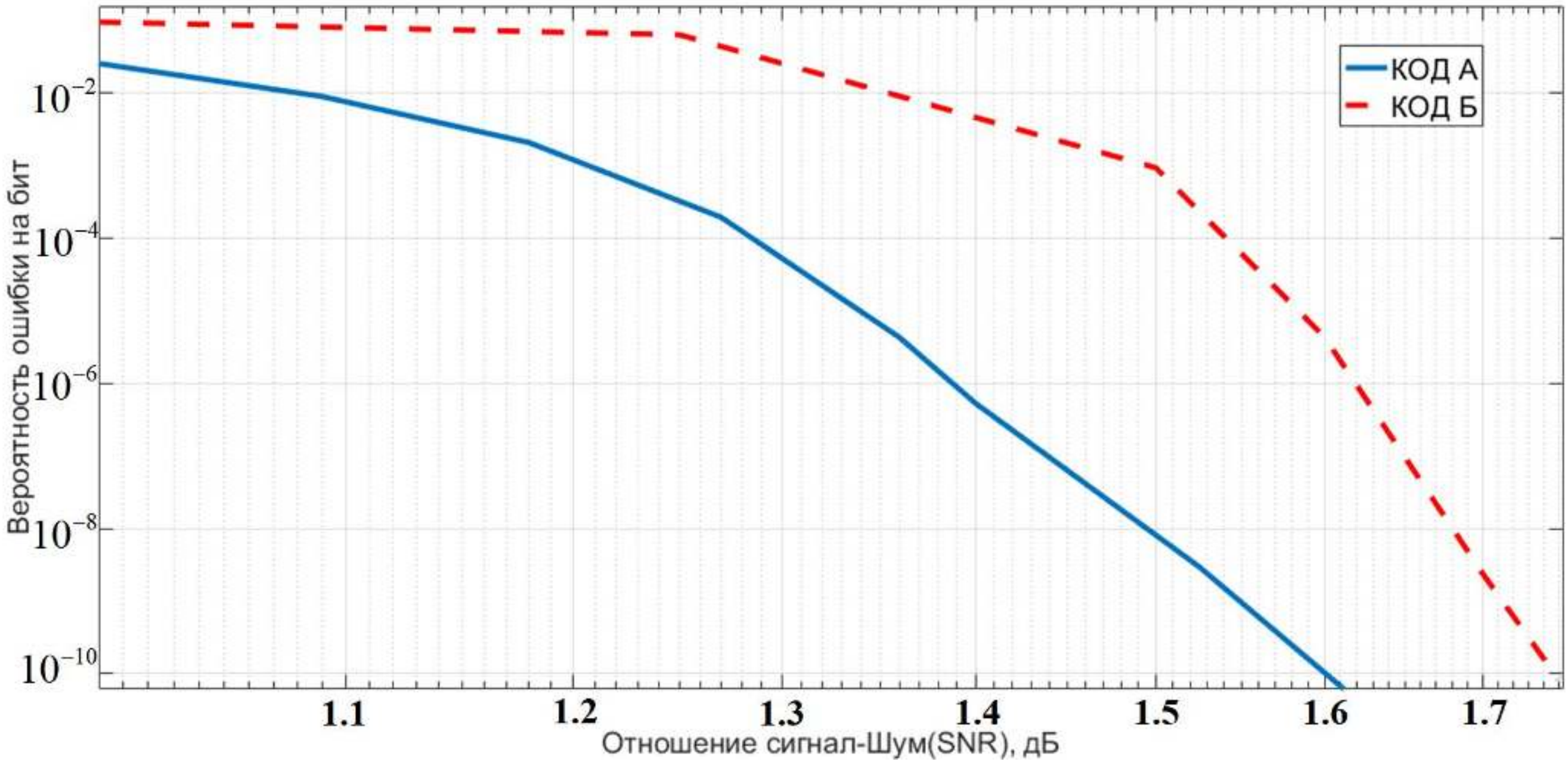

Рисунок 4.16 – Корректирующая способность построенного низкоплотностного кода в сравнении с F-LDPC низкоплотностным кодом компании TrellisWare

Таблица 4.5 – Надежность воспроизведения информации в накопителях архивной голографической памяти

| SNR, отношение сигнал-шум, дБ | TrellisWare код Б | Предложенный код А | Выигрыш по надежности |
|---|---|---|---|
| | $P_{BER}$ | $P_{BER}$ | $(P_{BER})_Б / (P_{BER})_А$ |
| 1,1 | $8 \times 10^{-2}$ | $9 \times 10^{-3}$ | 8,9 |
| 1,45 | $3 \times 10^{-3}$ | $7{,}2 \times 10^{-8}$ | 41670 |
| 1,6 | $5 \times 10^{-6}$ | $1{,}1 \times 10^{-10}$ | 45455 |

С целью демонстрации эффективности предложенной процедуры построения кодов наряду с протографом кода А (базовая матрица приведена на рис. 4.10) также были расширены протографы кодов В (базовая матрица приведена на рис. 4.8) и Д (базовая матрица приведена на рис. 4.9) с использованием метода имитации отжига (шаг 8). Лучшие квазициклические коды из полученных были выбраны путем имитационного моделирования, используя 30 итераций Normalize offset Min-Sum декодера с плавающей запятой (Layered scheduler), кривые помехоустойчивости представлены на рис. 4.18.

Рис 4.17 демонстрирует компромисс между началом «водопада» и «полкой» в кривой помехоустойчивости кода. Код B благодаря наилучшему порогу итеративного декодирования ($Eb/No =$ 0.8903) обеспечивает наилучший водопад, однако рано выходит на «полку» из-за недостаточного кодового расстояния ($d$ = 22). Код Д с кодовым расстоянием 24 и порогом итеративного декодирования $Eb/No =$ 1.0403 имеет «водопад» хуже, однако увеличение кодового расстояние понижает «полку» на порядок. Код А с кодовым расстоянием 54 и порогом итеративного декодирования $Eb/No =$ 1.1603 имеет худший из «водопадов», однако увеличение кодового расстояние значительно понижает «полку».

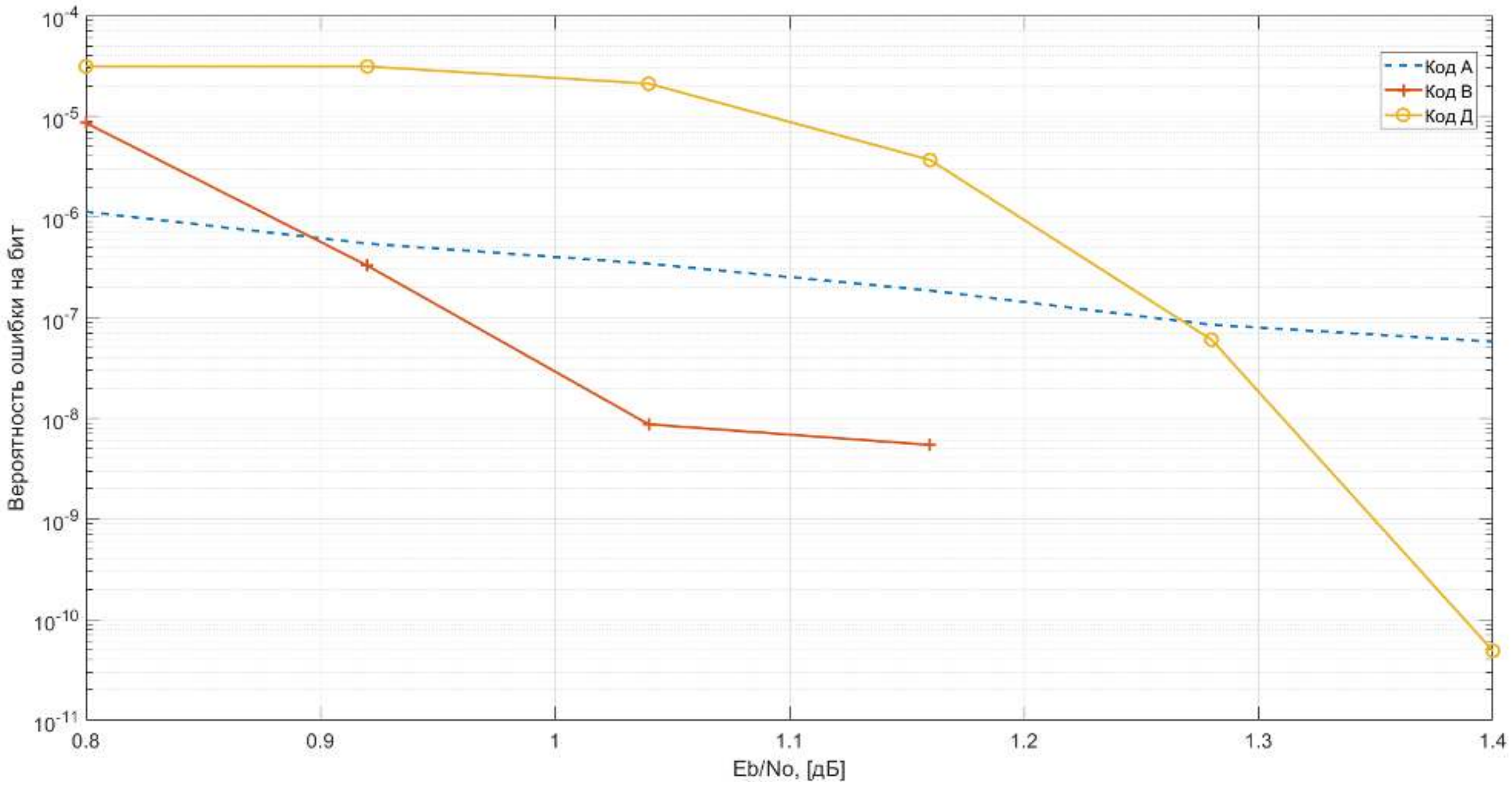

Рисунок 4.17 – Результаты моделирования двоичных квазициклических LDPC-кодов (А, В, Д) при 30 итерациях Adjusted Min-Sum декодера с плавающей запятой (Layered scheduler)

## 4.4 Выводы

Построение LDPC-кода для архивной голографической памяти осуществлялось с использованием следующих аппаратных средств: хост-компьютера и специализированного устройства поиска кратчайшего вектора в решетке.

Устройство, реализованное на ПЛИС, осуществляет поиск кратчайшего вектора в решетке, который является самой вычислительно сложной частью предложенных в диссертации методов оценки минимального кодового расстояния.

Созданное специализированное устройство поиска кратчайшего вектора в решетке, включающее операции модификации координатных компонентов вектора и блоков вычисления частичных сумм совместно с блоком модификации/вычисления приращений координат и его границ, отличающееся использованием регистровых стеков и параллельным выполнением мультипликативных операции в одном временном интервале. Устройство позволяет в подрешетки 512-размерности сократить количество DSP процессоров в 4 раза.

Разработанное специализированное устройство поиска кратчайшего вектора в решетке обеспечивает выигрыш по быстродействию в сравнении с программной реализацией в 33.93 раза для подрешетки 512-размерности для низкоплотных кодов.

В результате применения предложенных методов, алгоритмов и устройства был построен низкоплотностный код со скоростью 0.5 длины N=32000 для архивной голографической памяти, применение которого в декодере обеспечило повышение надежности воспроизведения информации от 8,9 раз при отношении сигнал-шум 1.1 дб в сравнение с F-LDPC кодом, рекомендованным стандартом HVD.

# ЗАКЛЮЧЕНИЕ

В диссертационной работе в рамках решения научно-технической задачи разработки методов, аппаратно-ориентированного алгоритма и специализированного устройства для построения квазициклических низкоплотностных кодов для декодеров, повышающих надежность воспроизведения информации в накопителях архивной голографической памяти, достигнуты следующие основные результаты:

1. Создан метод построения низкоплотностных кодов, отличающийся использованием в фазе расширения протографа жадного метода запрещенных коэффициентов и метода имитации отжига, обеспечивающий дистантные свойства кодов и их спектры связности для фильтрации кодов кандидатов, позволивший построить низкоплотностный код со скоростью 0.5 длины N=32000 для архивной голографической памяти, применение которого в декодере обеспечило повышение надежности воспроизведения информации от 8,9 раз при отношении сигнал-шум 1.1 дб в сравнение с F-LDPC кодом, рекомендованным стандартом HVD.

2. Разработан метод оценки кодового расстояния, основанный на вложение кода в решетку, отличающийся применением для поиска кратчайших векторов параллельным перебором линейных комбинаций базисных векторов решётки, а также применением на этапе ортогонализации параллельных методов QR-разложения матриц, применением метода ветвей и границ в скользящем окне по подрешетке 512-размерности, позволяющий ускорить нахождение кодового расстояния в 33,93 раза.

3. Разработан аппаратно-ориентированный алгоритм поиска кратчайшего вектора в решетке, отличающийся этапом распараллеливания вычисления координатных компонент с использованием зигзагообразного обхода Шнора элементов решетки, позволяющий оперативно получить необходимые индексы и ускоряющий поиск не менее чем в полтора раза.

4. Создано специализированное устройство поиска кратчайшего вектора в решетке, включающее операции модификации координатных компонентов

вектора и блоков вычисления частичных сумм совместно с блоком модификации/вычисления приращений координат и его границ, отличающееся использованием регистровых стеков и параллельным выполнением мультипликативных операции в одном временном интервале, позволяющееся в подрешетки 512-размерности сократить количество DSP процессоров в 4 раза.

**Рекомендации.** Результаты работы могут быть использованы при создании новых контроллеров архивной голографической памяти с повышенной надежностью хранения информации. Предложенные технические решения также могут быть использованы при разработке новых систем помехоустойчивого кодирования телекоммуникационных систем, например, для беспроводной связи (WI-FI, 6G) и оптических каналов передачи данных.

**Перспективы дальнейшей разработки темы.** Оценка влияния использования внешнего кода Рида-Соломона с мягкими решениями на повышение надежности считывания информации голографической памяти. Анализ повышения надежности при $P_{BER} < 10^{-10}$.

# СПИСОК ТЕРМИНОВ И ПРИНЯТЫХ СОКРАЩЕНИЙ

Низкоплотностный код, код Галаггера (LDPC, Low Density Parity Check Codes).

Низкоплотностный квазициклический код (QC-LDPC код, Quasi-Cyclic LDPC Codes), LDPC-код, состоящий из блочных матриц циркулянтов.

Граф-Таннера (Tanner-graph).

Геометрическая решетка (Решетка, Lattice).

АБГШ-канал (Additive white Gaussian noise).

Весовой спектр кода (Weigth Spectrum Enumerator).

Порог итеративного декодирования (Iterative threshold).

Фактор-Граф, граф факторизации вероятностей (Factor graph).

Субоптимальный расчет маргинальных вероятностей без учета циклов (loopy Belief Propagation, Belief Propagation, Message Passing, Sum-Product).

Min-sum декодер– аппаратно-ориентированная апроксимация субоптимального расчета маргинальных вероятностей без учета циклов, при высоких отношениях Сигнал-Шум Sum-Product эквивалентено Min-Sum.

Треппин-сет (Trapping set) – подграф, образованый циклом или их пересечением, обуславливающий ошибку при субоптимальном декодирование без учета циклов (loopy Belief Propagation).

Приближенное значение связности (ACE, Approximate Cycle Extrinsic Message Degree) цикла – локальное число инцидентности узлов, содержащихся в цикле соответствующей длины, без учета топологии внешних подграфов.

Спектр связности (Extrinsic Message Degree Spectrum, EMD Spectrum) – минимальное значение связности для массива циклов в графе с учетом глобальной структуры циклов.

ACE Spectrum – локальная аппроксимация спектра связности (EMD Spectrum).

Алгоритм разреза графа (cutset condition) - алгоритм, осуществляющий разбиение циклов путем назначения определенного значения символьному узлу. Позволяет улучшить порог итеративного декодирования графов низкоплотностные коды с неоднородной структурой факторов.

Низкоплотностные коды с неоднородной структурой факторов (Multi-edge Type LDPC, ME-LDPC). Графы, таких кодов содержат набор “надежных выколотых” символьных узлов, востанавливающих стирание за некоторое приемлемое число итераций (A4JA). Это позволяет улучшить порог итеративного декодирования, за счет увеличения числа итераций для сходимости кода.

Эволюция плотностей (Density Evolution) – метод оценки величины сигнал-шум (порога итеративного декодирования) при заданной вероятности ошибки.

Ковариационной Эволюции плотностей (Covariance Evolution)- обобщения метода Эволюцие плотностей на случай конечного числа символьных узлов в графе кода.

Гауссова аппроксимация аппроксимацию эволюции функции плотности вероятностей (Gaussian Approximation) - приближенный метод вычисления Эволюции плотностей, основанный на предположение о Нормальном распределении функции плотностей вероятности символьных узлов и вычисление соответсвующих стат. моментов.

Оценка порога итеративного декодирования при помощи графиков эволюции априорной и апостериорной взаимной информации (EXIT-charts) случайных величин и факторов для заданного распределения весов символов и проверок – ансамбля LDPC-кодов.

Оценка порога итеративного декодирования при помощи графиков эволюции априорной и апостериорной взаимной информации (Protograph EXIT-charts, PEXIT-charts) случайных величин и факторов для заданной базовой матрицы LDPC-кода.

Алгоритм расширения графа путем последовательного добавления ребер (Progressive edge growing, PEG-алгоритм).

Hill-Climbing («восхождение на вершину», поиск экстремума) – алгоритм итеративного расширения графа.

Guess-and-Test– алгоритм расширения графа путем случайного броска и проверки выполенения графом некоторых условий (обхват, спектра связность).

# СПИСОК ЛИТЕРАТУРЫ

## ПУБЛИКАЦИИ АВТОРА ПО ТЕМЕ ДИССЕРТАЦИИ

В рецензируемых научных журналах и изданиях, рекомендуемых ВАК:

# ПРИЛОЖЕНИЕ 1. Построенный MET QC LDPC-код (циркулянт, z=672)

Протограф предложенного кода со скоростью $RATE = \frac{VN - CN}{VN - PN} = 0.5$ (VN - 42 столбца/символа, CN - 22 строки/проверки, PN - 2 последних символа выколоты для улучшения порога cut-set condition в машинном обучение или Multi-Edge Type LDPC Code в помехоустойчивом кодирование), 40 итераций метода распространения доверия (layered планировщик) порог 0.55 дБ на уровне $P_{BER} = 10^{-12}$. Обхват кода 8(2864358 циклов), EMD: 6(7620),7(6858),8(15240),9(22098) …

| | | | | | | | | | | | | | | | | | | | | | | | | | | | | | | | | | | | | | | | | | |
|---|---|---|---|---|---|---|---|---|---|---|---|---|---|---|---|---|---|---|---|---|---|---|---|---|---|---|---|---|---|---|---|---|---|---|---|---|---|---|---|---|---|
| 3 | - | 0 | 6 | 7 | - | - | - | 1 | 2 | - | - | 384 | 0 | 2 | 351 | 5 | - | 6 | - | - | - | 3 | - | - | - | - | - | 5 | 4 | 157 | 7 | - | - | 5 | - | - | - | - | - | 223 | 422 |
| - | - | - | - | 27 | 4 | - | - | - | 2 | 0 | 3 | - | - | - | - | - | 3 | - | - | - | - | - | - | - | 443 | 705 | 621 | - | - | - | 14 | 32 | - | - | - | - | 214 | 383 | 134 | 139 | 357 |
| 1 | - | 8 | 10 | - | - | - | 9 | - | 116 | - | - | 15 | - | 516 | 458 | - | 462 | - | 28 | 43 | 466 | - | 66 | - | 7 | 663 | - | 93 | - | 1 | - | 0 | 490 | - | - | 46 | - | - | - | 86 | 326 |
| 311 | 17 | - | 0 | 2 | - | 236 | 535 | - | - | - | 185 | - | 5 | - | - | 7 | 401 | 524 | 583 | - | 208 | 584 | 46 | 686 | - | - | 450 | - | - | - | 210 | - | 201 | 526 | - | - | - | - | - | 142 | - |
| - | 40 | 697 | - | - | 13 | 8 | 685 | 7 | - | 68 | 434 | 466 | 489 | - | - | - | - | 605 | 125 | 522 | - | - | 31 | 689 | 71 | 750 | - | 137 | 567 | - | - | - | - | 586 | 3 | 129 | 327 | - | - | - | 563 |
| - | 449 | - | - | - | 74 | 565 | - | 558 | - | 153 | - | - | - | 283 | 1 | 655 | - | - | - | 2 | 738 | 318 | - | 1 | - | - | 555 | - | 72 | 3 | - | - | - | - | 62 | - | - | - | 282 | 486 | - |
| - | - | - | - | 351 | - | - | - | - | - | - | - | - | - | - | - | - | - | - | - | - | - | - | - | - | - | - | - | - | - | - | - | - | - | - | - | 449 | - | 392 | - | 749 | 626 |
| - | 87 | - | - | - | - | 534 | - | - | - | - | - | 272 | - | - | 168 | - | - | - | - | - | - | - | - | - | - | - | - | - | 278 | 416 | - | 166 | - | - | - | - | 674 | - | - | 300 | 137 |
| - | - | - | 613 | - | 496 | - | - | - | - | 337 | - | - | - | - | - | - | - | - | - | - | - | - | 76 | 376 | - | - | - | - | - | - | 277 | - | 499 | - | - | - | - | 295 | - | 732 | 511 |
| - | - | - | - | - | - | - | 390 | - | 611 | - | - | - | 635 | - | - | - | - | - | - | 378 | - | - | - | - | 246 | - | - | - | - | - | - | - | 509 | - | 665 | - | - | - | 657 | 475 | 50 |
| - | - | - | - | - | - | - | - | - | - | 217 | - | 229 | - | - | - | - | - | - | - | - | - | 627 | 199 | - | 223 | - | - | 0 | - | - | - | - | - | - | - | - | - | - | - | 711 | - |
| - | - | - | - | - | - | - | - | - | - | - | - | - | 677 | - | 378 | - | - | - | - | - | - | - | 666 | 567 | - | - | - | 601 | - | - | - | - | - | - | - | - | - | - | - | 144 | 273 |
| 217 | 36 | - | - | - | - | - | 128 | - | - | 24 | - | - | - | - | - | - | - | - | - | - | - | - | - | - | - | - | - | 671 | - | - | - | - | - | - | - | - | - | - | - | 100 | 108 |
| 665 | 541 | - | - | - | - | - | - | - | - | - | - | - | - | - | - | - | - | - | - | 0 | - | - | 286 | - | - | - | - | - | 263 | - | 114 | - | - | - | - | - | - | - | - | 513 | 448 |
| - | 15 | - | 592 | - | - | - | - | - | - | - | - | - | - | - | 59 | - | - | - | 586 | - | - | - | - | - | 406 | - | - | - | 293 | - | - | - | - | - | - | - | - | - | - | 702 | 734 |
| 447 | - | - | - | - | - | - | - | - | - | - | - | - | - | - | - | - | - | - | 620 | 208 | - | 489 | - | 185 | 3 | - | - | - | - | - | - | - | - | - | - | - | - | - | - | 384 | 554 |
| - | - | - | - | 143 | - | - | - | - | - | - | - | - | - | - | - | - | - | - | - | 23 | - | - | - | 349 | - | - | - | 682 | - | 163 | - | - | - | - | - | - | - | - | - | 426 | - |
| - | - | - | - | - | - | - | - | - | - | - | - | - | - | - | 413 | - | - | - | - | - | - | 177 | 473 | 367 | - | 116 | - | - | - | - | - | - | - | - | - | - | - | - | - | - | 639 |
| - | - | - | - | - | - | - | - | - | 597 | - | - | - | - | - | - | - | - | - | - | - | - | - | - | 107 | - | - | - | - | - | - | - | - | - | - | - | - | - | - | - | 301 | 376 |
| - | - | - | - | 427 | 710 | - | - | - | - | - | - | - | - | - | - | - | - | - | 486 | - | - | - | - | - | - | - | - | 698 | - | - | - | - | - | - | - | - | - | - | - | 615 | 516 |
| - | - | - | - | - | - | - | - | - | - | - | 593 | 629 | - | - | - | - | - | - | - | - | - | 513 | - | - | - | - | - | 445 | - | - | - | - | - | - | - | - | - | - | - | 578 | 198 |
| - | - | - | - | - | - | 710 | - | - | - | - | - | 88 | - | - | - | - | - | - | - | - | - | - | - | 200 | - | - | - | - | - | - | - | - | - | - | - | - | - | - | - | 274 | 42 |

ПРИЛОЖЕНИЕ 2. Акты о внедрении предложенных методов построения LDPC-кодов

*Перевод с английского языка на русский язык*

/логотип/
HUAWEI

Справка

Результаты диссертационного исследования Усатюк Василия Станиславовича:

1. Комбинаторная оптимизация свойств кода (вычисление кодового расстояния, вычисление весового спектра кода) и ансамблей графов (топологии циклов, метрик внешних для подграфов символьных весов (EMD), аппроксимация их величин (ACE).
2. Методы оптимизации кода и ансамблей графов, основанные на аппарате статистической физики – методе, основанном на графиках сходимости взаимной информации протографа (PEXIT-chart), Эволюции плотностей (Density Evolution) и Эволюции Ковариации (Covariance Evolution) применяются в компании Хуавэй (Huawei) при построении Низкоплотностных кодов (LDPC-codes).

Директор, Доктор */подпись/*
26-10-2016

Подпись г-на Ли Дзян
удостоверяю

КОПИЯ ВЕРНА
МЕНЕДЖЕР ПО ПЕРСОНАЛУ
Пивоваров А.С.
*/подпись/*

ООО «Техкомпания Хуавэй»
121614, Москва,
ул. Крылатская, д.17, корпус 2
ДЕПАРТАМЕНТ
УПРАВЛЕНИЯ
ПЕРСОНАЛОМ

Certificate

Results of Usatyuk Vasily Stanislavovich PhD theses research:

1. Combinatorial optimization of code (estimation of Hamming distance, enumerating of weight Spectrum) and graph ensembles (cycles topology, Extrinsic Message Degree – EMD, Approximate Cycle EMD – ACE);
2. Statistical physics optimization of code and graph ensembles based on protograph extrinsic information transfer chart – PEXIT chart, Density evolution and Covariance evolutional methods have application in Huawei's LDPC code construction.

Director, PhD

2016-10-26

Подпись г-на Ли Цзян удостоверяю

КОПИЯ ВЕРНА

МЕНЕДЖЕР ПО ПЕРСОНАЛУ

Пивоваров А.С.

ООО "Техкомпания Хуавэй" ДЕПАРТАМЕНТ УПРАВЛЕНИЯ ПЕРСОНАЛОМ Москва, ул. Крылатская, д. 17, корпус 2

***Перевод данного текста сделан мной, переводчиком Клепневым Виктором Игоревичем, верность перевода подтверждаю.***

Клепнев Виктор Игоревич

***Российская Федерация***

***Город Москва***

***Четвертого сентября две тысячи восемнадцатого года.***

*Я, Гришко Михаил Александрович, временно исполняющий обязанности нотариуса города Москвы Кузнецова Вячеслава Николаевича, свидетельствую подлинность подписи переводчика Клепнева Виктора Игоревича.*

*Подпись сделана в моем присутствии.*

*Личность подписавшего документ установлена.*

*Зарегистрировано в реестре: № 53/137-н/77-2018-* 23-543

*Взыскано государственной пошлины (по тарифу): 100 руб.*

*Уплачено за оказание услуг правового и технического характера: 200 руб.*

*М.А. Гришко*

*Всего прошнуровано, пронумеровано и скреплено печатью* 2 *лист(а)(ов)*

*ВРИО нотариуса*

УТВЕРЖДАЮ
Проректор по учебной работе
Юго-Западного государственного
университета
д.т.н., профессор
__________Локтионова О.Г.
«__» __________2018г.

# АКТ

об использовании результатов диссертационной работы на соискание ученой степени кандидата технических наук Усатюка Василия Станиславовича

Мы, ниже подписавшиеся, начальник учебно-методического управления, к.х.н., доцент Протасов В.В., заведующий кафедрой вычислительной техники, д.т.н., профессор Титов В.С., профессор кафедры вычислительной техники, д.т.н., доцент Чернецкая И.Е. составили настоящий акт о том, что результаты диссертационной работы Усатюка В.С. внедрены в учебный процесс по направлению подготовки 09.03.01 «Информатика и вычислительная техника», а именно:

– в курсе лекций и лабораторных работах по дисциплине «Защита информации» используются разделы диссертационной работы, связанные с разработкой методов и алгоритмов построения низкоплотностных кодов;

- при подготовке магистерских диссертаций используется раздел диссертационной работы, связанный с анализом дистантных свойств линейных кодов.

Начальник УМУ
к.х.н., доцент В.В. Протасов

Зав. кафедрой ВТ,
д.т.н., профессор В.С. Титов

Профессор кафедры ВТ
д.т.н., доцент И.Е. Чернецкая

УТВЕРЖДАЮ
Проректор по учебной работе
Юго-Западного государственного университета
д.т.н., профессор
________ Локтионова О.Г.
«06» 04 2022г.

# АКТ

об использовании результатов диссертационной работы на соискание ученой степени кандидата технических наук Усатюка Василия Станиславовича «Метод, аппаратно-ориентированный алгоритм и специализированное устройство построения для низкоплотностных кодов архивной голографической памяти»

Мы, ниже подписавшиеся, начальник учебно-методического управления, к.х.н., доцент Протасов В.В., и.о. декана факультета фундаментальной и прикладной информатики, к.т.н, доцент Таныгин М.О., и.о. заведующего кафедрой вычислительной техники, д.т.н., доцент Чернецкая И.Е составили настоящий акт о том, что результаты диссертационной работы Усатюка В.С. внедрены в учебный процесс по направлению подготовки 09.04.01 «Информатика и вычислительная техника», а именно:

– в курсе лекций и лабораторных работах по дисциплине «Схемотехника (элементная база перспективных ЭВМ)» используются разделы диссертационной работы, связанные с разработкой методов и алгоритмов построения и верификации специализированного устройства.

Начальник УМУ
к.х.н., доцент — В.В. Протасов

И.о. декана ФФиПИ,
к.т.н., доцент — М.О. Таныгин

И.о.зав. кафедрой ВТ,
д.т.н., доцент — И.Е. Чернецкая

# ПРИЛОЖЕНИЕ 3. Расширенные протографы

Пример с определением треппин-сетов. Код 4 на 36 с циркулянтом 278

| | | | | | | | | | | | | | | | | | | | | | | | | | | | | | | | | | | | |
|---|---|---|---|---|---|---|---|---|---|---|---|---|---|---|---|---|---|---|---|---|---|---|---|---|---|---|---|---|---|---|---|---|---|---|---|
| **236** | **176** | **256** | **107** | **239** | **125** | **154** | **153** | **199** | **210** | **168** | **260** | **277** | **68** | **12** | **15** | **101** | **108** | **191** | **246** | **117** | **157** | **134** | **62** | **173** | **73** | **171** | **197** | **184** | **148** | **81** | **44** | **56** | **32** | **203** | **268** |
| **277** | **237** | **177** | **257** | **108** | **240** | **126** | **155** | **154** | **200** | **211** | **169** | **261** | **278** | **69** | **13** | **16** | **102** | **109** | **192** | **247** | **118** | **158** | **135** | **63** | **174** | **74** | **172** | **198** | **185** | **149** | **82** | **45** | **57** | **33** | **204** |
| **230** | **278** | **238** | **178** | **258** | **109** | **241** | **127** | **156** | **155** | **201** | **212** | **170** | **262** | **279** | **70** | **14** | **17** | **103** | **110** | **193** | **248** | **119** | **159** | **136** | **64** | **175** | **75** | **173** | **199** | **186** | **150** | **83** | **46** | **58** | **34** |
| **34** | **231** | **279** | **239** | **179** | **259** | **110** | **242** | **128** | **157** | **156** | **202** | **213** | **171** | **263** | **280** | **71** | **15** | **18** | **104** | **111** | **194** | **249** | **120** | **160** | **137** | **65** | **176** | **76** | **174** | **200** | **187** | **151** | **84** | **47** | **59** |

42 на 7 кодовое расстояние 4 (верхняя оценка протографа 11) matrix21.txt

| | | | | | | | | | | | | | | | | | | | | | | | | | | | | | | | | | | | | | | | | | |
|---|---|---|---|---|---|---|---|---|---|---|---|---|---|---|---|---|---|---|---|---|---|---|---|---|---|---|---|---|---|---|---|---|---|---|---|---|---|---|---|---|---|
| **1** | **-1** | **14** | **20** | **2** | **-1** | **-1** | **-1** | **13** | **8** | **-1** | **-1** | **12** | **16** | **17** | **16** | **17** | **-1** | **7** | **-1** | **-1** | **-1** | **6** | **-1** | **-1** | **-1** | **-1** | **-1** | **16** | **15** | **10** | **25** | **-1** | **-1** | **9** | **-1** | **-1** | **-1** | **-1** | **-1** | **6** | **0** |
| **-1** | **-1** | **-1** | **-1** | **0** | **0** | **-1** | **-1** | **-1** | **10** | **6** | **2** | **-1** | **-1** | **-1** | **-1** | **-1** | **14** | **-1** | **-1** | **-1** | **-1** | **-1** | **-1** | **-1** | **16** | **14** | **1** | **-1** | **-1** | **-1** | **10** | **14** | **-1** | **-1** | **-1** | **-1** | **28** | **26** | **5** | **12** | **7** |
| **29** | **-1** | **15** | **12** | **-1** | **-1** | **-1** | **20** | **-1** | **29** | **-1** | **-1** | **15** | **-1** | **26** | **12** | **-1** | **15** | **-1** | **6** | **28** | **14** | **-1** | **27** | **-1** | **14** | **28** | **-1** | **4** | **-1** | **23** | **-1** | **11** | **2** | **-1** | **-1** | **26** | **-1** | **-1** | **-1** | **25** | **17** |
| **18** | **27** | **-1** | **0** | **10** | **-1** | **7** | **18** | **-1** | **-1** | **-1** | **2** | **-1** | **20** | **-1** | **-1** | **7** | **27** | **18** | **1** | **-1** | **7** | **1** | **21** | **0** | **-1** | **-1** | **27** | **-1** | **-1** | **-1** | **16** | **-1** | **8** | **23** | **-1** | **-1** | **-1** | **-1** | **-1** | **4** | **-1** |
| **-1** | **23** | **2** | **-1** | **-1** | **10** | **22** | **1** | **11** | **-1** | **1** | **26** | **1** | **10** | **-1** | **-1** | **-1** | **-1** | **24** | **11** | **16** | **-1** | **-1** | **12** | **2** | **13** | **6** | **-1** | **26** | **24** | **-1** | **-1** | **-1** | **-1** | **5** | **7** | **5** | **29** | **-1** | **-1** | **-1** | **20** |
| **-1** | **14** | **-1** | **-1** | **-1** | **18** | **22** | **-1** | **27** | **-1** | **10** | **-1** | **-1** | **-1** | **2** | **27** | **14** | **-1** | **-1** | **-1** | **26** | **18** | **22** | **-1** | **14** | **-1** | **-1** | **24** | **-1** | **7** | **13** | **-1** | **-1** | **-1** | **-1** | **13** | **-1** | **-1** | **-1** | **5** | **6** | **-1** |
| **-1** | **-1** | **-1** | **-1** | **0** | **-1** | **-1** | **-1** | **-1** | **-1** | **-1** | **-1** | **-1** | **-1** | **-1** | **-1** | **-1** | **-1** | **-1** | **-1** | **-1** | **-1** | **-1** | **-1** | **-1** | **-1** | **-1** | **-1** | **-1** | **-1** | **-1** | **-1** | **-1** | **-1** | **-1** | **-1** | **8** | **-1** | **28** | **-1** | **29** | **10** |

42 на 7 кодовое расстояние 5 (верхняя оценка протографа 11) matrix20.txt

| | | | | | | | | | | | | | | | | | | | | | | | | | | | | | | | | | | | | | | | | | |
|---|---|---|---|---|---|---|---|---|---|---|---|---|---|---|---|---|---|---|---|---|---|---|---|---|---|---|---|---|---|---|---|---|---|---|---|---|---|---|---|---|---|
| 8 | -1 | 17 | 11 | 8 | -1 | -1 | -1 | 2 | 1 | -1 | -1 | 24 | 4 | 0 | 7 | 24 | -1 | 0 | -1 | -1 | -1 | 3 | -1 | -1 | -1 | -1 | -1 | 11 | 23 | 2 | 22 | -1 | -1 | 24 | -1 | -1 | -1 | -1 | -1 | 3 | 7 |
| -1 | -1 | -1 | -1 | 2 | 0 | -1 | -1 | -1 | 6 | 10 | 28 | -1 | -1 | -1 | -1 | -1 | 10 | -1 | -1 | -1 | -1 | -1 | -1 | -1 | 8 | 20 | 23 | -1 | -1 | -1 | 15 | 10 | -1 | -1 | -1 | -1 | 19 | 25 | 8 | 0 | 16 |
| 13 | -1 | 27 | 27 | -1 | -1 | -1 | 1 | -1 | 14 | -1 | -1 | 3 | -1 | 12 | 3 | -1 | 11 | -1 | 25 | 20 | 21 | -1 | 19 | -1 | 1 | 27 | -1 | 17 | -1 | 4 | -1 | 4 | 2 | -1 | -1 | 25 | -1 | -1 | -1 | 6 | 1 |
| 0 | 10 | -1 | 2 | 7 | -1 | 1 | 20 | -1 | -1 | -1 | 7 | -1 | 12 | -1 | -1 | 20 | 14 | 4 | 2 | -1 | 7 | 26 | 12 | 29 | -1 | -1 | 1 | -1 | -1 | -1 | 17 | -1 | 3 | 14 | -1 | -1 | -1 | -1 | -1 | 14 | -1 |
| -1 | 25 | 28 | -1 | -1 | 10 | 9 | 3 | 1 | -1 | 15 | 2 | 18 | 0 | -1 | -1 | -1 | -1 | 3 | 25 | 11 | -1 | -1 | 18 | 3 | 26 | 2 | -1 | 0 | 11 | -1 | -1 | -1 | -1 | 19 | 21 | 11 | 0 | -1 | -1 | -1 | 7 |
| -1 | 3 | -1 | -1 | -1 | 5 | 22 | -1 | 7 | -1 | 17 | -1 | -1 | -1 | 12 | 17 | 12 | -1 | -1 | -1 | 7 | 10 | 26 | -1 | 24 | -1 | -1 | 2 | -1 | 21 | 8 | -1 | -1 | -1 | -1 | 5 | -1 | -1 | -1 | 19 | 19 | -1 |
| -1 | -1 | -1 | -1 | 26 | -1 | -1 | -1 | -1 | -1 | -1 | -1 | -1 | -1 | -1 | -1 | -1 | -1 | -1 | -1 | -1 | -1 | -1 | -1 | -1 | -1 | -1 | -1 | -1 | -1 | -1 | -1 | -1 | -1 | -1 | -1 | 7 | -1 | 26 | -1 | 10 | 23 |

42 на 7 кодовое расстояние 6 (верхняя оценка протографа 11) matrix26.txt

| | | | | | | | | | | | | | | | | | | | | | | | | | | | | | | | | | | | | | | | | | |
|---|---|---|---|---|---|---|---|---|---|---|---|---|---|---|---|---|---|---|---|---|---|---|---|---|---|---|---|---|---|---|---|---|---|---|---|---|---|---|---|---|---|
| 18 | -1 | 15 | 11 | 11 | -1 | -1 | -1 | 27 | 0 | -1 | -1 | 14 | 9 | 12 | 26 | 13 | -1 | 10 | -1 | -1 | -1 | 11 | -1 | -1 | -1 | -1 | -1 | 26 | 22 | 17 | 0 | -1 | -1 | 20 | -1 | -1 | -1 | -1 | -1 | 20 | 3 |
| -1 | -1 | -1 | -1 | 2 | 3 | -1 | -1 | -1 | 8 | 12 | 12 | -1 | -1 | -1 | -1 | -1 | 14 | -1 | -1 | -1 | -1 | -1 | -1 | -1 | 9 | 27 | 17 | -1 | -1 | -1 | 14 | 14 | -1 | -1 | -1 | -1 | 22 | 20 | 2 | 1 | 0 |
| 2 | -1 | 25 | 2 | -1 | -1 | -1 | 23 | -1 | 2 | -1 | -1 | 19 | -1 | 8 | 9 | -1 | 18 | -1 | 11 | 15 | 11 | -1 | 5 | -1 | 4 | 9 | -1 | 19 | -1 | 26 | -1 | 27 | 1 | -1 | -1 | 7 | -1 | -1 | -1 | 8 | 28 |
| 17 | 24 | -1 | 20 | 1 | -1 | 4 | 28 | -1 | -1 | -1 | 1 | -1 | 14 | -1 | -1 | 8 | 19 | 16 | 22 | -1 | 25 | 15 | 25 | 25 | -1 | -1 | 0 | -1 | -1 | -1 | 14 | -1 | 20 | 27 | -1 | -1 | -1 | -1 | -1 | 12 | -1 |
| -1 | 13 | 13 | -1 | -1 | 4 | 3 | 16 | 13 | -1 | 12 | 3 | 29 | 8 | -1 | -1 | -1 | -1 | 20 | 27 | 14 | -1 | -1 | 2 | 28 | 18 | 3 | -1 | 17 | 22 | -1 | -1 | -1 | -1 | 12 | 25 | 19 | 17 | -1 | -1 | -1 | 29 |
| -1 | 10 | -1 | -1 | -1 | 25 | 10 | -1 | 6 | -1 | 13 | -1 | -1 | -1 | 25 | 14 | 13 | -1 | -1 | -1 | 22 | 25 | 25 | -1 | 16 | -1 | -1 | 11 | -1 | 27 | 9 | -1 | -1 | -1 | -1 | 9 | -1 | -1 | -1 | 25 | 19 | -1 |
| -1 | -1 | -1 | -1 | 29 | -1 | -1 | -1 | -1 | -1 | -1 | -1 | -1 | -1 | -1 | -1 | -1 | -1 | -1 | -1 | -1 | -1 | -1 | -1 | -1 | -1 | -1 | -1 | -1 | -1 | -1 | -1 | -1 | -1 | -1 | -1 | 6 | -1 | 0 | -1 | 22 | 13 |

42 на 7 кодовое расстояние 7 (верхняя оценка протографа 11) matrix13.txt

| | | | | | | | | | | | | | | | | | | | | | | | | | | | | | | | | | | | | | | | | | | |
|---|---|---|---|---|---|---|---|---|---|---|---|---|---|---|---|---|---|---|---|---|---|---|---|---|---|---|---|---|---|---|---|---|---|---|---|---|---|---|---|---|---|---|
| 14 | -1 | 4 | 29 | 14 | -1 | -1 | -1 | 0 | 7 | -1 | -1 | 3 | 27 | 0 | 1 | 5 | -1 | 7 | -1 | -1 | -1 | 22 | -1 | -1 | -1 | -1 | -1 | 28 | 6 | 16 | 6 | -1 | -1 | 0 | -1 | -1 | -1 | -1 | -1 | 0 | 10 | 14 |
| -1 | -1 | -1 | -1 | 16 | 3 | -1 | -1 | -1 | 18 | 6 | 11 | -1 | -1 | -1 | -1 | -1 | 10 | -1 | -1 | -1 | -1 | -1 | -1 | -1 | 29 | 9 | 0 | -1 | -1 | -1 | 18 | 12 | -1 | -1 | -1 | -1 | 15 | 4 | 22 | 10 | 16 | -1 |
| 16 | -1 | 12 | 24 | -1 | -1 | -1 | 3 | -1 | 10 | -1 | -1 | 23 | -1 | 14 | 6 | -1 | 12 | -1 | 2 | 15 | 28 | -1 | 11 | -1 | 4 | 10 | -1 | 17 | -1 | 29 | -1 | 0 | 3 | -1 | -1 | 26 | -1 | -1 | -1 | 9 | 11 | 16 |
| 26 | 19 | -1 | 22 | 4 | -1 | 18 | 12 | -1 | -1 | -1 | 11 | -1 | 2 | -1 | -1 | 18 | 13 | 29 | 7 | -1 | 11 | 28 | 19 | 17 | -1 | -1 | 7 | -1 | -1 | -1 | 13 | -1 | 15 | 9 | -1 | -1 | -1 | -1 | -1 | 0 | -1 | 26 |
| -1 | 23 | 11 | -1 | -1 | 0 | 14 | 4 | 15 | -1 | 14 | 10 | 26 | 10 | -1 | -1 | -1 | -1 | 15 | 17 | 12 | -1 | -1 | 6 | 22 | 17 | 15 | -1 | 10 | 25 | -1 | -1 | -1 | -1 | 16 | 9 | 3 | 20 | -1 | -1 | -1 | 0 | -1 |
| -1 | 7 | -1 | -1 | -1 | 27 | 25 | -1 | 18 | -1 | 29 | -1 | -1 | -1 | 19 | 15 | 0 | -1 | -1 | -1 | 6 | 26 | 12 | -1 | 8 | -1 | -1 | 29 | -1 | 14 | 19 | -1 | -1 | -1 | -1 | 17 | -1 | -1 | -1 | 17 | 17 | -1 | -1 |
| -1 | -1 | -1 | -1 | 29 | -1 | -1 | -1 | -1 | -1 | -1 | -1 | -1 | -1 | -1 | -1 | -1 | -1 | -1 | -1 | -1 | -1 | -1 | -1 | -1 | -1 | -1 | -1 | -1 | -1 | -1 | -1 | -1 | -1 | -1 | -1 | 11 | -1 | 9 | -1 | 9 | 7 | -1 |

ПРИЛОЖЕНИЕ 4. Список исходных и бинарных файлов с использованными/реализованными и/или предложенными автором методами в процессе построения MET QC LDPC-кодов (Multi-Edge Type Quasi-Cyclic LDPC Codes):

1. Эволюции плотностей (Density Evolution, DE) или ее приближении (Gaussian Approximations, GA):
   -Density Evolution;- Relay Flat Fading DE; -Reciprocal-channel approximation; -Gaussian Approximation .
2. Графики взаимной информации проверочных и символьных узлов (аналог density evolution) для:
   -распределения весов , EXIT-Chart;
   -заданного протографа Protograph Exit-Chart (PEXIT-Chart).
3. Эволюция ковариации Covariance-Evolution.
4. Расширение протографа методом иммитации отжига с максимизацией обхвата(girth), спектра связности (EMD Spectrum), для построения квазициклического кода https://github.com/Lcrypto/Simulated-annealing-lifting-QC-LDPC.
5. Оценка спектра связности квазициклического LDPC-кода (MET QC LDPC) EMD Spectrum of QC-LDPC codes.
6. Адаптация кода по длине с оптимизации при помощи спектра связности EMD и оценки кодового расстояния floor scale modular lifting и их апроксимации ACE floor-scale-modular-lifting.
7. Верхняя оценка кодового расстояния MacKay-Vontobel-Smarandache: -Sequentuial ; -Parallel.
8. Нижняя оценка кодового расстояния Таннера Tanner lower bound.
9. Оценка кодового расстояния методом Брауэра-Циммерман требует Magma либо GAP.
10. Поиск кратчайшего базиса (Shortest Basis Problem) и кратчайшего вектора (Shortest Vector Problem) в решетке.
11. Параллельные методы ортогонализации базиса решеток Грамм-Шмидт, Гивенс и Хаусхолдер на GPU, Intel MKL;
12. Оценка кодового расстояния методом геометрии чисел.
13. Поиск треппин-сетов в коде и оценка его полки по FER(BLER) для заданного декодера методом распространения доверия Importance Sampling Trapping Set (TS) search and TS weighing, без спектрального метода для оценки BER.
14. Платформы для иммитационного моделирования Низкоплотностных кодов с планировщиками flooding, layered schedulers и декодерами: normalize min-sum; offset-min; sum-product;log-sum-product; self-corrected min-sum .
15. Платформа для моделирования Низкоплотностных кодов MET QC-LDPC из стандарта 5G. **По мере дальнейшей публикации статей часть исходных файлов будут раскрыты, а репозитории дополнены**.

ПРИЛОЖЕНИЕ 5. Скриншот результатов конкурса по Геометрии Чисел технического университета Да́рмштадт - Поиск кратчайшего вектора в циклической решетке (на идеалах кольца кругового многочлена). web.archive.org сентябрь 2013

In both cases you can enter the Hall of Fame if you are the first in dimension n to solve a challenge or your submitted vector is shorter than the shortest one in a dimension that was solved before.

**More Information**

1. T. Plantard, M. Schneider: Creating a Challenge for Ideal Lattices, IACR Cryptology ePrint Archive: Report 2013/039

## SVP HALL OF FAME

| Position | Dimension | Index | Seed | Euclidean norm | Contestant | Solution |
|---|---|---|---|---|---|---|
| 1 | 128 | 256 | 0 | 2959 | Tsukasa Ishiguro, Shinsaku Kiyomoto, Yutaka Miyake, Tsuyoshi Takagi | vec |
| 2 | 108 | 324 | 0 | 2669 | Tsukasa Ishiguro, Shinsaku Kiyomoto, Yutaka Miyake, Tsuyoshi Takagi | vec |
| 3 | 102 | 103 | 0 | 2670 | Usatyuk Vasiliy | vec |
| 4 | 100 | 202 | 0 | 2660 | Po-Chun Kuo, Po-Hsiang Hao | vec |
| 5 | 96 | 288 | 0 | 2493 | Tsukasa Ishiguro, Shinsaku Kiyomoto, Yutaka Miyake, Tsuyoshi Takagi | vec |

Complete list

## APPROX-SVP HALL OF FAME

| Position | Dimension | Index | Seed | Euclidean norm | Contestant | Solution |
|---|---|---|---|---|---|---|
| 1 | 478 | 958 | 0 | 487384 | Usatyuk Vasiliy | vec |
| 2 | 452 | 908 | 0 | 412796 | Usatyuk Vasiliy | vec |
| 3 | 424 | 856 | 0 | 367585 | Usatyuk Vasiliy | vec |
| 4 | 420 | 421 | 0 | 419016 | Michael Schneider | vec |
| 5 | 400 | 1212 | 0 | 283869 | Usatyuk Vasiliy | vec |

Complete list